%% file: paper.tex
\documentclass[a4paper,10pt]{elsarticle}
\usepackage{amsmath}
\usepackage{amssymb}
\usepackage{longtable}
\usepackage{array}
\usepackage{rotate}
\usepackage{geometry}
\usepackage{graphicx}
\usepackage{fancyhdr}
\usepackage{comment}
\usepackage{color}
\RequirePackage{lineno}

\begin{document}

\biboptions{sort&compress}




\title{Non-collider searches for stable massive particles}

\author[Liverpool]{S.~Burdin}
\author[King]{M.~Fairbairn}
\author[geneva]{P.~Mermod\corref{cor1}}
\author[Stockholm]{D.~Milstead}
\author[Alberta]{J.~Pinfold}
\author[Lancaster]{T.~Sloan}
\author[York]{W.~Taylor}
\address[Liverpool]{Department of Physics, University of Liverpool, Liverpool L69 7ZE, UK}
\address[King]{Department of Physics, King's College London, London WC2R 2LS, UK}
\address[geneva]{Particle Physics department, University of Geneva, 1211 Geneva 4, Switzerland}
\address[Stockholm]{Department of Physics, Stockholm University, 106 91 Stockholm, Sweden}
\address[Alberta]{Physics Department, University of Alberta, Edmonton, Alberta, Canada T6G 0V1}
\address[Lancaster]{Department of Physics, Lancaster University, Lancaster LA1 4YB, UK}
\address[York]{Department of Physics and Astronomy, York University, Toronto, ON, Canada M3J 1P3}

\begin{abstract}
The theoretical motivation for exotic stable massive particles (SMPs) and the results of SMP searches at non-collider facilities are reviewed. SMPs are defined such that they would be sufficiently long-lived so as to still exist in the cosmos either as Big Bang relics or secondary collision products, and sufficiently massive such that they are typically beyond the reach of any conceivable accelerator-based experiment. The discovery of SMPs would address a number of important questions in modern physics, such as the origin and composition of dark matter and the unification of the fundamental forces. This review outlines the scenarios predicting SMPs and the techniques used at non-collider experiments to look for SMPs in cosmic rays and bound in matter. The limits so far obtained on the fluxes and matter densities of SMPs which possess various detection-relevant properties such as electric and magnetic charge are given.
\end{abstract}

\maketitle
%
%


\tableofcontents
\newpage
\input{smp-intro}
\input{smp-theo}

\input{smp-matter}

\input{smp-trappedelectric}

\input{smp-trappedmagnetic}

\input{smp-cosmicmono}

\input{smp-compactobjects}

\input{smp-summary}

\bibliographystyle{utphys}
\bibliography{papercomp}

\end{document}

%% file: smp-intro.tex
\section{Introduction}

The field of elementary particle physics finds itself at an
interesting juncture. Collider experiments at the TeV scale have so far failed to falsify the Standard Model (SM)~\cite{PDG2012}. There
are, however, a number of reasons to suppose that hitherto unobserved
particles exist at or above the TeV scale. The requirement of
naturalness in the Higgs sector typically mandates the existence of
exotic particles. Similarly, there exist a number of theories of dark
matter involving heavy particles and/or new states of matter. Furthermore, progress in particle
physics has often been made by unexpected discoveries as high mass
scales are accessed, prompting new questions to be asked. Non-collider
experiments offer promising means to explore mass regions beyond that
available at colliders. This paper summarises the results of
non-collider searches for a specific class of new phenomena:
stable massive particles\footnote{The term {\it stable} is taken to mean that the object is sufficiently long lived that it can be directly observed rather than via its decay products. The term {\it massive} implies that the object mass is greater than $\mathcal{O}$(100~GeV). The term {\it particle} is taken in the broadest possible sense, including both fundamental and composite states of matter, of either microscopic or macroscopic dimensions.} (SMPs) which are not predicted within the SM and which can be directly observed via their strong and/or electromagnetic interactions in a detector.

SMPs are common features of theories beyond the SM~\cite{Fairbairn2007}. Examples include supersymmetric stable
particles and magnetic monopoles. SMP searches are thus an important component in the program of collider
experiments~\cite{Fairbairn2007}. The Large Hadron Collider (LHC) has
presently imposed constraints on production cross sections of exotic SMPs with masses up to $\sim1$~TeV~\cite{ATLAS2011a,ATLAS2011c,ATLAS2011b,DeRoeck2012a,DeRoeck2012b,CMS2012b,CMS2012c,ATLAS2012a,Mermod2013,CMS2013a,CMS2013d,ATLAS2013a,ATLAS2013b,ATLAS2013f,Bendtz2014}. Non-collider experiments are sensitive to SMPs with masses many orders of magnitude greater than this. The non-collider results form an impressive
body of work that is based on a range of experimental techniques. While some search methods are similar to those used in the collider community (e.g. scintillator-based searches), others (e.g., mass spectroscopy) are typically not used. One aim of this work is to summarise and interpret the non-collider searches in a consistent way, whilst highlighting the assumptions underpinning a given search.

This paper is organised as follows. Representative theories postulating SMPs with a range of different properties are outlined in Section~\ref{smp:theory}. In Section~\ref{smp_matter} the interactions of different SMPs are described, with the focus on those aspects that are most relevant for detection techniques. Sections~\ref{heavy_isotopes} and \ref{monopoles_matter} cover searches made for non-magnetically charged SMPs and magnetic monopoles trapped in matter. Cosmic ray searches for SMPs are covered in Section~\ref{cosmic}. Section~\ref{compact_objects} outlines searches for macroscopic composite objects.

%% file: smp-theo.tex
\section{Theory and cosmology of various kinds of SMPs}
\label{smp:theory}

In this section it is shown how SMPs feature in a range of theories of
physics beyond the SM. Putative SMPs are chosen such that
they represent the wide range of detection-relevant properties
(charge, mass, interactions) envisaged in the literature. The
theoretical motivation for a given type of SMP is described, along
with a description of the properties of the putative particle that are
most relevant for search techniques. The cosmological implications
arising from the existence of a given SMP are also discussed. The
current most mainstream model of cosmology consists of (arguably) high
scale inflation followed by reheating of the Universe at a high
temperature, i.e., within a few orders of magnitude of the Grand
Unified Theory (GUT) scale, followed by radiation domination until a
redshift of $z\sim 5000$. We will on occasion attempt to
identify regions of parameter space that seem to be ruled out by this
model of cosmology, being aware that although it is consistent with all
observations, we cannot yet know if the thermal history of the
Universe differs significantly from this simple model at early times.

\subsection{New particle states (elementary or composite)}

This subsection discusses many kinds of new stable massive particles that have been proposed and are relevant to cosmic-ray and astrophysics experiments (including heavy leptons and hadrons, fractionally charged particles, quark matter, exotic stars, and mirror matter), starting with general considerations on cosmology and dark matter. Particles that can arise as topological field configurations (magnetic monopoles, $Q$-balls and black holes) are discussed in the next subsections.

\subsubsection{General considerations}

Particles that are present in the early Universe with equal matter and
antimatter abundance start in thermal equilibrium, unless they have a
very weak self-interaction cross section. The particles would
therefore be created when other particles annihilate with each other
(e.g., $\gamma\gamma\rightarrow e^+e^-$) but the rate at which this
would occur is limited by the mass of the particle relative to the
temperature of the Universe at that time: $R_e\sim\exp(-m_e/T)$.  The
species in question would continue to follow the relic abundance set by
this Boltzmann suppression until it goes out of thermal equilibrium,
at which point whatever number of particles that are present per unit
entropy is ``frozen in''. At late times, entropy density corresponds to photon density, but if other species freeze out after the species in
question, they will dump their entropy into the photon bath and
further reduce the relic abundance of the initial massive particle~\cite{Kolb:1990vq}.

The current-day abundance of particles with matter/antimatter
asymmetry depends upon the rate at which the antimatter and matter
annihilate. If this happens rapidly then the anisotropy that is
initially left with the matter becomes entirely responsible for the
number of particles that are present in the plasma. If the matter and
antimatter are very inefficient in self-annihilation (due to a low
self-interaction cross section or a low number density) then the relic
abundance in the Universe today only depends upon the abundance at
freeze out, which happens at a relatively high temperature relative to
the particle mass.

There is a huge amount of evidence for dark matter in the Universe.  If dark matter
particles have no asymmetry and self-annihilate with a cross section
comparable to the electroweak scale, then we naturally end up with
approximately the right amount of dark matter in the Universe today
provided their masses are in the GeV--TeV range~\cite{Bergstrom2000}.
This is sometimes known as the WIMP miracle, where WIMP refers to
weakly interacting massive particles.

If, on the other hand, dark matter has a conserved quantum number that
is asymmetric, as it is for baryons, then the physics responsible for
obtaining the correct relic abundance today is completely different
and may be related to the asymmetry in the baryonic
sector~\cite{Hooper2005}.

This work will not be focusing on WIMP dark matter, which already has
many reviews~\cite{Strigari2013}. On the other hand, many of the SMPs considered here could provide alternative dark-matter candidates. This adds motivation to SMP searches. The suitability of a given SMP for dark matter will be discussed in each corresponding subsection.

\subsubsection{SMPs as heavy leptons and hadrons}{\label{sec:smpleptonhadron}}

A number of models provide motivation for stable objects with
elementary charge, as well as fractionally charged ($\pm
\frac{1}{3}e,\pm \frac{2}{3}e$) colour-triplet particles and neutral
colour-octet particles. Examples of such scenarios include models of
leptoquarks, extra dimensions and supersymmetry
(SUSY)~\cite{Fairbairn2007}. Here, SUSY is used as a paradigm
to motivate and outline the expected properties of heavy hadron- and
lepton-like objects, since phenomenological supersymmetric models
allow all of the aforementioned hypothesised SMPs.

Stable particles can arise as possible candidates for the lightest
supersymmetric particle or can be meta-stable due to, e.g., small mass
splitting or an approximately conserved $R$-parity quantum
number. While the hierarchy problem is often invoked as a means of
constraining supersymmetric particles to the TeV scale, the
requirement of naturalness in a theory remains an open question. The
need for a WIMP dark matter candidate is another typical argument for
TeV-scale SUSY. However, there are a number of dark matter
scenarios not requiring WIMPs. Supersymmetry should therefore also be
sought in experiments with a mass sensitivity far in excess of the TeV
scale.

There are a number of examples of supersymmetry scenarios predicting
meta-stable charged particles.  Stable slepton scenarios, leading to  SMPs with charge $\pm e$ that would
interact like a heavy lepton, arise when the lightest SUSY
particle is the gravitino and the next-to-lightest is a slepton. This
can occur in gauge-mediated scenarios~\cite{Giudice1999}.  The decay
of the slepton into the gravitino is suppressed because of the
gravitational scale couplings of the latter particle, leading to
sleptons that can have very long lifetimes depending upon the
gravitino mass~\cite{Buchmuller2006}.

In Split Supersymmetry~\cite{Giudice2004} the gluino may be stable,
giving rise to so-called $R$-hadrons. In this scenario, mesonic and
baryonic states are possible, along with the gluino-gluon state. Such
particles would be strongly interacting. Electromagnetic interactions
are also expected unless the lightest meson and baryon states to which
the others may decay are neutral. Observable meson and baryon states
may have electric charges $0, \pm e$ or $0,\pm e, 2e$. An antibaryon
with charge $-2e$, such as a gluino-$\bar{u}\bar{u}\bar{u}$ state, would be expected
to annihilate quickly in matter to become a $R$-meson. Stable squarks
would lead to $R$-hadron states with the same range of possible charges as for
the gluino $R$-hadrons.

The cosmological constraints on these particles are somewhat model
dependent. Charged particles would be in equilibrium in the early
Universe and would freeze out in the same way as other particles, such
as WIMPs. Their relic density depends upon their mass and charge.
Alternatively, if they are the result of the decay of other particles,
their abundance would be a result of different physics.

Theories with long-lived gluinos such as Split SUSY have interesting
cosmological phenomenologies because the gluino can only decay via
interactions involving squarks, which are much heavier, leading to
lifetimes that depend upon the squark mass to the fourth
power~\cite{Gambino2005}.  Such decays could lead to inconsistencies
with nucleosynthesis by photo-dissociating deuterium and other
isotopes if their lifetime is around 100~s; if they are longer lived
they can distort the thermal nature of the Cosmic Microwave Background
(CMB) or even add unacceptably to the cosmic gamma ray
background~\cite{Arvanitaki2005}.

Very similar constraints are valid for theories where the gravitino is
the lightest particle. Often a phase of entropy production is required
at late times to dilute the density of the next-to-lightest SUSY
particles before they decay to gravitinos, causing problems for light
element abundance in the process~\cite{Buchmuller2006}.

Stable negatively charged particles could bind with nuclei to form dense neutral objects which can act as dark matter~\cite{Khlopov2011,Khlopov2013}, for instance $X^-$p and $X^{--}$He. In most models, the viability of such dark matter candidates is strongly constrained by the scarcity of anomalous heavy isotopes (of hydrogen and helium in particular) in terrestrial matter, which are formed when positively charged particles combine with ordinary electrons (See Section~\ref{heavy_isotopes}).  However, low anomalous isotope abundances on Earth could be explained in the case where free positively charged particles are inherently suppressed in the early Universe~\cite{Khlopov2006b,Khlopov2011} or invoking a recombination mechanism~\cite{Fargion2005}. It was argued that a dark matter species such as $X^{--}$He cannot cause significant nuclear recoil in underground detectors based on nuclear recoil such as CDMS~\cite{CDMS2011}, XENON100~\cite{XENON1002012} and
LUX~\cite{LUX2014}, but could still provide an explanation for the positive results reported by DAMA~\cite{Bernabei2013} based on the mechanism of low-energy radiative
capture of $X^{--}$He by intermediate mass nuclei~\cite{Khlopov2013}. There exists an allowed range in which $X^{--}$He binding energy with sodium is in the interval 2--4~keV, which would lead to the observed annual modulation of the ionisation signal in the DAMA/NaI and DAMA/LIBRA experiments.

Searches for stable interacting particles at the LHC
typically exclude masses up $\sim 1$~TeV, whereas limits on the
lepton-like objects are more modest: $\sim 300$~GeV~\cite{ATLAS2011c,ATLAS2011b,CMS2012b,CMS2012c,CMS2013a,ATLAS2013b,ATLAS2013f}.
Should such objects exist they can become trapped in matter and form anomalous
isotopes, which provides a basis for non-collider searches for heavy
leptons and hadrons (see Section~\ref{heavy_isotopes}). In these searches
the new stable particles are often generically termed $X^0$ if they are electrically neutral
and $X^{\pm 1}, X^{\pm 2}$ if they carry single or double electric charge.
SMPs carrying electromagnetic charge which would be hitting the Earth have also been sought
at cosmic-ray facilities,  using, for example, time-of-flight techniques for signal identification (see Section~\ref{cosmic_ionisation_active}).

\subsubsection{Fractionally charged particles}
\label{fractional_charges}

The SM gauge group $SU(3)\times SU(2)\times U(1)$ does not on its own present any barrier to the existence of fractionally charged particles~\cite{Schellekens1990,Schellekens2013}. Anomaly cancellation does place restrictions on the charges that particles can have but if one were to introduce non-chiral particles or scalar fields there would be no constraint upon their charges.

It is therefore interesting that all observed colour-singlet particles
have integer charge. We will see in Section~\ref{theo_monopoles}
that charge quantisation arises in fundamental theories via the Dirac
argument but this on its own does not guarantee that charges have
values that are integer multiples of $e$. If the SM comes
from a Grand Unified Theory, such as $SU(5)$ or $SO(10)$, then there
are good reasons why charges must have the values we observe in the
SM, however, this is more or less by construction and has
no other motivation.

Physicists attempting to obtain the SM as a low-energy
limit of string theory find typically that this is not so simple as
one would like. For example, while it is possible to obtain the gauge
group $SO(10)$ in string theory~\cite{Lerche1987}, the representations
are such that the normal way of breaking the GUT gauge group to that
of the SM via the Higgs mechanism doesn't work. Instead,
string theorists break the gauge symmetry by turning on Wilson loops,
loops of flux in the compact space that act as a background
field~\cite{Wen1985,Athanasiu1988}. This results in a minimum possible
magnetic charge not of $2\pi/e$ but $k(2\pi/e)$, where $k$ is an
integer that depends on the group theory. This results in the
possibility of colour-singlet particles with charge $e/k$ existing in
the theory.

Obviously these particles are not around in great numbers today, otherwise we would have an interestingly different chemistry. There are ways for them to obtain large masses, and given that the characteristic mass scale in string theory is the string scale, which is usually close to the GUT or Planck scale, we might expect not to see these particles around
today. Nevertheless, there are of course constructions with lower string scales, and one can also imagine other extensions of the SM with fractionally charged particles. It is certainly worthwhile to constrain the existence of such objects experimentally.

One cosmological scenario involving fractionally charged particles is
that composite objects may exist with a large ($10^{12}$~GeV)
confinement scale. Such particles could in principle act as dark
matter~\cite{Birkel1998}. They could decay with lifetimes that would
be phenomenologically interesting for cosmic-ray experiments although
the idea that they could explain super-GZK events seems less
attractive since the AGASA excess has not been observed with the
Pierre-Auger experiment.

Fractionally charged particles can be directly sought in cosmic rays
using detector arrays sensitive to anomalous ionisation energy loss
(see Section~\ref{cosmic_ionisation_active}), anomalous Cherenkov
light emission (see Section~\ref{Cherenkov_light}), or anomalous
rigidity (see Section~\ref{cosmic_in_space}). Specific methods also
exist to search for fractional charge in matter: the levitometer and
the Millikan oil drop methods~\cite{Perl2001}. These are not described
here, but an up-to-date review can be found in Ref.~\cite{Perl2004}.

\subsubsection{Dark atoms and mirror matter}
\label{dark_atoms}

There is a large class of models where a whole new gauge group and spectrum of particles is assumed to exist and to be very similar to the SM. It may differ in just one respect, for example, it could have right-handed chirality as opposed to the left-handed nature of the SM. The particles of this gauge group can presumably possess all the complexity of our own physics, including chemistry. They are decoupled from our own matter unless additional interactions are assumed, and therefore the new gauge group is referred to as a ``dark sector''.

The simplest such models have ``shadow'' matter coupled to our own matter through kinetic mixing of our electromagnetism with the electromagnetism of the dark sector. Electromagnetically bound states between shadow particles of opposite charge form ``dark atoms''. If we refer to the electromagnetic tensor of our electromagnetic field as $F_{\mu\nu}$ and assume that the shadow particles are charged under a different U(1) symmetry, which we denote $G_{\mu\nu}$, then the mixing between them will be given by (see, e.g., Ref.~\cite{Holdom1986})
\begin{equation}
\mathcal{L}_{gauge}=-\frac{1}{4}F_{\mu\nu}F^{\mu\nu}-\frac{1}{4}G_{\mu\nu}G^{\mu\nu}-\frac{\chi}{2}F_{\mu\nu}G^{\mu\nu},
\end{equation}
where the final term is the kinetic mixing term. The strength of the mixing is controlled by the dimensionless parameter $\chi$. In the special case where the hidden sector is isomorphic with the SM --- which reduces the number of parameter and enhances the symmetry of the theory --- the shadow particles are called ``mirror matter''. If the left and right-handed chiral fermions are interchanged in the mirror sector, then the latter simply arises from restoring space-time parity symmetry as a fundamental symmetry of nature.

The intriguing possibility that dark matter could be formed of dark atoms has been studied extensively over the past few decades~\cite{Kobzarev1966,Blinnikov1982,Blinnikov1983,Kolb1985,Carlson1987,Khlopov1989,Hodges1993,Berezhiani1995b,Mohapatra2000,Foot2004,Okun2007,Kaplan2010,Kaplan2011,Behbahani2011,Cline2012,Cline2013,CyrRacine2013,Cline2014,Foot2014}. A wide variety of complex cosmological and astrophysical effects can arise from such models. One simple and theoretically well-motivated hypothesis is that of mirror matter, as discussed above. Assuming a kinetic mixing between the photon and the mirror photon of the order of $\chi\sim 10^{-9}$ and specific initial conditions in the early Universe provides a very rich phenomenology which does not seem to be in obvious conflict with experiments~\cite{Foot2004,Foot2014}. Although mirror matter would naively not qualify as a valid dark matter candidate due to its strong self-interaction cross section, a more careful analysis shows that it could still potentially account for structure formation and bullet cluster observations for certain values of kinetic mixing~\cite{Foot2014}. In addition, models of dark atoms can help reducing the tension between the observed DAMA/LIBRA annual modulation and other underground nuclear recoil experiments such as CDMS, XENON and LUX. For instance, interactions of mirror particles via kinetic mixing introduces a differential cross section for dark matter nucleon scattering which makes the detection efficiency very sensitive to the energy threshold used in the experiment~\cite{Foot2008}. Likewise, another simple hidden sector model for dark atoms can accommodate inelastic collisions which can be seen in DAMA while evading constraints from other experiments~\cite{Kaplan2010}.

Interactions between large aggregates of mirror matter and ordinary matter (which can be quite significant in the presence of kinetic mixing) are discussed in Section~\ref{interactions_mirror}. The presence of bodies made of dark atoms gravitationally bound to the Solar System could hypothetically lead to a number of anomalies in the behaviour of some meteoroids~\cite{Foot2002,Foot2003a}. Such phenomenology is discussed further in Section~\ref{meteoroids}.

\subsubsection{Strange quark matter (strangelets and nuclearites)}\label{sec:strangelets}

It has been suggested that there is a more stable configuration of
quarks than nuclei, namely {\it strange quark matter} or SQM, where
there are roughly equal numbers of $u$, $d$ and $s$
quarks~\cite{Bodmer1971,Chin1979,Farhi1984}.  This seems
counter-intuitive as $s$ quarks have a mass of around 100~MeV so
replacing either an $u$ or $d$ quark, which both have masses less than
around 5~MeV, with an $s$ quark seems energetically unfavourable.
However, the chemical potential due to the Pauli-Exclusion principle
in bulk quark matter is significantly more than the $s$ mass, so it is
energetically more favourable to have the three species in
co-existence. Strange quark matter could be stable for baryon numbers $A$
ranging from a few to $10^{57}$~\cite{DeRujula1984}, beyond which
so-called strange stars collapse into black holes. Strange quark matter with
$A\simeq10^3$ is often called a ``strangelet", while large clumps are
often termed ``quark nuggets" or ``nuclearites". Strange quark matter would
have a nuclear density $\sim5\cdot10^{14}$~g/cm$^3$. The charge $Z$
carried by a lump of SQM is predicted to be significantly
lower than in the case of nuclei, approximately $Z\sim0.3A^{2/3}$.

Note that the transition to this phase from the phase where quarks are
confined in nuclei is very much suppressed.  A baryon with only a
single $s$ quark is a hyperon, which decays quickly. The only possible
way for a nucleus to make a transition to strange quark matter is for
many of the $u$ and $d$ quarks to simultaneously undergo weak
interactions such that they convert into strange quarks, at which point
the large freeing up of chemical potential will make the configuration
energetically stable.  Since the probability of this happening is
astronomically small, normal baryonic matter is stable.

Lumps of SQM can therefore only be produced either in high-energy
collisions at particle accelerators or in the early Universe, the
latter being more favourable since presumably a large amount of
entropy would make it easier to produce such objects.  In particular,
if the Quantum Chromodynamics (QCD) phase transition is first order,
the dynamics of bubble nucleation are such that quark matter lumps
would form at that stage and shrink and cool, moving on the
equation-of-state diagram from a high temperature to a zero
temperature, high chemical potential configuration.

The current status of neutron star observations and theory is rather
interesting, with the vast majority of neutron stars having masses
just above 1.44~$M_\odot$, but recent observations suggest the
existence of stars with much higher masses, around
2~$M_\odot$~\cite{Clark2002,Demorest2010,Antoniadis2013}. Studies of
the properties of hyperons have lead people to believe that these will
play a role in the nuclear equation of state inside neutron
stars~\cite{SchaffnerBielich2008} but this would lower the maximum
mass to less than 1.5~$M_\odot$.  Possible solutions to this include
quark stars~\cite{Weissenborn2011}, but the situation with regards to
observations and theory is still fluid (see
Section~\ref{old_neutron_stars}).

A mechanism for the production of quark nuggets would be the collision and fragmentation of quark
stars~\cite{Madsen1988}, which could in principle lead to a rate of
$10^{-10}~M_\odot$yr$^{-1}$ in the Galaxy~\cite{Madsen2005}, although the absolute expected rate is very
uncertain.

It is also possible that dark matter is made up of antiquarks in
certain schemes of Charge-Parity (CP) violation in QCD. In these models the antiquark sector is sufficiently different so as to form quark nuggets during the QCD phase transition, while the normal quarks form baryons~\cite{Oaknin2005}. Such antiquark nuggets would have very similar phenomenologies to quark nuggets, and besides potentially explaining dark matter and matter-antimatter asymmetry, they provide an interesting phenomenology of diffuse cosmic photon emission~\cite{Forbes2008,Forbes2010} (see Section~\ref{diffuse_emission}).

Despite the fact that quark matter nuggets are strongly interacting,
they do not conflict with astrophysical observations as cold
dark-matter candidates. The ratio of self-interaction cross section to
mass is constrained to be less than $\sim1$~cm$^2$/g by gravitational
lensing observations of merging galaxy
clusters~\cite{Markevitch2004}. Although this constraint would prevent,
e.g., an elementary positively charged particle $X^+$ with mass less than
$10^6$~GeV from constituting dark matter because they would form
heavy hydrogen atoms with atomic-size cross sections, macroscopic
objects with nuclear density (such as nuclearites) would satisfy it by
several orders of magnitude.

The techniques used for searching for nuclearites depend on the
targeted mass range~\cite{DeRujula1984}: up to roughly 1~g ($10^{24}$~GeV), they can be sought in matter as an anomalous abundance of
heavy isotopes (Section~\ref{heavy_isotopes}) or in cosmic rays using
ionisation arrays (Section~\ref{cosmic_ionisation}), while for higher
masses, one can look for signatures of superheavy compact objects,
e.g., in the form of seismic signals (Section~\ref{compact_objects}).

\subsubsection{Fermionic exotic compact stars}
\label{compact_stars}

In Section~\ref{sec:strangelets} it was discussed how strange quark matter could be bound by strong interactions and possibly form stable configurations even for low aggregate masses. More generically, objects composed of fermions can be bound by gravity in a dense phase of matter if the mass is sufficiently large: in that case, a high fermionic degeneracy pressure holds the star up against gravitational collapse. Well known examples are white dwarfs and neutron stars, where electrons and neutrons support the degeneracy pressure, respectively. In each case, stability analysis predicts an allowed mass window; for instance, white dwarfs with masses beyond $\sim 1.5~M_{\odot}$ (the Chandrasekhar limit) collapse into neutron stars, and neutron stars with masses below $\sim 0.2~M_{\odot}$ and beyond $\sim 3~M_{\odot}$ are unstable~\cite{Colpi1993,Narain2006}. Also, any compact star with mass larger than $\sim 20~M_{\odot}$ would collapse into a black hole.

The same principles can be applied to massive agglomerates of exotic
fermions. Several authors have shown that such objects can be stable
in mass ranges that depend on assumptions made about the fermion mass
and interaction
strength~\cite{Narain2006,Dietl2012,Hansson2005,Horvath2007}. For
instance, simple models of a degenerate Fermi gas of fermions with
mass at the TeV scale predict a stability mass range between
$10^{-10}M_{\odot}$ and $10^{-3}M_{\odot}$, corresponding to radii
between 1~$\mu$m and 100~m~\cite{Narain2006,Dietl2012}. Another interesting model is that of a degenerate gas of hypothetical fermions beyond the quark level (hereafter termed preons)~\cite{Hansson2005,Horvath2007}. Despite the fact that no theory can reconcile preon dynamics at a high energy scale with low quark and lepton masses, minimal assumptions about fermionic properties of preons, their low mass, and the compositeness scale, address the features of preon stars in a fairly generic way~\cite{Sandin2005,Horvath2007}. A bag model, motivated in this case by the expectation that quark and lepton masses come mainly from interactions, allows for stable preon stars with masses in the range $10^{-7}-10^{-3}M_{\odot}$ and radii in the range $0.01-1000$~cm (for a density of the order of $10^{23}$~g/cm$^3$)~\cite{Sandin2005,Horvath2007,Sandin2007}.

Stellar collapse is unlikely to be a viable mechanism for exotic compact star formation~\cite{Horvath2007}, leaving a phase transition at the early stage of the Big Bang as the only alternative. While such a phase transition does not conflict with cosmology~\cite{Nishimura1987}, it is unclear if sufficient pressure can be provided to form massive compact stars, and the details of potential mechanisms that could achieve that feat depend on the considered model of exotic particles forming the star~\cite{Horvath2007}.

The only constraints on the abundance of compact massive objects in the mass ranges discussed here were obtained with searches relying on gravitational lensing. Abundances are limited to less than $\sim10\%$ of the dark matter density by femtolensing in the mass range $2\cdot 10^{-17}M_{\odot}-10^{-13}M_{\odot}$ and by microlensing in the mass range $2\cdot 10^{-9}M_{\odot}-2 M_{\odot}$ (see Section~\ref{grav_lensing}). The remaining mass range windows remain unconstrained, and it is conceivable for exotic compact stars to constitute dark matter if they have masses either exclusively in unconstrained ranges or distributed over a wide mass range that only partly overlaps with those probed by gravitational lensing searches.

\subsection{Magnetic monopoles}
\label{theo_monopoles}

The idea that a magnetic monopole could exist can be attributed to
Pierre Curie in 1894~\cite{Curie1894}. But the first persuasive
physics case, made within the framework of quantum mechanics, was put
forward by Dirac in 1931~\cite{Dirac1931,Dirac1948}. He noted that, if
monopoles exist, then electric charge must be quantised, as is
observed in nature. No other explanation for this fundamental
principle was known at the time. In 1984, `t~Hooft~\cite{tHooft1974}
and Polyakov~\cite{Polyakov1974} demonstrated the {\it necessity} of
monopoles in Grand Unified Theories. Previously, Dirac had only
demonstrated the {\it consistency} of magnetic monopoles with
electrodynamics and quantum theory. To date, no generally convincing
experimental evidence for monopoles has been seen. However, there are
sound theoretical reasons for believing that the magnetic monopole
must exist.

\subsubsection{Electric-magnetic duality}

Maxwell's theory of classical electrodynamics relates the electric and
magnetic fields to each other and to the motions of electric
charges. The four basic equations of Maxwell's theory utilise
fundamental electric charges but they do not allow isolated magnetic
charges, i.e., magnetic monopoles. However, the existence of magnetic
charges would fully symmetrise Maxwell's theory. These generalised
Maxwell's equations possess a symmetry, under the duality
transformation, that mixes the electric ($\mathbf{E}$) and magnetic
($\mathbf{B}$) fields:
 \begin{equation}
 \mathbf{E} + i\mathbf{B} = e^{i\phi}(\mathbf{E} + i\mathbf{B}).
 \end{equation}
But, this duality symmetry is broken if free magnetic charges do not
exist in nature. It is easy to incorporate magnetic charges into
classical electrodynamics by simply introducing a magnetic charge
density ($\rho_{M}$) into Gauss's law for magnetism:
$\mathbf{\nabla}\cdot\mathbf{B} = \rho_{M}$.  However, this would
require the introduction of a vector potential $\mathbf{A}$, which is
related to the magnetic field by $\mathbf{B} = \mathbf{\nabla} \times
\mathbf{A}$. Unfortunately, a smooth vector potential would
automatically render the magnetic field sourceless --- seemingly
eliminating the possibility of magnetic charge.

\subsubsection{The Dirac quantisation condition}

In 1931 Dirac, perhaps motivated by the enhanced symmetry of the
Maxwell's equations that would result from the existence of a magnetic
charge, published the first quantum theory of a magnetic
monopole~\cite{Dirac1931,Dirac1948}. Dirac noted that because
$\mathbf{A}$ is not an observable quantity, it does not have to be
smooth. He envisioned the magnetic monopole as the end of an
infinitely long, infinitesimally thin solenoid. Along the solenoid,
the magnetic field is undefined; this line of singularity is known as
the Dirac string. Thus, Dirac's magnetic monopole can only exist if
the Dirac string is impossible to detect. One can imagine an electron
transported along a closed path enclosing the solenoid string. The
electron's wave function would acquire a phase, as in the
Aharonov-Bohm effect~\cite{Aharonov1959}. But, if the wave function
only acquires a ``trivial'' phase then the string would not be
detected. In this case the vector potential $\mathbf{A}$ describes a
localised magnetic charge, surrounded by the magnetic
Coulomb field with a singularity at the origin.

We can express this test more mathematically by supposing that a
point-like monopole with charge $g$ sits at the origin with magnetic
field $\mathbf{B} = g(\hat{r}/r^{2})$ and the solenoid lies along the
negative $z$ axis. Using spherical coordinates and in the appropriate
gauge, the non-zero component of the vector potential is $A_{\phi} =
g(1 - cos\theta)$, where $A_{\phi}$ is defined by $\mathbf{A}\cdot
d\mathbf{r} \equiv A_{\phi}d\phi$. The electron's wave function
acquires a trivial phase, leaving the Dirac string undetected, if
\begin{equation}
e^{-ie\oint \mathbf{A}\cdot d\mathbf{r}} =  e^{-i4\pi eg} = 1.
\label{eq:aphase}
\end{equation}
This gives rise to Dirac's quantisation condition, $eg = n/2$. Importantly, using this relation we see that Dirac had explained the quantisation of electric charge on the condition that at least one magnetic monopole exists.

The minimum allowed magnetic charge is called the Dirac charge, $g_{D}
= 1/(2e)$. In natural units the fine structure constant $\alpha =
e^{2}\approx 1/137$. Thus, $g_{D} = e/2\alpha \approx 137e/2 $. The
Dirac quantisation condition requires all magnetic charges to be
integer multiples of $g_{D}$. Note that the magnetic charge is large
compared to the electric charge, i.e., $g_D=n 68.5e$. So, a relativistic
magnetic monopole with a minimum magnetic charge would ionise $\sim$ 4700
times more than a particle with a single electric charge. Likewise,
the strength of the magnetic Coulomb force between two elementary
single charges is $\sim$ 4700 times that of the Coulomb force. The magnetic
counterpart of the fine structure constant is $\alpha_m\approx 430$.
In the SM, the quantum of electric charge is $e = 1$
($=1.6\times10^{-19}$~C). But, in the quark model the smallest unit of
electric charge is $\frac{1}{3}$, suggesting that, according to the Dirac
condition, the fundamental unit of magnetic charge is
3g$_{D}$. However, this possibility has been discounted due to the
fact that quarks are
confined~\cite{Preskill1984,tHooft1976,Corrigan1976}. Presumably, this
prohibition will need to be revisited if free quarks are ever
observed.

In 1969 Schwinger hypothesised a particle with electric and magnetic charge, called a Dyon\footnote{Unless made clear by the context, the term {\it monopole} is used to refer to both magnetic monopoles and dyons in this paper.}~\cite{Schwinger1966,Schwinger1968,Schwinger1969,Yock1969,DeRujula1978,Fryberger1981}
. In this case the quantisation condition is
\begin{equation}
e_{1}g_{2} - e_{2}g_{1}  = n,
\label{eq:qcSchwinger}
\end{equation}
where we have considered two dyons with electric and magnetic charges
($e_{1}, g_{1}$) and ($e_{2}, g_{2}$). Each dyon is unable to detect
the string of the other if and only if Eq.~\ref{eq:qcSchwinger} is
satisfied. The allowed charges of dyons are restricted by the Dirac
quantisation condition. Most arguments made for monopoles may be
readily extended to dyons, and some gauge models, such as SU(5),
predict the existence of both monopoles and dyons. Schwinger argued
that generalising the Dirac condition for a dyon restricts the
magnetic charge of the dyon to be even with a smallest value
2g$_{D}$~\cite{Schwinger1966,Schwinger1968,Schwinger1969}. Dyons
should be heavier than electrically neutral monopoles. Consequently,
they can decay to monopoles via a process such as Dyon$^{\pm}$
$\rightarrow$~Monopole$ + e^{\pm} + X$.

\subsubsection{GUT monopoles}
\label{GUT_mono}

A symmetry-breaking phase transition in the early Universe may give
rise to topological defects, such as magnetic monopoles; this idea is
known as the Kibble mechanism~\cite{Kibble:1976sj}.  In 1974,
't~Hooft~\cite{tHooft1974} and Polyakov~\cite{Polyakov1974}
independently discovered a magnetic monopole solution --- within the
framework of the Georgi-Glashow model~\cite{Georgi1974} --- generated
by an SU(2) gauge symmetry breaking to U(1). In this broken phase, the
theory has a spherically symmetric ``hedgehog'' solution with magnetic
charge equal to the Dirac charge. This 't~Hooft-Polyakov monopole is
not point-like and has a finite mass of the order of the Grand Unification scale. From far away it looks like a
Dirac monopole. In contrast with the Dirac monopole, 't~Hooft-Polyakov
monopole theory does not necessitate the introduction of a source of
magnetic charge, rather, here it is due to the topological charge, i.e., twists or knots
in the vacuum expectation value of a field. The absence of any singularity in the description of the
't~Hooft-Polyakov monopole makes it mathematically preferable to the
Dirac monopole.


In some GUT theories, with several stages of symmetry breaking, lighter monopoles may arise. A good example is provided by the Pati-Salam model~\cite{Pati1974} based on the SO(10) gauge group:
\begin{equation}
SO(10) \dfrac{M_{1}}{\rightarrow}  SU(4) \times SU(2) \times SU(2)
 \dfrac{M_{2}}{\rightarrow} SU(3) \times SU(2) \times U(1).
 \end{equation}
In this theory one has a heavy monopole with mass $\propto M_{1}$ and
single Dirac charge and a lighter monopole with mass $\propto M_{2}$
and a double Dirac charge~\cite{Lazarides:1980cc}. The mass of the
lighter monopole is $\sim10^{15}$~GeV. There are even lighter
monopoles (M$\sim10^{8}$~GeV) in the SU(15)
model~\cite{Frampton1989,Frampton1990}. Models with fermions in
bifundamental representations can naturally lead to family unification
as opposed to family replication. Such models~\cite{Kephart2007}
utilising the symmetry group $SU(4)\times SU(3)\times SU(3)$ can
give rise to a multiply charged magnetic monopole with mass as low as
$\sim10^{7}$~GeV.

Magnetic monopole solutions also exist within the framework of
electroweak theory. Notably, such monopoles would be expected to have
a mass set by the electroweak (EW) scale. Contrary to earlier work,
which asserted that the Weinberg-Salam model could not admit
monopoles~\cite{Vachaspati1992,Barriola1994}, it has been established
that monopole solutions are possible~\cite{Cho1997,Yang1998}. These
lower-mass gauge EW monopoles could be present in cosmic rays but also
may be producible at the LHC.

It has also been proposed that monopoles belonging to a dark sector~\cite{Bruemmer:2009ky} would, via mixing, appear as objects with small electric charge. In this scenario, monopoles with masses greater than several hundred TeV represent a contribution to the measured dark matter relic density~\cite{Khoze:2014woa}.


\subsubsection{Monopole catalysis of proton decay}
\label{monopole_catalysis}

The possibility that a GUT monopole could catalyse a baryon number
violating process was suggested as early as 1980~\cite{Dokos1980}. The
central core of a GUT monopole retains the original symmetry and
contains the fields of the superheavy gauge bosons that mediate baryon
number violation. Within this core the forces of the universe are
still indistinguishable from one another and the quarks and their
leptons are, in this domain, the same particles. Thus, it is not
unreasonable to expect that baryon number conservation could be
violated in baryon-monopole scattering. However, it was originally
thought that the cross section of this process would be of the order
of the tiny geometrical cross section of the monopole core ($\sim
10^{-58}$~cm$^{2}$).

 \begin{figure}[htbp]
   \centering
   \includegraphics[width=0.4\linewidth]{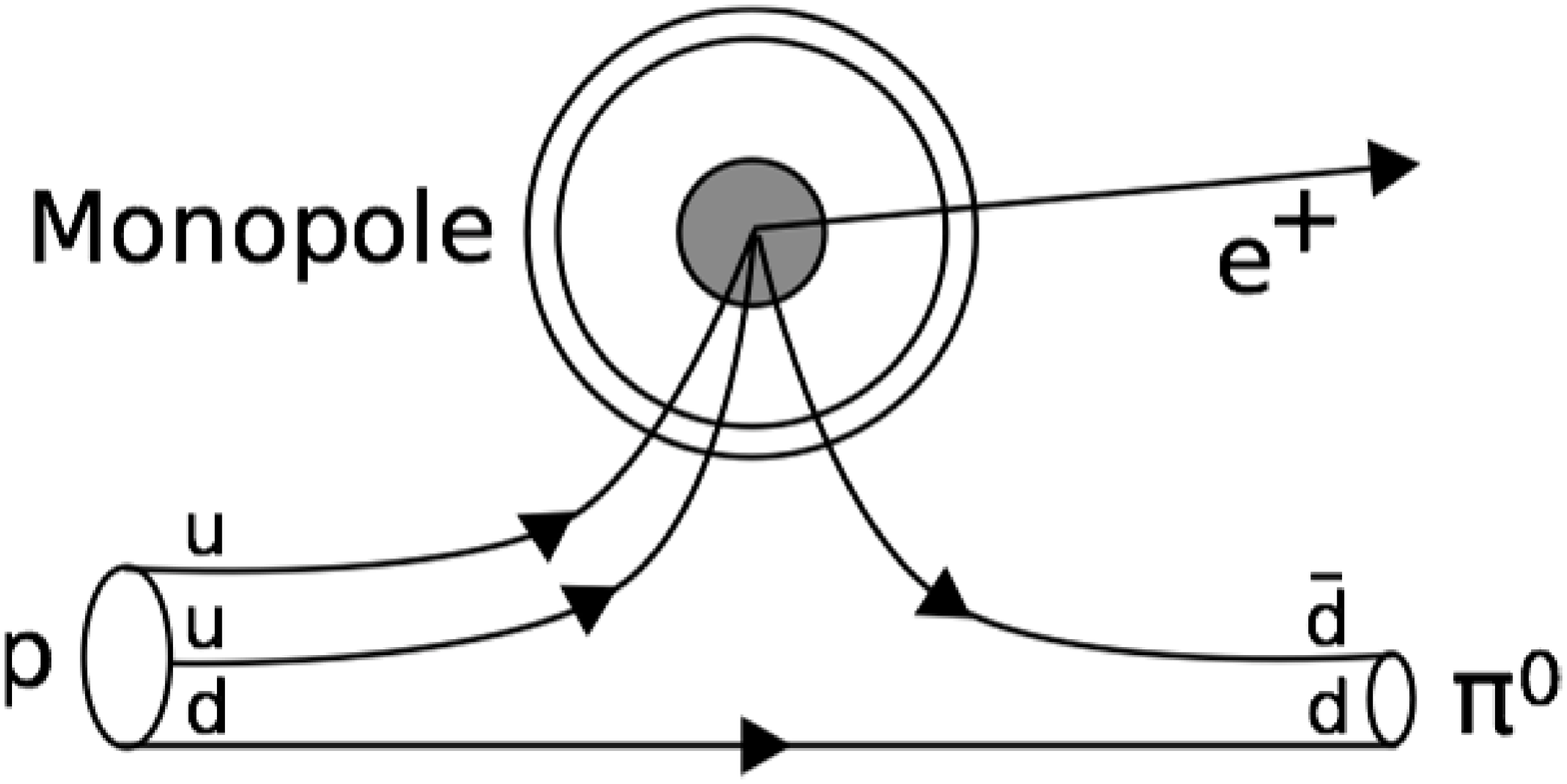} 
   \caption{A depiction  of a proton decay into a positron and a neutral pion
catalysed by a GUT monopole.}
   \label{fig:catalysis}
\end{figure}

Later studies by Rubakov~\cite{Rubakov1981,Rubakov1982} and Callan~\cite{Callan1982,Callan1983} concluded that these processes
are not suppressed by powers of the gauge boson mass. Instead,
catalysis processes such as p~+~Monopole$\rightarrow$e$^{+} +
\pi^{0}$, pictured in Fig.~\ref{fig:catalysis}, could have strong
interaction rates. An explanation for a potentially large monopole catalysis
cross section is the following. The monopole core should be surrounded
by a fermion-antifermion condensate. Some of the condensate will have
baryon number violating terms extending up to the confinement
region. The increase in size of this region gives rise to the
essentially geometric cross-section: $\sigma_{B}\beta \sim10^{-27}$~cm$^{2}$, where $\beta =v/c$.

However, there are theoretical uncertainties in this arena and it is
not certain that strong catalysis is a general feature of all GUT
theories. It may be that catalysis does occur but at considerably
lower rates, as is discussed
elsewhere~\cite{Walsh1984,Rubakov1984}. For example, it has been
proposed~\cite{Arafune1983,Rubakov1984} that the monopole catalysis
cross section could have a 1/$\beta^{2}$-dependence: $\sigma\sim
(1$~GeV)$^{-2}/\beta^{2}$, at least for sufficiently low
monopole-proton relative velocities. It should also be noted that
intermediate mass monopoles arising at later stages of symmetry
breaking, such as the doubly charged monopoles of the SO(10) theory,
do not catalyse baryon number violation.

\subsubsection{Relic abundance of topological defects --- The Kibble Mechanism and inflation}
\label{relic_abundances}

In the hot Big Bang model of cosmology, the coherence length of fields
is limited by the distance that signals can travel --- there is no
possibility of coherence between two points that are further away than
the size of the particle horizon at any given time.  We expect
monopoles from non-Abelian gauge theories where there is some internal
three-dimensional field space, which can be labelled $\phi_a$,
where $a=1,2,3$.  Monopoles will form in such theories if they also
possess degenerate minima at $|\phi_a|=v$, which will have the topology
in field space of the surface of a sphere.

If we go back to the very early Universe, finite temperature
corrections drive the field to $\phi_a=0$ such that all coordinates
are equal and the symmetry is unbroken.  We can map directions in
field space using, for example, $\phi_1=|\phi_a|\cos\theta$,
$\phi_2=|\phi_a|\sin\theta\cos\phi$ and $\phi_3=|\phi_a|\sin\theta\sin\phi$. Then, as the Universe
expands, the finite temperature corrections go away and the field
moves towards some point in its classical minimum
$\langle|\phi_a|\rangle=v$ but in some random ($\theta,\phi$)
direction.

In each different horizon volume region of the Universe, different
values of $\theta$ and $\phi$ are chosen at random.  Then, as the
Universe expands, these different regions come into causal
contact with each other. The energy associated with the kinetic term
in the Lagrangian will try to smooth out the field so that it moves
continuously between such regions but there will be on average one
defect per horizon volume at symmetry breaking which, if the vacuum
has the topology of the surface of a sphere, will be point-like --- a
monopole~\cite{Kibble1980}.

The precise density of monopoles in the Universe today depends upon
the symmetry breaking scale and the radius in field space $v$ at which
the fields will sit in their minimum, but the generic models for GUT
monopoles suggest a relic abundance of
\begin{equation}
\Omega_{monopole}\sim 10^{11}\left(\frac{T_{GUT}}{10^{14}~{\rm
    GeV}}\right)^3\left(\frac{m_{monopole}}{10^{16}~{\rm GeV}}\right),
\end{equation}
or a number density of $10^{-21}$~cm$^{-3}$, i.e., one monopole per
each (100~km)$^3$ cube. This is huge and would correspond to a Universe very different to the
one we live in.

This is in fact one of the motivating factors for cosmological
inflation (the problem was the primary motivation for Guth's original
work~\cite{Guth1981}). To solve this monopole problem you would only require
a dilution factor of $10^{13}$. However inflation also solves the
horizon problem and to do that, we expect a minimum expansion factor
during inflation of $e^{50}$, resulting in a $10^{65}$ increase in
volume, which would reduce the concentration
of these monopoles to a small value, so if inflation
happens after monopoles are formed, they are effectively diluted such
that the number density of a $10^{16}$~GeV monopole would be something
not very far in orders of magnitude of one per Universe.

Hence, monopoles formed by the Kibble Mechanism are either very
abundant or completely absent in the simplest models depending upon
whether or not they form before or after inflation. It is possible to
introduce freedom into almost every part of this story. We don't know
the mass of the monopole or the scale of inflation, although we do
know that even extreme models of inflation with strange thermal
histories (e.g., long dust phase before reheating) require volume dilution
factors of $10^{50}$ in order to solve the horizon problem. Below we
discuss two examples of scenarios in which the conclusions reached
above are circumvented.

\subsubsection*{Multiple stages of inflation}

It is possible that multiple stages of inflation could occur. This would
mean that much of the expansion required to alleviate the horizon
problem would occur in an initial stage of inflation, after which the
Universe could in principle thermalise again, monopoles could form and
then a second stage of inflation would reduce their abundance and make
them observable without being problematic. It is important to note that
constraints on the tensor-to-scalar ratio $r_T$ (the ratio between tensor
gravitational waves and scalar density perturbations in the primordial
plasma) in the CMB tells us that the maximum energy scale of inflation
corresponds to a potential with energy density around $(10^{15-16} {\rm
GeV})^4$~\cite{Planck2014,BICEP22014}. Therefore, the absolute
maximum temperature after inflation must be less than $T=10^{16}$~GeV
(and in most models quite a lot less since the inflaton doesn't decay
immediately).  Because of this, only monopoles that
correspond to a symmetry that is broken at temperatures considerably
less than $10^{16}$~GeV could be produced after a period of inflation,
placing pressure on GUT monopoles that traditionally correspond to
symmetries that are only restored at $10^{16}$~GeV. It should be noted
that the presence of massive particles far beyond the reach of the LHC
could change the running of the gauge couplings and consequently lead
to Grand Unification occurring at a lower energy scale, although many
particles are usually required to achieve a considerable shift in $M_{GUT}$.

\subsubsection*{Stochastic fluctuations created during inflation}

Unless fine tuning occurs and one or more of the couplings of a theory
are many orders of magnitude different from unity, then for a given
potential associated with some symmetry breaking all the relevant
scales associated with that potential, e.g., the expectation value,
the scale of symmetry breaking and the mass (curvature) of the
potential, should be within a couple of orders of magnitude of each
other in energy. For example, the Higgs mass is 125~GeV while the
vacuum expectation of the Higgs field is 246~GeV.

We consider a situation where the mass scale associated with the
symmetry-breaking potential leading to a monopole is much less than
the Hubble scale during inflation, i.e., less than $10^{13}$~GeV. Then
the field will be effectively light and will be given stochastically
different values in different regions of the Universe due to quantum
fluctuations.
When inflation ends, the {\it absolute maximum} temperature that can be created
corresponds roughly to the quarter power of the inflaton energy
density, which is much larger than the Hubble expansion rate at the
end of inflation, so if the inflaton were to reheat efficiently the
temperature corrections to the potential would erase any memory of the
quantum fluctuations. However, if the inflaton is extremely weakly
coupled to the rest of the particle spectrum, it could then take a
long time to reheat and when it does so, the temperature may in
principle be so low as to not erase the stochastically generated
values of the monopole field. Once the Hubble expansion rate drops to the point at which the monopole ﬁeld can start to classically roll down its potential, stochastic values of the ﬁeld that are picked up during inﬂation will lead to the initial conditions for the monopole being different in different regions and will determine the number of topological defects left over. The final number will be
highly dependent upon the physics of the monopole field and the
physics of the inflaton field. However this could only work for
relatively light topological defects.

\subsubsection{Monopoles in extra dimensions and in string theory}

There are a huge number of particles in higher dimensional theories which may posses magnetic charge.  It was pointed out by \cite{Sorkin1983,Gross1983} that monopoles in Kaluza Klein (KK) theories are expected to exist when the U(1) of electromagnetism is obtained by the symmetries of a closed loop in the compact dimensions.  This result is expected to hold in more realistic compactifications, although of course chirality remains a problem in such theories.


In string theory there are a variety of monopoles, in fact the aforementioned KK monopoles are also thought to be solutions of the wider theory \cite{Banks1988}.  The same authors also pointed out that when one moves away from the self-dual compactifications radius where the symmetry between KK and winding modes leads to an enhanced SU(2) symmetry then the higgsing of the massless degrees of freedom at that point can give rise to monopoles exactly in the same manner as the 't~Hooft-Polyakov monopole.

Five branes can couple to the magnetic potential associated with the anti-symmetric tensor field $B_{\mu\nu}$ and upon their dimensions lying along the six compact dimensions can appear as magnetic monopoles in the 3+1 dimensional theory \cite{Callan1991,Gauntlett1993}.  There are many such examples in different string theories with different compactifications and for each compactification upon a (typically complicated) compact manifold, there will be many different ways of wrapping extended magnetically charged objects resulting in point like monopoles in the 3+1D world.

In modern string theory, fundamental open strings have their ends attached to D-branes embedded in the higher dimensional space.  In some scenarios the gravity only part of the bulk can be pierced by lower-dimensional D-brane defects~\cite{Ellis2000,Ellis2004,Ellis2005,Ellis2008}. In our world, these particles would appear as point-like ``D-particles''~\cite{Ellis2000,Ellis2004,Ellis2005,Ellis2008,Shiu2004}, non-perturbative stringy objects with masses of order m$_{D}\sim$ M$_{s}$/g$_{s}$, where g$_{s}$ is the string coupling.
The lightest D-particle (LDP) is stable, because it is the lightest state carrying its particular charge. Electrically and magnetically charged states are possible but normally the lightest state is assumed to be neutral to evade cosmological constraints (they are not observed in the same way magnetic monopoles are not observed) however there could be some aspect of cosmology which prevents them from being ruled out, such as inflation or a late time injection of entropy.  An important difference between the D-matter states and magnetic monopoles is that D-particles could have perturbative magnetic couplings.

The masses of all of these objects will be typically set by the string scale and/or the compactification scale of the compact space.  Since these are objects in which gauge fields can propagate, the constraints on their scale is set by the fact that 3+1D quantum field theory works well up to the multi-TeV energies of the LHC.  However typical theoretical descriptions usually set the compactification scale close to the GUT scale $10^{16}$~GeV although this can be reduced to lower energies, in particular when some of the extra dimensions are gravity only.  In those scenarios, the strongest constraint is the cosmology of the (now light) KK graviton modes associated with the large gravity only extra dimensions and the string scale and/or the inverse radius of the compact space (and hence the masses of the monopoles) can be brought close to the discovery reach of the LHC~\cite{Fairbairn2002}.

\subsubsection{Monopole acceleration in galactic magnetic fields}
\label{monopoles_in_galaxy}

Here we discuss the behaviour of magnetic monopoles in the Galaxy,
where they are assumed to drift under the influence of the magnetic
and gravitational fields. The acceleration of monopoles in galactic
magnetic fields is a well studied
subject~\cite{Goto1963,Osborne1970,Wick2003}.

An upper limit on the flux of galactic monopoles of $\sim
10^{-16}$~cm$^{-2}$sr$^{-1}$s$^{-1}$ --- the so-called ``Parker bound''
--- is obtained based on the argument that fluxes above this value
would decrease the energy stored in the galactic magnetic field faster
than it could be generated by the galactic dynamo
effect~\cite{Parker1970,Parker1971}. This limit was revised to account
for the monopole mass dependence~\cite{Turner1982}. An even tighter
bound was obtained for monopole masses below $10^{17}$~GeV by considering the
survival and growth of a small galactic seed field~\cite{Adams1993}.

The rapid acceleration occurring in galactic magnetic fields causes
monopoles with the Dirac charge or higher to acquire kinetic energies
of the order of $10^{13}$~GeV, and thus monopoles with masses of the
order of $10^{13}$~GeV or lower would quickly become
relativistic~\cite{Wick2003}. This estimate takes into account
possible multiple passages of monopoles through the galactic plane.

Below we report the results of calculations of the influences of the
galactic magnetic and gravitational fields on monopoles. Our
calculations use the improved knowledge of the galactic magnetic field
and galactic mass distribution given in
Ref.~\cite{Jansson2012,Jansson2012b}. The random galactic magnetic
field component is neglected since it should average to
zero~\cite{GF1}. The lifetime of a monopole in the Galaxy is assumed
to be the time for it to reach a radius of 25~kpc, beyond which it is
assumed to have escaped from the Galaxy. Throughout, we assume that
monopoles interact with matter predominantly through their
gravitational and electromagnetic fields and that strong interaction
and ionisation effects are not significant.

\begin{figure}[ht]
\centering
\includegraphics[width=0.7\linewidth]{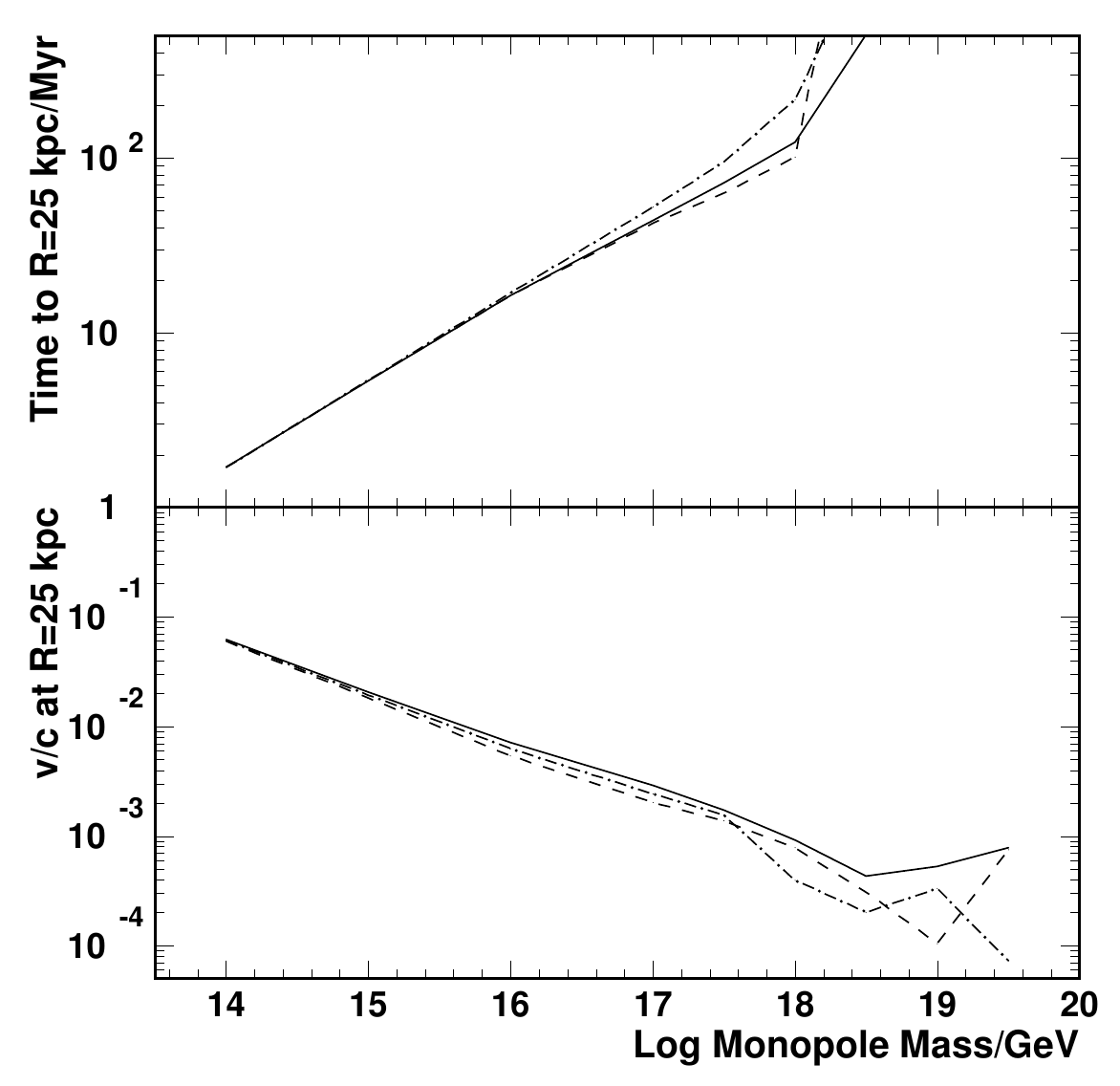}
\caption{\label{MMlive} Top: Time for a monopole in the Galaxy
  to reach radius 25~kpc as a function of monopole mass. Each curve is
  for a monopole starting at the same galactic coordinates as the
  Solar System (radius 8.5~kpc). The solid (dashed) curve is for a
  monopole starting with the same (opposite) velocity as the Solar
  System. The dash-dotted curve is for a monopole starting at rest at
  the position of the Solar System. Bottom: The velocities at
  radius 25~kpc in the same three designations as above as a function of
  monopole mass.}
\end{figure}

Figure~\ref{MMlive} shows the monopole lifetime for several different
initial conditions. The outcome is found to be approximately independent of the
initial conditions, and a linear fit to the data gives a lifetime of
0.17$~M^{1/2}$~Myr with the monopole mass $M$ in GeV. Monopoles with
masses below $10^{17.5}$~GeV have short lifetimes compared to the age
of the Galaxy ($\sim$4.5~Gyr) and are thus quickly expelled from the
Galaxy. This is consistent with the results reported in
Ref.~\cite{Turner1982}. Such short lifetimes will severely reduce the
concentration of relatively low mass monopoles released within the
Galaxy (e.g., ejected in supernova explosions). For $M>10^{18}$~GeV
monopoles remain gravitationally bound to the Galaxy. However, they
tend to spend a major fraction of their dwell time at radii that are
much larger than the solar orbit radius. Hence, again their flux in
the vicinity of the Earth will be limited. Furthermore, since they
have large kinetic energies, their ranges in matter will be longer
than the planetary or star radii. They would thus be unlikely to stop
in matter. They could still have bound to matter before galaxies were
formed, e.g., during cosmic nucleosynthesis~\cite{Bracci1984}. A
monopole trapped in a larger rigid structure (such as an interstellar dust grain)
would have to drag the whole object with it during acceleration and
would not reach high velocities.

Monopoles ejected from other galaxies into extra-galactic space are
likely to have similar velocities to those ejected from our own
Galaxy. They could enter our Galaxy. Their large kinetic energy would
overcome the potential energy due to the galactic magnetic fields so
that they could be detected in Earth-bound detectors. Hence if a
naturally occurring monopole is eventually detected in flight it will
likely be of extra-galactic origin.

\subsubsection{Monopole phenomenology}
\label{monopole_pheno}

Monopole searches have been performed at colliders, in cosmic rays and
in matter since the 1950s, and are still continuing today. A wide
variety of techniques can be employed to search for monopoles. These
include induction techniques with superconducting magnetometers to
detect magnetic charge trapped in matter
(Section~\ref{induction_technique}) or in flight
(Section~\ref{cosmic_induction}), detector arrays based on ionisation
energy loss (Section~\ref{cosmic_ionisation}), detection of Cherenkov
radiation in water or ice to probe relativistic monopoles
(Section~\ref{cosmic_cherenkov}), and nucleon-decay detectors for
probing the monopole catalysis reaction
(Section~\ref{cosmic_catalysis}). A summary of current experimental
limits on monopoles in a wide range of hypotheses and detection
techniques is provided in Table~\ref{tab:summary_monopoles} and
in Figs.~\ref{fig:betagamma} and~\ref{fig:mass} (Section~\ref{cosmic}). Interactions of monopoles with matter
are discussed in Section~\ref{interactions_magnetic}.

As discussed above, very massive monopoles ($10^{18}$~GeV or higher) would form galactic halos. It is conceivable that they could contribute to dark matter. Oppositely charged monopoles can also form a neutral bound state termed ``monopolium''. In an SU(5) GUT model, monopolium can have a lifetime ranging from several days to $10^{11}$ years depending on its diameter, and would be copiously formed in the early Universe at the time of nucleosynthesis given that a sufficient amount monopole defects were produced at an earlier epoch~\cite{Hill1983}. More generally, a simple way to allow for stable monopolium is to postulate the existence of at least two kinds of monopoles which could not decay into one another nor annihilate with each other (just as electrically charged fermions can exist either as leptons or hadrons). Monopolium with lifetime comparable to or greater than the age of the Universe could play the role of dark matter, similarly to the dark atoms discussed in Section~\ref{dark_atoms}. Monopole-antimonopole annihilations within relic monopolia could also give rise to ultra-high energy cosmic rays~\cite{Bhattacharjee1995}. A scenario of monopoles existing within a dark sector have also been put forward to explain dark matter (see Section~\ref{GUT_mono}).

\subsection{Field configurations other than monopoles}

A number of theories predict the existence of objects with a more
extended structure than that expected of the exotic leptons and
hadrons outlined in Section~\ref{sec:smpleptonhadron}.

\subsubsection{$Q$-balls}
\label{theo_Qballs}

$Q$-balls are topological objects that are stable due to a conserved
angular momentum in an internal space (essentially, a charge) combined
with a lack of lighter particles in the spectrum into which the
charged object could decay. Consider a theory with a two-dimensional
internal space of scalar fields $\phi_1$ and $\phi_2$ with a potential
that is only a function of $|\phi|=\sqrt{\phi_1^2+\phi_2^2}$.  We
assume that $|\phi|$ is constant within a (real space) sphere of
radius $R$ and zero outside that sphere.  If the vector $\phi_i$ is
rotating around the internal SO(2) symmetry with a specific angular
frequency $\omega$, then the conserved charge $Q$ is given by
\begin{equation}
Q=\int d^3 x
\left[\phi_1\partial_0\phi_2-\phi_2\partial_0\phi_1\right]=\frac{4\pi}{3}
R^3\omega|\phi|^2, \label{qcharge}
\end{equation}
whereas the energy $E$ is given by
\begin{equation}
E=\frac{4\pi}{3}R^3\left(\frac{1}{2}\omega^2|\phi|^2+V\right),
\label{qenergy}
\end{equation}
where $V$ is the potential of $\phi$.  Using Eq.~\ref{qcharge} to
replace $\omega$ in Eq.~\ref{qenergy}, one can see that there is a
radius $R$ at which the energy is minimised, and at this minimum
$E=Q\sqrt{2V/\phi^2}$. This field configuration, or ``$Q$-ball'', can
decay by emitting the basic scalar particle of the Lagrangian, which
has unit charge and mass
$m^2=\left.\partial^2V/\partial\phi^2\right|_{\phi=0}$. However the
energy per unit charge of the $Q$-ball is $E/Q=\sqrt{2V/\phi^2}$, so
that if the mass of the scalar $m> \sqrt{2V/\phi^2}$, the $Q$-ball
will not be able to decay into these scalar
particles~\cite{Lee1974,Friedberg1976,Friedberg1977,Coleman1985}.

Unlike topological monopoles, $Q$-balls are dynamic configurations of
the vacuum that may or may not exist depending upon the history of the
Universe. Like topological monopoles, $Q$-balls can form with very
many different mass scales depending upon the theory in question. Also
the charge in question can vary; it could be electric charge or it
could, for example, be baryon number.

Many potentials have characteristic logarithmic one-loop corrections that allow $m>\sqrt{2V/\phi^2}$. In SUSY there are examples of $Q$-balls in both gravity-mediated~\cite{Kasuya2000} and gauge-mediated~\cite{Kasuya2000b} SUSY-breaking scenarios. For typical values of the parameters in these models, $Q$-balls must have rather large masses, much higher than the electroweak scale, in order to be stable~\cite{Kusenko1998,Kusenko1998b}.

The interactions of $Q$-balls with matter depend upon their properties, such as electric charge and ability to catalyse nucleon decay, as discussed in Section~\ref{interactions_Qballs}.

\subsubsection{Primordial black holes}
\label{black_holes}

Black holes are gravitationally collapsed objects whose energy content
is located in such a small region that they are shrouded by an event
horizon.  While we are perhaps most familiar with astrophysical black
holes at the centres of galaxies or at the end point of stellar
evolution, it is possible that much smaller black holes were created
in the early Universe. These are called primordial black holes, or
PBHs.

PBHs are hypothesised to have formed in density fluctuations in the
early Universe~\cite{Zeldovich1967,Carr1974,Carr1975} with masses
typically peaked near the mass enclosed in the particle horizon at
that epoch. There are a large number of theoretical scenarios
describing PBH formation~\cite{Carr2010,Frampton2010} including
specific types of
inflation~\cite{Carr1993,Ivanov1994,Yokoyama1998,Kanazawa2000,Leach2000,Chongchitnan2007,Nozari2007,Saito2008,Lin2013},
phase transitions, for example, from bubble
collisions~\cite{Hawking1982,Kodama1982,Berezin1983,La1989,Moss1994,Konoplich1998},
and the collapse of cosmic
strings~\cite{Hogan1984,Hawking1989,Polnarev1991,Caldwell1996,MacGibbon1998,Hansen2000,Nagasawa2005}
and necklaces~\cite{Matsuda2006,Lake2009}. Consequently, PBHs can span
a huge mass range.

While classically, nothing including light can escape from a black hole, it is now generally accepted that these objects gradually decay
due to quantum effects. Hawking's analysis~\cite{Hawking1974,Hawking1975} (inspired by earlier work by Beckenstein~\cite{Bekenstein1973}) showed that quantum fields in curved space-times develop non-zero creation operators, leading to the production of particles from the vacuum at a rate determined by the local space-time curvature.  While a fully self-consistent treatment of the problem is still elusive, it seems necessary to accept that this happens from the point of view of thermodynamics. It is assumed that the energy associated with the particles comes from the gravitational field such that the mass of the black hole slowly leaks away.

Although PBHs with mass less than 10$^{14}$~g would have completely evaporated by now, they would nevertheless have contributed to the development of the early Universe. For example,
PBHs would affect baryogenesis~\cite{Turner1979,Barrow1991,Upadhyay1999,Bugaev2003}; generate neutrinos~\cite{Bugaev2002,Bugaev2009} or hypothetical particles~\cite{Green1999,Lemoine2000,Khlopov2006}; swallow monopoles~\cite{Izawa1984,Stojkovic2005a}; and destroy domain walls \cite{Stojkovic2005b}.


The masses of primordial black holes that survive to the current day is likely to be larger than $10^{15}$~g. Such a black hole would have a Schwarzchild radius of $10^{-15}$~m so it would be about the size of a proton. To investigate the detectability of such objects it is interesting to estimate their expected electric charge. The temperature of a small black hole is related to its mass by
\begin{equation}
T_{Hawking}=\left(\frac{1.058\times 10^{16}}{M_{BH}}\right){\rm MeV},
\end{equation}
so that black holes with masses less than around $10^{16}$~g would be
able to emit electrons and positrons and would do so, resulting in a
net charge of order 1~\cite{Page1976b}.  Much heavier black holes
could in principle have much larger charges and would not be able to
emit particles to get rid of this excess charge but there is no
particular reason why this would occur. For black holes with masses
less than $10^{18}$~g the cross section for absorbing a proton is
greater than that for absorbing an electron due to the wave-function
overlap being suppressed so the black hole might classically absorb
protons more efficiently while not being able to balance this net
charge by radiating positrons~\cite{Islam1980}. In this small mass
range, it seems difficult to imagine a net charge greater than around
500$e$ at the absolute most.

PBHs are constrained due to their indirect effects on nucleosynthesis, fluctuations in the cosmic microwave background, and galaxy structure formation~\cite{Mack2007,Carr2010,Pani2013}, as well as a variety of dynamical arguments~\cite{Carr1999}. Another strong requirement comes from the mere existence of old neutron stars~\cite{Capela2013}. PBHs must be small enough to escape detection but also large enough not to be evaporating today, otherwise we would detect the gamma-ray emission associated with their demise. Direct and indirect experimental limits on PBH abundances inferred from impacts on Earth and astrophysical observations are discussed further in Section~\ref{compact_objects}. All these constraints put together indicate that PBHs cannot be the dominant constituent of dark matter in the Universe.

Small black holes not only interact gravitationally with the surrounding matter, but also electromagnetically if they posses electric charge. This is discussed in Section~\ref{interactions_black_holes}.

\subsubsection{Black hole remnants}
\label{black_hole_remnants}

The end point of the evolution of black holes is unknown, as the objects will shrink to the size associated with the fundamental length scale of quantum gravity.
The Hawking description for the quantum thermal emission of particles from black holes ceases to apply as the black hole reaches the Planck mass, $M_{Pl} = [\hbar c/G]^{1/2}$ = $10^{19}$~GeV = $2\cdot 10^{-5}$g, since the size of such a black hole is comparable to its Compton wavelength.



There are cogent arguments for the existence of black hole remnants,
i.e., assuming that the black hole does not radiate away to ordinary
stable particles and vacuum. A generic problem for black holes is how
information stored behind the horizon can be emitted via Hawking
radiation. Either this occurs, in which case there needs to be some
blurring of space-time locality near the horizon, or it does not, in
which case whatever is left at the end point of black hole evaporation
must contain all the information associated with the particles that came together to form the black hole in the first place. The latter situation has led to the idea that Hawking evaporation stops at the last minute, leaving behind a stable kernel. One such possibility is based on the ``Generalised Uncertainty Principle'' (GUP)~\cite{Adler2001,Carr2013}. In this approach the usual uncertainty relation is replaced by $\Delta x > \hbar/\Delta p + l_{Pl}^{2}\Delta p/\hbar$, where the second term accounts for self-gravity effects. This results in the evaporation ceasing at the Planck mass, leaving a stable remnant.
Further arguments for the existence of remnants have been predicated on black holes with axionic charge~\cite{Bowick1988}, the modification of the Hawking temperature due to magnetic monopoles~\cite{Lee1992} or quantum hair~\cite{Coleman1991}.  The coupling of a dilaton field to gravity can also yield relics, with detailed features depending on the dimension of space-time~\cite{Torii1993}.

The Planck mass remnants of vaporising PBHs would be expected to
contribute to the dark matter budget of the
Universe~\cite{MacGibbon1987}. This possibility has been explored for
a number of inflationary
scenarios~\cite{Barrow1992,Alexeyev2002,Barrau2004,Chen2005,Nozari2005}. If
reheating occurs at temperature $T_{R}$ and the relics have mass
$\kappa M_{Pl}$, then the requirement that the PBH remnant density is  less than the dark matter density $\Omega_{CDM}$ implies $ \beta'(M) < 2 \times 10^{-28} \kappa^{-1} (M/M_{Pl})^{3/2}$, for the mass range $(T_{R}/T_{Pl})^{-2}  < M/M_{Pl} < 10^{11} \kappa^{2/5}$~\cite{Carr1994}. It is possible to construct scenarios in which Planck mass PBH relics are capable of producing the observed dark matter density and baryon asymmetry of the Universe.


Assuming that PBH remnants ionise matter, e.g., because they carry electromagnetic charge, constraints on their cosmic flux can be inferred from ionisation array experiments (Section~\ref{cosmic_ionisation} and Table~\ref{tab:summary_monopoles}). However, none of these searches has been explicitly interpreted for PBH remnants.

\subsection{Summary}
\label{theo_summary}

Table~\ref{tab:hypotable} summarises the properties of SMPs that have been proposed and which are discussed in this Section. It shows the approximate SMP mass range expected from theoretical scenarios. Where an upper SMP mass limit is not prescribed by theory,  the Planck mass $M_{PL}$ is assigned. The main search methods are also shown.

\begin{table}
\begin{center}
\begin{tabular}{|l|c|c|c|}
    \hline
    SMP                  & mass range     & \multicolumn{2}{c|}{main search method}     \\
                         &                & in cosmic rays        & in matter                \\
    \hline
    $X^-$, $X^{--}$      & $<M_{PL}$      & time-of-flight        & mass spectrometry            \\
                         &                &                       & radiochem. properties       \\
    $X^+$, $X^{++}$      & $<M_{PL}$      & time-of-flight        & mass spectrometry              \\
                         &                &                       & atomic spectroscopy                  \\
    $X^0$                & $<M_{PL}$      & nuclear recoil        & mass spectrometry             \\
    fractional charge    & $<M_{PL}$      & low ionisation        & levitometer                     \\
                         &                & high rigidity         & oil drop                      \\
    dark atom            & $<M_{PL}$      & nuclear recoil        & anomalous meteoroids   \\
    monopole             & $<M_{PL}$      & time-of-flight        & induction               \\
                         &                & high ionisation       &                           \\
                         &                & Cherenkov             &                              \\
                         &                & nucleon decay         &                             \\
    charged $Q$-ball     & $<M_{PL}$      & high ionisation       & ---                        \\
    neutral $Q$-ball     & $<M_{PL}$      & nucleon decay         & ---                          \\
    quark matter         & $<10^{33}$~g   & spectrometry          & mass spectrometry            \\
                         &                & high ionisation       & heavy-ion activation       \\
                         &                & seismic signals       &                            \\
                         &                & grav. lensing         &                            \\
    fermion star         & $10^{23}-10^{30}$~g & grav. lensing    & ---                      \\
    prim. black hole     & $10^{15}-10^{35}$~g & grav. lensing    & ---                      \\
                         &                & bursts                &                              \\
    black-hole remnant   & $<M_{PL}$      & ionisation    & ---                      \\

  \hline
\end{tabular}\caption{\label{tab:hypotable} Summary of the properties of a range of SMPs that have been proposed in the literature. Note that $M_{PL}$ refers to the Planck scale}
\end{center}
\end{table}

Many SMPs provide dark matter candidates which are superheavy in most of the cases. The charged SMPs in bound states which involve sometimes ordinary particles can give neutral dark matter constituents. Some constituents could remain charged and still satisfy the dark matter candidates requirements~\cite{Chuzhoy2009}.

A wide range of experimental techniques must be employed to investigate the whole spectrum of different SMP types. These will be described in detail in Sections~4$-$7.

%% file: smp-matter.tex
\section{Interactions of SMPs with matter}
\label{smp_matter}

\subsection{Ionisation energy loss for electrically charged SMPs}
\label{interactions_electric}

For high velocity SMPs, the energy loss in matter of density
$\rho$~g/cm$^3$ will be dominated by interactions with the atomic
electrons, which can be considered as quasi-free at such
velocity. Here, high velocity is such that the maximum kinetic energy
transferred in a collision with an electron, $2m_e V v_F$, is
much greater than the energy required to excite the atoms in the
material. Here, $m_e$ is the electron mass, $V$ the velocity of the
SMP and $v_F\sim 2\times10^6$m/s is the Fermi velocity of the
electrons when treated as belonging to a Fermi gas. In this regime,
the stopping power is given by the familiar Bethe-Bloch equation for a
particle moving with velocity $\beta=V/c$ and
$\gamma=\sqrt{1/(1-\beta^2)}$~\cite{PDG2012}:
\begin{equation}
-\frac{1}{\rho}\frac{dE}{dx}=4\pi \frac{Z N_A}{A} \frac{z^2e^4}{m_e c^2 \beta^2}\bigg(\frac{1}{2}\ln{\frac{2 m_e c^2 \beta^2 \gamma^2 T_{max}}{I^2}} - \beta^2
\bigg),
\label{BB}
\end{equation}
where $z$ is the charge number on the SMP, $e$ is the electronic
charge, $Z$ and $A$ are the atomic number and mass of the material,
respectively, $N_A$ is Avogadro's number, $T_{max}=2m_e c^2 \beta^2
\gamma^2$ is the maximum kinetic energy that can be transferred in a
collision with a stationary electron and $I$ is the average ionisation potential
of the atoms of the material. The latter is roughly parameterised as
$I(Z)=(12 Z+7)$~eV for $Z \leq 13$ and $I(Z)=(9.76 Z + Z^{-0.19})$~eV
for $Z > 13$~\cite{Lewin1977} (more recent values of $I$ can be found
in Ref.~\cite{PDG2012}). Note that we have assumed that the SMP of
mass $M$ is sufficiently heavy that terms in powers of $m_e/M$ are
negligible. In addition, we assume that the SMP moves with velocity $\beta
\gamma < 10$ so that density effects in the material can be neglected.

For lower velocities, $\beta < \alpha z^{1/3}/(1+\alpha z)$, where
$\alpha \cong 1/137$ is the fine structure constant, the atomic
binding energy is such that the atomic electrons can no longer be
treated as quasi-free. Here, the Bethe-Bloch formula breaks down. The
stopping power was derived in this velocity regime in a series of papers by
Lindhard {\it et al.}~\cite{Lindhard1961}. They treat the atomic
electrons as a degenerate Fermi gas and include the energy loss due to
collisions with the nucleus. The stopping power then becomes
\begin{equation}
-\frac{1}{\rho}\frac{dE}{dx}=8\pi \hbar c \frac{N_A}{A} a_0 \frac{z Z}{(z^{2/3}+Z^{2/3})^{3/2}} \beta,
\label{LS}
\end{equation}
where $a_0$ is the Bohr radius of the hydrogen atom.

The velocity range between the two described by Eqs.~\ref{BB} and
\ref{LS} is usually joined smoothly by a cubic polynomial with
coefficients chosen by solving the four simultaneous equations
obtained by equating the values of $dE/dx$ and $d(dE/dx)/d\beta$ at
the two boundaries~\cite{Lewin1977}.

The region of applicability of the Bethe-Bloch formula is well tested and the formula is found to be accurate to within a few percent. The lower velocity region described by Eq.~\ref{LS} is found to agree with heavy ion measurements~\cite{Ormrod1963,Ormrod1965} to within 10\% over most of the velocity range but with occasional larger disagreements~\cite{Lewin1977}.

\subsection{Other sources of energy loss for electrically charged SMPs}

In addition to a point-like electrically charged SMP, a number of
putative composite SMP candidates exist in the literature. The
presence of inner structure influences the scattering cross sections
and energy loss as the SMP propagates through matter. In this section,
interactions of the most widely sought composite SMPs are described.

\subsubsection{Hadronising SMPs}
\label{interactions_Rhadrons}

As has been seen in Section~\ref{sec:smpleptonhadron}, if the requirement of a SUSY WIMP candidate is relaxed, then SUSY admits a range of supersymmetric particles that can
provide SMP candidates~\cite{Fairbairn2007}. Non-WIMP LSPs are one way in which SMPs can arise in SUSY scenarios. Properties of putative states formed by hadronising massive coloured particles,
such as $R$-hadrons in the context of SUSY, have been extensively
studied~\cite{Kraan2004,deBoer2008,Mackeprang2010,Farrar2011}. Of
particular relevance to this work are the predicted mechanisms through
which $R$-hadrons scatter. From considerations of perturbative QCD,
the heavy particle would largely act as a spectator, with the light
quarks being the active participants. Based on analogies with low energy
hadron-hadron data, a cross section of order 10~mb would be expected
for scattering in matter. Such interactions would also be
characterised by low energy loss ($\sim1$~GeV) per collision. Charge exchange reactions are also expected and, based on estimates of the low lying $R$-hadron mass states, $R$-hadrons containing gluinos and sbottom-type quarks would be expected to become neutral after around ten interactions. Following repeated scattering, a $R$-hadron with a net positive charge would only be expected for a hypothesis of stable stop-like quarks. Owing to the approximate spin-independence of the mass hierarchies of states containing a heavy coloured object, the aforementioned results are also valid for non-SUSY exotic hadrons that could arise.

\subsubsection{Mirror matter}
\label{interactions_mirror}

The simplest model of dark atoms is that of mirror particles with exactly the same properties as Standard Model particles but with $V+A$ weak interactions instead of $V-A$ (see Section~\ref{dark_atoms}). A mirror atom would interact only very weakly with ordinary matter. In addition to gravitational interactions, this can also happen through a tiny kinetic mixing $\chi$ between the photon and the mirror photon: mirror charged particles would then act as ordinary particles with charge $\chi e$. However, mirror atoms would interact among themselves in the same way as ordinary atoms do, thus allowing (in certain conditions) for the formation of macroscopic objects such as mirror stars, mirror dust, mirror rocks, etc.~\cite{Blinnikov1982}. The large number of mirror atoms present in mirror bodies would result in strong interactions with ordinary matter. Using a kinetic mixing parameter $\chi=10^{-9}$ (allowed by current constraints~\cite{Foot2014}), a mirror matter body impacting the Earth would experience a significant drag force proportional to the total number of molecules within the body~\cite{Foot2003a}. The repulsion force between a mirror body and ordinary matter could be sufficient to counteract the Earth's gravitational force (thus allowing for mirror fragments to stay on the Earth's surface), and conversely ordinary dust impacting a mirror body could be slowed down and remain embedded inside the body or accumulate on its surface.

\subsubsection{Strange quark matter}
\label{interactions_nuclearites}

As mentioned in Section~\ref{sec:strangelets}, strange quark matter
would comprise roughly the same number of up, down and strange
quarks. SQM systems can be stable for baryon numbers between those of
ordinary nuclei to those of neutron stars ($A~\sim 10^{57}$). Two
specific types of strange quark matter are typically sought: strangelets and nuclearites.

Strangelets are small ($A<10^7$) SQM systems. The energy loss in matter is expected to resemble that of heavy ions. However, strangelets would have a far lower charge-to-mass ratio than heavy ions. Typical values are $\frac{Z}{A} \sim 0.1$ (for $A \ll 10^3$) and $\frac{Z}{A} \sim 8A^{-\frac{2}{3}}$ (for $A \gg 10^3$)~\cite{Heiselberg:1993dc}. Anomalous charge-to-mass ratios in the cosmic radiation can be sought at balloon or space-borne spectrometers (see Section~\ref{cosmic_in_space}). There are large uncertainties in the propagation of strangelets in matter (e.g. the atmosphere). In one approach, the strangelet mass is reduced following collisions with molecules~\cite{Wilk1996}, whereas another approach~\cite{Banerjee2000} posits that the mass and charge increase following the accretion of nucleons.

Nuclearites are large ($A>10^7$) and are electrically neutral atom-like systems of nuclear density. They would be expected to possess a electron cloud around the core which, for $A>10^{15}$, would be largely contained within the nuclear matter. They are generally assumed to have a typical galactic velocity of $\beta \sim 10^{-3}$. The dominant types of interactions in matter would be elastic and inelastic collisions. The energy loss~\cite{DeRujula1984} is given by
\begin{equation}
\frac{dE}{dx}=-{\sigma}{\rho}\beta^2,
\end{equation}
where $\rho$ is the material density, $v$ is the nuclearite speed and $\sigma$ is the nuclearite's geometric cross section. Different expected values for $\sigma$ can be found in Ref.~\cite{DeRujula1984}. The energy loss is such that nuclearites would be visible. Interactions in transparent matter (e.g., scintillators and water) have been used in searches. It would be expected that nuclearites moving slowly through water would give rise to a thermal shock, emitting black-body radiation and producing a large visible photon yield. Energy loss leading to damage in plastic track detectors is also expected (see Section~\ref{cosmic_ionisation}). Nuclearites with masses of the order of 1~g or larger would melt rock and produce acoustic/seismic signals when colliding with the Earth or the Moon (see Section~\ref{seismic}).

\subsubsection{$Q$-balls}
\label{interactions_Qballs}

Some extensions to the SM predict $Q$-balls as non-topological soliton solutions (see Section~\ref{theo_Qballs}). Two classifications are typically made. $Q$-balls can be supersymmetric electrically charged solitons (SECS) or supersymmetric electrically neutral solitons (SENS), each with different scattering properties. The SENS variety largely interacts in matter via the catalysis of nucleon decay leading to pion production and has been sought at nucleon decay experiments~\cite{SuperKamiokande2007}. Interactions of a SECS in matter could lead to the $Q$-balls obtaining a net positive charge in the interior of the particle although the net charge of the object could be neutralised by the acquisition of a surrounding electron cloud. The $Q$-ball would then possess an atomic size ($\sim 10^{-10}$m) with a heavy nucleus. Interactions of $Q$-balls would then proceed via electromagnetic and nuclear processes, with the latter being due to reactions between the $Q$-ball core and nuclei in the medium. As $Q$-ball interactions would resemble those of nuclearites (Section~\ref{interactions_nuclearites}), cosmic-ray experiments typically search for both types of exotic objects~\cite{MACRO2000,SLIM2008b}. More details of such searches are given in Section~\ref{cosmic_ionisation}.

\subsubsection{Small black holes}
\label{interactions_black_holes}

Hawking radiation is the dominant energy loss mechanism for black
holes with masses smaller than $10^{15}$~g passing through
matter. Other than Hawking radiation, there are two primary energy
loss mechanisms for small black holes: absorption, and ionisation from
the gravitational pull. Absorption losses result from the increase in
mass of the black hole as atoms cross within its Schwarzschild radius,
which, for a black hole the mass of a planet, a mountain and a person,
is the size of a golf ball, a neutron, and a neutrino,
respectively~\cite{Greenstein1984}. More important for detection
purposes are the ionisation losses induced at large distance by the
gravitational pull, which, for black-hole masses larger than
$10^{15}$~g, always results in a large transfer of kinetic energy from
the black hole to the atoms of the detector
materials~\cite{Greenstein1984}. Such deformations could potentially
be observed in the form of high-ionisation tracks, tubes of melted
rock, sound waves, and seismic signals~\cite{Khriplovich2008} (see
Section~\ref{seismic}).

Much lighter neutral black-hole remnants (Section~\ref{black_hole_remnants}) should interact very weakly with matter, since they neither radiate nor produce significant ionisation in their wake.

\subsection{Energy loss for magnetic monopoles}
\label{interactions_magnetic}

As with electrically charged particles
(Section~\ref{interactions_electric}), the major contribution to the
energy loss of magnetically charged particles with velocities $\beta
\gamma < 10^4$ is interactions with atomic electrons and nuclei.
Such interactions transfer small amounts of energy at each collision so that the stopping power
$dE/dx$, and hence the range, is well defined.

For velocities $\beta \gamma < 10^4$, the stopping power of monopoles
in material is estimated by analogy to the treatment of the stopping
power for charged particles in matter. For velocities greater than
this, energy losses are dominated by stochastic processes, in which it
is possible to lose a large amount of energy at each collision.

\subsubsection*{$10^{-1} < \beta\gamma < 10^4 $}

At these velocities the atomic electrons can be treated as quasi-free
and a version of the Bethe-Bloch formula is
appropriate~\cite{Ahlen1975,Ahlen1978}. However, since the force
between the monopole and an electron is the velocity-dependent Lorentz
force (rather than the velocity-independent Coulomb force, as in the
case of charged SMPs), the $1/\beta^2$ term cancels. The modified
Bethe-Bloch formula for magnetic monopoles is then~\cite{Ahlen1975,Ahlen1978}
\begin{equation}
-\frac{1}{\rho}\frac{dE}{dx}=4 \pi \frac{ZN_A}{A} \frac{g^2 e^2}{m_ec^2}\bigg(\ln \frac{2m_ec^2\beta^2 \gamma^2}{I} +\frac{K(|g|)}{2} -\frac{1}{2} - B(|g|)\bigg).
\end{equation}
Here, $g$ and $e$ are the magnetic pole strength of the monopole and electronic charge, respectively,
and $I$ is the mean ionisation potential of the atoms of the material (see Section~\ref{interactions_electric}). The terms $K(|g|)$ and $B(|g|)$
are numerical factors that depend on the magnetic monopole pole strength. They are given by $B(|g|)=$0.248 (0.672), $K(|g|)=$0.406 (0.346) for magnetic pole strength 137$e$/2 (137$e$). The density effect correction
is again neglected. (Details of this correction are given in Refs.~\cite{Ahlen1975,Sternheimer1971}).

\subsubsection*{$10^{-3} < \beta \gamma < 10^{-2}$}

The situation is less definite for monopoles at lower
velocities. Indeed, no theory exists for the intermediate region
$10^{-2} < \beta < 10^{-1}$. The region $\beta < 10^{-2}$ is discussed
quantitatively in papers by a number of
authors~\cite{Ahlen1975,Ahlen1978,Ahlen1982,Ahlen1983,Ficenec1987,Drell1983,Bracci1983,Derkaoui1998}. A
comprehensive treatment is given in Ref.~\cite{Derkaoui1998}.

Ahlen and Kinoshita~\cite{Ahlen1982} used the theory of
Lindhard~\cite{Lindhard1961} as applied to slowly moving charged
particles to compute the stopping power of lower velocity magnetic
monopoles in the range \mbox{$10^{-3} < \beta < 10^{-2}$}. This gives the stopping power in nonconductors as
\begin{equation}
-\frac{1}{\rho}\frac{dE}{dx}=2\pi \frac{ZN_A}{A} \frac{g^2 e^2 \beta}{m_ec v_F}\bigg(\ln \frac{2m_e v_F a_0}{\hbar}-0.5\bigg),
\label{dedx2}
\end{equation}
where $v_F=(\hbar/m)(3 \pi^2 ZN_A/A)^{1/3}$ is the Fermi velocity.

A further correction is necessary for the conduction electrons in
metallic solids (such as in the Earth's core); see
Ref.~\cite{Derkaoui1998} for details. Furthermore, the intense
magnetic field of magnetic monopoles is expected to shift the atomic
energy levels by the Zeeman effect. The shifts in energy are so great
that the levels can cross, allowing transitions between them. Such
transitions cause the monopole to lose energy~\cite{Drell1983}. This
was shown to increase the stopping power by an order of magnitude
compared to the results from Lindhard theory for hydrogen and helium
for $\beta > 3\cdot10^{-3}$~\cite{Drell1983}. However, the effect has not
been studied for other materials and is not taken into account in
Eq.~\ref{dedx2}, implying uncertainties in it.

\subsubsection*{$\beta \gamma < 10^{-3}$}

In this regime the energy loss occurs mainly through elastic collisions with atoms, mediated by interactions between the magnetic dipole moment of the atom and the magnetic field of the monopole. This leads to differences in stopping power for paramagnetic and diamagnetic materials.

The atoms of paramagnetic substances have a fixed, permanent magnetic dipole moment $\chi=g_J \sqrt{J(J+1)}\mu_B$, where $\mu_B$ is the Bohr magneton, $g_J$ is the Lande splitting factor and $J$ is the total angular momentum quantum number of the atom. The potential energy of a monopole in the field of a magnetic dipole follows a $1/r^2$ behaviour. For scattering in such a potential, the stopping power becomes
\begin{equation}
\frac{dE}{dx} = \int_0^{E_{max}} N E \frac{d\sigma}{dE} dE =
\chi \frac{\hbar ge \mu_0 N}{m}  \frac{<J_z>}{\sqrt{J(J+1)}},
\end{equation}
i.e.,
\begin{equation}
\frac{dE}{dx} = \chi \mu_B \mu_0 g N  \frac{<J_z>}{\sqrt{J(J+1)}},
\end{equation}
where $N$ is the atomic density, $J_z$ and $J$ are the angular momentum quantum numbers of the atom and $E_{max}$ is the maximum energy transferred to the atom in a single collision (corresponding to an atomic velocity of $2\beta$). Note that this value is independent of the monopole velocity.

For diamagnetic materials the induced dipole moment depends on the
distance between the monopole and the dipole. The scattering of a
monopole from such an atom is more complicated and it is necessary to
consider the detailed atomic structure. This is described in detail in
Ref.~\cite{Derkaoui1998}.

\subsubsection*{$\beta\gamma > 10^{4}$}

At very large velocities ($\beta \gamma > 10^4$), stochastic processes such as bremsstrahlung, pair production and hadron production dominate the energy loss of a monopole. Photonuclear interactions by  strong interactions are the largest contributor of these for materials with moderate values of atomic mass (such as rock), with a smaller contribution from pair production. In contrast pair production is the dominant process in materials with high atomic number $Z$. The contribution from bremsstrahlung is small and is usually neglected.

Such stochastic processes mean that the energy loss per unit path length is dominated by either single or small numbers of collisions with the possibility of a large energy transfer in each collision. The stopping
power and the range are therefore not well defined. Therefore, experiments in this velocity region must compute the probability of the loss of a monopole in traversing the material surrounding the detector and correct for it in the detection efficiency.

Searches for ultra-relativistic monopoles in this energy regime have
been made by, e.g., the RICE~\cite{RICE2008} and ANITA~\cite{ANITA2011} collaborations. To compute the detection efficiency they use the interaction cross sections for stochastic processes computed for high velocity muon and tau leptons~\cite{Dutta2001}.  To allow for the stronger coupling due to the large effective charge of a monopole, the cross sections are multiplied by the square of the equivalent Dirac charge ($137e/2$), i.e., by 4700.

It should be noted that the cross sections for interactions of charged leptons by these stochastic processes were derived from perturbation theory, which requires small coupling constants. This is quite reliable for charged leptons. However, for magnetic monopoles the coupling constant is very large and perturbation theory is almost certainly invalid. Better non-perturbative calculations are desirable for these processes. For example, treating the monopole as a flux of virtual photons by the Weizs\"acker-Williams method could lead to a non-perturbative calculation of the cross sections for stochastic
processes~\cite{Kim1973}.

\subsubsection{Range of monopoles in matter}
\label{range_monopoles_matter}

The range was computed by integrating the stopping power:
\begin{equation}
R=\int_{E_{min}}^E \frac{dE}{dE/dx}.
\end{equation}
In principle the value of the minimum kinetic energy $E_{min}$ should be zero. However, the stopping power is unknown for velocities $\beta < 10^{-5}$. Here, $E_{min}$ was taken to be the kinetic energy corresponding to $\beta=10^{-6}$ and the stopping power was extrapolated to this value from the curves in Ref.~\cite{Derkaoui1998}. The loss of range due to this approximation is quite small for paramagnetic materials such as iron since the stopping power becomes roughly independent of velocity at very low velocity. However, the loss leads to considerable uncertainty for low velocity monopoles in diamagnetic materials such as silicon since the stopping power falls strongly with velocity.

\begin{figure}[ht]
\centering
\includegraphics[width=0.7\linewidth]{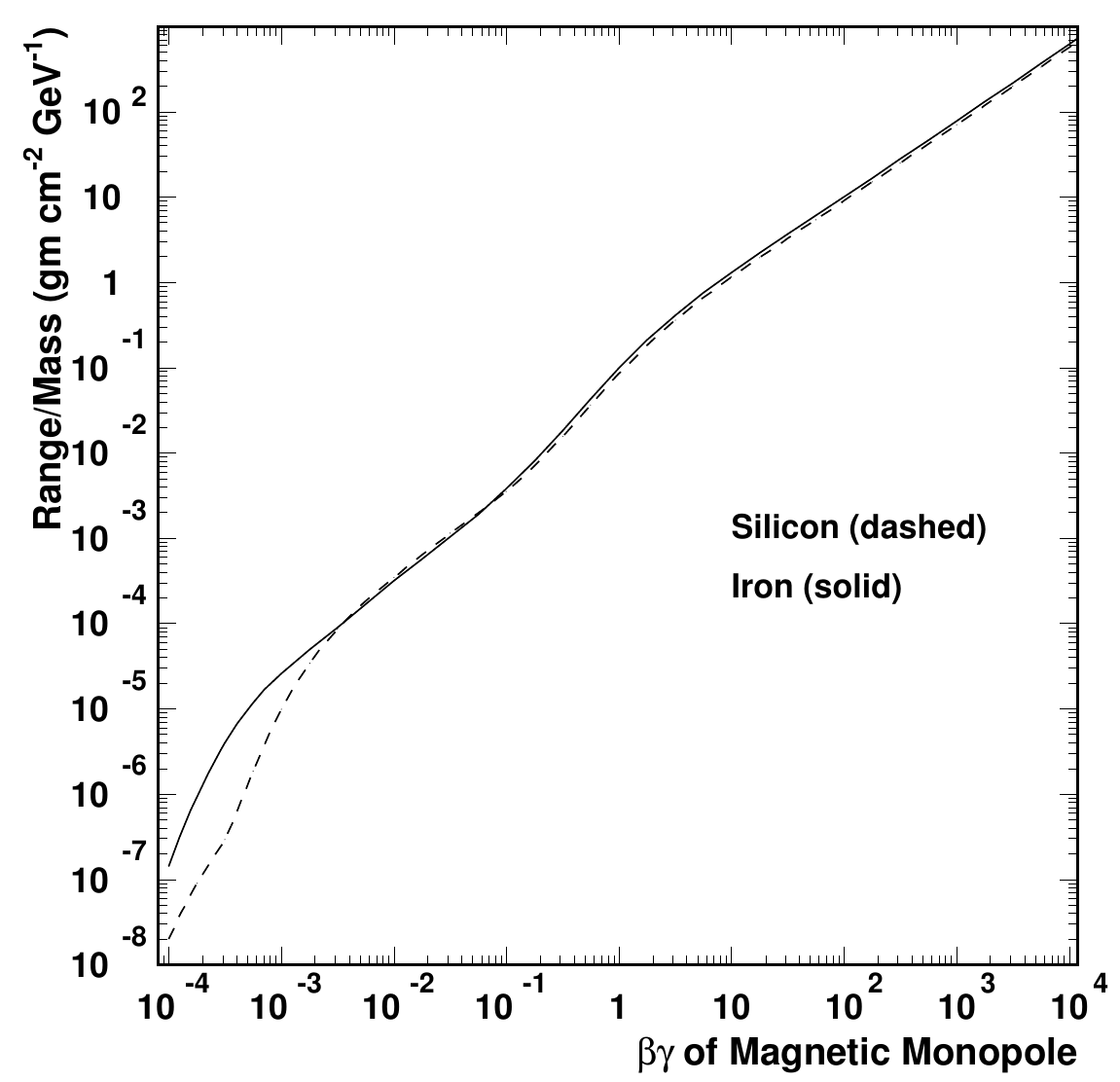}
\caption{\label{fig3}The ratio of range/mass for Dirac monopoles in iron and silicon as a function of monopole velocity factor $\beta\gamma$. The range is computed from the stopping powers given in Ref.~\cite{Derkaoui1998} and it is defined as the thickness of material to slow the monopole to a speed $\beta < 10 ^{-6}$. }
\end{figure}

The monopole range/mass ratio is independent of the mass. The computed values of this ratio, using the values of stopping power from Ref.~\cite{Derkaoui1998}, are shown in Fig.~\ref{fig3} as a function of the velocity factor $\beta \gamma=P/M$ for a monopole of mass $M$ and momentum $P$. For values of $\beta \gamma>10$ the variation is linear following the form
\begin{equation}
\frac{R}{M}=k(\beta \gamma)^{0.909},
\end{equation}\label{eq:rm}
with the constant $k$=0.136 (0.150) for silicon (iron). The ranges for
silicon for $\beta \gamma < 10^{-3}$ are rather uncertain due to the
sensitivity to the assumed minimum kinetic energy, $E_{min}$. However, they are relatively insensitive to this assumption for higher values of $\beta \gamma$.


For dyons (particles with both magnetic and electric charges) the
stopping power is increased relative to that for a magnetic charge
alone and depends on the electric charge
strength~\cite{Derkaoui1998}. The stopping power also depends
quadratically on the magnetic pole strength. For example, if the
fundamental unit of charge were $\frac{e}{3}$ (i.e., the $d$ quark) then the
magnetic pole strength could be three times stronger than the Dirac
monopole strength described above. In this case the stopping power
would be increased by an order of magnitude compared to that for a
Dirac monopole. There would then be a corresponding decrease in the
computed range.

\begin{figure}[ht]
\centering
\includegraphics[width=0.7\linewidth]{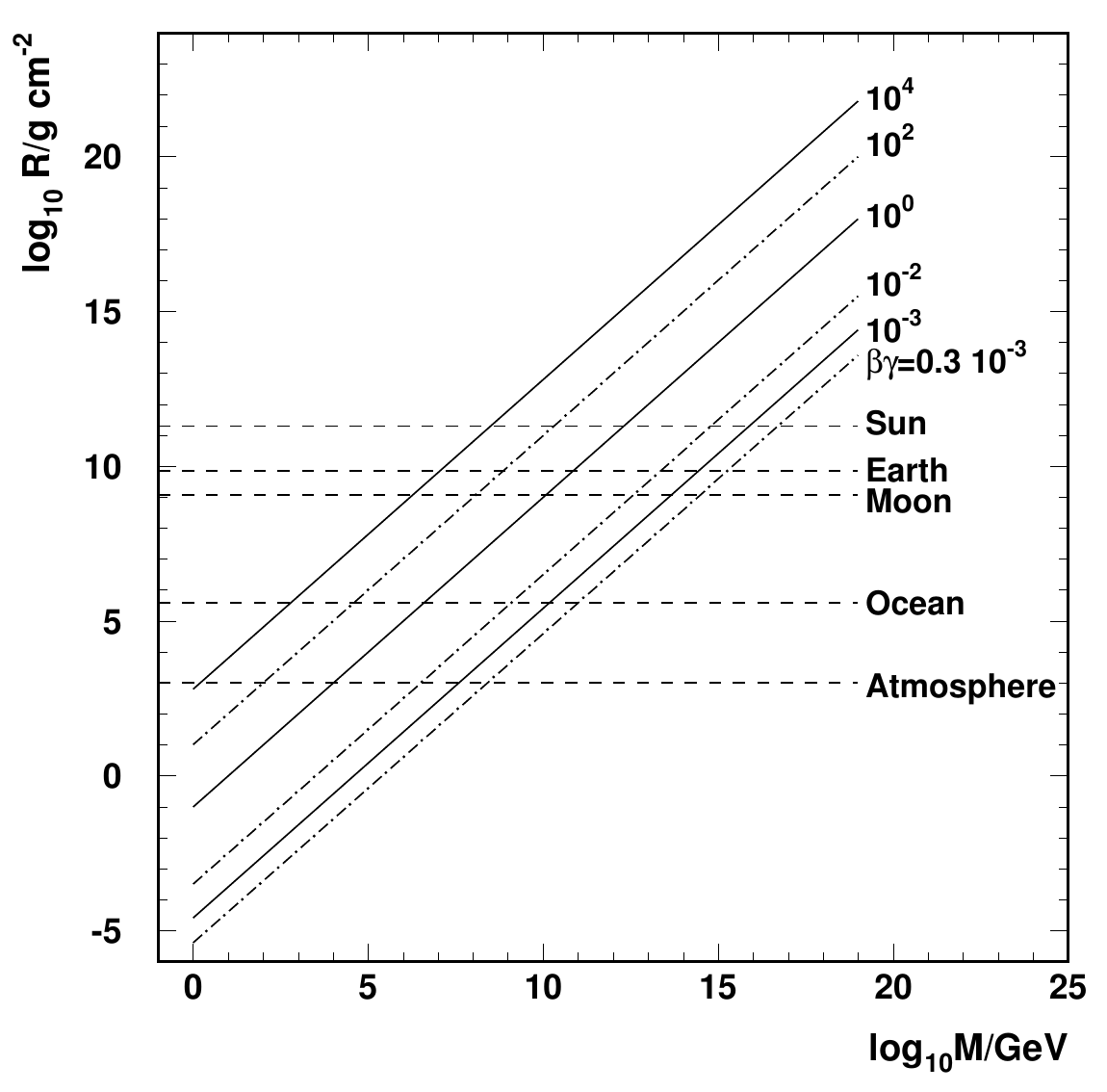}
\caption{\label{fig4} The ranges of Dirac magnetic monopoles as a
  function of the monopole mass for different values of $\beta
  \gamma$, for velocities below which stochastic processes make little
  contribution to the monopole stopping power. The horizontal lines show the
  widths of various possible stopping bodies for the magnetic monopole.}
\end{figure}

An alternative way to present the range is shown in
Fig.~\ref{fig4}. This shows the range of a Dirac monopole (charge
$137e/2$) as a function of mass for different values of $\beta \gamma$
up to the value where energy loss from stochastic processes becomes
significant. The horizontal lines show the sizes of various objects
that act as targets in which monopoles could potentially stop.

\subsubsection{Monopole binding}
\label{monopole_binding}

An understanding of monopole binding in matter is important in view of the many searches have been undertaken for trapped monopoles (see Section~\ref{monopoles_matter}).

Classical electrodynamics ensure that a monopole would be attracted to the surface of a ferromagnetic material by a mirror image force (in the same way an electrically charged object sticks to the surface of a conductor), resulting in a binding energy of at least 50 eV~\cite{Kittel1977}. Ferromagnetic materials will thus trap monopoles regardless of assumptions about the interactions within the material.

When coming to rest in matter, monopoles would be expected to bind to atoms or nuclei through interactions with dipole magnetic moments. The chemical potential of the atom-monopole molecule allows for bound states with $\sim 1.5$ eV binding energy~\cite{Ruijgrok1983,Ruijgrok1984}. The binding of monopoles to atomic nuclei has been discussed by several authors~\cite{Sivers1970,Bracci1983,Milton2006}. All show that the binding energy of a monopole to an spin-1/2 nucleus is strong, with values varying from several keV to MeV. Nuclear form factors affect the binding~\cite{Olaussen1985} but the binding energy and capture cross section remain large. Monopoles that slow down in matter are therefore likely to become attached to an atomic nucleus in the material, tearing it from its lattice. The slowing down will then continue until the nucleus-monopole system settles into thermal equilibrium with the lattice. Here it is likely to be at least as tightly bound as the original nucleus. Once a monopole becomes bound to a nucleus inside an atomic lattice, it is likely to be very difficult to dislodge it~\cite{Milton2006}.

In the presence of monopoles in the early stages of the Universe, bound states of monopoles and nuclei would likely be formed~\cite{Bracci1983,Bracci1984}. The possibility that cosmic monopoles could be attached to a proton or a nucleus when impacting the Earth needs therefore to be considered in searches.

%% file: smp-trappedelectric.tex
\section{Searches for heavy isotopes of matter}
\label{heavy_isotopes}

Several basic hypotheses motivate searches for SMPs trapped in
matter. Should SMPs be produced in
cosmic-ray interactions or exist as cosmic relics, then an observable concentration could
accumulate at the surface of the Earth. In this respect the Earth can be
regarded as an SMP-absorber over its $\sim3\cdot10^9$ year
lifetime. Depending on the properties of the SMP, it can form
heavy isotopes of the known elements. An analogous example of a
production process for an SMP is given by that of cosmic-ray-induced
tritium (T). This radioactive heavy isotope is observed as an approximately
uniformly distributed concentration (of size $\sim10^{-17}$ with
respect to H$_2$O). SMPs can also arise as relics from the Big Bang
that were present at the time of formation of the Earth.

A basic and frequently sought SMP is an object with the elementary
charge $+e$, denoted
$X^+$~\cite{Alvager1967,Muller1977,Boyd1978b,Smith1979,Smith1982,Hemmick1990,Verkerk1992,Yamagata1993} (see Section~\ref{sec:smpleptonhadron}). Depending
on its interaction properties, this object would act as a heavy
positively charged nucleon and can thus form heavy isotopes. The
simplest case would be for $X^+$ to form a {\it heavy hydrogen}
nucleus, which would capture an electron, leading to an atom with
similar chemical properties as observed for hydrogen. Anomalous water
molecules H$X$O would then be formed.  Objects with neutral, higher or
negative charges, i.e., $X^0, X^{++}, X^{-}$, would behave similarly
and form putative heavy isotopes, e.g., He ($X^{0}ppnn$,
$X^{++}$,$X^{-}pppnn$) and Li ($X^{0}pppnnn$,$X^{-}ppppnnnn$). In the case of a negatively charged SMP,
$X^-$, electromagnetic binding would be expected to occur with a
nucleus. The Bohr orbit would typically lead to the SMP effectively
existing inside the nucleus. If such an SMP were to bind with a
nucleus of atomic number $Z$, it would lead to an atom with similar
chemical properties as expected for the element with atomic number
$Z-1$. A neutral SMP, $X^0$, could bind were it to interact strongly
in the manner of a neutron.

Fractionally charged objects (see Section~\ref{fractional_charges}) with charges $\pm \frac{2e}{3}$ and $\pm \frac{e}{3}$ (commonly referred to as free quarks in the literature) are also hypothesised. Should they bind to atoms, they would lead to the unique property of atoms that are not electrically neutral. This feature is often used in searches~\cite{Perl2001}. Searches for fractional charge in matter are not described here, but an up-to-date review can be found in Ref.~\cite{Perl2004}.

In this section the different experimental techniques are catalogued~\cite{Lu:2004pk} and a description given of selected searches. A comprehensive summary of the results of the body of work performed in this area is given in Table~\ref{tab:summary_trapped_electric} in Section~\ref{smp-trapped-summ}.

\subsection{Experimental techniques}

The range of techniques used to search for SMPs trapped in matter
include mass spectrometry, Rutherford back-scattering, atomic spectroscopy, gamma-ray
emission and radiochemical analysis. Each of these techniques is
described below. It should be noted that the various techniques are
complementary. When taken together, the different
technologies typically provide a sensitive mass range
spanning the GeV scale up to $\sim10^{8}$~GeV.

\subsection{Choice of materials}

A wide range of materials have been used. These are selected
according to their availability, the potential of SMPs to
stop in them, and the underlying hypotheses that anticipate stopped
SMPs with potentially visible or enhanced concentrations. The following material types have been explored:

\begin{itemize}

\item {\bf Water.} Owing to its near ubiquity, samples of water have been frequently used~\cite{Smith1979,Smith1982,Verkerk1992,Yamagata1993}. Furthermore, the most simple SMP, $X^+$, would be expected to readily form anomalous water. Another advantage to using water is that enrichment factors of up to $\sim10^{12}$ have been achieved for heavy particles via electrolysis and/or centrifugation. However, uncertainty remains owing to the circulation of heavy SMPs in water. Some searches have sought SMPs at large ocean depths~\cite{Yamagata1993}.

\item {\bf Rare gases.} Helium is naturally depleted from the Earth's atmosphere due to its escape into the exosphere, which would ensure enhanced concentrations of anomalously massive He isotopes~\cite{Boyd1991}. Likewise, the concentration of stable radon isotopes in air would be enhanced due to the short lifetime of natural Rn (a few days)~\cite{Holt1976}.

\item {\bf Non-gaseous elements}. A number of searches have taken place for low SMP concentrations among elements such as Be, O, C, N, Na, Fe and Pb~\cite{Middleton1979,Dick1984,Dick1986,Nitz1986,Norman1987,Norman1989,Hemmick1990}.

\item {\bf Lunar rocks}. Unlike the aforementioned material types, searches with moon rock are limited due to sample availability. Such searches~\cite{Stevens1976,Han2009} are motivated by the absence of large-scale chemical and geological changes over eons in the upper layers of lunar surface. Searches in lunar soil thus permit the conversion  concentration limits into cosmic flux limits, while the complexity of SMP propagation in the Earth's atmosphere and oceans leads to more unambiguous interpretations~\cite{Banerjee2000}. Strangelet flux limits from searches in lunar samples are competitive with searches at balloon or space-borne spectrometers~\cite{Perillo1998,Han2009} (see Section~\ref{cosmic_in_space} and Fig.~\ref{fig:nuclearite_flux_limits}).

\item {\bf Meteorites}. Searches in meteorites~\cite{Jones1989} are attractive because meteoric material has generally not been subjected to chemical differentiation. Massive SMPs which were trapped in the matter from which the Solar System condensed would likely have sunk to planetary cores during planet formation and therefore would be expected to be mostly absent from the Earth's and Moon's crusts. As discussed in Section~\ref{stellar_monopoles}, the same argument also motivates dedicated searches for monopoles in meteorites.

\end{itemize}

Somewhat surprisingly, no searches have been made for electrically charged SMPs trapped in matter at high-energy colliders.

\subsection{Mass spectrometers}

\begin{figure}[htbp]
\centering
\includegraphics[width=0.725\textwidth]{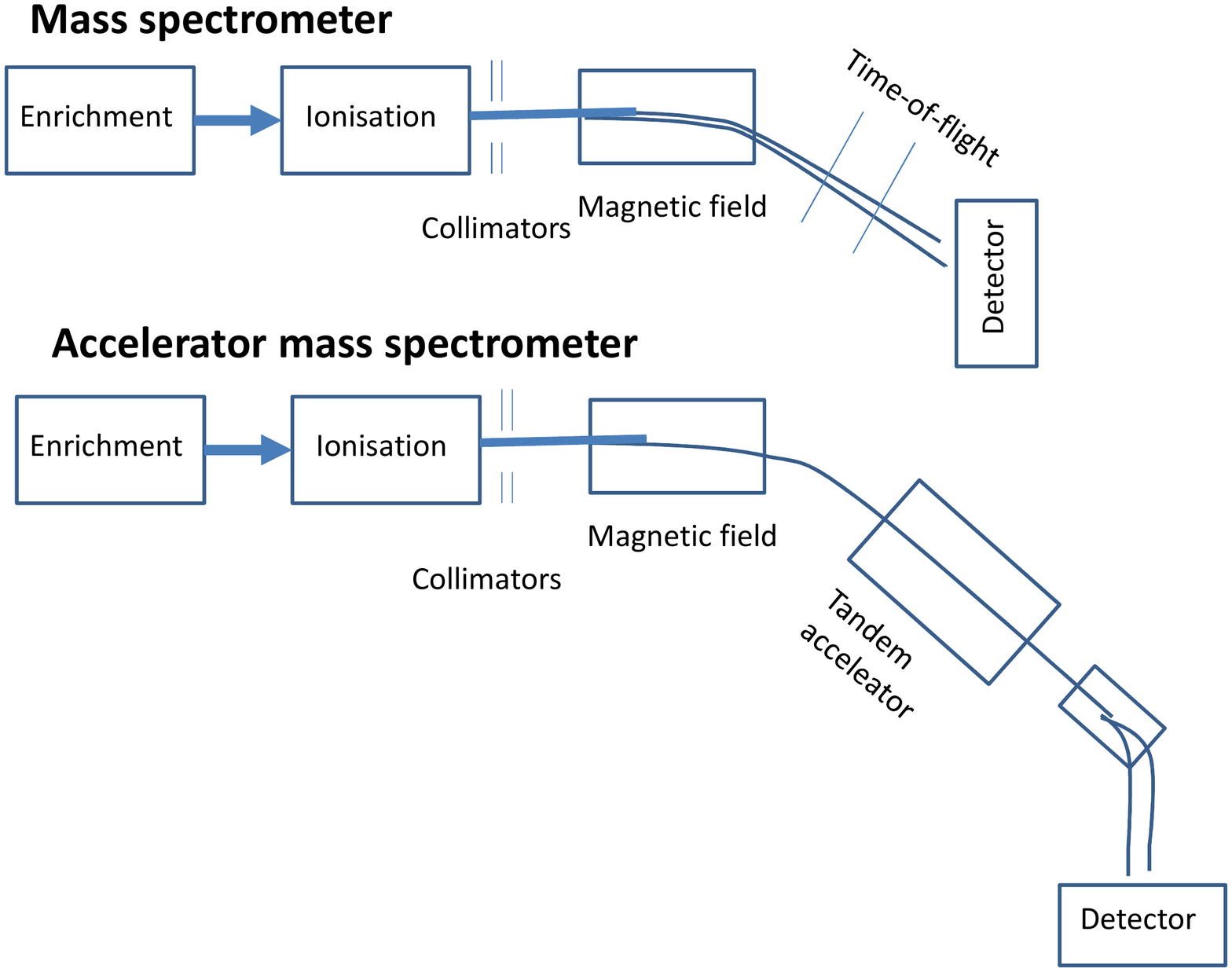}
\vspace{-0.35cm}
\caption{Schematic diagram of the various steps in the experiment.}
\label{fig:spectrometeroverview}
\end{figure}

Most searches have used the mass spectrometer (MS) method~\cite{Alvager1967,Muller1977,Middleton1979,Smith1979,Smith1982,Hemmick1990,Yamagata1993,Vandegriff1996,Han2009,Javorsek2001a}. The results typically do not depend on the modelling of the SMP scenario, but rather on the hypothesis of binding leading to an anomalously high isotope mass. A spectrometer-based experiment is schematically sketched in
Fig.~\ref{fig:spectrometeroverview}. Following enrichment for SMPs (e.g., via electrolysis), ionised samples are collimated, focused and ultimately separated by a magnetic field prior to passing to a detection system. Accelerator mass spectrometry (AMS) is an extension of the MS method in which the ions, following passage through an initial separating magnet, are subsequently accelerated in a tandem accelerator and ultimately separated by a further magnet before emerging onto a detection system. In both cases, time-of-flight systems can also be utilised as a means of further suppressing backgrounds. Lower and upper mass limits are typically determined by the experimental configuration. For example, the detection system can employ counters at various locations corresponding to ranges of SMPs with specific masses. SMPs with masses up to ~$10^4$~GeV have so far been sought.

The most stringent limits on the concentration of SMPs was published in Ref.~\cite{Smith1979,Smith1982}. A sample of enriched heavy water (D$_2$O) of size $2\cdot10^{-5}$~l was extracted from a starting water sample of size $10^8$~l using multi-step electrolysis. It was estimated that around 10\% of SMPs with charge $+e$ would survive the enrichment process. This step is estimated to correspond to an enrichment factor of $6\cdot 10^{11}$. The sample was then processed within a mass spectrometer. No heavy candidates were observed. This allowed an abundance limit of $\frac{X}{D}\sim10^{-16}-10^{-17}$ to be calculated for the sample. Taking into account the enrichment factor, this limit is reduced to $\frac{X}{D}\sim10^{-28}-10^{-29}$. This limit applies to a mass range of $\sim$10$-$1000~GeV.  A complementary means of establishing a limit is via a comparison of the densities of the enriched and unenriched samples. Since the densities are consistent within an accuracy of 10\% a limit of $\frac{X}{H}\sim 10^{-14}-10^{-17}$ can be inferred. This limit nominally extends up to a mass of $\sim10^{6}$~GeV, although the electrolysis theory used above may not be valid for higher masses, and gravitational effects may have depleted the original surface water sample from very massive SMPs.

\subsection{Atomic spectroscopy}

Searches made using atomic spectroscopy~\cite{Dick1984,Dick1986,Verkerk1992,Mueller2004,Lu2005} are  sensitive to any type of SMP as long as it can bind to form a heavy anomalous isotope. The excitation spectra of this isotope would be similar in structure to that of the lighter isotopes albeit at shifted frequencies.
The frequency shift $v_{MS}$ of a given transition between an isotope of mass $M$ and one of infinite mass is given by

\begin{equation}
\delta v_{MS} = - \frac{F_{MS}}{M}
\end{equation}\label{eq:massshift}

Here, the coefficient $F_{MS}$ depends on the given transition. Since the mass shifts depend on $\frac{1}{M}$ atomic spectroscopy searches can in principle extend to arbitrarily large masses. In practice, factors such as resolution and gravitational effects lower the mass sensitivity.

Searches for anomalous atomic transitions due to SMPs have been made with hydrogen extracted from sea water~\cite{Verkerk1992}, sodium~\cite{Dick1984,Dick1986}, and helium taken from air~\cite{Mueller2004}.  A brief description of Ref.~\cite{Mueller2004} is given below to highlight the underlying principles of this type of search.

In Ref.~\cite{Mueller2004} doubly charged SMPs which would form Helium-like particles were sought. Heavy helium would be expected to behave in a similar manner to the other noble gases, such as neon and argon, and remain in the atmosphere, while the light helium atoms would be depleted. Air samples, as used in this search, therefore provide a promising way to look for heavy helium. The line shape of the well studied transition $1s2s$ $^3S_{1} \rightarrow 1s2p$ $^3P_{2}$ was used in the search.
Following a purification procedure, helium from air was exposed to a RF-discharge to populate the $1s2s$ $^3S_{1,F=3/2}$ and $^3S_{1,F=1/2}$ states. Frequency-modulation saturation spectroscopy with a laser was then performed to look for transitions arising from the depopulation of this state.

\begin{figure}
\begin{center}
\includegraphics[width=0.725\textwidth]{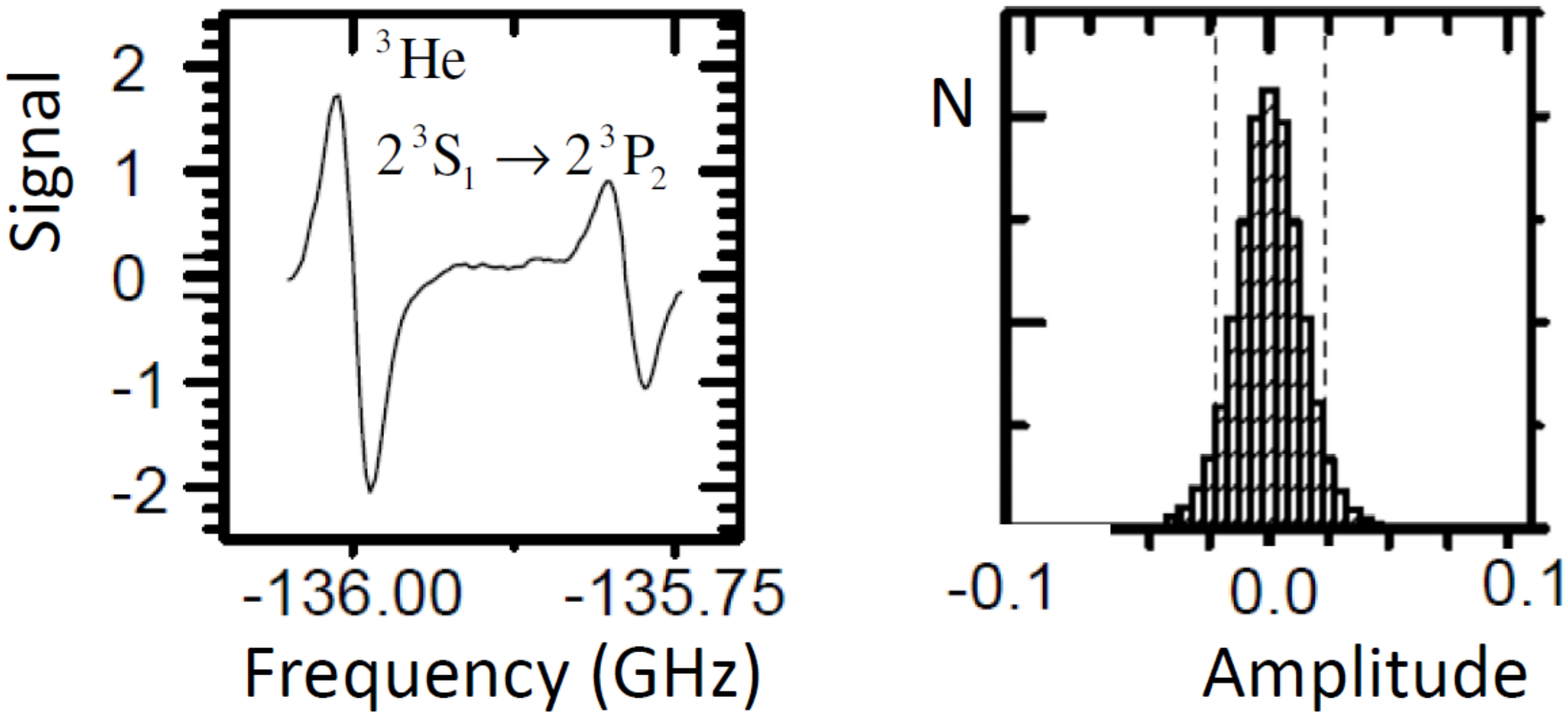}
\end{center}
\vspace{-3.5cm}
\caption{Left: the measured line spectra corresponding to the
$^3$He transitions
 2$^3$S$_{1,F=3/2} \rightarrow$ 2$^3$P$_{2,F=5/2}$ and
2$^3$S$_{1,F=1/2} \rightarrow$ 2$^3$P$_{2,F=3/2}$. The former (latter) transition corresponds to the stronger (weaker) signal. Right: Distribution of the amplitude, in fixed windows of frequency, of a signal possessing the lineshape given in the left plot. Data outside of the dashed lines correspond to 5\% of the total sample of amplitude measurements. The Figure is from Ref.~\cite{Mueller2004}}
\label{fig:laserheliumraw}
\end{figure}

Figure~\ref{fig:laserheliumraw} (left) shows the line shape for the $^3$He transitions
 2$^3$S$_{1,F=3/2} \rightarrow$ 2$^3$P$_{2,F=5/2}$ and
2$^3$S$_{1,F=1/2} \rightarrow$ 2$^3$P$_{2,F=3/2}$. The former (latter) transition corresponds to the stronger (weaker) signal.
The $^3$He deexcitation lineshape was used as a template to look for signals of heavy isotope transitions at anomalous frequencies. The best fit of the template to the data for a given frequency interval was used to give the frequency-dependent amplitude of a possible heavy isotope signal. The distribution of amplitudes is given in Figure~\ref{fig:laserheliumraw} (right). From these data it was concluded, to a 95\% confidence level, that no transition with an amplitude larger than $7.9\times10^{-2}$ times the $^3$He amplitude is present over the studied range in frequency. The sensitive mass region extends up to $10^4~$~GeV, beyond which Doppler shift effects may change the signal shape.
The data correspond to a mass-dependent limit on the concentration of SMPs of type $X^{++}$ per nucleus in the Solar System of $10^{-17}-10^{-13}$.

\subsection{Radiochemical properties}

Searches for negatively charged particles with masses less than $\sim 10^5$~GeV were performed under the assumption that they would bind to positively charged nuclei and doing so reduce the nuclear charge by one unit~\cite{Turkevich1984,Norman1987,Norman1989}. For instance, an $X^-$ particle could bind to nitrogen to form an isotope $_6(^{14}$N$X^-)$: assuming only electromagnetic interaction between the SMP and the nucleus, the bound state would have the nuclear properties of nitrogen but the chemical properties of carbon and thus would exist in a carbon sample~\cite{Turkevich1984}. The same principle was also employed using samples of manganese (for probing Fe$X^-$)~\cite{Norman1987} and thallium (for probing Pb$X^-$)~\cite{Norman1989}. In the carbon search~\cite{Turkevich1984}, a sample of reactor grade graphite was irradiated with neutrons such that $_5(^{14}$C$X^-)$ would be formed in an analogous way to the reaction of $^{14}$N with neutrons to form $^{14}$C. After irradiation, the sample was dissolved by chemical processes in which boron atoms were separated and impurities which could give rise to radioactive activity were removed. The decay properties of the nuclei within the sample was then studied. The particle $_5(^{14}$C$X^-)$ would be expected to have the chemistry of boron but also emit $\beta$ particles with energies around $\sim 0.8$~MeV; due to coulomb effects, the half-life would be shorter than ordinary $^{14}$C, around 15 years. No such decay was observed, and limits of the abundance of $X^-$ per nucleon of less than $\sim 2 \cdot 10^{-15}$ were obtained.

\subsection{Gamma rays following neutron absorption}

A search for anomalous very tightly bound states of nuclei was made in~\cite{Holt1976}. An atmospheric xenon sample was prepared in which it was assumed that the anomalous radon would also be present. This was exposed to a thermal neutron beam and high energy photons in the energy range $20-250$~MeV were sought. The hypothesis here is that radon-like heavy nuclei would absorb hadronic matter and release energy in the form of particles. A radon-like element was considered owing to its large expected abundance in the atmosphere and its enhanced concentration due to the absence of stable isotopes of radon. No evidence for anomalous gamma-ray emission was found and a limit of $10^{-29}$ per atom of silicon was set on the terrestrial concentration of anomalous radon-like isotopes of matter.

\subsection{Rutherford back scattering}

The Rutherford back scattering method is based on the same principle used to discover the atomic nucleus, namely, that back scattering will occur due to interactions of a light object in the vicinity of a stationary heavy object. At GSI in Germany, a $^{238}$U beam impinged on a number of samples, such as meteorites~\cite{Polikanov1991}. The beam energy (1.4~MeV per nucleon) was lower than the expected Coulomb barrier in order to ensure elastic collisions. Exotic particles with masses as high as $10^7$~GeV were sought; the upper mass limit arises in the case of strangelets due to the potential screening of a nuclear charge by electrons. In addition, a $^{244}$Pu target was used. Since $^{244}$Pu has a finite half-life (~$8\cdot10^7$ years), anomalous particles would be enriched (with a factor $10^{17}$) with respect to terrestrial Pu atoms.

\subsection{Heavy-ion activation}

One method which relies on the strangelet hypothesis is to irradiate the sample with a heavy-ion beam and monitor for photon emission in the GeV range which would be expected in a model of strange quark matter~\cite{Farhi1985}. One search applied this method to samples of nickel ore, meteorites, and lunar soil~\cite{Perillo1998}.

\subsection{Summary}\label{smp-trapped-summ}

\begin{table}
\hspace{-1.5cm}
\begin{tabular}{|c|c|c|c|c|c|} \hline
    Sample & Method & SMP  & Mass range (GeV) & Concentration limit & Ref.   \\
\hline D$_2$O & MS  & $X^+$   & $6-16$ &     $3\cdot 10^{-18}$ (H)      & \cite{Alvager1967} \\
\hline D$_2$O & MS  & $X^+$ & $6-350$ & $4\cdot 10^{-22}-2\cdot 10^{-21}$ (H)  & \cite{Smith1979} \\
\hline D$_2$O & MS/TOF  & $X^+$ & $12-1200$ &  $2\cdot 10^{-29}-2\cdot 10^{-28}$ (H)  & \cite{Smith1982} \\
\hline H$_2$O & spectroscopy  & $X^+$ & $10^4-10^7$ &  $6\cdot 10^{-15}$ (H)  & \cite{Verkerk1992} \\
\hline H$_2$O & MS/TOF  & $X^+$ & $5-1600$ &   $4\cdot 10^{-17}$ (H)  &  \cite{Yamagata1993} \\
\hline H,D    & MS      & $X^+$ & $0.3-8$ & $2\cdot 10^{-19}-10^{-13}$ (H) & \cite{Muller1977} \\
\hline D,Li,Be,  &  &  & & &   \\
B,C,O,F & MS & $X$ & $10^2-10^4$ &  $10^{-23}-10^{-9}$ (nucleon) &  \cite{Hemmick1990} \\
\hline He,Ar  & MS &  $X^{++}$, $X$ & $42-82$ & $2 \cdot 10^{-12}-2 \cdot 10^{-11}$ (nucleus)  & \cite{Vandegriff1996}\\
\hline He & spectroscopy & $X^{++}$, $X$ & $20-10^4$ & $10^{-17}-10^{-12}$ (nucleus) & \cite{Mueller2004} \\
\hline Na  &  spectroscopy & $X$ & $10^2-10^5$  & $7\cdot 10^{-12}$ (nucleon)  &   \cite{Dick1984,Dick1986} \\
\hline lunar soil & MS & $X$   &  $42-70$  & $6\cdot 10^{-17}-5\cdot 10^{-12}$ (nucleus) &   \cite{Han2009} \\
\hline O & MS & $X$   & $20-54$ & $10^{-16}$ (nucleon) &  \cite{Middleton1979} \\
\hline chondrites, &  &  & & &   \\
       nickel ore,                       & activation & strangelet & $2\cdot 10^{3}-10^8$ & $4 \cdot 10^{-17}-10^{-13}$ (nucleon) & \cite{Perillo1998}  \\
       lunar soil &  &  & & &   \\
\hline Au & MS & $X$  &  $<1.7 \cdot 10^3$  &  $6 \cdot 10^{-12}-10^{-8}$ (nucleus) & \cite{Javorsek2001a} \\
\hline meteorite & MS & $X$   & $<647$  &  $6 \cdot 10^{-9}-10^{-8}$ (nucleus) & \cite{Javorsek2001a} \\
\hline Rn   & gamma & X &  $\sim 150$    &   $10^{-29}$ (nucleus)       & \cite{Holt1976} \\
\hline C    & radiochemistry & $X^-$   & $<10^5$ &  $2\cdot 10^{-15}$ (nucleon) & \cite{Turkevich1984}    \\
\hline Fe   & radiochemistry & $X^-$   &  $<10^4$  & $2 \cdot 10^{-12}$ (nucleon) & ~\cite{Norman1987}\\
\hline Pb   & radiochemistry &  $X^-$   &  $<10^4$  & $3 \cdot 10^{-13}$ (nucleon) & \cite{Norman1989} \\
\hline Pu,  &  &  & & &   \\
    meteorites, & back scattering & $X$ & $4\cdot 10^2-10^7$ & $10^{-17}-10^{-11}$ (nucleon) & \cite{Polikanov1991} \\
       sediments & & & & &  \\
\hline
\end{tabular}
  \caption{\label{tab:summary_trapped_electric} Summary of published concentration limits (95\% confidence level) obtained in experiments probing for heavy isotopes of matter.}
\end{table}

A number of inventive searches utilising a range of techniques has been performed for exotic isotopes in matter. A summary of concentration limits is presented in Table~\ref{tab:summary_trapped_electric}. Many experiments have been specifically interpreted for strangelets~\cite{Blackman1989,Klingenberg2001,Lu2005,Finch2006} although the majority can be regarded as searches for generic SMPs with specific masses and electric charges which bind to form superheavy nuclei.

 It is noteworthy that the mass sensitivity of this body work extends up to $\sim 10^8$~GeV, well beyond that achievable at any present or planned collider experiment. The mass sensitivity in the type of search presented in this section is typically determined by the details of experimental set-up. One area of possible future improvement would therefore be the design of experiments sensitive to SMP scenarios (see Section~\ref{theo_summary}) with a higher mass sensitivity than previously achieved.

%% file: smp-trappedmagnetic.tex

\section{Searches for magnetic monopoles in matter}
\label{monopoles_matter}

\begin{figure}[tb]
\begin{center}
\includegraphics[width=0.75\linewidth]{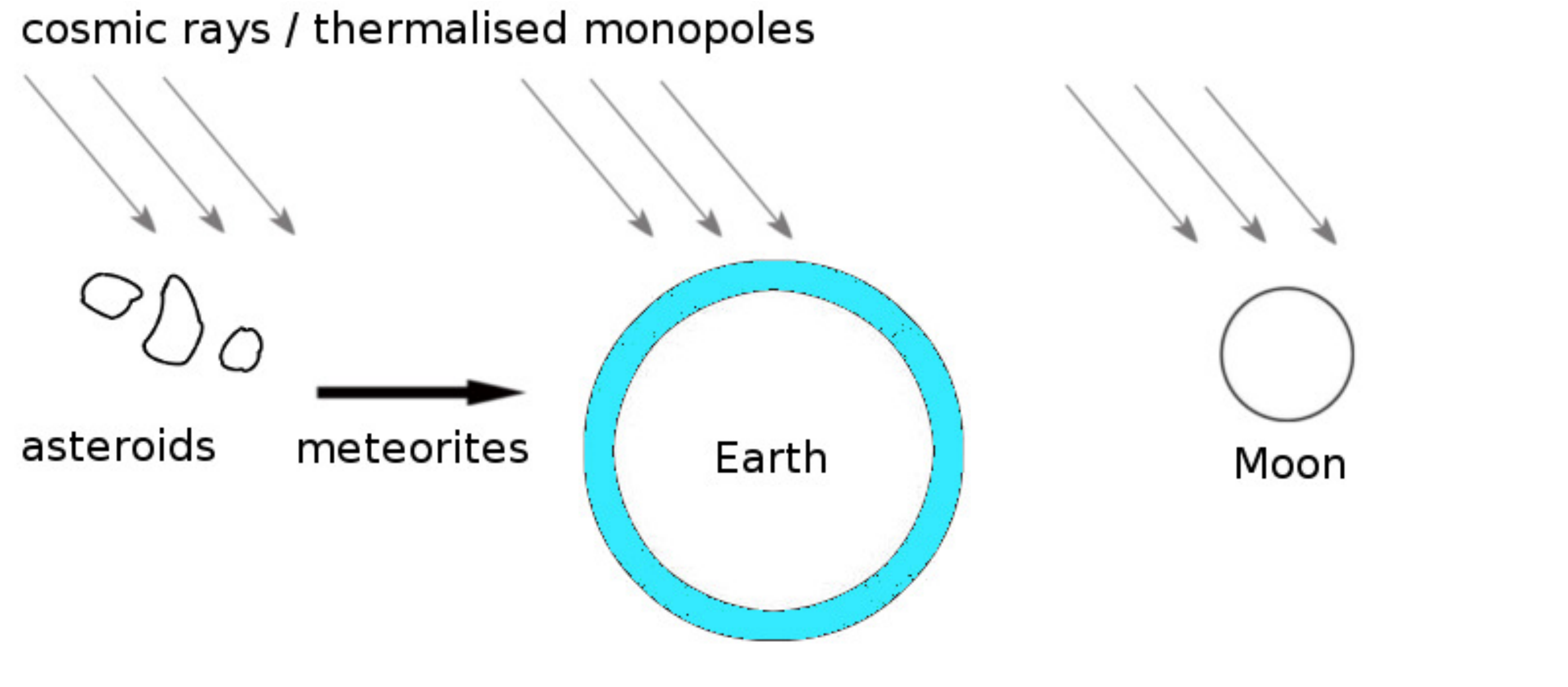}
\caption{Early (prior to 1983) searches for monopoles in matter were based on the hypothesis that monopoles may be created as secondary products in high-energy cosmic ray collisions, or may themselves be a component of the cosmic rays. They were assumed to possess a mass around the GeV-TeV scale, in which case they would quickly thermalise and remain trapped in asteroids, in the Moon's surface, and in the Earth's atmosphere, oceans and surface rocks.}
\label{fig:cosmic_ray_impact}
\end{center}
\end{figure}

As was seen in Section~\ref{theo_monopoles}, magnetic monopoles are SMPs well worth looking for, and they would be expected to bind to matter. Short reviews of monopole searches in matter were given in Refs.~\cite{Carrigan1983,PDG2012}. This section provides an updated, more complete picture and puts the different investigations into their context. Searches for magnetically charged particles in bulk matter broadly fall into four categories depending on the assumptions made:

\begin{itemize}

\item Generic searches for SMPs bound in matter such as the heavy isotope searches described in Section~\ref{heavy_isotopes} directly apply to monopoles and dyons if corresponding assumptions are made on the binding to nuclei or nucleons and enrichment factors. In the interpretation of such results, monopoles are equivalent to $X^0$ particles and dyons are equivalent to electrically charged $X$ particles. The search techniques listed below, which are specific for monopoles, allow to probe much larger quantities of material. However, contrarily to generic searches, monopole-specific studies have generally not exploited enrichment techniques such as electrolysis, centrifuging, or the natural high abundance of exotic isotopes in atmospheric rare gases --- even if, in principle, they could.

\item Searches based on the premise of relatively light and quickly thermalised monopoles, as illustrated in Fig.~\ref{fig:cosmic_ray_impact}. Most of these searches used the \textit{extraction technique}, where the property of a magnetic charge (typically the Dirac charge) was exploited to attempt to extract monopoles from a sample and then accelerate them with a magnetic field through a detection device (see Section~\ref{extraction_technique}). In the 1970s, researchers started to use the more powerful \textit{induction technique}, in which an induced persistent current after passage of a sample through a superconducting coil is sought (see Section~\ref{induction_technique}).

\item Searches based on the premise of supermassive monopole relics
  produced at the early stages of the Big Bang, trapped in matter either before or after the formation of the Solar System. They are more recent and use the induction technique.

\item Indirect searches based on the changes in some of the properties of celestial bodies that could be expected if they contained large numbers of monopoles.

\end{itemize}

A number of samples that were exposed to high-energy collisions at accelerators have also been analysed to search for monopoles with masses up to several hundred GeV~\cite{Carrigan1973,Kalbfleisch2000,Kalbfleisch2004,H12005,DeRoeck2012b,Bendtz2014}. A review of collider searches prior to the LHC is given in Ref.~\cite{Fairbairn2007}.

This section starts with a short description of the experimental strategies and techniques. Then it reviews searches for monopoles in the atmosphere, surface outcrops, deep-oceans deposits, ocean water, moon rocks, meteorites, crust-derived rocks and ores, and mantle-derived rocks. Finally, constraints obtained from indirect effects of monopoles inside the Earth and the Moon are presented.

\subsection{Experimental strategies}

Choices of samples and experimental arrangements are motivated by the
target that the experimenter has in mind, which in turn is influenced by theoretical models.

\subsubsection{Low-mass monopoles}

Between the early 1960s and the late 1970s, accelerator facilities did
not reach sufficient energies to exclude monopoles with masses beyond
the GeV scale. Masses above $10^4$~GeV were seldom given consideration in the design of the experimental
setup and in the interpretation the results during this time. The
choice of sample was likewise in a large part motivated by the
assumption of a relatively low mass. One leading scenario considered
to interpret the results was secondary production of monopole pairs in
the collision between high-energy cosmic rays and astronomical
bodies. Due to the dramatic decrease in cosmic-ray flux with energy,
indirect cosmic-ray production is most relevant if the monopole has a
relatively low mass (in the GeV range). A discussion of how search
results interpreted in terms of limits on the secondary monopole
production cross section by incident high-energy cosmic rays relate to
the tight cross section limits obtained at recent hadron collider
experiments for monopoles in the relevant mass range (up to the TeV
scale~\cite{Fairbairn2007,ATLAS2011c}) is given in
Section~\ref{cosmic_prod}. In a second scenario, TeV-mass cosmic
monopoles impacting the Earth, the Moon or asteroids would thermalise
in the Earth's atmosphere, in the Moon's surface and inside the
asteroid, respectively. This model is illustrated in
Fig.~\ref{fig:cosmic_ray_impact} and explains why early searches for monopoles in bulk matter focused on samples consisting of air, meteorites, sediments, and rocks from the Moon's surface.

\subsubsection{High-mass monopoles}
\label{stellar_monopoles}

From the 1980s, guided by the indication of a high monopole mass in
Grand Unification theories (see Section~\ref{GUT_mono}) as well as the non-observation of free monopoles with masses up to several hundred GeV at collider experiments, more consideration was given to massive cosmic monopoles, which would not easily be stopped in matter (see Section~\ref{range_monopoles_matter}). Due to an expected penetration depth exceeding that of the planetary crust, searches relying on samples from the Earth and Moon surfaces would have only  faint chances of finding cosmic monopoles with masses beyond $\sim 10$~TeV.

Massive monopoles could still have bound to nuclei before the Solar
System was formed, e.g., during Big Bang nucleosynthesis or via
trapping inside galactic dust and stars. These so-called
\textit{stellar} monopoles are assumed to have already been trapped in
matter before planetary differentiation occurred. Stellar monopoles
would be recycled in supernovae along with the heavy nuclei to form
new stars and planets, remaining trapped near the cores of the
astronomical bodies if the mass is of the order of the Grand Unification scale. 
They would be expected to sink down during planet formation. Thus, they would be mostly absent from planetary crusts.

Meteorites provide a way to probe stellar monopoles: iron meteorites
might originate from the cores of planetary bodies, and chondrites
(which constitute about 86\% of the meteorites on Earth) are believed
to be made of undifferentiated material from the primary solar
nebula. Another way is to search in polar volcanic rocks. Stellar
monopoles inside the Earth would tend to accumulate at a point along
the magnetic axis where the downwards gravitational force is equal to
the upwards force exerted by the Earth's magnetic field. An
equilibrium above the core-mantle boundary for a monopole with the
Dirac charge corresponds to the condition that the mass should be less
than $4\cdot 10^{14}$~GeV. Assuming a binding of monopoles to nuclei,
from this point, the solid mantle convection would be expected to
slowly bring up monopoles to the surface. Over geologic time, it can
be surmised that monopoles of a wide range of masses and charges would tend to accumulate in the mantle beneath the geomagnetic poles~\cite{Bendtz2013a}. Stellar monopole searches are discussed in Section~\ref{meteorites} (meteorites) and in Section~\ref{polar_rocks} (polar volcanic rocks). 

\subsection{Experimental techniques}

\subsubsection{Extraction technique}
\label{extraction_technique}

\begin{figure}[tb]
\begin{center}
\includegraphics[width=0.33\linewidth]{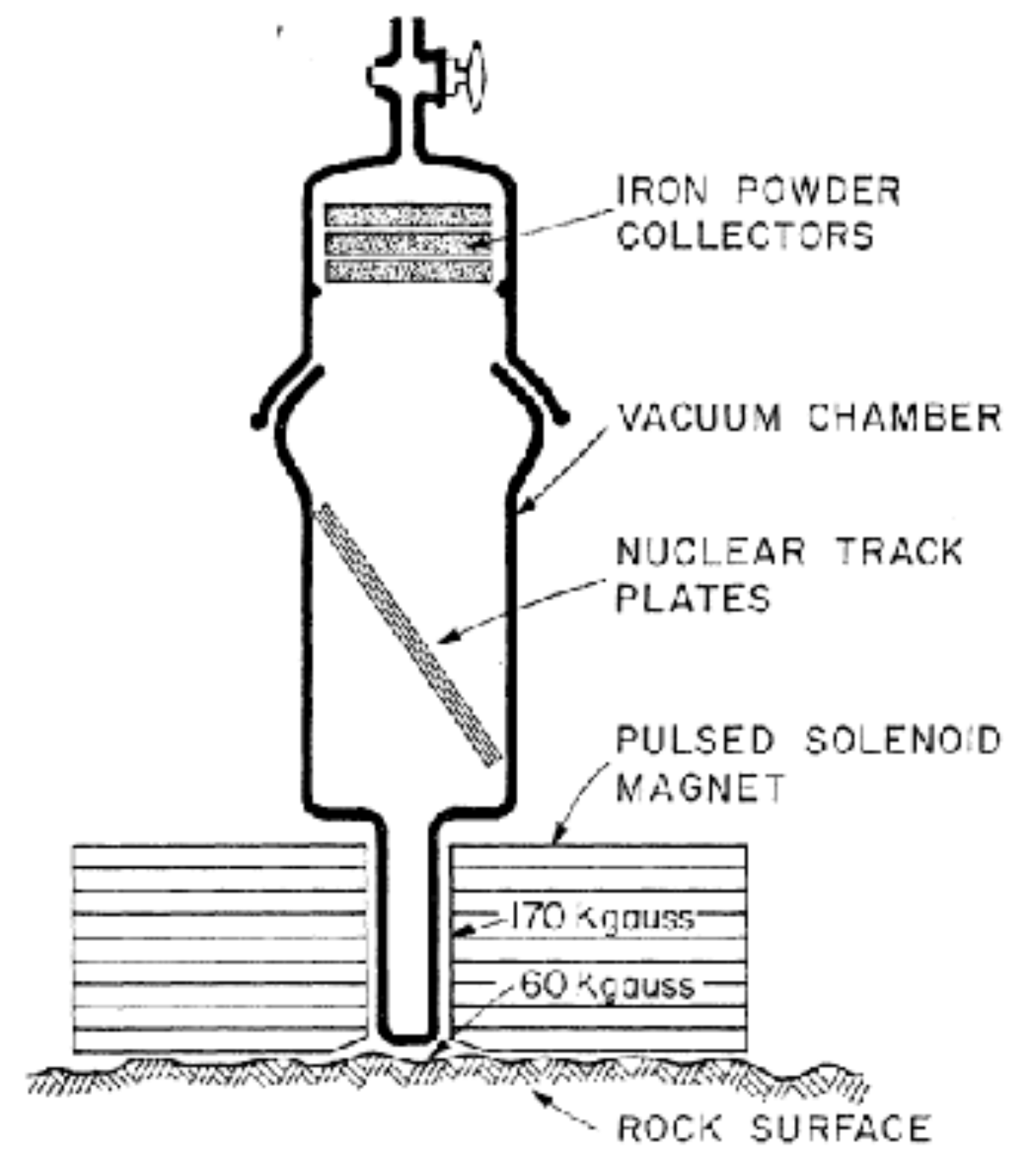}
\includegraphics[width=0.66\linewidth]{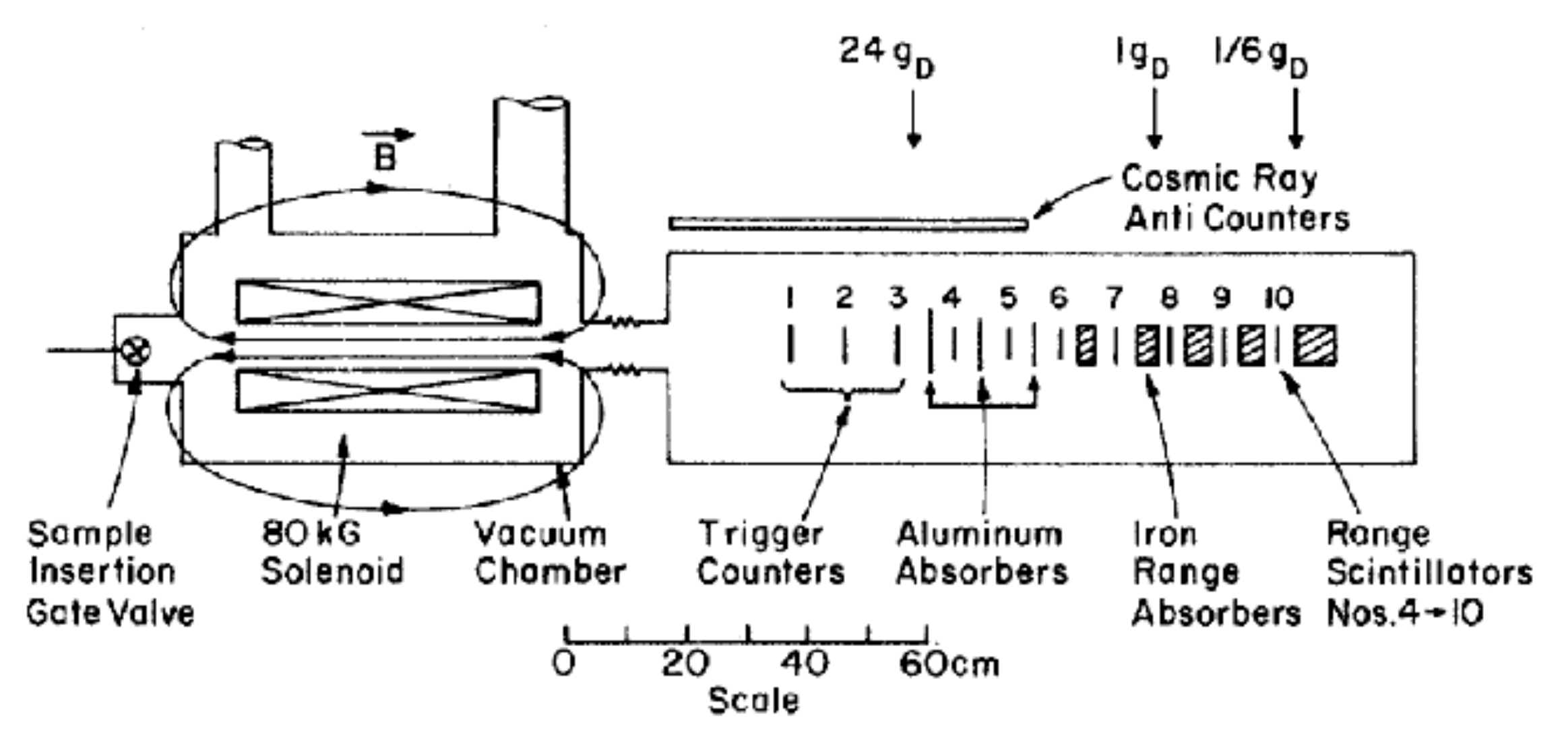}
\caption{Examples of experimental apparatuses used for monopole searches with the extraction technique. It is assumed that monopoles of a given polarity present in the sample surface would be dragged out and accelerated by the strong magnetic field and identified by detectors sensitive to high ionisation energy losses. The device on the left was designed to be portable to be carried to a site where it can scan rock outcrops (Figure from Ref.~\cite{Goto1963}). The device on the right was designed for samples from the Fermilab accelerator and was also used in a search for monopoles trapped in air and seawater (Figure from ref.~\cite{Carrigan1976}).}
\label{fig:extraction}
\end{center}
\end{figure}

Monopoles trapped in matter can be extracted by applying a magnetic
field to a sample that is heated and
evaporated~\cite{Petukhov1963}. Even at room temperature, the
application of a strong (5~T or higher) magnetic field could be
enough to pull monopoles out of a solid sample surface (magnetic extraction technique)~\cite{Goto1963,Fleischer1969a,Fleischer1969b,Kolm1971,Carrigan1976}. Monopoles present at thermal velocities in the Earth's atmosphere may also be collected using extended magnetic fields~\cite{Malkus1951,Carithers1966,Bartlett1981}. In all these experiments, a magnetic field would accelerate the extracted monopole to an appropriate velocity, and an array typically made of scintillators or nuclear-track detectors would identify the high ionisation energy loss expected from a monopole (examples of experimental apparatuses are shown in Fig.~\ref{fig:extraction}). Such a procedure thus involves three steps (extraction, acceleration, detection) and efficiency uncertainties are associated with each of these steps. These efficiencies generally depend on the assumed monopole mass and charge and often drop to zero above a certain mass value.

Extraction searches often considered ferromagnetic materials and assumed a monopole binding to the material surface through the mirror image force~\cite{Goto1958} (see Section~\ref{monopole_binding} for a discussion of monopole binding to matter). If strong binding to nuclei occurs, the available magnetic fields could be insufficient for extraction~\cite{Sivers1970}. In Ref.~\cite{Goto1963} it is pointed out that a binding to nuclei would not affect the results of searches with the extraction technique because, for the sample thicknesses considered in this search, the magnetic field would be strong enough to extract the whole atom or even clusters of atoms from the material surface. However, this claim has been questioned on the basis that monopoles bound to nuclei would remain stuck if there is no available nearby site for the atom to jump to in a solid lattice~\cite{Milton2006}.

\subsubsection{Induction technique}
\label{induction_technique}

In the 1970s, with the development of superconducting magnetometers, the extraction technique was gradually abandoned in favour of the induction technique, which provides substantial advantages: there are fewer steps involved (no extraction nor acceleration are needed), the measurement is directly sensitive to the magnetic charge and does not depend on the monopole mass, and larger solid sample volumes can be analysed faster.

The relevant quantity measured by the magnetometer is the persistent
current, defined as the difference between the output after and
before passage of a sample through the sensing region. For any sample
containing only magnetic dipoles, the induced currents cancel
out to zero after passage. The persistent current is directly
proportional to the magnetic charge contained in a sample, and a
consistent non-zero persistent current for multiple passages is an
unmistakable signature of a monopole. The calibration constant
translating persistent current into magnetic charge (in units of
$g_D$) can be determined using either a convolution method with a
sample of well-known dipole magnetic moment or a long thin solenoid
that mimics a magnetic pole~\cite{DeRoeck2012b}. The minimum magnetic
charge that can be resolved depends on the intrinsic magnetisation of
the samples, as magnetised samples are more likely to provoke flux
jumps~\cite{DeRoeck2012b,Bendtz2013a}. A resolution well below the
Dirac charge can be achieved even with highly magnetised samples after demagnetisation or crushing~\cite{DeRoeck2012b,Bendtz2013a} or by placing samples in a mumetal container~\cite{Jeon1995}.

\subsection{Atmosphere}

\begin{figure}[tb]
\begin{center}
\includegraphics[width=0.4\linewidth]{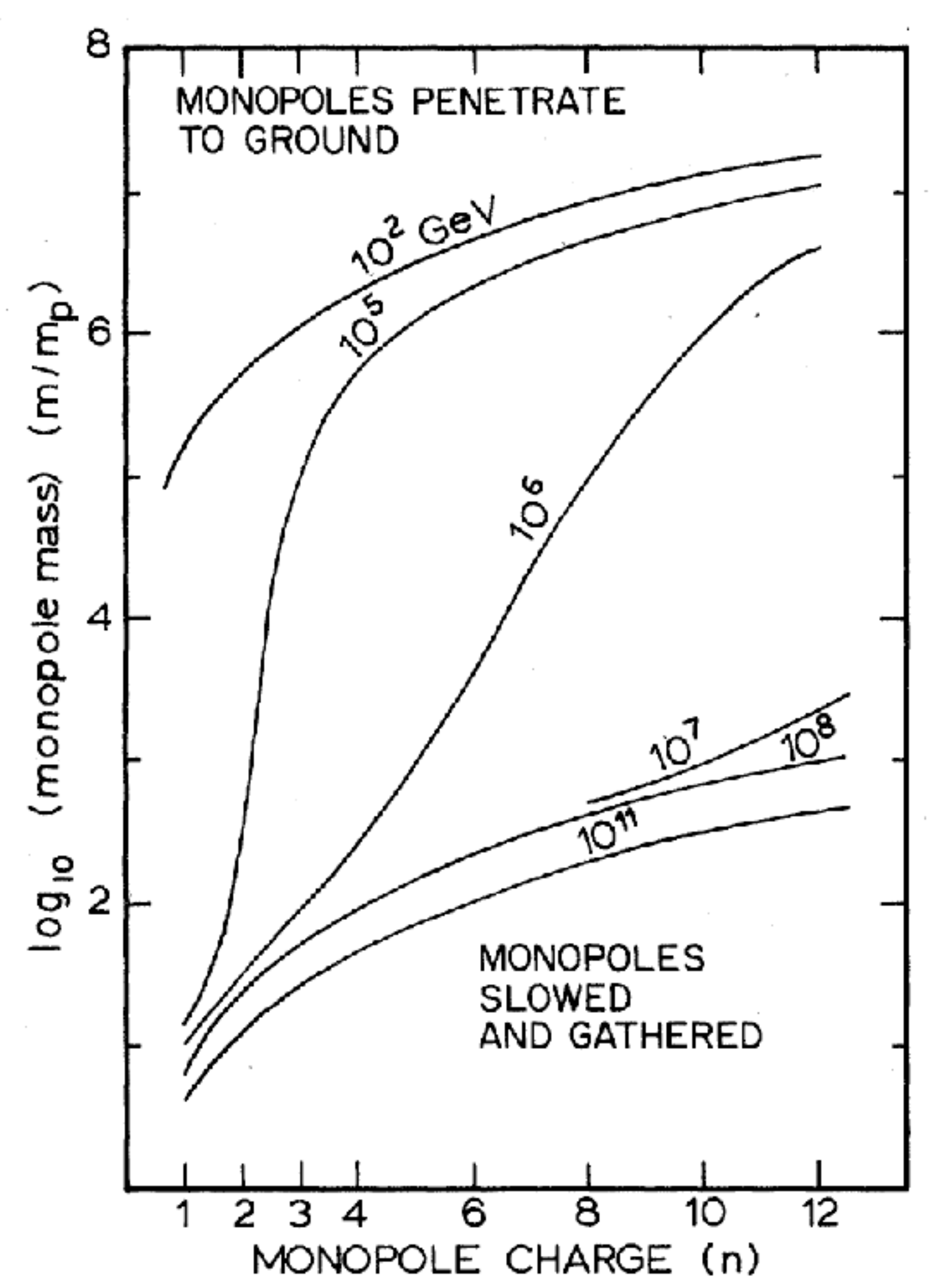}
\caption{Mass and charge regions for cosmic monopoles to be stopped in the Earth's atmosphere. Incident monopoles with mass and charge such that they lie below the curves would be stopped and gathered, where the different curves correspond to different kinetic energies. The calculations are made using three energy-loss mechanisms: ionisation, bremsstrahlung, and pair production (see Section~\ref{interactions_magnetic}). If monopoles are also allowed to stop in the oceans, the curves are shifted towards higher masses by about one order of magnitude in mass~\cite{Fleischer1969a}. The figure is from Ref.~\cite{Bartlett1981}.}
\label{fig:stopping_atmosphere}
\end{center}
\end{figure}

Monopoles supposedly present in the atmosphere at thermal velocities would drift under the force exerted by the geomagnetic field. If they carry a magnetic charge of the right polarity, they may be gathered through a detector using the stronger field of a large magnet. This approach was initiated by Malkus in 1951~\cite{Malkus1951} and developed in subsequent decades with more powerful magnetic fields~\cite{Carithers1966,Bartlett1981}. The latest search~\cite{Bartlett1981} covered an effective collecting area of 24~000~m$^2$ and set a 90\% confidence-level limit of 
$4.4\cdot 10^{-16}$~cm$^{-2}$sr$^{-1}$s$^{-1}$ in the flux of incoming positive-sign monopoles with mass, charge and energy such as they would be stopped in the atmosphere. Estimated ranges for which this condition is satisfied are indicated in Fig.~\ref{fig:stopping_atmosphere}.

Assuming that monopoles thermalise in the atmosphere, they could be found on the surface of magnetite outcrops on the Earth's surface, which act as attraction sites through the mirror image force~\cite{Goto1963}. The portable magnetic extraction device shown in Fig.~\ref{fig:extraction} (left) was taken to a site in the Adirondack Mountains (USA) to scan outcrop ferromagnetic rocks likely to have been exposed since glacial recession. An area of 1000~cm$^2$ was searched for positively charged monopoles. Estimating that monopoles would remain near the rock surface for 100 to 1000 years before being eroded away, the negative results imply a cosmic monopole flux limit of about 
$1.6\cdot 10^{-14}$~cm$^{-2}$sr$^{-1}$s$^{-1}$ for magnetic charge above 1~$g_D$ and mass below 10~GeV (assuming the highest energies).

\subsection{Oceans}

Monopoles that thermalise either in the atmosphere or in ocean water
would likely accumulate in deep-ocean
deposits~\cite{Fleischer1969a,Fleischer1969b,Kolm1971}. The main
advantage with such samples is the large collection time: some samples
remain exposed and grow slowly on the ocean floor under periods of the
order of one million years. Some samples have ferromagnetic properties
and can thus act as attractors of monopoles. Monopoles may thermalise
down to the considered depth, which allows to probe higher masses and
energies than experiments conducted above sea level (for the same
energy, a factor 10 on the mass is gained as compared to the curves
shown in Fig.~\ref{fig:stopping_atmosphere}; see also
Section~\ref{range_monopoles_matter} and Fig.~\ref{fig4}). Four manganese nodules of about
2-cm radius each, collected from the floor of the Drake Passage
(Southern Ocean) at 3 km depth, were analysed with a magnetic
extraction device~\cite{Fleischer1969a}. This search was extended
using a new device to analyse 8~kg of samples from a 2.5-cm-thick ferromanganese crust formed over 16 million years in the Mid-Atlantic ridge~\cite{Fleischer1969b}. Another similar search was performed on over 1600~kg of sediments from deeper ocean (4.4~km depth on average)~\cite{Kolm1971}. Assuming that the extraction technique is efficient, these searches set a combined upper limit of 
$4.8\cdot 10^{-19}$~cm$^{-2}$sr$^{-1}$s$^{-1}$ for cosmic monopoles
with magnetic charge (positive or negative) of magnitude in the range
$1-60~g_D$ stopping in the atmosphere or ocean. Assuming the highest
energies, this corresponds to a mass below 130~GeV. 

A new search in 1985 analysed vast amounts of terrestrial material with the induction technique~\cite{Kovalik1986}. Probed samples included 498~kg of blueschists, 145~kg of ferromanganese nodules, and 180~kg of seawater. The blueschists are rocks from Southern California that originated as deep-sea sediments about 150 million years ago and show evidence of having been buried $20-40$~km in depth and held below the Curie temperature of the ferromagnetic minerals present. No monopoles were found in a total of 643~kg of rocks and although the study does not provide quantitative interpretations in terms of cosmic flux limits, qualitatively two conclusions can be drawn: the earlier negative results quoted above from deep-sea deposit searches are confirmed with the more robust induction technique, and the flux of cosmic monopoles that would be stopped after traversing $20-40$~km of the Earth's crust is directly constrained.

In another experiment~\cite{Carrigan1976}, 1630 litres of seawater and 68~000 litres of air were pumped into the magnetic extraction device shown in Fig.~\ref{fig:extraction} (right).  The search is based on the assumption that monopoles remain trapped within the atmosphere and ocean fluids. In a simple model where the only mechanism that would reduce the monopole density in air and water is monopole-antimonopole recombination, the search sets limits of the order of 
$1.6\cdot 10^{-16}$~cm$^{-2}$sr$^{-1}$s$^{-1}$ for monopoles of both polarities with magnetic charge in the range $1-24~g_D$. It must be stressed that the assumption made to derive this limit (all monopoles remain in the water) is orthogonal to the one made in ocean-bottom searches (all monopoles are trapped in the deposits).

\subsection{Moon rocks}

In the early 1970s, Moon rock samples returned from the Apollo
missions (19.8~kg in total) were analysed with the induction
technique~\cite{Eberhard1971,Ross1973}. These samples are considered
to have an average exposure time of 500 million years. Another
advantage is that the monopole fate after stopping can be assessed in
a robust manner due to the lack of atmosphere and magnetic field and
the stable lunar geology. Under the influence of cosmic-ray and
micrometeorite impacts, the surface of the Moon is assumed to have
been mixed completely and uniformly down to a depth of 4~m since the crystallisation of the magma ocean. The search derives a flux limit of 
$6.4\cdot 10^{-19}$~cm$^{-2}$sr$^{-1}$s$^{-1}$ for cosmic monopoles that would stop within 4~m depth of the Moon's surface. This corresponds to about the same amount of material as the Earth's atmosphere, so that Fig.~\ref{fig:stopping_atmosphere} can be used to roughly indicate the regions in charge and mass to which this result applies.

\subsection{Meteorites}
\label{meteorites}

Several searches used small meteoric samples~\cite{Petukhov1963,Goto1963,Kolm1971,Eberhard1971}. The mother asteroids are estimated to have average irradiation times of the order of 500 million years. However, only monopoles with very low mass and energy would thermalise and stop inside them. The possibility that monopole relics of higher masses could have been already trapped in the solar nebula (stellar monopoles, see Section~\ref{stellar_monopoles}) was not given consideration before a search in 1995 that probed a total of 331~kg of material with the induction technique, including iron ores and sedimentary rocks, but also notably 112~kg of meteorites~\cite{Jeon1995}. The search sets a limit on the monopole density in meteoric material of less than $2.1\cdot 10^{-5}$/gram at 90\% confidence level. The limit for chondrites is $2.5\cdot 10^{-5}$/gram, which can be directly interpreted as a limit on the stellar monopole density in the Solar System. One important assumption needed for drawing this conclusion is that the monopole was not dislodged by large accelerations exerted upon impact of the meteor. Conservatively assuming ferromagnetic binding only, the authors claim that this should not happen for monopoles with a mass below 10$^{17}$~GeV.

\subsection{Polar volcanic rocks}
\label{polar_rocks}

As discussed in Section~\ref{stellar_monopoles}, mantle-derived rocks at high latitude can also be used to search for stellar monopoles. In 2013, 23.4~kg of volcanic rock samples from a large variety of locations in arctic and antarctic areas were analysed with the induction technique~\cite{Bendtz2013a}. The probed polar volcanic rocks represent $\sim 4$ times less material than used in the meteorite search described above. However, for monopole mass and charge satisfying the criterion for a position above the core-mantle boundary ($M\cdot g_D/g<4\cdot 10^{14}$~GeV), this difference is compensated for by an increase in monopole concentration of roughly a factor six in polar mantle-derived rocks (assuming whole-mantle convection). It results that, in a simple model a limit of $1.6\cdot 10^{-5}$/gram at 90\% confidence level can be set on the monopole density in the matter averaged over the whole Earth. This can be interpreted as a limit on the stellar monopole density in the Solar System, a slightly better limit than obtained with the meteorite search. However, for low values of the magnetic charge, the mass range for which monopoles may be found at the Earth's surface is more constrained. Also, the search relies on the assumption of a binding to nuclei. 

In addition to constraining stellar monopoles, the matter density limit can also be interpreted for cosmic monopoles with energy and mass such as they would stop inside the Earth. This yields 2.2$\cdot 10^{-14}$~cm$^{-2}$sr$^{-1}$s$^{-1}$ --- a pretty weak limit compared with direct searches with detector arrays (Section~\ref{cosmic}) due to the fact that the monopoles are diluted into a very large volume.

\subsection{Iron furnace}

Another notable search claims to be able to set constraints on the
presence of monopoles in over 137 tons of iron ore
material~\cite{Ebisu1987}. A SQUID-based magnetometer was placed
directly under an iron ore processing furnace. A monopole would lose
its ferromagnetic binding as soon as the ore is heated above the Curie
temperature. In the search, it is assumed that the monopole has a high
mass (10$^{16}$~GeV) and begins to fall down under gravitational
influence. Given that the ore is moving on a conveyor belt,
assumptions about the speed at which monopoles drift inside the ore
determine the amount of material that is probed. If the ferromagnetic
image force is the only binding mechanism, monopoles are expected to
begin a free fall as soon as the temperature reaches the Curie
point. In the case where monopoles also bind to nuclei, the argument
is made that monopoles would still drift through the ore with
diffusion coefficients taken as those of hydrogen or carbon. However,
the paper does not address the case of a monopole attached to a much
larger atom, e.g., bound to an iron nucleus. Nor does it explain
how a monopole-atom system would be supposed to drift through the
conveyor belt. The limits set in this search rely heavily on the
assumption of a quick diffusion through large amounts of material, an
assumption that does not seem to be well motivated, as it has been
advocated that monopoles bound to nuclei that are part of a solid
would be very difficult to budge~\cite{Milton2006}. Another point
that the search fails to elucidate is the plausibility of the
presence of Grand Unification monopoles inside the Earth's crust. Cosmic monopoles with masses of the order of 10$^{16}$~GeV would likely traverse the Earth's crust without stopping (see Section~\ref{range_monopoles_matter}); and stellar monopoles of similar mass would not be found in the crust but rather stay inside the Earth's core~\cite{Carrigan1980}.

\subsection{Indirect searches}

In 1980, Carrigan considered the case of Grand Unification monopoles inside the Earth. The two opposite magnetic charges will accumulate at the two positions for which the magnetic force equals the gravitational force, which happen to lie inside the Earth's liquid core for a monopole with Dirac charge and $M\sim 10^{16}$~GeV~\cite{Carrigan1980}. Carrigan reasoned that whenever a geomagnetic field reversal occurs (every 0.1 to 1 million years), monopoles and antimonopoles would have a chance to meet and annihilate during the $\sim$1000 years duration of the reversal, thus creating extra heat. Under the assumptions of $M\sim 10^{16}$~GeV, a stable dipole field when there is no reversal, and the presence of both monopoles and antimonopoles within the Earth, the non-observation of the extra heat results in an upper limit on the monopole density of $6\cdot 10^{-5}$/gram~\cite{Carrigan1980}. It also corresponds to a limit of 8$\cdot 10^{-14}$~cm$^{-2}$sr$^{-1}$s$^{-1}$ on the flux of cosmic monopoles with energy and mass such that they would stop inside the Earth.

A constraint on the sensitivity of monopoles inside the Moon was obtained in 1983 using magnetometer observations aboard Explorer 35 orbiting the Moon~\cite{Schatten1983}. This is done by measuring the radial component of the magnetic field emanating from the Moon with a high precision. A limit of $6\cdot 10^{-9}$ monopoles/gram was obtained, however, it only constrains the excess of one polarity over the other one; no effect is expected if equal amounts of north and south monopoles are present.

\subsection{Summary}

In this section, searches for monopoles in various samples such as
meteorites and a variety of Earth and Moon rocks were
reviewed. These constrain monopole concentrations in matter to be
below 1 per $\mathcal{O}$ few hundred kg. Given that massive cosmic
monopoles are unlikely to be stopped by the crustal layers of large
planetary bodies, today it would seem that searches with colliders and
large cosmic detector arrays have much better sensitivities, as will
be seen in Section~\ref{cosmic} and summarised in
Fig.~\ref{fig:mass}. However, this does not apply to primordial
monopoles trapped in matter before the formation of planetary bodies
(i.e., stellar monopoles). The most promising places to search for high-mass stellar monopoles would be in meteorites and polar volcanic rock samples. If large amounts  ($>100$~kg) of such samples can be analysed using high-efficiency magnetometers, existing limits can be improved in the future. Another improvement would come from probing samples of primary material gathered directly in space, whose monopole content would be guaranteed not to be altered by impacting the Earth or by geological processes. In this respect, returned samples from asteroids or comets would be extremely valuable.

%% file: smp-cosmicmono.tex
\section{Searches at cosmic ray facilities}
\label{cosmic}


This section reviews detection techniques and searches for cosmic SMPs --- presumably produced in energetic events through the history of the Universe, travelling through space, and impacting the Earth. The techniques described below rely on assumptions about the interaction of SMPs with matter (discussed in Section~\ref{smp_matter}). An extensive review for the techniques based on ionisation energy loss was given in 1986 by Groom~\cite{Groom1986}. Results are most commonly interpreted as flux limits for monopoles with the Dirac charge in a given velocity range. Often the same limits can be reinterpreted for other types of objects, and sometimes dedicated searches are made with the same kind of detectors for an interpretation in a scenario of, e.g., nuclearites or $Q$-balls.

\subsection{Induction detectors}
\label{cosmic_induction}

The induction technique (described in Section~\ref{induction_technique}) can be used to detect the passage of one specific type of cosmic SMP: the magnetic monopole. A monopole may come with an attached proton or heavier nucleus, may carry electric charge of its own, or may be moving too slowly to be detected in a scintillator or proportional chamber. However, it still carries a magnetic charge (presumably an integer multiple of $g_D$) and as such will always induce a predictable persistent current when passing through a superconducting loop. Therefore, induction detectors are considered the most robust means of identifying a monopole. When searching for monopoles in cosmic rays, several loops in coincidence are needed to avoid spurious signals. A major drawback is the difficulty and cost of building large superconducting devices with the required sensitivity, and the fact that such devices cannot be easily operated in any environment. Induction experiments are usually built in magnetically-shielded rooms at ground level, with an effective loop area up to a square meter.

A number of searches using apparatuses with two loops~\cite{Incandela1984,Ebisu1985,Incandela1986,Ebisu1987}, three loops~\cite{Cabrera1983a,Caplin1985,Caplin1986,Cromar1986,Gardner1991}, eight loops~\cite{Huber1990,Huber1991} and a closed box~\cite{Bermon1985,Bermon1990} can be summarised by a combined flux limit of $3.8\cdot 10^{-13}$~cm$^{-2}$sr$^{-1}$s$^{-1}$. This limit applies to magnetic charges equal to or above the Dirac charge regardless of particle mass and velocity. They provide the only direct constraints on the monopole flux for $\beta<10^{-4}$, although indirect limits based on galactic magnetic field survival (for masses below $10^{17}$ GeV) or the mass density of the Universe (for higher masses) are more powerful~\cite{Turner1982} (see Section~\ref{monopoles_in_galaxy}).

\subsection{Ionisation arrays}
\label{cosmic_ionisation}

Active detectors employ time-of-flight and/or pulse length to
discriminate between slow-moving SMPs and light-speed backgrounds. The
ionisation signal threshold provides sensitivity to SMPs within a
given range of velocity and charge (either electric or
magnetic). Passive detectors rely only on ionisation and generally
have a higher threshold, appropriate for magnetic monopoles or highly charged particles.

\subsubsection{Active detectors}
\label{cosmic_ionisation_active}

Devices capable of recording the passage of charged particles, like scintillators and gaseous detectors, can be used to search for SMPs. Cosmic-ray backgrounds --- in general, mostly muons from secondary interactions --- can be rejected using one or both of two methods. The first method relies on the low speed expected from very massive SMPs and is based on either a time-of-flight measurement or the measurement of the width of a pulse in a thick slab of detector. The second method relies on an anomalously high ionisation energy loss, as would be expected, e.g., from a monopole or an electrically charged SMP at low velocity (see Section~\ref{interactions_electric}). When considering monopoles, the two methods are complementary, since timing allows to detect slow monopoles that would ionise just enough to produce a detectable signal ($\beta\sim 10^{-3}$), while discriminating against high ionisation is efficient for the detection and identification of fast monopoles ($\beta\gtrsim 10^{-2}$) (see Section~\ref{interactions_magnetic}). Monopoles in the velocity range $10^{-4}<\beta<10^{-3}$ usually do not undergo sufficient energy losses to ionise matter, although they may still excite atoms which would emit scintillating light~\cite{Ahlen1983,Ficenec1987}. One technique to extend the sensitivity of gaseous detectors to monopoles in this velocity range is to exploit the Drell effect~\cite{Drell1983}: in a mixture of helium and CH$_4$, for instance, a low-velocity monopole can excite helium atoms, which then ionise CH$_4$ molecules through atomic collisions (Penning effect)~\cite{Jesse1964}.

The state-of-the-art ionisation-array experiment is embodied by the
MACRO detector, which operated in the late 1990s at the Grand Sasso
laboratory at an average depth of 3600 meters water
equivalent~\cite{MACRO1994,MACRO1997,MACRO2002a}. MACRO spanned a
total area of 918 m$^2$ and used a combination of liquid
scintillators, streamer tubes and nuclear-track detectors. The active
detectors operated during 4.3 years (the nuclear-track detectors,
discussed in Section~\ref{cosmic_ionisation_passive}, were exposed
for a longer time). A scintillator data analysis with tight
time-of-flight selection set an upper flux limit of $2.5\cdot
10^{-16}$~cm$^{-2}$sr$^{-1}$s$^{-1}$ for slow-moving monopoles in the
range $10^{-4}<\beta<4\cdot 10^{-3}$. Another scintillator data
analysis using pulse height as a main discriminant set a limit of
$2.2\cdot 10^{-16}$~cm$^{-2}$sr$^{-1}$s$^{-1}$ in the range
$10^{-3}<\beta<0.1$. The MACRO streamer tubes were filled with 73\%
helium and 27\% n–pentane, allowing exploitation of the Drell and
Penning effects and resulting in a limit of $2.8\cdot 10^{-16}$~cm$^{-2}$sr$^{-1}$s$^{-1}$ in the range $10^{-4}<\beta<5\cdot
10^{-3}$. Combined limits lie around $1.4\cdot 10^{-16}$~cm$^{-2}$sr$^{-1}$s$^{-1}$ in the range $10^{-4}<\beta<0.1$ for a bare
Dirac monopole, which would reach the detector~\cite{MACRO2002a}. This
is the most stringent constraint to date in this $\beta$ range of
special relevance to Grand Unification monopoles. It should be noted that a velocity $\beta=10^{-4}$ corresponds to a mass $M=10^{21}$~GeV if one assumes a kinetic energy of $10^{13}$~GeV, as would be expected from acceleration in galactic magnetic fields~\cite{Wick2003} (see Section~\ref{monopoles_in_galaxy}).

Prior to MACRO, ionisation arrays were built on a smaller scale (most of them during the 1980s) at various locations using a variety of different techniques. Many searches were performed with scintillator arrays using either time-of-flight techniques or a high ionisation threshold (or both) at sea level~\cite{Mashimo1982,Bonarelli1982,Bonarelli1983,Kajino1984a,Tarle1984,Liss1984,Barish1987} and underground~\cite{Alekseev1982,Mashimo1983,Groom1983,Kawagoe1984,Tsukamoto1987,Shepko1987} (most notably with the BAKSAN telescope~\cite{Alekseev1982}), covering very wide velocity ranges. Numerous searches were also made with gaseous detector arrays underground~\cite{Bartelt1983,Krishnaswamy1984,Soudan21992,L3C2009} (most notably the SOUDAN-2 nucleon-decay detector~\cite{Soudan21992}) and at sea level~\cite{Ullman1981,Kajino1984a,Hara1986,Masek1987,Buckland1990} (the most extensive published in 1990~\cite{Buckland1990}), where experiments exploiting the Drell effect were sensitive to $\beta$ down to $10^{-4}$ for monopoles. One underground experiment in the Mont Blanc Tunnel was also sensitive to the low-velocity range using plastic streamer tubes~\cite{Battistoni1983}.

A summary of combined flux limits obtained with specific techniques at specific locations for monopoles in various $\beta$ ranges is given in Table~\ref{tab:summary_monopoles}.

In the experiments where ionisation energy loss was used as a discriminant~\cite{Mashimo1982,Bonarelli1982,Bonarelli1983,Mashimo1983,Kajino1984a,Kawagoe1984,Krishnaswamy1984,Hara1986,Masek1987,Tsukamoto1987,Buckland1990,MACRO1997,MACRO2002a}, the limits also apply to electrically charged particles with high $Z/ \beta$ which would reach the detector. In the analyses where time-of-flight is used as the only  discriminant~\cite{Ullman1981,Alekseev1982,Battistoni1983,Groom1983,Mashimo1983,Kawagoe1984,Krishnaswamy1984,Liss1984,Tarle1984,Barish1987,Tsukamoto1987,Shepko1987,MACRO1994,MACRO1997,MACRO2002a,L3C2009}, the limits extend to SMPs with low values of electric charges. For instance, the scintillator slow-monopole analysis of MACRO, with an upper flux limit of $2.5\cdot 10^{-16}$~cm$^{-2}$sr$^{-1}$s$^{-1}$ for $10^{-4}<\beta<4\cdot 10^{-3}$, generally applies to light yields corresponding to that of a muon or above~\cite{MACRO2002a}. MACRO also explicitly set a limit of $6.1\cdot 10^{-16}$~cm$^{-2}$sr$^{-1}$s$^{-1}$ for fractionally charged particles, with $0.2e<q_e<0.7e$ and $0.25<\beta<1$~\cite{MACRO2000b,MACRO2004}. Similar previous searches for fractionally charged particles with ionisation detectors underground~\cite{Kawagoe1984,Aglietta1994,L3C2009} and on a mountain top~\cite{DeLise1965} were performed, yielding limits not as stringent as MACRO.

MACRO also used liquid scintillators to search for nuclearites with velocities down to $\beta\sim 10^{-5}$~\cite{MACRO1992,MACRO2000}. Sensitivity to the $\beta\sim 10^{-4}$ range allows to probe nuclearites that would have been gravitationally trapped in the Solar System. Combined cosmic nuclearite flux limits from the MACRO scintillator detector using various analysis techniques can be summarised as $4.1\cdot 10^{-15}$~cm$^{-2}$sr$^{-1}$s$^{-1}$ for $2\cdot 10^{-5}<\beta<10^{-3}$, $3.6\cdot 10^{-16}$~cm$^{-2}$sr$^{-1}$s$^{-1}$ for $10^{-3}<\beta<0.1$, and $4.4\cdot 10^{-15}$~cm$^{-2}$sr$^{-1}$s$^{-1}$ for $0.1<\beta<1$.

Another interesting scenario is that of a stable strongly interacting massive particle (see Sections~\ref{sec:smpleptonhadron} and \ref{interactions_Rhadrons}). A global analysis of direct and indirect constraints identified allowed regions in the particle-proton scattering cross section versus mass plane~\cite{Starkman1990}. Events featuring delayed coincidences of recoil energies from nuclear elastic scattering on sodium and/or iodine were sought using the DAMA detector setup~\cite{Bernabei1999}. The negative results significantly improved the limits for masses up to $4\cdot 10^{16}$ GeV, leaving unconstrained regions at very high masses as well as at a mass around $\sim 1$~TeV.

\subsubsection{Plastic nuclear-track detectors}
\label{cosmic_ionisation_passive}

Passive solid-state nuclear-track detector arrays are frequently used
to search for highly ionising particles. The detector generally
consists of thin foils of polymers whose molecular bonds are broken
by the passage of a particle that ionises above a certain threshold
--- typically corresponding to $Z/\beta>5$ for CR39 or $Z/\beta>50$ for flexible Makrofol polycarbonate. When immersed in a chemical solution, the damaged locations are attacked more rapidly than the undamaged regions. After etching, they are revealed in the form of micrometer-sized cone-shaped pits, visible by optical means. Such a technique has the advantage to be devoid of electronics and thus easy to deploy in all kinds of environments. An efficient calibration method for nuclear-track detectors is to place them behind a target that is bombarded with high-energy heavy ions, producing a variety of fragments of various charges, which are identified from the characteristics of the etch-pit cones they produce. Modern scanning and automatic pattern recognition make this technique even more attractive, allowing for efficient coverage of large areas at relatively low cost. The most studied polymer is CR39, used in large-area experiments such as OHYA~\cite{Orito1991}, MACRO~\cite{MACRO2002a} and SLIM~\cite{SLIM2008a}. A magnetic monopole of Dirac charge with trajectory perpendicular to a CR39 foil would be expected to be recorded if it has a velocity $\beta>2\cdot 10^{-3}$ or in the narrow range where elastic recoil of atoms is important ($2\cdot 10^{-5}<\beta<2\cdot 10^{-4}$)~\cite{Derkaoui1999}.

Early nuclear-track detector arrays deployed at sea level~\cite{Fleischer1971,Bartlett1981,Doke1983} set an overall flux limit of $5.2\cdot 10^{-15}$~cm$^{-2}$sr$^{-1}$s$^{-1}$. Larger arrays were deployed underground~\cite{Nakamura1987,Orito1991,MACRO1997,MACRO2002a}, most notably a 2.1 years exposure of 2000~m$^2$ array of CR39 at the OHYA mine near Tokyo~\cite{Orito1991} and a 9.5 years exposure of a 1263~m$^2$ (of which 845.5~m$^2$ were analysed) array of CR39 and Lexan as a part of MACRO~\cite{MACRO1997,MACRO2002a}. Assuming full acceptance --- considering also SMPs which traverse the Earth before reaching the detector --- the limit from the MACRO nuclear-track detectors is $1.5\cdot 10^{-16}$~cm$^{-2}$sr$^{-1}$s$^{-1}$ and the limit from OHYA is $3.2\cdot 10^{-16}$~cm$^{-2}$sr$^{-1}$s$^{-1}$.

\begin{figure}[tb]
\begin{center}
\includegraphics[width=0.6\linewidth]{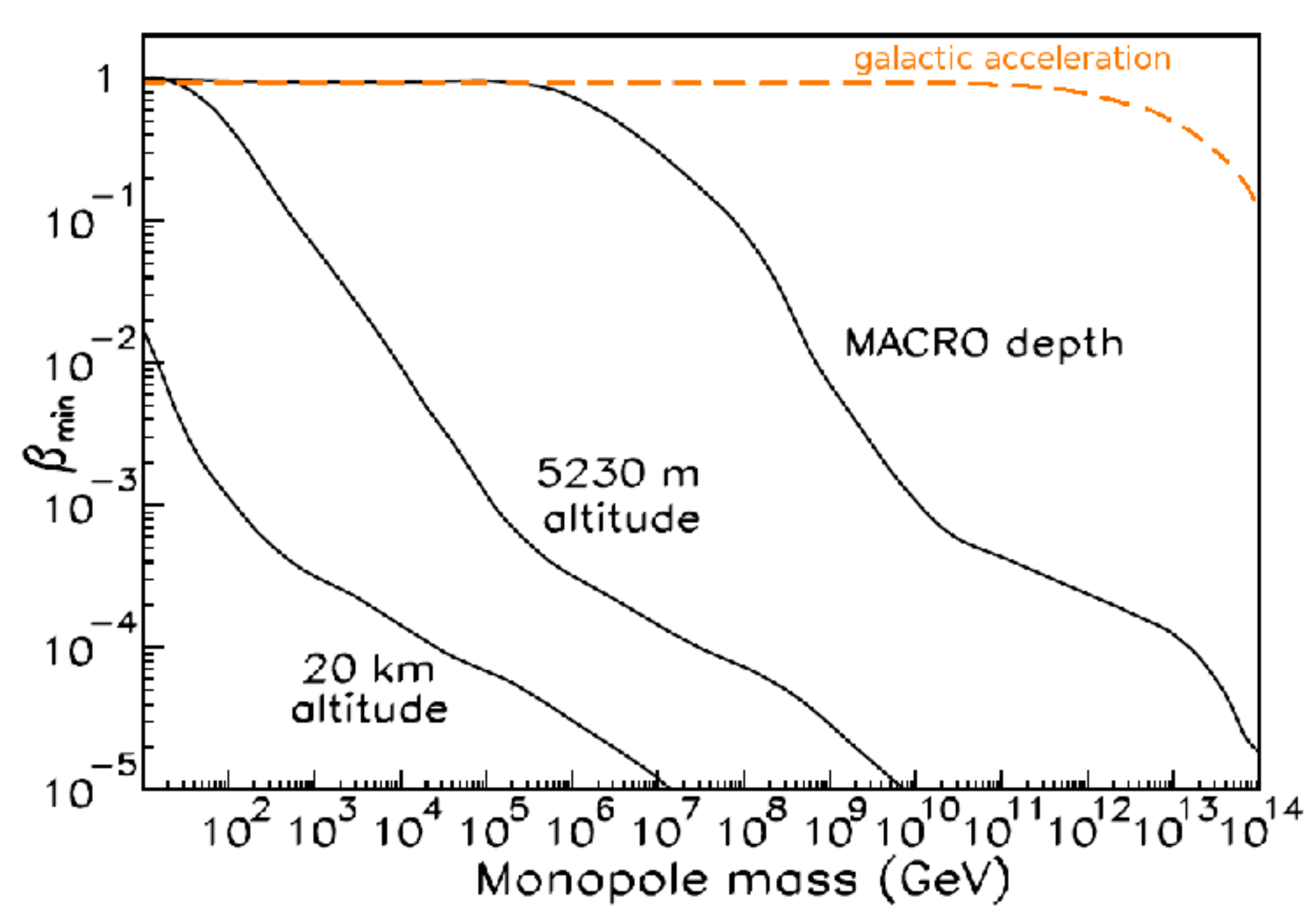}
\caption{Ranges in mass and velocity for which a monopole with one Dirac charge coming from above would reach a detector at various depths~\cite{SLIM2008a}. The altitude level of 5230 m corresponds to that of the SLIM detector~\cite{SLIM2008a}. The dashed line indicates the typical velocity expected from a monopole due to its acceleration by galactic magnetic fields~\cite{Wick2003}. The Figure is taken from ref.~\cite{SLIM2008a} upon which the dashed line is superimposed.}
\label{fig:monopole_altitudes}
\end{center}
\end{figure}

Nuclear-track detectors placed on balloons~\cite{Price1975,Price1978}
or on mountain
tops~\cite{Barwick1983,Price1984b,Kinoshita1981,SLIM2008a} are
sensitive to monopoles that do not possess sufficient energy to reach
the ground or underground (see the discussion in
Section~\ref{range_monopoles_matter}). This is also true for detectors
placed directly in space~\cite{Shirk1978,Weaver1998} although in that
case more sophisticated techniques are needed to separate exotic
highly ionising particles from ordinary ions in the cosmic radiation (see Section~\ref{cosmic_in_space}). Fig.~\ref{fig:monopole_altitudes} (from Ref.~\cite{SLIM2008a}) shows the velocity above which a monopole of a given mass would reach a detector in a given location. Superimposed on the figure is the velocity expected if the monopole has an energy of $10^{13}$~GeV, as expected from acceleration in galactic magnetic fields~\cite{Wick2003} (Section~\ref{monopoles_in_galaxy}). Assuming such energies, the mass range of cosmic monopoles that is accessible to high-altitude experiments but inaccessible to underground experiments such as MACRO is between 20 and 1000~GeV. Also, high-altitude experiments can probe secondary monopoles (in a similar mass range) produced by high-energy cosmic-ray collisions in the upper atmosphere. Another way to efficiently search for such low-mass monopoles, discussed in Section~\ref{monopoles_matter}, is to look for trapped monopoles in samples of air, water, rock surface, and deep-sea deposits. The relation between low-mass cosmic monopole searches and collider searches is further discussed in Section~\ref{cosmic_prod}. Among the high-altitude cosmic-ray experiments, SLIM at Mt. Chacaltaya (5230~m) was the most sensitive, with an area of 400~m$^2$~\cite{SLIM2008a}, providing a monopole flux limit of $6.5\cdot 10^{-16}$~cm$^{-2}$sr$^{-1}$s$^{-1}$ in the range $0.03<\beta<1$.

Nuclearites are expected to produce etch-pit cones in nuclear-track detectors (see Section~\ref{interactions_nuclearites}), and thus results that have been obtained for monopoles with large CR39 arrays at OHYA, MACRO and SLIM --- with a combined flux limit at the level of $10^{-16}$~cm$^{-2}$sr$^{-1}$s$^{-1}$ --- are believed to also be valid for nuclearites~\cite{Nakamura1990,Orito1991,MACRO1992,MACRO2000,SLIM2008b}. The mass range to which the limits apply depend on the detector depth and the nuclearite velocity. Assuming a typical galactic velocity ($\beta\sim 2\cdot 10^{-3}$), all detectors have full acceptance to nuclearites with $M\gtrsim5\cdot 10^{22}$~GeV, which would be expected to traverse the Earth. For very large ($M\gtrsim 100$ g) objects the effect of strong heating on detector sensitivity remains unclear~\cite{MACRO1992}. At such velocities, the minimum nuclearite mass for reaching the detector from above is $M\gtrsim10^{14}$~GeV at MACRO depth, $M\gtrsim10^{12}$~GeV at OHYA depth, $M\gtrsim10^{11}$~GeV at sea level, and $M\gtrsim10^{10}$~GeV at SLIM altitude~\cite{MACRO2000,SLIM2008b}. Combined limits on the nuclearite flux from MACRO, OHYA and SLIM are reported in Fig.~\ref{fig:nuclearite_flux_limits}.

As was discussed in Section~\ref{interactions_Qballs}, $Q$-balls can also undergo significant energy losses in matter and give rise to signatures similar to that of nuclearites of the same mass. However, energy loss predictions strongly depend on $Q$-ball properties other than the mass (such as electric charge $Z_Q$) and on model assumptions (e.g., SUSY hypotheses), making the interpretation more difficult. Within one particular model of $Q$-ball interactions, it was found that CR39 nuclear-track detectors are sensitive to $Z_Q\geq 3$~\cite{Arafune2000}. Based on such an interpretation, the nuclearite flux limits from MACRO, OHYA and SLIM mentioned above can also be applied to $Q$-balls of roughly the same masses.

\subsubsection{Ancient mica}
\label{mica_searches}

Track forming by very highly ionising particles is a also general feature for non-conducting minerals~\cite{Fleischer1965}, and the etching technique can be advantageously applied to ancient rocks with geological exposure times such as mica and obsidian, although the signal threshold for crystals and glasses is generally higher than for plastics. Fossil amber has a chemical composition similar to plastics but was found to have a very high threshold, even higher than minerals~\cite{Uzgiris1971}. Exposure of a sample to high temperature can cause thermal fading, an important effect to assess in a search. The effective time of exposure of a sample of known uranium concentration can be intrinsically determined by measuring the density of fossil tracks from spontaneous uranium fission: in that way it was shown that selected mica crystals can efficiently record tracks for at least 600 million years~\cite{Fleischer1964a}. The detection threshold for mica lies around $Z/\beta \sim 130$~\cite{Fleischer1967}. A difficulty arises when trying to interpret null results for particles that provide ionisation close to this threshold. The analysis of tracks produced by ion beams of MeV energies support the ion-spike explosion model~\cite{Fleischer1967}, in which the formation of etch pits in mica depends on the number of ions produced in the vicinity of the particle: the ions repel each other and are displaced, producing lattice defects. However, the response to a new material cannot be reliably predicted from first principles and has to be studied empirically.

Searches using ancient crystals (such as mica) or glasses (such as
obsidian) as nuclear-track detectors have the advantage of very long
exposure times. Searches were performed with $\sim 1600$~cm$^2$ of
mica samples with recording age varying from 185 to 900 million
years~\cite{Fleischer1969c,Price1984a,Price1986,Ghosh1990}. In these
searches, high-quality transparent mica was cleaved into $\sim100$~$\mu$m sheets and etched in hydrofluoric acid, and a signal was sought
in the form of etch-pit cones aligned on several surfaces. The results
were interpreted for low-velocity monopoles that capture a nucleus on
their way to the rock layer, in which case a track is guaranteed due
to a high ionisation energy loss for velocities in the range $5\cdot
10^{-3}<\beta<0.5$ and electromagnetic nuclear collisions in the range
$4\cdot 10^{-4}<\beta<5\cdot
10^{-3}$~\cite{Price1984a,Price1986,Ghosh1990}. In this
interpretation, flux limits at the level of $10^{-18}$~cm$^{-2}$sr$^{-1}$s$^{-1}$ have been set in a narrow $\beta$ range
around $10^{-3}$, with large model uncertainties on the probability
that the monopole captures a nucleus in the Earth's crust and the
monopole-nucleus bound pair reaches the mica. Here we chose to factor
out the model dependence and merely quote flux limits assuming that
SMPs reach the mica and produce a detectable track: this yields a
combined limit of $5.5\times 10^{-20}$~cm$^{-2}$sr$^{-1}$s$^{-1}$. If
one assumes that the ion-spike explosion model holds for mica, bare
monopoles with $\beta>0.1$ carrying three Dirac charges or more are
guaranteed to be detected~\cite{Fleischer1969c}, without further
assumptions. High-velocity ($\beta>0.9$) monopoles with two Dirac
charges lie just above the threshold, and singly charged relativistic ($\gamma>2000$) monopoles could possibly produce tracks if they undergo sufficiently local energy losses. Within such assumptions, as can be seen in Fig.~\ref{fig:betagamma}, searches in ancient mica provide the best limit for large parts of the velocity range.

These strong limits from ancient mica in the flux of highly ionising particles also apply to nuclearites with masses $M\gtrsim 10^{14}$~GeV~\cite{DeRujula1984,Price1988}. That provides the best means of testing the hypothesis that nuclearites could be a major dark matter component in this mass range. As can be seen in Fig.~\ref{fig:nuclearite_flux_limits}, current mica limits constrain nuclearite dark matter for masses up to $10^{26}$~GeV~\cite{Price1988}. For a mass of 1 kg or higher, effect from rock heating may become significant and the size of the nuclearite becomes the same order as the size of the etch-pit cones sought in monopole searches in mica~\cite{DeRujula1984}. A search for large nuclearites in mica would thus require special care with the way a signal is identified.

One particular model of $Q$-ball interactions predicts that mica detectors are also sensitive to $Q$-balls with $Z_Q\geq 10$~\cite{Arafune2000}. Based on such an interpretation, the nuclearite flux limits from ancient mica mentioned above can also be applied to $Q$-balls of roughly the same masses.

\subsection{Water or ice neutrino observatories}
\label{cosmic_cherenkov}

Relativistic charged particles travelling through an homogeneous and transparent medium like water or ice emit detectable Cherenkov radiation. Detectors equipped with arrays or strings of photomultiplier tubes for Cherenkov-light detection include SuperKamiokande, ANTARES and IceCube, and in the future KM3NeT~\cite{KM3NeT2011}. Other detectors, such as RICE and ANITA, are equipped with radio antennas for detection of coherent radio Cherenkov emission from dense electromagnetic showers. Some SMPs would produce distinctive signatures in these detectors, such as a large yield in the case of a relativistic monopole. Neutrino detectors are optimised for a large acceptance and a long duty cycle, features which are also great advantages for cosmic SMP searches.

\begin{figure}[tb]
\begin{center}
\includegraphics[width=0.44\linewidth]{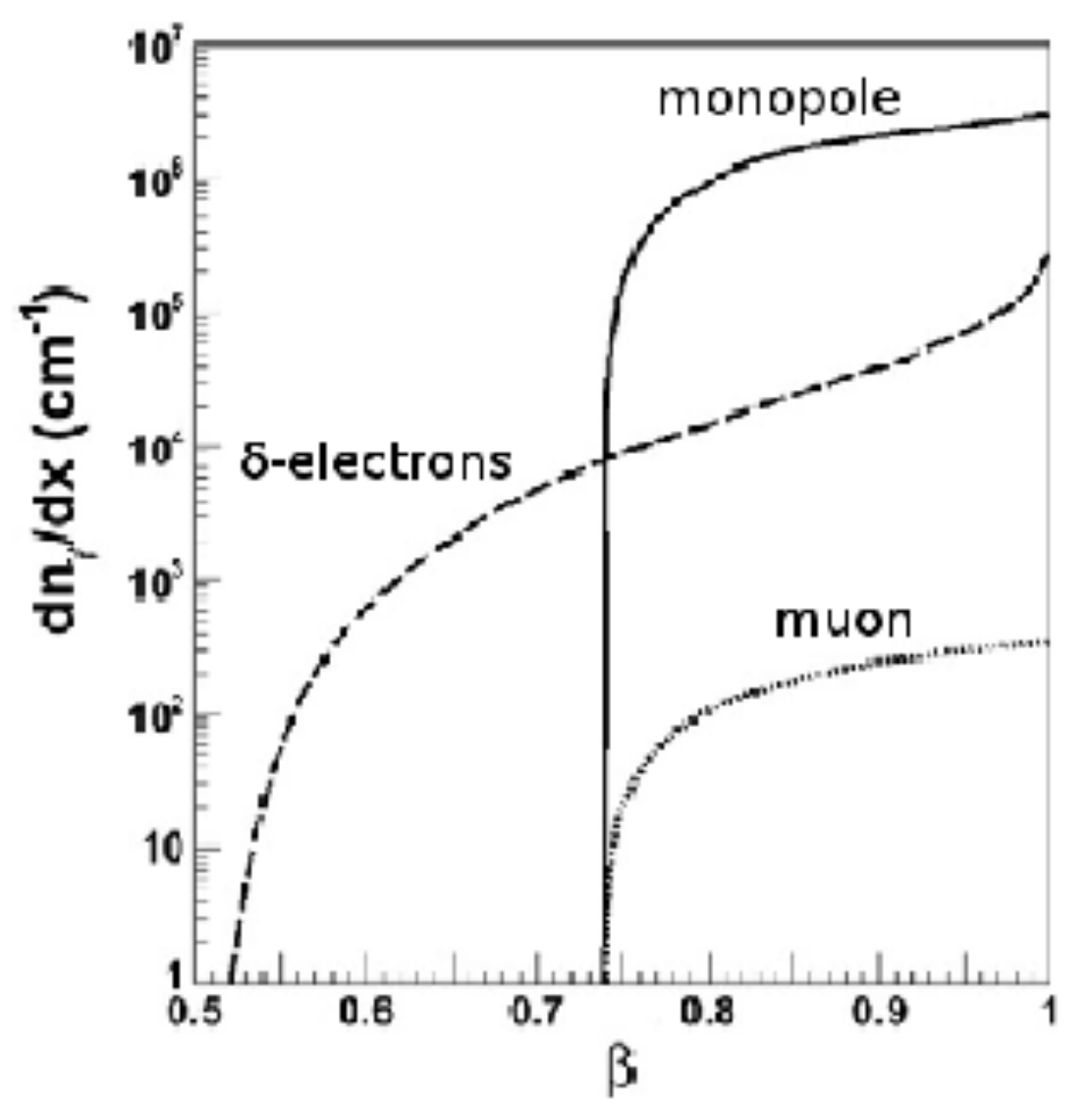}
\includegraphics[width=0.55\linewidth]{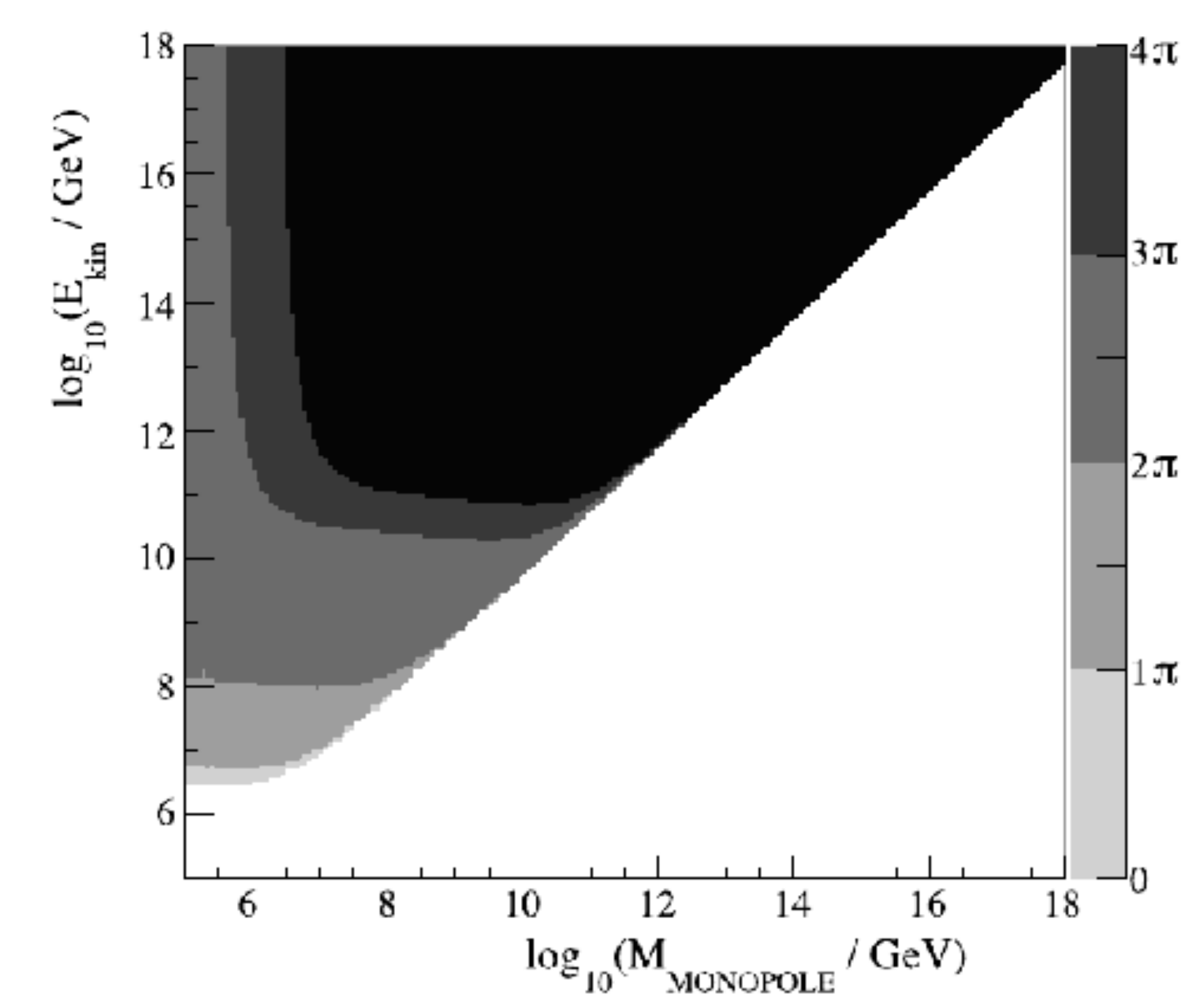}
\caption{Left: Number of Cherenkov photons in the $300-600$~nm wavelength range emitted per cm in sea water, with contributions from a monopole with $g=g_D$ (solid line) and from $\delta$ electrons produced along its path (dashed line), as a function of the monopole velocity $\beta$ (Figure from ref.~\cite{ANTARES2012}). Direct Cherenkov emission by a muon is also shown (dotted line). Right: Angular acceptance of a typical neutrino telescope for relativistic magnetic monopoles as a function of mass and kinetic energy (Figure from ref.~\cite{AMANDA2010}). The grey shaded areas represent the regions for which monopoles are capable of reaching the detector with $\beta\geq 0.75$. The black area corresponds to the region of full acceptance, where monopoles are still relativistic after traversing the full diameter of the Earth.}
\label{fig:monopoles_ice}
\end{center}
\end{figure}

\subsubsection{Cherenkov light emission}
\label{Cherenkov_light}

SMPs that carry a high electromagnetic charge (such as monopoles) would provide striking Cherenkov signatures in water or ice. As illustrated in the left plot of Fig.~\ref{fig:monopoles_ice} in the case of a Dirac monopole, copious amounts of visible Cherenkov light would be emitted for $\beta>0.75$, and additional light exceeding muon backgrounds would be produced by Cherenkov radiation from $\delta$ electrons along the monopole path for velocities down to $\beta=0.6$ (and, to a lesser extent, by water luminescence down to $\beta=0.001$~\cite{BAIKAL2008}). Large backgrounds from cosmic muons necessitate the implementation of high thresholds. This can be improved by considering up-going events in which the SMP is required to traverse the Earth before reaching the detector.

Searches have been performed at the Baikal~\cite{BAIKAL1995,BAIKAL2008}, AMANDA~\cite{AMANDA2010}, ANTARES~\cite{ANTARES2012} and IceCube~\cite{IceCube2013} neutrino telescopes. ANTARES searched for up-going SMPs and set flux limits between $10^{-16}$ and $10^{-17}$~cm$^{-2}$sr$^{-1}$s$^{-1}$ in the velocity range $0.625<\beta<1$ assuming a Dirac monopole~\cite{ANTARES2012}. The IceCube detector very recently set a more stringent limit of $3\cdot 10^{-18}$~cm$^{-2}$sr$^{-1}$s$^{-1}$ in the velocity range $0.8<\beta<1$. The IceCube analysis has a ionisation threshold low enough to accept all monopoles coming from below the horizon: this corresponds to the black area in the right plot of Fig.~\ref{fig:monopoles_ice}. For down-going signals, the threshold is adjusted as a function of the incoming angle to reject the much larger backgrounds, at the cost of rejecting most of the signal (assuming a Dirac monopole) for polar angles below 75$^{\circ}$. Despite this low selection efficiency, this allows to probe monopoles with lower masses and lower velocities, illustrated by the grey shades in Fig.~\ref{fig:monopoles_ice}. The kinetic energy gained by monopoles due to acceleration in galactic magnetic fields is expected to be of the order of $10^{13}$~GeV (possibly up to a maximum of $10^{16}$~GeV)~\cite{Wick2003} (see Section~\ref{monopoles_in_galaxy}). This corresponds to a mass range between $10^7$ and $10^{13}$~GeV for ensuring a good acceptance to up-going Dirac monopoles (possibly up to $10^{16}$~GeV for the highest energies), and extending the search to down-going monopoles pushes the lower mass reach to $10^{5}$~GeV (a discussion of the range of monopoles in matter is given in Section~\ref{range_monopoles_matter}).

The Kamiokande-II detector performed a search for fractionally charged particles, distinguished from unit charge particles by the lower intensity of the emitted Cherenkov light~\cite{Kamiokande1991}. The obtained flux limits are at the level of $2\cdot 10^{-15}$~cm$^{-2}$sr$^{-1}$s$^{-1}$ --- a little bit weaker than MACRO results on fractional charges~\cite{MACRO2004}.

It has been pointed out that the characteristics of black-body radiation light from the thermal shock along the passage of a nuclearite can be used to set interesting nuclearite flux constraints at neutrino telescopes such as ANTARES~\cite{Pavalas2010}. However, a full study including background estimate and systematic uncertainties remains to be done.

\subsubsection{Radio emission}

High-energy neutrinos interacting in a non-conducting medium produce charged particle showers. As a shower propagates the number of electrons increases relative to the
number of positrons due to photons in the shower interacting with atomic
electrons. This leads to coherent emission of Cerenkov radiation of
wavelength similar to the transverse size of the shower: the
Askaryan effect~\cite{ask1,ask2}.
High-energy neutrino experiments based on this principle have been conducted, using data from the FORTE satellite~\cite{Lehtinen2004}, Earth-based radio telescopes pointed at the Moon~\cite{Jaeger2010,Buitink2010}, the ANITA balloon-borne radio antenna array, which hovered over the Antarctic landscape for one month~\cite{ANITA2010}, and the RICE radio antenna array buried in ice at the South Pole~\cite{Kravchenko2012}. Ultrarelativistic charged SMPs moving through a medium undergo dramatic radiative losses, and the induced electromagnetic showers create coherent Cherenkov radiation in the radio frequency domain. Magnetic monopoles with masses below $\sim 10^{13}$~GeV would be accelerated to relativistic speeds and therefore would produce bright electromagnetic showers that could be detected by radio neutrino telescopes~\cite{Wick2003}.

A search for high-energy neutrinos with RICE found no bright event candidates and this result was reinterpreted as flux limits between $10^{-18}$ and $10^{-19}$~cm$^{-2}$sr$^{-1}$s$^{-1}$ for magnetic monopoles in the velocity range $10^7<\gamma<10^{12}$~\cite{RICE2008}. In the future, the ARA radio antenna is expected to improve the RICE sensitivity by approximately two orders of magnitude~\cite{Kravchenko2012}. ANITA did a similar analysis where monopole selection was made by requiring a row of four successive trigger events within 500~ns. Further waveform analysis of the four remaining candidates showed that the data were consistent with backgrounds, and monopole flux limits between $10^{-17}$ and $10^{-19}$~cm$^{-2}$sr$^{-1}$s$^{-1}$ were set in the range $10^9<\gamma<10^{13}$~\cite{ANITA2011}. The RICE and ANITA velocity ranges correspond to masses in the range $10^3<M<10^7$~GeV assuming galactic acceleration. At such very high velocities, only searches with ancient mica (Section~\ref{mica_searches}) can reach better sensitivities than searches with the radio Cherenkov technique, assuming that the energy loss is sufficiently high to exceed the mica threshold for track formation.

\subsection{Gravitational-wave detectors and acoustic sensors}

Gravitational-wave detectors are extremely sensitive to tiny
vibrations and were shown to be able to detect comic-ray
showers~\cite{Yamamoto2008}. They can thus also be used to search for
SMPs that would induce vibrations, such as monopoles and
nuclearites~\cite{Allega1983,Bernard1984}. In a study comparing the
suitability of different kinds of gravitational-wave detectors for SMP detection, it was shown that resonant-bar detectors should be $30-300$ times more sensitive than interferometers, due to a larger volume and a higher efficiency~\cite{Yamamoto2008}. Searches performed at sea level using resonant-bar detectors quoted nuclearite flux limits at the level of $10^{-12}$~cm$^{-2}$sr$^{-1}$s$^{-1}$ for nuclearite masses $M\gtrsim 10^{11}$~GeV~\cite{Liu1988,Astone1993,Astone2013}. While less restrictive than the limits obtained by larger arrays like MACRO and SLIM (Section~\ref{cosmic_ionisation}), these results were obtained using a completely different technique, with different assumptions underlying the detector response expectations.

If a large amount of energy is nearly instantaneously produced within a small volume of matter, thermal expansion is predicted to result in a pressure pulse detectable with acoustic sensors~\cite{Nahnhauer2012}. This principle has been used to search for particle cascades, which would be induced by neutrinos with very high energies impacting the sea~\cite{Kurahashi2010}. Ambitious projects with strings of acoustic detectors spanning very large volumes of water or ice have been proposed~\cite{Nahnhauer2012}. Analysis techniques designed for very high energy neutrinos can be adapted to select for massive objects that would undergo very high energy losses~\cite{Danaher2012}. The volumes considered --- of the order of 100~km$^3$ --- could make such a future search very competitive.

\subsection{Air-shower observatories}
\label{cosmic_airshower}

Very large air showers may excite air molecules to the
extent that the emitted blue fluorescence light can be detected by
ground-based telescopes using focusing mirrors and
photomultipliers. The air fluorescence technique was pioneered by the
Fly's Eye detector~\cite{HiRes2004} and is now commonly used in
combination with ground arrays at observatories, such as the Pierre
Auger and Telescope Array observatories, whose aim is to study ultra-high energy cosmic rays. In principle, this technique could also be used in water, although large air showers are usually initiated high in the atmosphere. Fluorescence would probably only be detectable in water in very rare cases where high-energy neutrino interactions or exotic SMPs produce very high ionisation (such as high-velocity monopoles). To our knowledge, the fluorescence technique has never been used in underwater studies.

Cosmic-ray detector arrays with a large surface collection area, such as the Pierre Auger observatory in Argentina, can be sensitive to very low fluxes. Fluorescence detectors can be used to measure shower characteristics. SMPs would produce showers with exploitable signatures, e.g., strong electromagnetic component, long time span, narrow showers, and deep showers. The best studied case is that of $R$-hadrons (see Sections~\ref{sec:smpleptonhadron} and \ref{interactions_Rhadrons}), motivated by the possibility that they could be responsible for the observed ultra-high-energy cosmic-ray events~\cite{Albuquerque1998,Berezinsky2001,Gonzalez2005,Anchordoqui2007,Albuquerque2009}. A generic search for deviations from expected air-shower shapes could be used to hint at the presence of $R$-hadrons, monopoles, $Q$-balls, strangelets, and micro-black holes~\cite{Auger2012}, although such studies have not been published yet.

\subsection{Balloon and space particle telescopes}
\label{cosmic_in_space}

Cosmic-rays particles with high charges or anomalous charge-to-mass ratios can potentially be identified at experiments aboard satellites or space stations.

Experiments in space with energy loss capabilities (using, e.g., nuclear-track detectors) include Ariel-6~\cite{Fowler1987}, HEAO-3~\cite{Binns1989}, Slylab~\cite{Shirk1978} and Trek~\cite{Westphal1998,Weaver1998} and together constrained the flux of particles with $Z\geq 100$ to be less than $2\cdot 10^{-12}$~cm$^{-2}$sr$^{-1}$s$^{-1}$. This limit applies to nuclearites with masses $M\gtrsim 1000$~GeV and is actually stronger than limits inferred from searches for heavy isotopes in matter in this mass range (see Section~\ref{heavy_isotopes}).

The AMS spectrometer on the International Space Station has the capability to identify anomalous particle rigidities and has been shown to have the potential to largely surpass the sensitivities of previous experiments (including matter searches) for strangelets in the mass range $10\leq M\leq 1000$~GeV~\cite{Sandweiss2004,Finch2006}. The Bess balloon-borne experiment searched for fractionally charged particles in cosmic rays and set a limit of $4.5\cdot 10^{-7}$~cm$^{-2}$sr$^{-1}$s$^{-1}$ for charge $2/3e$~\cite{BESS2008}. Results from AMS for both strangelets and fractional charges are expected in the near future~\cite{Sbarra2003,Sandweiss2004,Finch2006}.

\subsection{Catalysis of nucleon decay}
\label{cosmic_catalysis}

SMPs can be hypothesised to catalyse nucleon decay reactions when
traversing matter (see Section~\ref{monopole_catalysis}). This
reaction would be expected in the case of a slow-moving Grand Unification monopole via what is known as the Rubakov-Callan effect~\cite{Rubakov1981,Callan1982,Rubakov1983,Callan1983}. Although the catalysis cross section $\sigma_c$ is highly uncertain, the range $1<\sigma_c<1000$~mb is well motivated and worth searching for.

\subsubsection{Underground nucleon-decay detectors}

Underground experiments have been used to search for monopole catalysis of nucleon decay~\cite{Bosetti1983,Errede1983,Battistoni1983,Krishnaswamy1984,Kajita1985,Stone1985,Bartelt1987,IMB1994,BAIKAL1998,MACRO2002b,IceCube2014}. The simplest thing to do is to reinterpret proton decay results for a monopole inducing no more than one single decay during its crossing of the detector. Taking the IMB detector as a typical example, this applies to the cases of a large catalysis cross section combined with a low velocity ($\beta<0.1$ for $\sigma_c=10$~mb, or $\beta<2\cdot 10^{-5}$ for $\sigma_c=100$~mb) or a small cross section, with the best sensitivity for $\sigma_c=1$~mb with a velocity-independent flux limit of $3\cdot 10^{-13}$~cm$^{-2}$sr$^{-1}$s$^{-1}$~\cite{IMB1994}.

The sensitivity to cross sections exceeding 1 mb can be considerably enhanced with the use of a dedicated analysis where multiple (typically two or three) nucleon decays are sought in a given time window. This strategy for searching for nucleon decay catalysis by cosmic monopoles was first used in 1983 with a purpose-built water Cherenkov detector~\cite{Bosetti1983}. The search, which set a limit of 3$\cdot 10^{-12}$~cm$^{-2}$sr$^{-1}$s$^{-1}$ (assuming $5\cdot10^{-4}<\beta<0.05$ and $\sigma_c\sim 100$~mb), is still the only one to have been performed at sea level. The technique was then used at a number of underground detectors, including SOUDAN~\cite{Bartelt1987}, IMB~\cite{IMB1994}, BAIKAL\cite{BAIKAL1998}, MACRO~\cite{MACRO2002b}, and IceCube~\cite{IceCube2014}. Depending on experimental constraints such as timing of the successive decays, these searches probe various  $\sigma$ ranges between $10$ and $2\cdot 10^{5}$~mb. Together, these result in a combined limit at the level of $10^{-16}$~cm$^{-2}$sr$^{-1}$s$^{-1}$ on the monopole flux (typically assuming $10^{-5}<\beta< 10^{-2}$ and $\sigma_c\sim 100$~mb).

Catalysis of nucleon dissociation into quarks resulting in pion emission by electrically neutral $Q$-balls was sought at the SuperKamiokande detector using a similar technique~\cite{SuperKamiokande2007}. This constrained the neutral $Q$-ball flux to below $10^{-15}$~cm$^{-2}$sr$^{-1}$s$^{-1}$ for catalysis cross sections in the range $1-200$~mb.

\subsubsection{Neutrinos from the Sun}

Another approach is to measure the flux of neutrinos with energy around 45~MeV from the Sun and infer the rate of SMP-induced nucleon decays happening at the centre of the Sun. The efficiency with which the Sun captures monopoles striking its surface depends on the monopole mass and velocity (see Section~\ref{range_monopoles_matter}). Calculations of monopole stopping power in the Sun show that $M=10^{16}$~GeV and $\beta=10^{-3}$ yield a stopping length of the order of a solar radius~\cite{Meyer1985}. However, it should be noted that if we assume that monopoles acquire energies of the order of $10^{13}$~GeV by acceleration in galactic magnetic fields, they will stop when traversing the Sun only if they have masses much below the GUT scale. Monopoles with such masses which arise in theories with an intermediate mass scale below the GUT scale are not expected to catalyse nucleon decay (see Section~\ref{monopole_catalysis}). At the SuperKamiokande neutrino telescope such an analysis was used to set a limit $6\cdot 10^{-24}\left(\frac{\beta}{10^{-3}}\right)^2$~cm$^{-2}$sr$^{-1}$s$^{-1}$ on the magnetic monopole flux, assuming monopoles with mass and velocity such that they stop in the Sun, with nucleon decay catalysis cross section larger than 1~mb~\cite{Kajita1985,SuperKamiokande2012}.

\subsubsection{Neutron stars}

Cosmic monopoles striking the surface of a neutron star would lose sufficient energy to be captured and thus would accumulate inside the neutron star at a rate proportional to the monopole flux. Assuming monopole catalysis of nucleon decay, observations of the energy emitted by old or young neutron stars have been used to set indirect constraints on the monopole flux. This can be done using the total X-ray emission per neutron star~\cite{Kolb1982}, the contribution of old neutron stars to diffuse X-ray background~\cite{Dimopoulos1982b,Kolb1984}, serendipitous searches for soft X-ray point sources~\cite{Kolb1982}, and the luminosities of old radio pulsars~\cite{Freese1983b}. These different methods make slightly different assumptions and together give a bound of the order of $10^{-23}$~cm$^{-2}$sr$^{-1}$s$^{-1}$ on the flux of monopoles, where the nucleon decay catalysis cross section times the nucleon-monopole relative velocity is of the order of 0.1~mb~\cite{Kolb1984}. These limits apply to monopoles with masses $M \geq 10^{14}$~GeV, below which monopoles would probably not reach the surface of the neutron star due to deflection by magnetic fields. They also rely on the assumption that monopole annihilations inside the neutron star can be neglected. The limit can be improved by several orders of magnitude if one assumes that monopoles are also captured in the matter that will later form the neutron star, although this applies only to lower monopole masses ($M\leq 10^{17}$~GeV), and the fate of the monopole after capture is uncertain~\cite{Freese1983b,Kolb1984}.

\subsection{Discussion of candidate events}
\label{cosmic_candidates}

Historical claims were made that exotic events have been observed in cosmic rays. None of the cases was confirmed.

\begin{itemize}

\item A very highly ionising nuclear-track candidate was observed by Price \textit{et al.} in a balloon-flight experiment in 1973. It was first interpreted as evidence for a magnetic monopole~\cite{Price1975}. A subsequent paper by the same authors refuted the monopole interpretation, while still claiming that the track is not compatible with any known particle~\cite{Price1978}. The event remains a mystery, as no further specimen has been recorded since.

\item In 1982, Cabrera reported the detection of an event consistent with a Dirac monopole at one of the first cosmic-ray induction experiments~\cite{Cabrera1982}. A similar event was also seen by Caplin \textit{et al.} in 1986~\cite{Caplin1986,Guy1987}. In both cases, the flux change was observed in only one loop, thus not allowing to rule out a spurious jump (see Section~\ref{induction_technique}). The lesson was learned that such experiments need at least two loops in coincidence to detect an unambiguous signal. Later induction searches (See Section~\ref{cosmic_induction}) placed monopole flux limits much below what is needed to produce the observed events.

\item Certain classes of observed cosmic-ray air showers appear to have anomalous features which have been tentatively interpreted within exotic scenarios involving SMPs~\cite{Shaulov1996,Chen1997}. The most studied are the so-called ``centauro'' events at Mt. Chacaltaya in 1980~\cite{Lattes1980}, whose alternative interpretations were recently refuted in favour of more mundane explanations~\cite{Kopenkin2003,Ohsawa2004,Kopenkin2006,Kopenkin2008}. In general, the need for exotic physics does not appear to be convincing due to the limited information made available by the detectors. It is still very worthwhile to continue to try to confirm reported anomalies in cosmic-ray experiments, and search for new ones.

\item The underground observation of several $\alpha$-particles of
  anomalously long range was tentatively interpreted as nuclear
  fission induced by Grand Unification monopoles~\cite{Anderson1983}, even if non-exotic explanations cannot be ruled out.

\item In 1990 a balloon-borne experiment at an atmospheric depth of 9~g/cm$^2$ with exposure $6\cdot 10^8$~cm$^2$sr$\cdot$s reported two events with anomalous rigidity, corresponding to $Z\sim 14$ and $A\sim 350$~\cite{Saito1990}. The characteristics of the events are not compatible with known elements but fit nicely with those expected from strange quark matter~\cite{Saito1990}. This result is in mild conflict with matter searches~\cite{Hemmick1990}. In another balloon-borne chamber experiment, an event was found with $Z/\beta\sim 40$ and high $\beta$, which could not be reconciled with a high penetration depth~\cite{Ichimura1993}. New data by AMS is expected to shed new light on these phenomena in the near future~\cite{Sandweiss2004,Finch2006}.

\end{itemize}

\subsection{Connecting cosmic-ray and collider monopole production}
\label{cosmic_prod}

Cosmological SMP relic abundances are highly dependent on the models and can generally be made as low as one wishes depending on the scale of inflation, the annihilation rate, and the extent to which SMPs bind to matter (see Section~\ref{relic_abundances}). At present, cosmic flux limits are therefore very difficult to relate to direct cross section limits at colliders. However, high-energy cosmic rays impacting the Earth's atmosphere would still be expected to produce SMPs by the same mechanisms as at hadron colliders, at a rate which sharply decreases with mass. This allows for a direct comparison between flux limits and collider constraints.

Below we convert flux limits into cross section limits in the case of magnetic monopoles, making the following assumptions:

\begin{itemize}

\item The cosmic-ray primary particles are mainly protons with only a small fraction of heavier nuclei.
\item A parameterisation of the cosmic-ray flux, $\phi(E)$, at primary energy $E$ is given by
\begin{equation}
\phi(E)=1.05~10^4~E^{-2.6}\rm{(m^2~sec~sr~GeV)}^{-1}~~~(E<4.9~10^6)\rm{GeV}
\end{equation}
\begin{equation}
\phi(E)=6.01~10^7~E^{-3.16}\rm{(m^2~sec~sr~GeV)}^{-1}~~~(E > 4.9~10^6)\rm{GeV}
\end{equation}
from a fit to the primary spectrum~\cite{PDG2012}.
\end{itemize}

It follows from these assumptions that the differential rate for monopole production is given by
\begin{equation}
\frac{d^2N}{dEd\Omega}=\int_x e^{-x/\lambda}~dx~
\sigma(M,E)~N_A~\phi(E)~(\rm{GeV~cm}^2 \rm{~s~sr})^{-1}
\label{integral}
\end{equation}
where $x$ is the thickness of the atmosphere penetrated by the cosmic ray, $N_A$ is Avogadro's number, $\lambda$ is the mean free path of the cosmic ray in the upper atmosphere, $\sigma(M,E)$ is the production cross section for a monopole of mass $M$ by a cosmic-ray primary of energy $E$ and $\phi(E)$ is the cosmic-ray flux at energy $E$ given above.


Here $\lambda$ is taken to be 30~g/cm$^{2}$, assuming that after the first collision the primary will have insufficient energy to produce a monopole pair. Hence the cross section lower limit for primaries between energies $E_1$ and $E_2$ from the flux limit $L$ is given by
\begin{equation}
\sigma < \frac{L}{\lambda N_A \int_{E_1}^{E_2} \phi(E)dE}.
\label{limit}
\end{equation}
The lower energy limit, $E_1$ is taken as the maximum of the production threshold ($\sqrt{s}=2M$) and the energy needed for the monopole to penetrate to the experiment depth. The upper energy integration limit is taken to be infinite (the result is not sensitive to this upper limit).

The penetration depth of a monopole of a particular energy is deduced from the parameterisation of the range given in Section~\ref{range_monopoles_matter} and Equation~\ref{eq:rm}. The primary energy to produce a monopole of this energy is deduced under the assumption that the monopole has negligible momentum in the centre-of-mass frame.

The best limit is from the ancient mica~\cite{Fleischer1969c,Price1984a,Price1986,Ghosh1990} (Section~\ref{mica_searches}). However, the depth throughout geological time is unknown but thought to be less than 10~km \cite{Fleischer1969c}, thus primaries must be assumed to give monopoles enough energy to reach depths greater than 10~km. This constraint means that mica cross section limits are valid for masses above $\sim 5$~TeV.

For masses in the range $1<M<5$ TeV the best limit is from IceCube~\cite{IceCube2013} (Section~\ref{Cherenkov_light}). Monopoles of this mass have sufficient energy to reach the detector depth in IceCube.

For the lowest masses, the SLIM high-altitude experiment gives the best limit~\cite{SLIM2008a} (Section~\ref{cosmic_ionisation_passive}).

\begin{table}
\begin{tabular}{|c|c|c|c|c|c|}
\hline
monopole mass (TeV) & 0.316 & 1.0 & 3.16 & 10.0 & 31.6 \\\hline
Experiment & SLIM & IceCube& IceCube & mica & mica \\ \hline
Mean value of $\sqrt{s}$ (TeV) & 0.86 & 8.9 & 8.63 & 27.3 & 86.3 \\ \hline
$\sigma$ (cm$^2$)  &  0.86~10$^{-32}$ & 0.48~10$^{-30}$ & 0.42~10$^{-30}$ &
0.24~10$^{-29}$ & 0.35~10$^{-27}$ \\ \hline
\end{tabular}
\caption{The best cross-section lower limits for various monopole masses, obtained from cosmic flux limits. }
\label{tab:sigmas}
\end{table}

\begin{figure}[htb]
\begin{center}
\hspace{-7mm}
\includegraphics[width=0.725\textwidth]{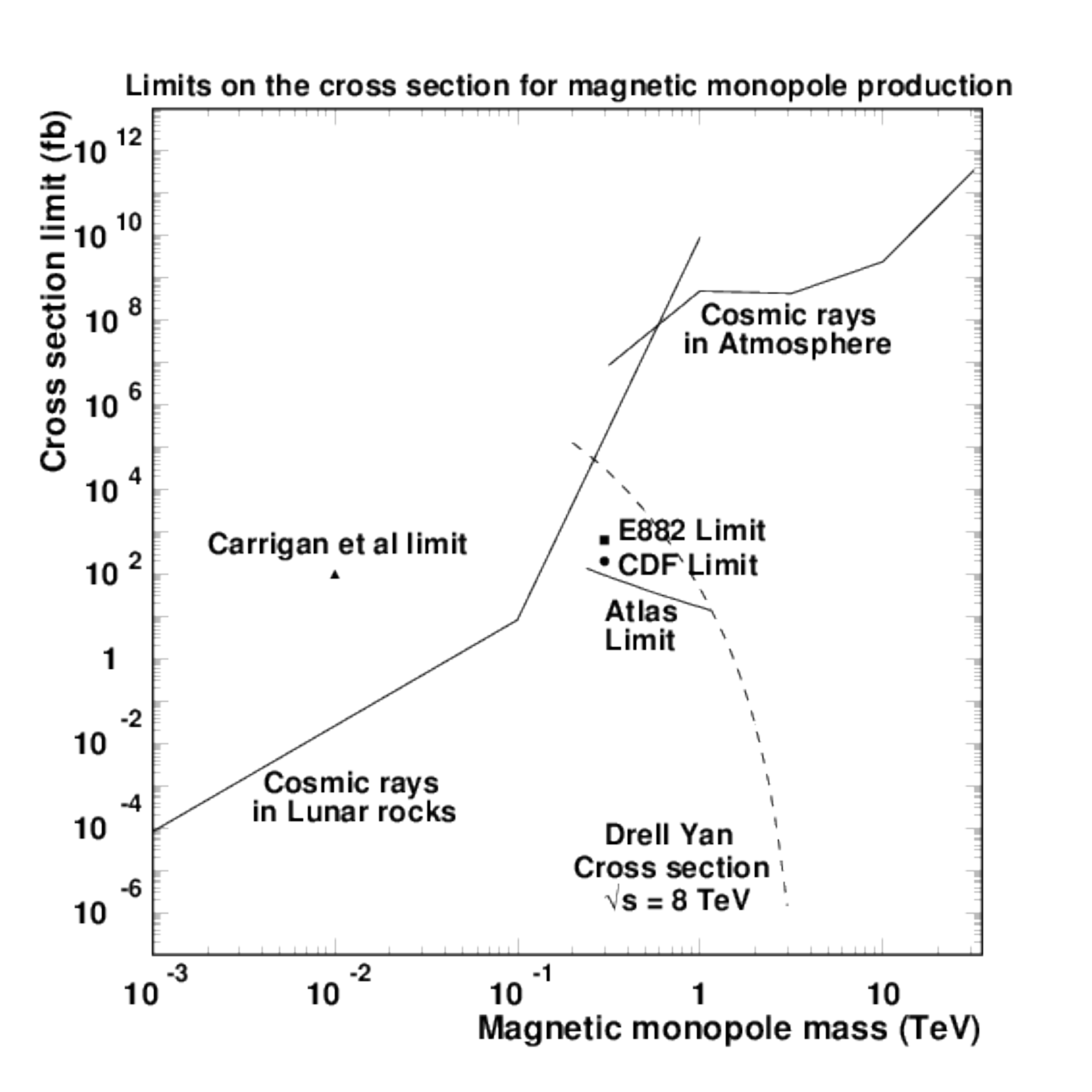}
\caption{The cross section limit from cosmic-ray production of monopoles (this work). Limits from accelerator-based experiments~\cite{Carrigan1973,Bertani1990,CDF2006,ATLAS2012a} are also shown together with the cross section for Drell-Yan production of monopole pairs~\cite{Milton2006} at the LHC (2012) energy. }
\label{fig:sigmas}
\end{center}
\end{figure}

The best cross section limits are summarised in Table~\ref{tab:sigmas} and plotted in Figure~\ref{fig:sigmas}. These are compared with the limit determined by accelerator-based experiments~\cite{Carrigan1973,Bertani1990,CDF2006,ATLAS2012a} (the latest and most stringent one from ATLAS~\cite{ATLAS2012a}) and with the theoretical production cross section via the Drell Yan process \cite{Milton2006}. Note that the cosmic-ray primary spectrum falls rapidly with energy (approximately as $1/E^3$). Hence the mean value of the $\sqrt{s}$ to produce each mass is only of order 40\% above the threshold energy. The mean values of the $\sqrt{s}$ are given in the table for each mass. The cross section for monopole production is expected to rise rapidly with energy.

Thanks to high luminosities, modern colliders provide the best constraints in the mass range up to half the collision energy. Cross section limits for monopole production by cosmic rays are weaker but they extend to much higher masses and they are relatively model independent in terms of the kinematic distributions of produced monopoles. A model dependence comes in from the assumption that the monopoles are
nearly at rest in the centre of mass system so that one can compute the energy in the lab frame. At masses below 1 TeV, monopoles need a large energy to penetrate through the atmosphere, and experiments at high altitude become more attractive. Searches in matter are the most constraining for masses below 100 GeV.

\subsection{Summary}
\label{cosmic_summary}

\begin{figure}[tb]
\begin{center}
\includegraphics[width=0.9\linewidth]{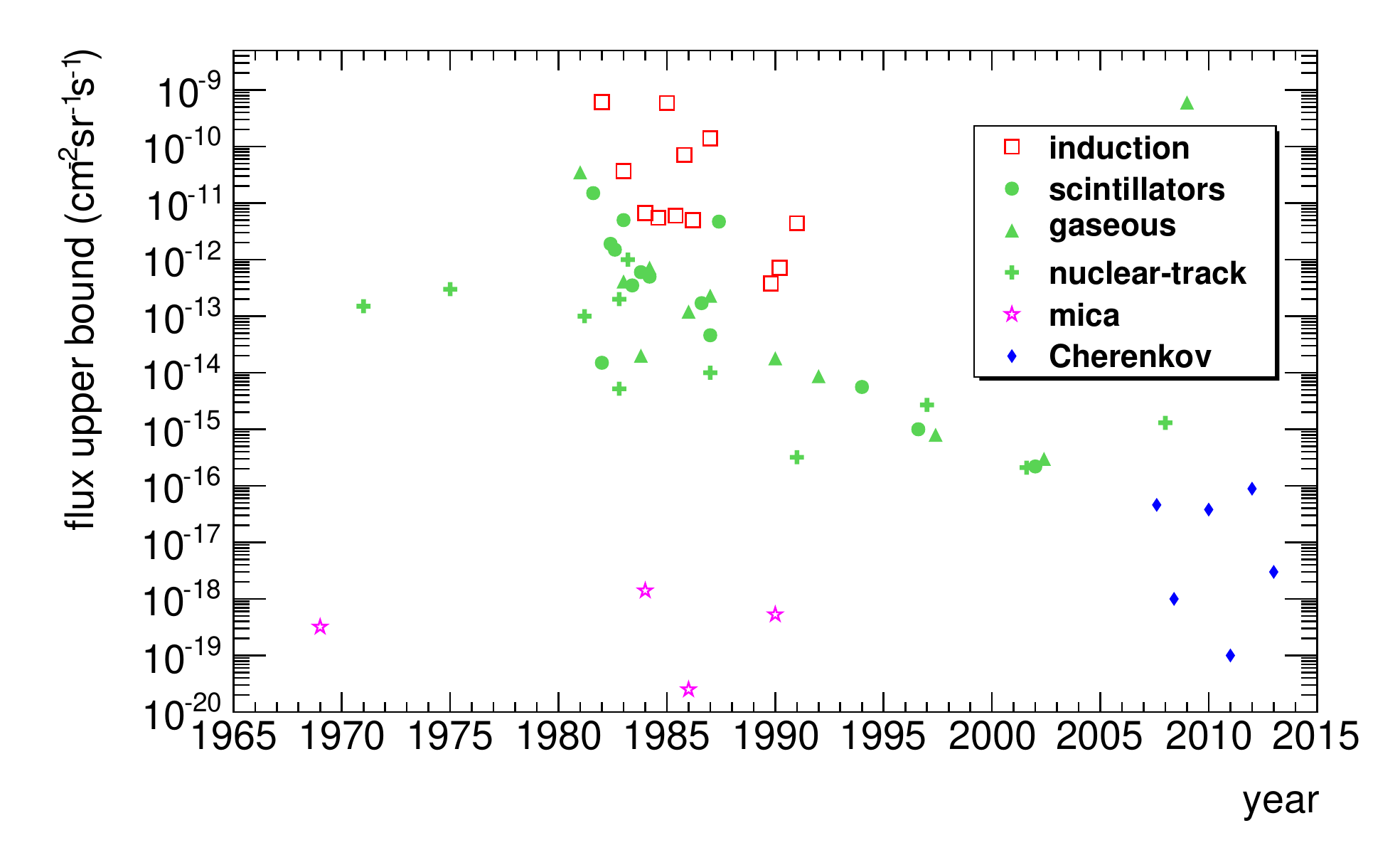}
\caption{Single-experiment monopole flux limits (90\% confidence level) versus year of publication. The limits assume monopoles with $g\geq g_D$ with mass and energy such that they would possess a velocity in the sensitivity range of the detector (see Fig.~\ref{fig:betagamma}). The mica limits are guaranteed if the monopole has $g\geq 2g_D$.}
\label{fig:time}
\end{center}
\end{figure}

\begin{figure}[tb]
\begin{center}
\includegraphics[width=0.95\linewidth]{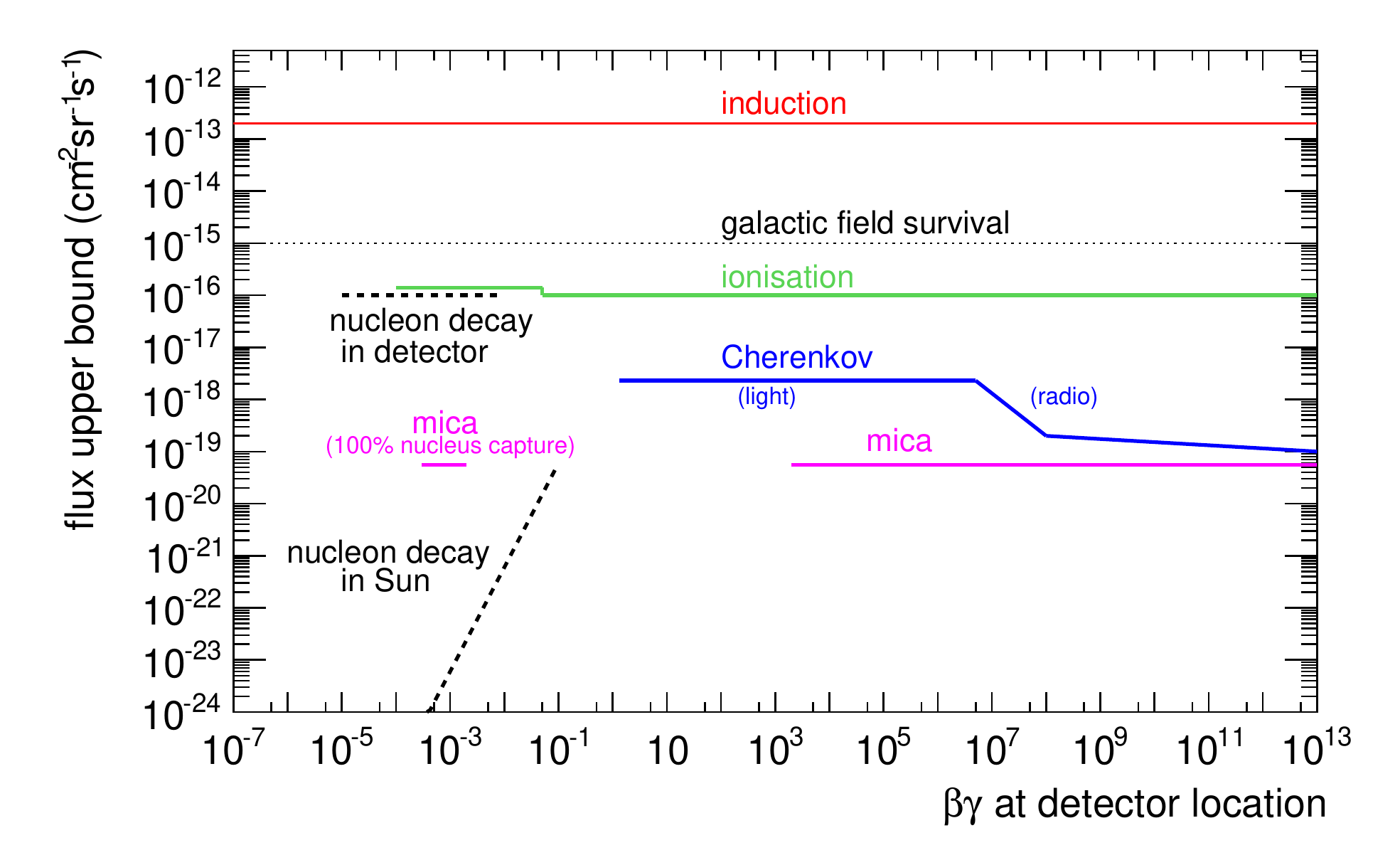}
\caption{Summary of combined cosmic monopole flux limits (90\% confidence level) as a function of velocity times Lorentz factor. The limits assume monopoles with $g\geq g_D$ with mass and energy such that they would possess the given velocity after reaching the detector. The high-velocity mica limit applies to $g\geq 2g_D$. The limits based on nucleon decay in detectors assume catalysis cross sections $\sigma_c$ of the order of 100~mb.}
\label{fig:betagamma}
\end{center}
\end{figure}

\begin{figure}[tb]
\begin{center}
\includegraphics[width=0.95\linewidth]{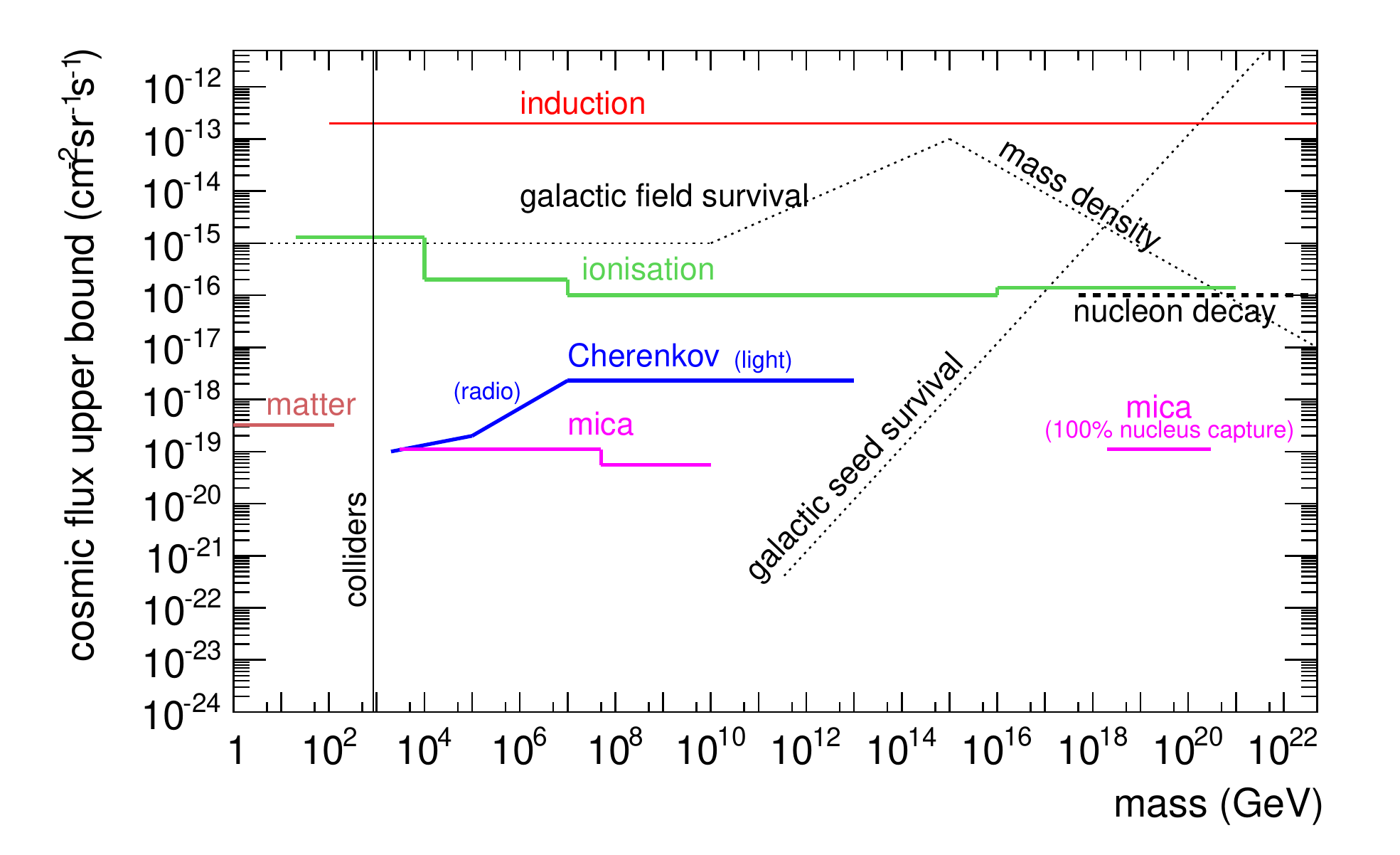}
\caption{Summary of combined cosmic monopole flux limits (90\% confidence level) as a function of mass. The limits assume monopoles with $g\geq g_D$ which possess an energy of $10^{13}$~GeV (expected after acceleration by galactic magnetic fields~\cite{Wick2003}). The high-velocity mica limit applies to $g\geq 2g_D$. The limits based on nucleon decay in detectors assume catalysis cross sections $\sigma_c$ of the order of 100~mb.}
\label{fig:mass}
\end{center}
\end{figure}

Experiments capable of directly probing the flux of massive interacting particles can be based on a very large variety of techniques, such as magnetic induction, ionisation signals in active or passive detectors, Cherenkov emission, vibrations, extended air showers, and space-borne spectrometers. A review of constraints and assumptions for each search was given, mostly within models of magnetic monopoles, exotic electrically charged massive particles, $Q$-balls, and nuclearites. The limited detector areas make the detection of fluxes much lower than that $\sim1$ particle per km$^2$ per year difficult. In the case of massive compact objects (such as nuclearites), lower fluxes can still be probed using dedicated techniques, as will be discussed in Section~\ref{compact_objects}.

The monopole hypothesis is the most prevalent. The evolution of flux limits obtained since 1970 (within the assumptions of each experiment) as a function of time is presented in Fig.~\ref{fig:time}. This shows that the current focus is on obtaining very strong constraints for relativistic monopoles with Cherenkov techniques at large neutrino telescopes. Given the impressive advances in modern automatic microscope technologies, the possibility of new search with large nuclear-track detector arrays and ancient mica should also be kept in mind. A summary of direct and indirect constraints on the flux of cosmic monopoles is given in Table~\ref{tab:summary_monopoles} and in Figs.~\ref{fig:betagamma} and ~\ref{fig:mass}. While Fig.~\ref{fig:betagamma} presents combined flux limits as a function of velocity assuming that a monopole of such velocity (with charge $\geq g_D$) traverses the detector, Fig.~\ref{fig:mass} presents the same results as a function of monopole mass assuming a monopole energy of $10^{13}$~GeV (expected after acceleration by galactic magnetic fields~\cite{Wick2003}, see Section~\ref{monopoles_in_galaxy}).

\input{smp-summarytablemono}

The experiments based on ionisation energy loss are generally also sensitive to composite objects such as electrically charged $Q$-balls and nuclearites (nuclearite flux limits are reported in Fig.~\ref{fig:nuclearite_flux_limits}). One interesting challenge is to probe the possibility that such objects could constitute dark matter, in particular this is not excluded for masses above $\sim 100$~g. A more extensive search with ancient mica can potentially probe the $100-10^5$~g mass range, while seismic or gravitational techniques would be needed to probe higher masses (see Section~\ref{compact_objects} and Fig.~\ref{fig:nuclearite_flux_limits}).

As current models cannot accurately predict monopole production and fate during the long and complex history of the Universe, the fact that monopoles evaded detection both in cosmic rays and in matter for wide ranges of monopole masses does not preclude their existence. The monopole density could have been diluted down to quantities below current sensitivities by cosmic inflation (see Section~\ref{relic_abundances}). There are also scenarios in which monopoles would be copiously produced in the early Universe but still almost invisible to experiments. One such scenario is the confinement of monopoles into long-lived monopolium states (see Section~\ref{monopole_pheno}).
In another scenario monopoles belong to a hidden sector and, through mixing, manifest themselves as objects with small electric charges (see Section~\ref{GUT_mono}). It can also be speculated that most monopoles which were produced bound to matter early, perhaps during nucleosynthesis; Section~\ref{monopole_binding} discusses  monopole binding. If they are also very massive, they could subsequently have remained trapped in the cores of stars and dead-star remnants.


Accelerator experiments are complementary as they allow for a controlled laboratory environment in which new particles, if they exist in the accessible mass range, are more or less guaranteed to be produced and detected at a high rate. An example of this complementarity was presented here with a comparison between monopole production cross-section limits at colliders and in atmospheric interactions by high-energy cosmic rays (Fig.~\ref{fig:sigmas}).

%% file: smp-summarytablemono.tex
{\footnotesize
\LTcapwidth=\textwidth
\begin{longtable}{|l|l|l|c|c|}
    \hline
    technique            & location     & reference & velocity range                       & limit (cm$^{-2}$sr$^{-1}$s$^{-1}$)                          \\
    \hline
    induction            & sea level    &
    ~\cite{Cabrera1982,Cabrera1983a,Incandela1984,Bermon1985,Ebisu1985,Caplin1985,Incandela1986,Caplin1986,Cromar1986,Ebisu1987,Bermon1990,Huber1990,Huber1991,Gardner1991}
                                            & all $\beta$                          & 2$\cdot 10^{-13}$                        \\
    scintillator         & underground   &
    ~\cite{Alekseev1982,Mashimo1983,Groom1983,Kawagoe1984,Tsukamoto1987,Shepko1987,MACRO1994,MACRO1997,MACRO2002a}
                                            & $10^{-4}<\beta<0.1$                  & 2.5$\cdot 10^{-16}$                        \\
                         &              &
    ~\cite{Alekseev1982,Mashimo1983,Kawagoe1984,Tsukamoto1987}
					    & $0.1<\beta<0.5$                      & 1.3$\cdot 10^{-14}$                        \\
                         &              &
    ~\cite{Mashimo1983,Kawagoe1984,Tsukamoto1987}
				            & $0.5<\beta<1$                        & 1.3$\cdot 10^{-13}$                        \\
                         & sea level    &
    ~\cite{Barish1987,Tarle1984,Liss1984}
				            & $3\cdot10^{-4}<\beta<10^{-3}$        & 7.8$\cdot 10^{-13}$                        \\
                         &              &
    ~\cite{Mashimo1982,Bonarelli1982,Bonarelli1983}
					    & $10^{-3}<\beta<0.5$                  & 2.9$\cdot 10^{-13}$                        \\
    plastic streamer tube & underground  &
    ~\cite{Battistoni1983}
                                            & $10^{-4}<\beta<2\cdot10^{-3}$        & 4.6$\cdot 10^{-12}$                        \\
    gaseous detector     & underground  &
    ~\cite{Krishnaswamy1984,MACRO1997,MACRO2002a}
                                            & $10^{-4}<\beta<5\cdot10^{-3}$        & 2.8$\cdot 10^{-16}$                        \\
                         &              &
    ~\cite{Bartelt1983,Krishnaswamy1984,Soudan21992,L3C2009}
                                            & $5\cdot10^{-3}<\beta<1$              & 6$\cdot 10^{-15}$                          \\
                         & sea level    &
    ~\cite{Ullman1981,Kajino1984a,Hara1986,Masek1987,Buckland1990}
                                            & $10^{-4}<\beta<1$                    & 1.4$\cdot 10^{-14}$                        \\
    nuclear-track detector & underground  &
    ~\cite{Nakamura1987,Orito1991,MACRO1997,MACRO2002a}
                                            & $2\cdot 10^{-5}<\beta<2\cdot 10^{-4}$  & 3$\cdot 10^{-16}$                        \\
                         &              &
    ~\cite{Nakamura1987,Orito1991,MACRO1997,MACRO2002a}
                                            & $2\cdot 10^{-3}<\beta<0.02$  & 2$\cdot 10^{-16}$                        \\
                         &              &
    ~\cite{Nakamura1987,Orito1991,MACRO1997,MACRO2002a}
                                            & $0.02<\beta<1$               & $10^{-16}$                        \\
                         & sea level    &
    ~\cite{Fleischer1971,Bartlett1981,Doke1983}
                                            & $0.04<\beta<1$                       & 5.2$\cdot 10^{-15}$                        \\
                         & high altitude&
    ~\cite{Price1984b,SLIM2008a}
                                            & $3\cdot10^{-5}<\beta<2\cdot10^{-4}$  & 3$\cdot 10^{-14}$                          \\
                         &              &
    ~\cite{Price1975,Price1978,Kinoshita1981,Barwick1983,Price1984b,SLIM2008a}
                                            & $0.03<\beta<1$                       & 6.5$\cdot 10^{-16}$                        \\
                         & ancient mica     &
    ~\cite{Fleischer1969c,Price1984a,Price1986,Ghosh1990}
                                            & $4\cdot10^{-4}<\beta<0.5$            & $5.5\cdot 10^{-20}$~
    $^($\footnote{Assumes that the monopole picks an Al or Mg nucleus while traversing the rock on its way to the mica layer.}$^)$ \\
                         &              &
    ~\cite{Fleischer1969c,Price1984a,Price1986,Ghosh1990}
                                            & $\gamma>2000$                        & $5.5\cdot 10^{-20}$~
    $^($\footnote{Assumes that bare ultrarelativistic monopoles with $g=g_D$ induce tracks in mica. Valid for $|g|=2g_D$, $\beta>0.9$ and $|g|\geq 3g_D$, $\beta>0.1$.}$^)$ \\
    Cherenkov light      & deep water/ice &
    ~\cite{ANTARES2012}
                                            & $0.6<\beta<0.8$                      & 5$\cdot 10^{-17}$                          \\
                         &              &
    ~\cite{BAIKAL2008,AMANDA2010,ANTARES2012,IceCube2013}
                                            & $0.8<\beta<1$                        & 2.3$\cdot 10^{-18}$                          \\
    Cherenkov radio      & South Pole   &
    ~\cite{RICE2008,ANITA2011}
                                            & $10^7<\gamma<10^{13}$                & 2$\cdot 10^{-19}$                          \\
    nucleon decay        & underground  &
    ~\cite{Battistoni1983,Errede1983,Krishnaswamy1984,Kajita1985,Stone1985,Bartelt1987,IMB1994,BAIKAL1998,MACRO2002b}
                                            & $10^{-5}<\beta<10^{-2}$        & $10^{-16}$
    $^($\footnote{Assumes a nucleon decay catalysis cross section $\sigma=100$ mb.}$^)$ \\
                         & deep ice         &
    ~\cite{IceCube2014}
                                            & $\beta\sim 10^{-3}$                & $10^{-17}$
    $^($\footnote{Assumes a nucleon decay catalysis cross section $\sigma=1000$ mb.}$^)$ \\
                         & sea level    &
    ~\cite{Bosetti1983}
                                            & $5\cdot10^{-4}<\beta<0.05$           & 3$\cdot 10^{-12}$
    $^($\footnote{Assumes a nucleon decay catalysis cross section $\sigma=100$ mb.}$^)$ \\
                         & in the Sun       &
    ~\cite{Kajita1985,SuperKamiokande2012}
                                            & $\beta<0.1$                          & $6\cdot 10^{-24}\left(\frac{\beta}{10^{-3}}\right)^2$
    $^($\footnote{Assumes that monopoles impacting the Sun remain trapped, and a nucleon decay catalysis cross section $\sigma>1$ mb.}$^)$ \\
    old neutron stars                       & galaxy     &
    ~\cite{Kolb1982,Dimopoulos1982b,Freese1983b,Kolb1984}
                                            & all initial $\beta$                  &   $10^{-23}$
    $^($\footnote{Assumes nucleon decay catalysis cross section times nucleon-monopole relative velocity of the order of 0.1~mb, 100\% capture probability for monopoles incident on the neutron star, and negligible effect from monopole-antimonopole annihilations.}$^)$ \\
    collection (air) & sea level  &
    ~\cite{Malkus1951,Carithers1966,Bartlett1981}
                                            & stop in atmosphere               & $4.1\cdot 10^{-16}$
    $^($\footnote{Assumes monopoles with positive charge.}$^)$                          \\
    extraction (deposits) & ocean bottom &
    ~\cite{Goto1963,Fleischer1969a,Fleischer1969b,Kolm1971,Carrigan1976}
                                            & stop in ocean                    & 4.8$\cdot 10^{-19}$
    $^($\footnote{Assumes that monopoles would be efficiently extracted from the solid lattice of the samples by applying a strong magnetic field --- this has been criticised in Ref.~\cite{Milton2006}. However, Ref.~\cite{Kovalik1986} analysed a variety of deep-sea deposits with the induction technique, and although not explicitly quoted by the authors, this negative result provides a robust limit of similar magnitude.}$^)$ \\
    matter induction         & Moon rocks       &
    ~\cite{Eberhard1971,Ross1973}
                                            & stop on Moon surface             & 6.4$\cdot 10^{-19}$                           \\
                             & polar rocks  &
    ~\cite{Bendtz2013a}
                                            & stop inside Earth                & 2.2$\cdot 10^{-14}$
    $^($\footnote{Assumes $M\cdot g_D/g<4\cdot 10^{14}$~GeV, whole-mantle convection, and monopole binding to nuclei.}$^)$    \\
    Earth heat               & Earth core   &
    ~\cite{Carrigan1980}
                                            & stop inside Earth                & 8$\cdot 10^{-14}$
    $^($\footnote{Assumes $4\cdot 10^{14} < M < 10^{17}$~GeV, a stable dipole geomagnetic field in the Earth's core when there is no reversal, and the presence of both monopoles and anti-monopoles.}$^)$    \\

    magnetic field survival  & galaxy       &
    ~\cite{Turner1982}
                                            & all $\beta$                          & $10^{-15}$                                 \\
                             & galactic seed&
    ~\cite{Adams1993}
                                            & all $\beta$                          & $1.2\cdot 10^{-33}M$
   $^($\footnote{This formula requires the monopole mass ($M$) to be expressed in GeV. The bound assumes that the current galactic magnetic field grew starting from a small value, and is valid for $M>3\cdot 10^{11}$~GeV. For smaller masses, another formula applies, dependent on the value of the field seed~\cite{Adams1993}.}$^)$                                            \\
    mass density             & Universe     &
    ~\cite{Turner1982}
                                            & all $\beta$                          & $5.4\cdot 10^4 \frac{\beta}{M}$
  $^($\footnote{This formula requires the monopole mass ($M$) to be expressed in GeV and assumes a uniform distribution.}$^)$          \\
  \hline
  \caption{\label{tab:summary_monopoles} Summary of combined flux limits for cosmic monopoles with $g\geq g_D$ that would reach the detector and possess a velocity in the given range. Experiments with sensitivity to low speeds ($\beta<0.04$) can probe very high masses ($M>10^{15}$~GeV). Searches for monopoles that stopped in the atmosphere, in oceans and on the Moon surface probe only low masses ($M<100$~GeV). Specific assumptions needed for the limits to be valid are specified as footnotes.}
\end{longtable}
}

%% file: smp-compactobjects.tex
\section{Searches for macroscopic composite objects}
\label{compact_objects}

In this section we discuss composite objects so massive that the flux expected if they constituted dark matter is unconstrained by cosmic array detectors. This roughly corresponds to masses larger than 100~g. Possible candidates include strangelet nuclearites (see Section~\ref{sec:strangelets}), fermionic exotic compact stars (see Section~\ref{compact_stars}), and primordial black holes (see Section~\ref{black_holes}). Cosmic fluxes for such objects can be constrained using specific techniques such as gravitational lensing. In addition, such objects could be gravitationally captured during star formation~\cite{Capela2013b} and remain trapped in the cores of stars, planets and asteroids, or form meteoroids. In such a scenario, their presence could be revealed in the form of anomalies in the properties of Solar System bodies.


\subsection{Gravitational lensing}
\label{grav_lensing}

Massive compact objects with masses of the order of that of planets could be detected through occasional magnification of light from distant sources by gravitational lensing.

\subsubsection{Microlensing}

Experiments such as EROS and MACHO have been monitoring temporal variation in fluxes from millions of background stars~\cite{Dalcanton1994,MACHO1998}. In such measurements, the observed frequency of brightening expected when a dark object passes close to the line of sight which allows the exclusion of objects in the mass range $4\cdot 10^{26}<M<4\cdot 10^{33}$~g as a major component of the dark matter halo~\cite{MACHO1998}. Recent results obtained with the Kepler satellite extend this mass range down to $4\cdot 10^{24}$~g~\cite{Griest2013}. These limits apply to any type of compact object, including primordial black holes and nuclearites. They are reported in Fig.~\ref{fig:nuclearite_flux_limits}.

\subsubsection{Femtolensing of gamma-ray bursts}

Interference effects between lensed images of a pointlike astrophysical source can be exploited to identify asteroid-sized compact objects, a technique called femtolensing~\cite{Gould1992}. Gamma-ray sources are appropriate because the time delay induced by the lens would be comparable to the oscillation period of a gamma ray. Gamma-ray burst sources can be very distant and still sufficiently bright: since the probability to have a candidate in the line of sight scales with distance, this is an additional advantage.

The femtolensing technique was applied on gamma-ray bursts to constrain the abundance of compact objects with masses in the range $10^{17}<M< 10^{21}$~g~\cite{Marani1999,Barnacka2012}. Recent results using the Fermi satellite set limits on the contribution to dark matter at the level of 8\% for $5\cdot 10^{17}<M<2\cdot 10^{20}$~g~\cite{Barnacka2012}. These limits apply to compact objects whose physical radius is smaller than the Einstein ring radius, which in the search above is the case for e.g. primordial black holes, exotic fermionic stars, and nuclearites. They are reported in Fig.~\ref{fig:nuclearite_flux_limits}.

\subsection{Meteoroids}
\label{meteoroids}

When colliding with the Earth or other planets, new massive objects would give rise to signatures such as anomalous meteors, seismic signals, and fossil traces. For a given mass density, the impact rate is proportional to the speed and inversely proportional to the mass of the object. The impact rates for various bodies in the Solar System are summarised in Table~\ref{tab:dark_object_fluxes} assuming a speed corresponding to galactic rotation (250~km/s) and a density corresponding to the local density of dark matter ($\sim 10^{-24}$~g/cm$^3$). Also, some Solar System bodies could be made partly or entirely of extended objects in a new state of matter. If dressed by ordinary matter (e.g. by gravitational interaction), such objects could be observed as asteroids and comets with anomalous properties.

\begin{table}
\begin{tabular}{|l|c|c|c|c|c|}
  \hline
  object mass (g)  & Eros & Moon & Earth & Jupiter & Sun \\
  \hline
  1         & $10^4$     & $3\cdot 10^{8}$   & $4\cdot 10^{9}$  & $4\cdot 10^{11}$  & $5\cdot 10^{13}$ \\
  $10^{3}$  & 10         & $3\cdot 10^{5}$   & $4\cdot 10^{6}$  & $4\cdot 10^{8}$   & $5\cdot 10^{10}$ \\
  $10^{6}$  & $10^{-2}$  & $3\cdot 10^{2}$   & $4\cdot 10^{3}$  & $4\cdot 10^{5}$   & $5\cdot 10^{7}$ \\
  $10^{9}$  & $10^{-5}$  & 0.3               & 4                & $4\cdot 10^{2}$   & $5\cdot 10^{4}$ \\
  $10^{12}$ & $10^{-8}$  & $3\cdot 10^{-4}$  & $4\cdot 10^{-3}$ & 0.4               & 50 \\
  $10^{15}$ & $10^{-11}$ & $3\cdot 10^{-7}$  & $4\cdot 10^{-6}$ & $4\cdot 10^{-4}$  & $5\cdot 10^{-2}$ \\
  $10^{18}$ & $10^{-14}$ & $3\cdot 10^{-10}$ & $4\cdot 10^{-9}$ & $4\cdot 10^{-7}$  & $5\cdot 10^{-5}$ \\
  \hline
\end{tabular}
  \caption{\label{tab:dark_object_fluxes} Expected number of collisions per year between an object of given mass and various bodies in the Solar System, assuming that the object constitutes 100\% of the local dark matter density and that it possesses a typical galactic velocity.}
\end{table}

\subsubsection{Seismic detection}
\label{seismic}

It has been argued that extended exotic compact objects would manifest a number of signatures such as anomalous earthquakes when traversing a rocky planetoid~\cite{DeRujula1984,Bernard1984,Khriplovich2008,Luo2012,Rafelski2013}. For instance, nuclearites with $M>1000$~g would heat and break rock to such an extent that they would produce a characteristic seismic track if at least seven stations are used~\cite{Herrin1996}.

Absence of evidence for epilinear seismic signals on Earth allowed to exclude nuclearites as the dominant dark matter component in the mass range $10^5 < M < 3\cdot 10^8$~g~\cite{Herrin2006}. In addition, the measurement of the total amount of seismic energy obtained with the five seismic stations implanted on the Moon by the Apollo astronauts were interpreted to set conservative limits on the rate of nuclearite impacts, concluding that the flux does not exceed one tenth of the dark matter density in the mass range $5\cdot 10^4 < M < 10^6$~g~\cite{Herrin2006}. These limits are reported in Fig.~\ref{fig:nuclearite_flux_limits}. The same authors pointed out that the larger area on Earth leads to eight times as many detectable collisions, while the Moon has the advantage of much lower backgrounds thanks to a lower seismic activity and the lack of noise from winds and waves. A sufficient number of seismometers judiciously placed on the Moon would allow to detect objects with masses two orders of magnitude lower than on Earth, potentially down to the kg level~\cite{Herrin2006}. As this mass range has not been explored by any other means (existing searches in ancient mica reach up to 160~g~\cite{Price1986}, see Section~\ref{mica_searches}), this would motivate plans for seismic nuclearite searches as a part of any future trip to the Moon~\cite{Herrin2006,Banerdt2006,Herrin2007} or any other seismologically quiet planetary body.

Seismic detection techniques should also be sensitive to impacts and punctures from compact objects other than nuclearites. Matter inside compact exotic fermion stars (see Section~\ref{compact_stars}) would have a density over 10 orders of magnitude higher than quark matter, leading to cross sections $\sim 10^6$ smaller for the same mass (neglecting gravitational effects)~\cite{Sandin2007}. As a consequence, fermion star fluxes below what would saturate the dark matter density would be extremely difficult to detect using seismic signals on the Earth or the Moon. In the case of black holes, the interactions with matter are dominated by gravitational acceleration and Hawking radiation~\cite{Greenstein1984,Khriplovich2008} (see Section~\ref{interactions_black_holes}). If one does not invoke exotic scenarios (see Section~\ref{black_holes}), only primordial black holes with $M>10^{15}$~g would remain today. Assuming that black holes with $M\sim 10^{16}$ g constitute dark matter, that corresponds to a flux of $10^{-34}$ cm$^{-2}$sr$^{-1}$s$^{-1}$, or about four collisions with Earth every $~\sim 10^7$ years~\cite{Danaher2012,Luo2012}. The consideration of the excitation of seismic waves by non-colliding close encounters enhances the expected event rate by factor 100~\cite{Luo2012}. This corresponds to 1 event every 25000 years.

\subsubsection{Meteors}

Compact objects with masses such that they can occasionally impact
larger rocky bodies (see Table~\ref{tab:dark_object_fluxes}) would
exhibit salient features such as punctures and exit
wounds~\cite{Rafelski2013}. Such objects could possibly be dressed in
loosely bound ordinary matter and thus look like meteors when traversing the atmosphere, although with unusual stability and absence of normal impactor material~\cite{Rafelski2013}. Anomalous impact events on Earth presenting one or several of these features have been subject of much speculation. For instance, an alternative interpretation for the 1908 Tunguska explosion involves an incident black hole~\cite{Jackson1973}, although this hypothesis has been disproved based on the lack of observation of an exit event~\cite{Beasley1974}. The Tunguska event and other anomalous events were also proposed to be caused by meteors made of mirror matter~\cite{Foot2002,Foot2003a} (see Section~\ref{dark_atoms}). However, even if details of such events are associated with mysteries which are the subject of speculation and scientific debate, conventional models involving ordinary meteors cannot be disproved.

The energy dissipated by a nuclearite with $M>1$~g when passing through rock would be enough to melt the rock in its wake, producing a macroscopic track or pipe~\cite{DeRujula1984}. If such fossil tracks in ancient rock could be identified, they would be interesting indications of the passage of exotic compact objects. Such evidence could be confirmed if the track is associated with high concentrations of fullerenes --- exceptionally long-lived molecules which are not expected to be created in significant amounts by processes other than violent heating~\cite{Collar1999}.

Meteor-like events with anomalously high velocities relative to Solar-System meteors can be identified using cosmic-ray air Cherenkov telescopes. If observed, such events would be signatures of, e.g., nuclearites in the mass range $10^{-3}-10^3$~g~\cite{Porter1985,Gorham2012}. It was also suggested that high-velocity transits could be identified by detecting the emitted thermal radio signal with, e.g., balloon-borne instruments such as ANITA~\cite{Gorham2012}.

An even more promising observatory for future anomalous meteor searches is the proposed JEM-EUSO telescope, to be installed on board the International Space Station~\cite{Bertaina2014}. This instrument is designed to have a very large field of view from above the Earth's atmosphere and to be able to detect events of very short duration. Being space-borne, it has also the advantage to be immune from adverse weather conditions. Exotic meteors such as nuclearites would be easily distinguished from normal meteors using a variety of criteria such as velocity, light profile, and fireball profile (e.g., events developing at low heights, or moving upwards). It is estimated that only 24 hours of JEM-EUSO operation would allow to constrain the flux of nuclearites with $M>1$ g at the level of $10^{-20}$~cm$^{-2}$sr$^{-1}$s$^{-1}$~\cite{Bertaina2014}. That is ten times better than the limit obtained from existing searches in ancient mica (see Section~\ref{mica_searches} and Fig.~\ref{fig:nuclearite_flux_limits}), thus potentially allowing to close the gap of nuclearite masses between those constrained by ancient mica analyses and those constrained by seismic detection.

\subsubsection{Comets}

An intriguing proposition is that the major component of comets would be in the form of extended invisible exotic matter, e.g., dark atoms/mirror matter~\cite{Foot2003a} (see Section~\ref{dark_atoms}). A small kinetic mixing between the photon and the mirror photon ($\chi\sim 10^{-9}$) could possibly result in a mirror body being dressed by ordinary dust and behave like comets do (see Section~\ref{interactions_mirror}). This speculative idea is inspired by the observation of fewer returning comets than expected (so-called ``comet fading problem''), indicating that they often disappear after their first passage near the Sun~\cite{Levison2002}. Detailed observations of sun-grazing comets such as ISON in 2013 showed that they often disintegrate near perihelion without leaving any visible debris~\cite{Knight2014}. Another puzzling observation obtained from the Deep Impact space probe collision with comet Tempel 1 in 2005 is that the comet material has an extremely high porosity~\cite{AHearn2007}. Even if more standard explanations cannot be ruled out at present, these three facts --- comet fading, disintegration without debris, and high porosity --- could be explained if comets are hypothesised to be mostly made of extended (few km) aggregates of invisible matter interpenetrated by loosely bound ordinary matter. Once all ordinary matter has evaporated, such objects would become completely dark.

Comets are complex objects and it seems clear that many aspects of comet composition and behaviour are still poorly understood. The data to be collected by the Rosetta spacecraft, which successfully made a rendezvous with comet 67P/Churyumov–Gerasimenko in August 2014, will probably be very valuable in that respect.

\subsubsection{Asteroids}

A very low rate of low-diameter ($<100$ m) craters relative to the rate of large craters is observed on asteroid Eros. It has been proposed that this could be explained if most impactors were mostly made of mirror matter~\cite{Foot2003a}. Later is was shown, however, that seismic shaking provoked by impacts on the asteroid can effectively erase small craters~\cite{Thomas2005,OBrien2009}. In such conditions, it seems difficult to draw strong conclusions about the nature of the impactors.

Rotation speeds of asteroids with irregular shapes are known to be either increased or decreased over time due to the Yarkovsky-O'Keefe-Radzievskii-Paddack (YORP) effect~\cite{Lowry2007,Taylor2007,Kaasalainen2007}, explaining the observed broadening in spin rate distributions for asteroids under 10~km in diameter~\cite{Taylor2007}. There is a limit in rotation speed beyond which centrifugal forces will cause the asteroid to disintegrate. In a "rubble pile" model (assuming low tensile strength), it is predicted that no asteroids would survive rotation speeds higher than 2.2 rotations per hour, and indeed a sharp barrier is observed at this value for asteroids in the size range $1-10$~km~\cite{Harris2006}. However, a significant fraction of small asteroids ($<100$~m in diameter) are observed to spin dramatically faster than this barrier~\cite{Harris2006,Eubanks2014}. This could be explained if asteroids possessed a core of ultradense matter (such as quark matter): in the case of small asteroids, the ultradense core would constitute a large fraction of the total mass and a negligible fraction of its volume, allowing for surface material to remain gravitationally bound at much higher rotation speeds than would be possible otherwise~\cite{Eubanks2014}. However, the possibility of a high tensile strength (as opposed to loose regolith) in the case of small asteroids cannot be ruled out using current data~\cite{Harris2006}.

\subsection{Earth core radiography}

The possibility that the Earth would possess a small core of ultradense matter (e.g. of nuclear density if constituted of quark matter) can potentially be tested directly using neutrinos as probes. A measurement of the absorption of atmospheric neutrinos with energies of tens of TeV when they pass through Earth is capable of revealing its density distribution~\cite{GonzalezGarcia2008}. Using the  IceCube detector, 10 years of data would provide at least $3\sigma$ separation in averaged density between core and mantle~\cite{GonzalezGarcia2008,Hoshina2012}, and the prospects will become better at future large-scale neutrino observatories. Another method is to use the much higher flux of atmospheric neutrinos at lower energies ($5-10$~GeV) and exploit matter effects of neutrino oscillations to extract the electron density. The future Hyper-Kamiokande experiment would have the capability to perform such an analysis~\cite{HyperKamiokande2011}.

\subsection{Indirect searches}
\label{compact_indirect}

The abundance of specific types of compact objects such as nuclearites and black holes can be constrained based on the effects they would be expected to have on the Earth, the Sun, neutron stars, and cosmic rays.

\subsubsection{Geothermal effects}


Constraints from geothermal energy budget of the Earth can be used to limit the flux times radiation efficiency of compact objects. In a model of antiquark matter nuggets~\cite{Oaknin2005} (see Section~\ref{sec:strangelets}), this constrains the flux to be below what would be required to match the dark matter density for masses below $\sim 3$~g~\cite{Gorham2012}. Volcanoes could also result from the intense heat after passage of large compact objects~\cite{Rafelski2013}.

\subsubsection{Existence of old neutron stars}
\label{old_neutron_stars}

If strange matter exists in a stable state, a collision between a lump of strange matter and a neutron star would be expected to seed the conversion of the neutron star into a strange star~\cite{Olinto1987}. It was argued that pulsars which are observed to exhibit glitches must be neutron stars~\cite{Alpar1987}, and hence the cosmic flux of strange matter must be so small as to not convert all neutron stars into strange stars~\cite{Madsen1988,Friedman1991}. In a model where strange star mergers produce ejecta of strange matter, this argument seems to rule out the existence of strange stars~\cite{Friedman1991}. However, it was shown that the amount of ejecta in strange star mergers can be arbitrarily small depending on the model~\cite{Bauswein2009}. Also, the argument does not hold for nuclearites made of antiquark matter, which would simply annihilate in a collision with a neutron star.

A primordial black hole can be gravitationally captured in a close encounter with a neutron star and subsequently disrupt it by rapid accretion. Recent studies~\cite{Capela2013,Pani2014} constrain the primordial black hole density to be at least 10 times lower than the dark matter density for masses in the range $10^{18}<M<10^{24}$~g by considering the existence of old neutron stars in regions of high dark matter density. This indirect result applies to a window which was unconstrained by direct searches, indicating that primordial black holes are not the dominant constituent of dark matter in the universe.

\subsubsection{Solar oscillations}

Stars have been proposed to serve as seismic detectors for primordial black holes or similar objects of very large mass~\cite{Kesden2011}. The gravitational field of a primordial black hole would induce solar oscillations. Such a signal is predicted to be well separated from backgrounds using current solar observatories for $M>10^{21}$~g~\cite{Kesden2011}. As can be seen in Table~\ref{tab:dark_object_fluxes}, assuming a dark matter density and galactic speed this corresponds to a rate of one event every $\sim 10^{8}$ years, too low to be observable. However, oscillation signatures of black holes could also be studied in a large number of stars much larger than the Sun, potentially greatly increasing the event rate~\cite{Kesden2011}.

\subsubsection{Dynamical effects}

If sufficiently abundant, very massive compact objects would disrupt or disturb astronomical systems. In particular, disruption of binary stars or open star clusters constrains the mass of dark-matter objects in the galactic disk to be below $\sim 10^{35}$~g~\cite{Carr1999}. Similarly, disruption of globular clusters and heating of the galactic disk constrains the mass of dark halo objects to be below $\sim 10^{39}$~g, and tidal distortion of galaxies constrains dark objects in galaxy clusters to have a mass below $\sim 10^{43}$~g~\cite{Carr1999}.

\subsubsection{Galactic and cosmic diffuse radiation}
\label{diffuse_emission}

Both the observed 511~keV line from the galactic plane~\cite{Prantzos2011} and MeV-energy continuum emission have been interpreted as resulting from electron annihilations with positron from the interior of quark antimatter nuggets, specifically in the model of Refs.~\cite{Oaknin2005,Forbes2010} (see Section~\ref{sec:strangelets}). Despite the large gap in energy, spectra and fluxes predicted by this model seem to be in good agreement with observations in both cases~\cite{Forbes2008}. The diffuse cosmic radio emission excess measured by ARCADE 2 was also explained in the same model by considering the thermal evolution of antiquark matter nuggets after recombination~\cite{Lawson2013}. One should note that the latter can be equally well reproduced with a model of WIMP dark matter~\cite{Fornengo2011}.

\subsubsection{Bursts from black-hole evaporation}

PBHs (Section~\ref{black_holes}) can be detected indirectly through the products of PBH evaporation that reach the Earth as part of the cosmic-ray flux.

In the present epoch PBHs  with  mass $\sim$10$^{15}$~g  would be contributing  photons with energy of the order of 100~MeV to the cosmological $\gamma$-ray background. The  observational limit on this background strongly suggests that the density of 10$^{15}$~g PBHs cannot exceed  approximately  10$^{-8}$ times the critical density~\cite{Page1976}. Also, as PBHs are  dynamically cold, it is reasonable to expect that some of them have clustered within the galactic halo, contributing to the galactic $\gamma$-ray background measured, for instance, by EGRET~\cite{Lehoucq2009}.

PBHs would also contribute to the presence of positrons or antiprotons in cosmic rays~\cite{Carr1976,Turner1982b,Halzen1991} detectable by, for example, AMS~\cite{AMS2013}. Evaporating PBHs are also a contending explanation for gamma ray bursts~\cite{Cline2011}, radio bursts~\cite{Rees1977} and the 511~keV annihilation line radiation originating from the Galaxy's centre~\cite{Okele1980,Bugaev2009}.

The possibility that evaporating PBHs could produce ultra high energy cosmic rays with energy $\gtrsim$10$^{20}$ eV has been studied \cite{Halzen1991,Barrau2000}. This emission results from direct production of fundamental constituents and from hadronisation of quarks and gluons. This idea has been criticised with a claim that the energy of the primary particles would be reduced by interactions with an optically thick photosphere generated around the PBH~\cite{Heckler1997}.  However, these arguments were refuted in another study~\cite{MacGibbon2008}. A contribution to the cosmic-ray energy spectrum above the GZK cutoff~\cite{Greisen1966,Zatsepin1966} could be detected by the AUGER or the Telescope Array experiments via air shower production. While more statistics are needed to settle the matter definitively, to date the cosmic energy spectrum at the highest energies appears to be consistent with GZK suppression expected when primaries originate from distant sources such as Active Galactic Nuclei~\cite{Auger2013}.

Another unique signature for PBH production of cosmic rays would be short bursts of ultra high energy gamma radiation from the evaporating PBH. Signals lasting for of the order of one second have been unsuccessfully sought in the past by experiments such as CYGNUS~\cite{Alexandreas1993}. It was argued, however, that the burst duration should be shorter, and a technique to search for microsecond burst was proposed~\cite{Krennrich2000}.

\subsection{Summary}

\begin{figure}[tb]
\begin{center}
\includegraphics[width=0.95\linewidth]{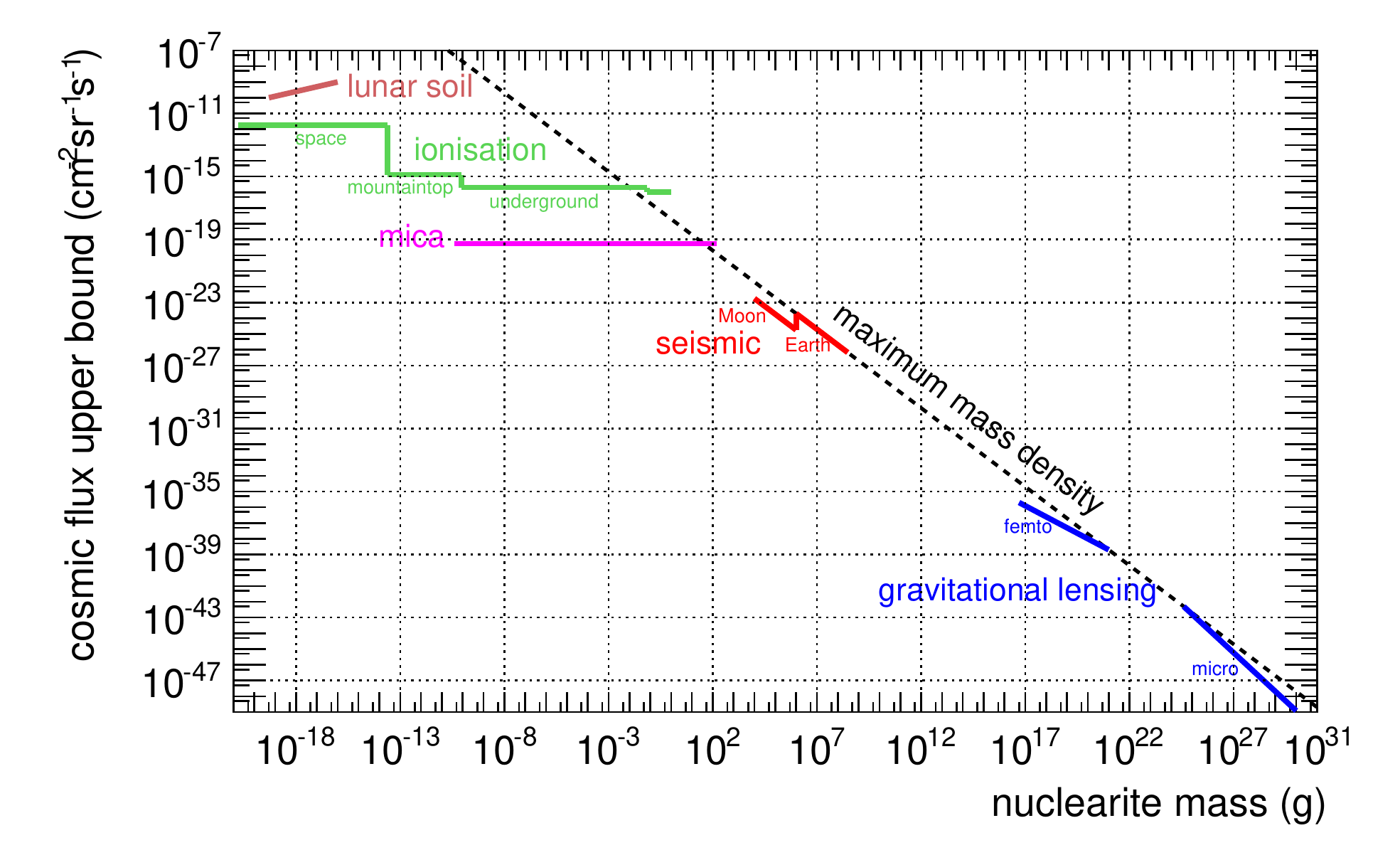}
\caption{Summary of combined nuclearite flux limits (90\% confidence level) as a function of mass, assuming typical galactic velocities. The dashed line represents the flux expected if 100\% of the dark matter density is composed of nuclearites of a given mass.}
\label{fig:nuclearite_flux_limits}
\end{center}
\end{figure}

A review was given on constraints or hints on rare massive ($M\gtrsim 100$~g) extended objects, using various methods such as gravitational lensing, impacts on Earth or on the Moon, anomalous meteoroid properties, and indirect detection of emitted radiation. In the case of nuclearites (Section~\ref{sec:strangelets}), in Fig.~\ref{fig:nuclearite_flux_limits} such constraints are shown together with limits from cosmic-ray detectors (Section~\ref{cosmic}) over many orders of magnitude for the nuclearite mass. It is evident that wide mass ranges remain unconstrained, leaving open the possibility that nuclearites (or other similar composite objects) could constitute dark matter~\cite{Alam1999}. Constraints on primordial black holes as dark matter are much more stringent due to the expectation of evaporation by the Hawking radiation process; these are extensively discussed and summarised in Ref.~\cite{Carr2010}.

%% file: smp-summary.tex
\section{Summary and outlook}

There is strong theoretical motivation for the existence of exotic stable massive particles (SMPs) such as long-lived SUSY particles, strangelets, and magnetic monopoles. SMPs with masses far beyond the reach of  accelerators could still have been produced in the early Universe or during other high-energy events taking place during the cosmos' history. They could be freely roaming the galaxy and detectable at cosmic-ray facilities, or they could have stopped and remained trapped in matter. The breadth of phenomenological scenarios considered in this paper include scenarios of SMPs which would bind to matter to form  samples containing anomalously heavy isotopes or magnetic charge, the possibility of direct detection of SMPs which impinge the Earth or the Moon, and direct and indirect effects of SMPs of macroscopic dimensions. Within these scenarios, a large number of searches using a variety of complementary detection techniques were described, discussed, and interpreted within the context of theoretical models which predict SMPs and their properties.

The cosmos offers a wealth of observable physics phenomena to be studied and unique possibilities to unravel fundamental mysteries which perhaps cannot be tackled at Earth-based laboratories. The nature of dark matter remains to be elucidated, unknown states of matter could exist inside compact stars, and cosmic events encompass energy ranges which extend many orders of magnitude beyond the electroweak scale. The body of work described in this paper, while very impressive, represents a modest effort compared with what could be done in the future by performing experiments on a larger scale, making use of new technologies, and employing human imagination to perform original measurements and explore unknown territories. There are important gaps in the experimental coverage of SMPs in some ranges of masses and properties, and there are many ways in which existing sensitivities could be improved. The discovery of an SMP would have a profound impact on our understanding of particle physics.

\section*{Acknowledgements}

We gratefully acknowledge the following people who have helped us in the preparation of part of this paper by providing useful comments and suggestions: Marshall Eubanks, Teresa Montarulli, Michele Weber. 

%% file: paper.bbl
\providecommand{\href}[2]{#2}\begingroup\raggedright\begin{thebibliography}{100}

\bibitem{PDG2012}
J.~Beringer {\em et~al.} {\em Phys. Rev. D} {\bfseries 86} (2012) 010001.

\bibitem{Fairbairn2007}
M.~Fairbairn, A.~Kraan, D.~Milstead, T.~Sjostrand, P.~Skands, and T.~Sloan {\em
  Phys. Rept.} {\bfseries 438} (2007) 1,
  \href{http://arxiv.org/abs/0611040}{{\ttfamily arXiv:0611040 [hep-ph]}}.

\bibitem{ATLAS2011a}
{ATLAS Collaboration}
  \href{http://dx.doi.org/10.1016/j.physletb.2011.03.033}{{\em Phys. Lett. B}
  {\bfseries 698} (2011) 353}, \href{http://arxiv.org/abs/1102.0459}{{\ttfamily
  arXiv:1102.0459 [hep-ex]}}.

\bibitem{ATLAS2011c}
{ATLAS Collaboration} {\em Phys. Lett. B} {\bfseries 701} (2011) 1,
  \href{http://arxiv.org/abs/1103.1984}{{\ttfamily arXiv:1103.1984 [hep-ex]}}.

\bibitem{ATLAS2011b}
{ATLAS Collaboration} {\em Phys. Lett. B} {\bfseries 703} (2011) 428,
  \href{http://arxiv.org/abs/1106.4495}{{\ttfamily arXiv:1106.4495 [hep-ex]}}.

\bibitem{DeRoeck2012a}
A.~De~Roeck, A.~Katre, P.~Mermod, D.~Milstead, and T.~Sloan {\em Eur. Phys. J.
  C} {\bfseries 72} (2012) 1985,
  \href{http://arxiv.org/abs/1112.2999}{{\ttfamily arXiv:1112.2999 [hep-ph]}}.

\bibitem{DeRoeck2012b}
A.~De~Roeck, H.-P. {H\"achler}, A.~M. Hirt, M.~Dam~Joergensen, A.~Katre,
  P.~Mermod, D.~Milstead, and T.~Sloan {\em Eur. Phys. J. C} {\bfseries 72}
  (2012) 2212, \href{http://arxiv.org/abs/1206.6793}{{\ttfamily arXiv:1206.6793
  [physics.ins-det]}}.

\bibitem{CMS2012b}
{CMS Collaboration} {\em JHEP} {\bfseries 1208} (2012) 026,
  \href{http://arxiv.org/abs/1207.0106}{{\ttfamily arXiv:1207.0106 [hep-ex]}}.

\bibitem{CMS2012c}
{CMS Collaboration} {\em Phys. Lett. B} {\bfseries 713} (2012) 408,
  \href{http://arxiv.org/abs/1205.0272}{{\ttfamily arXiv:1205.0272 [hep-ex]}}.

\bibitem{ATLAS2012a}
{ATLAS Collaboration} {\em Phys. Rev. Lett.} {\bfseries 109} (2012) 261803,
  \href{http://arxiv.org/abs/1207.6411}{{\ttfamily arXiv:1207.6411 [hep-ex]}}.

\bibitem{Mermod2013}
P.~Mermod {\em Proceedings of 48th Rencontres de Moriond, Very High Energy
  Phenomena in the Universe} (2013) ,
  \href{http://arxiv.org/abs/1305.3718}{{\ttfamily arXiv:1305.3718 [hep-ex]}}.

\bibitem{CMS2013a}
{CMS Collaboration} {\em JHEP} {\bfseries 07} (2013) 122,
  \href{http://arxiv.org/abs/1305.0491}{{\ttfamily arXiv:1305.0491 [hep-ex]}}.

\bibitem{CMS2013d}
{CMS Collaboration} {\em Phys. Rev. D} {\bfseries 87} (2013) 092008,
  \href{http://arxiv.org/abs/1210.2311}{{\ttfamily arXiv:1210.2311 [hep-ex]}}.

\bibitem{ATLAS2013a}
{ATLAS Collaboration} {\em Phys. Lett. B} {\bfseries 722} (2013) 305,
  \href{http://arxiv.org/abs/1301.5272}{{\ttfamily arXiv:1301.5272 [hep-ex]}}.

\bibitem{ATLAS2013b}
{ATLAS Collaboration} {\em Phys. Lett. B} {\bfseries 720} (2013) 277,
  \href{http://arxiv.org/abs/1211.1597}{{\ttfamily arXiv:1211.1597 [hep-ex]}}.

\bibitem{ATLAS2013f}
{ATLAS Collaboration} {\em Phys. Rev. D} {\bfseries 88} (2013) 112003,
  \href{http://arxiv.org/abs/1310.6584}{{\ttfamily arXiv:1310.6584 [hep-ex]}}.

\bibitem{Bendtz2014}
K.~Bendtz, A.~Katre, D.~Lacarr{\`e}re, P.~Mermod, D.~Milstead, J.~Pinfold, and
  R.~Soluk {\em Proceedings of the 14th ICATPP Conference} (2014) ,
  \href{http://arxiv.org/abs/1311.6940}{{\ttfamily arXiv:1311.6940
  [physics.ins-det]}}.

\bibitem{Kolb:1990vq}
E.~W. Kolb and M.~S. Turner
{\em Front.Phys.} {\bfseries 69} (1990) 1.

\bibitem{Bergstrom2000}
L.~Bergstrom {\em Rept. Prog. Phys.} {\bfseries 63} (2000) 793,
  \href{http://arxiv.org/abs/hep-ph/0002126}{{\ttfamily arXiv:hep-ph/0002126
  [hep-ph]}}.

\bibitem{Hooper2005}
D.~Hooper, J.~March-Russell, and S.~West {\em Phys. Lett. B} {\bfseries 605}
  (2005) 228, \href{http://arxiv.org/abs/hep-ph/0410114}{{\ttfamily
  arXiv:hep-ph/0410114 [hep-ph]}}.

\bibitem{Strigari2013}
L.~Strigari {\em Phys. Rept.} {\bfseries 531} (2013) 1,
  \href{http://arxiv.org/abs/1211.7090}{{\ttfamily arXiv:1211.7090
  [astro-ph.CO]}}.

\bibitem{Giudice1999}
G.~Giudice and R.~Rattazzi {\em Phys. Rept.} {\bfseries 322} (1999) 419,
  \href{http://arxiv.org/abs/hep-ph/9801271}{{\ttfamily arXiv:hep-ph/9801271
  [hep-ph]}}.

\bibitem{Buchmuller2006}
W.~Buchmuller, K.~Hamaguchi, M.~Ibe, and T.~Yanagida {\em Phys. Lett. B}
  {\bfseries 643} (2006) 124,
  \href{http://arxiv.org/abs/hep-ph/0605164}{{\ttfamily arXiv:hep-ph/0605164
  [hep-ph]}}.

\bibitem{Giudice2004}
G.~Giudice and A.~Romanino {\em Nucl.Phys. B} {\bfseries 699} (2004) 65,
  \href{http://arxiv.org/abs/hep-ph/0406088}{{\ttfamily arXiv:hep-ph/0406088
  [hep-ph]}}.

\bibitem{Gambino2005}
P.~Gambino, G.~Giudice, and P.~Slavich {\em Nucl. Phys. B} {\bfseries 726}
  (2005) 35, \href{http://arxiv.org/abs/hep-ph/0506214}{{\ttfamily
  arXiv:hep-ph/0506214 [hep-ph]}}.

\bibitem{Arvanitaki2005}
A.~Arvanitaki, C.~Davis, P.~Graham, A.~Pierce, and J.~Wacker {\em Phys. Rev. D}
  {\bfseries 72} (2005) 075011,
  \href{http://arxiv.org/abs/hep-ph/0504210}{{\ttfamily arXiv:hep-ph/0504210
  [hep-ph]}}.

\bibitem{Khlopov2011}
M.~Khlopov {\em Mod. Phys. Lett. A} {\bfseries 26} (2011) 2823,
  \href{http://arxiv.org/abs/1111.2838}{{\ttfamily arXiv:1111.2838
  [astro-ph.CO]}}.

\bibitem{Khlopov2013}
M.~Khlopov {\em Int. J. Mod. Phys. A} {\bfseries 28} (2013) 1330042,
  \href{http://arxiv.org/abs/1311.2468}{{\ttfamily arXiv:1311.2468
  [astro-ph.CO]}}.

\bibitem{Khlopov2006b}
M.~Khlopov {\em JETP Lett.} {\bfseries 83} (2006) 1.

\bibitem{Fargion2005}
D.~Fargion and M.~Khlopov {\em Grav. Cosmol.} {\bfseries 19} (2005) 219,
  \href{http://arxiv.org/abs/hep-ph/0507087}{{\ttfamily arXiv:hep-ph/0507087
  [hep-ph]}}.

\bibitem{CDMS2011}
{CDMS Collaboration} {\em Phys. Rev. Lett.} {\bfseries 106} (2011) 131302,
  \href{http://arxiv.org/abs/1011.2482}{{\ttfamily arXiv:1011.2482
  [astro-ph.CO]}}.

\bibitem{XENON1002012}
{XENON100 Collaboration} {\em Phys. Rev. Lett.} {\bfseries 109} (2012) 181301,
  \href{http://arxiv.org/abs/1207.5988}{{\ttfamily arXiv:1207.5988
  [astro-ph.CO]}}.

\bibitem{LUX2014}
{LUX Collaboration} {\em Phys. Rev. Lett.} {\bfseries 112} (2014) 091303,
  \href{http://arxiv.org/abs/1310.8214}{{\ttfamily arXiv:1310.8214
  [astro-ph.CO]}}.

\bibitem{Bernabei2013}
R.~Bernabei, P.~Belli, F.~Cappella, V.~Caracciolo, S.~Castellano, R.~Cerulli,
  C.~Dai, A.~d’Angelo, S.~d’Angelo, A.~Marco, H.~He, A.~Incicchitti,
  H.~Kuang, X.~Ma, F.~Montecchia, D.~Prosperi, X.~Sheng, R.~Wang, and Z.~Ye
  {\em Eur. Phys. J. C} {\bfseries 73} (2013) 2648,
  \href{http://arxiv.org/abs/1308.5109}{{\ttfamily arXiv:1308.5109
  [astro-ph.GA]}}.

\bibitem{Schellekens1990}
A.~Schellekens {\em Phys. Lett. B} {\bfseries 237} (1990) 363.

\bibitem{Schellekens2013}
A.~Schellekens {\em Rev. Mod. Phys.} {\bfseries 85} (2013) 1491,
  \href{http://arxiv.org/abs/1306.5083}{{\ttfamily arXiv:1306.5083 [hep-ph]}}.

\bibitem{Lerche1987}
W.~Lerche, D.~Lust, and A.~Schellekens {\em Nucl. Phys. B} {\bfseries 287}
  (1987) 477.

\bibitem{Wen1985}
X.-G. Wen and E.~Witten {\em Nucl. Phys. B} {\bfseries 261} (1985) 651.

\bibitem{Athanasiu1988}
G.~Athanasiu, J.~Atick, M.~Dine, and W.~Fischler {\em Phys. Lett. B} {\bfseries
  214} (1988) 55.

\bibitem{Birkel1998}
M.~Birkel and S.~Sarkar {\em Astropart. Phys.} {\bfseries 9} (1998) 297,
  \href{http://arxiv.org/abs/hep-ph/9804285}{{\ttfamily arXiv:hep-ph/9804285
  [hep-ph]}}.

\bibitem{Perl2001}
M.~Perl, P.~Kim, V.~Halyo, E.~Lee, I.~Lee, D.~Loomba, and K.~Lackner {\em Int.
  J. Mod. Phys. A} {\bfseries 16} (2001) 2137,
  \href{http://arxiv.org/abs/0102033}{{\ttfamily arXiv:0102033 [hep-ex]}}.

\bibitem{Perl2004}
M.~Perl, E.~Lee, and D.~Loomba {\em Mod. Phys. Lett. A} {\bfseries 19} (2004)
  2595.

\bibitem{Holdom1986}
B.~Holdom {\em Phys. Lett. B} {\bfseries 166} (1986) 196.

\bibitem{Kobzarev1966}
I.~Kobzarev, L.~Okun, and I.~Pomeranchuk {\em Sov. J. Nucl. Phys.} {\bfseries
  3} (1966) 837.

\bibitem{Blinnikov1982}
S.~Blinnikov and M.~Khlopov {\em Sov. J. Nucl. Phys.} {\bfseries 36} (1982)
  472.

\bibitem{Blinnikov1983}
S.~Blinnikov and M.~Khlopov {\em Sov. Astron.} {\bfseries 27} (1983) 371.

\bibitem{Kolb1985}
E.~Kolb, D.~Seckel, and M.~Turner {\em Nature} {\bfseries 314} (1985) 415.

\bibitem{Carlson1987}
E.~Carlson and S.~Glashow {\em Phys. Lett. B} {\bfseries 193} (1987) 168.

\bibitem{Khlopov1989}
M.~Y. Khlopov, G.~Beskin, N.~Bochkarev, L.~Pustylnik, and S.~Pustylnik {\em
  Sov. Astron.} {\bfseries 35} (1991) 21.

\bibitem{Hodges1993}
H.~Hodges {\em Phys. Rev. D} {\bfseries 47} (1993) 456.

\bibitem{Berezhiani1995b}
Z.~Berezhiani, A.~Dolgov, and R.~Mohapatra {\em Phys. Lett. B} {\bfseries 375}
  (1996) 26, \href{http://arxiv.org/abs/hep-ph/9511221}{{\ttfamily
  arXiv:hep-ph/9511221 [hep-ph]}}.

\bibitem{Mohapatra2000}
R.~Mohapatra and V.~Teplitz {\em Phys. Rev. D} {\bfseries 62} (2000) 063506,
  \href{http://arxiv.org/abs/astro-ph/0001362}{{\ttfamily
  arXiv:astro-ph/0001362 [astro-ph]}}.

\bibitem{Foot2004}
R.~Foot {\em Int. J. Mod. Phys. D} {\bfseries 13} (2004) 2161,
  \href{http://arxiv.org/abs/astro-ph/0407623}{{\ttfamily
  arXiv:astro-ph/0407623}}.

\bibitem{Okun2007}
L.~Okun {\em Phys. Usp.} {\bfseries 50} (2007) 380,
  \href{http://arxiv.org/abs/hep-ph/0606202}{{\ttfamily arXiv:hep-ph/0606202
  [hep-ph]}}.

\bibitem{Kaplan2010}
D.~Kaplan, G.~Krnjaic, K.~Rehermann, and C.~Wells {\em JCAP} {\bfseries 1005}
  (2010) 021, \href{http://arxiv.org/abs/0909.0753}{{\ttfamily arXiv:0909.0753
  [hep-ph]}}.

\bibitem{Kaplan2011}
D.~Kaplan, G.~Krnjaic, K.~Rehermann, and C.~Wells {\em JCAP} {\bfseries 1110}
  (2011) 011, \href{http://arxiv.org/abs/1105.2073}{{\ttfamily arXiv:1105.2073
  [hep-ph]}}.

\bibitem{Behbahani2011}
S.~Behbahani, M.~Jankowiak, T.~Rube, and J.~Wacker {\em Adv. High Energy Phys.}
  {\bfseries 2011} (2011) 709492,
  \href{http://arxiv.org/abs/1009.3523}{{\ttfamily arXiv:1009.3523 [hep-ph]}}.

\bibitem{Cline2012}
J.~Cline, Z.~Liu, and W.~Xue {\em Phys. Rev. D} {\bfseries 85} (2012) 101302,
  \href{http://arxiv.org/abs/1201.4858}{{\ttfamily arXiv:1201.4858 [hep-ph]}}.

\bibitem{Cline2013}
J.~Cline, Z.~Liu, and W.~Xue {\em Phys. Rev. D} {\bfseries 87} (2013) 015001,
  \href{http://arxiv.org/abs/1207.3039}{{\ttfamily arXiv:1207.3039 [hep-ph]}}.

\bibitem{CyrRacine2013}
F.-Y. Cyr-Racine and K.~Sigurdson {\em Phys. Rev. D} {\bfseries 87} (2013)
  103515, \href{http://arxiv.org/abs/1209.5752}{{\ttfamily arXiv:1209.5752
  [astro-ph.CO]}}.

\bibitem{Cline2014}
J.~Cline, Z.~Liu, G.~Moore, and W.~Xue {\em Phys. Rev. D} {\bfseries 89} (2014)
  043514, \href{http://arxiv.org/abs/1311.6468}{{\ttfamily arXiv:1311.6468
  [hep-ph]}}.

\bibitem{Foot2014}
R.~Foot {\em Int. J. Mod. Phys. A} {\bfseries 29} (2014) 1430013,
\href{http://arxiv.org/abs/1401.3965}{{\ttfamily arXiv:1401.3965
  [astro-ph.CO]}}.

\bibitem{Foot2008}
R.~Foot {\em Phys. Rev. D} {\bfseries 78} (2008) 043529,
  \href{http://arxiv.org/abs/0804.4518}{{\ttfamily arXiv:0804.4518 [hep-ph]}}.

\bibitem{Foot2002}
R.~Foot and T.~Yoon {\em Acta Phys. Polon. B} {\bfseries 33} (2002) 1979,
  \href{http://arxiv.org/abs/astro-ph/0203152}{{\ttfamily
  arXiv:astro-ph/0203152 [astro-ph]}}.

\bibitem{Foot2003a}
R.~Foot and S.~Mitra {\em Astropart. Phys.} {\bfseries 19} (2003) 739,
  \href{http://arxiv.org/abs/astro-ph/0211067}{{\ttfamily
  arXiv:astro-ph/0211067 [astro-ph]}}.

\bibitem{Bodmer1971}
A.~Bodmer
{\em Phys.Rev. D} {\bfseries 4} (1971) 1601--1606.

\bibitem{Chin1979}
S.~Chin and A.~Kerman {\em Phys. Rev. Lett.} {\bfseries 43} (1979) 1292.

\bibitem{Farhi1984}
E.~Farhi and R.~Jaffe {\em Phys. Rev. D} {\bfseries 30} (1984) 2379.

\bibitem{DeRujula1984}
A.~De~Rujula and S.~Glashow {\em Nature} {\bfseries 312} (1984) 734.

\bibitem{Clark2002}
J.~Clark, S.~Goodwin, P.~Crowther, L.~Kaper, M.~Fairbairn, N.~Langer, and
  C.~Brocksopp {\em Astron. Astrophys.} {\bfseries 392} (2002) 909,
  \href{http://arxiv.org/abs/astro-ph/0207334}{{\ttfamily
  arXiv:astro-ph/0207334 [astro-ph]}}.

\bibitem{Demorest2010}
P.~Demorest, T.~Pennucci, S.~Ransom, M.~Roberts, and J.~Hessels {\em Nature}
  {\bfseries 467} (2010) 1081, \href{http://arxiv.org/abs/1010.5788}{{\ttfamily
  arXiv:1010.5788 [astro-ph.HE]}}.

\bibitem{Antoniadis2013}
J.~Antoniadis, P.~Freire, N.~Wex, T.~Tauris, R.~Lynch, {\em et~al.} {\em
  Science} {\bfseries 340} (2013) 6131,
  \href{http://arxiv.org/abs/1304.6875}{{\ttfamily arXiv:1304.6875
  [astro-ph.HE]}}.

\bibitem{SchaffnerBielich2008}
J.~Schaffner-Bielich {\em Nucl. Phys. A} {\bfseries 804} (2008) 309,
  \href{http://arxiv.org/abs/0801.3791}{{\ttfamily arXiv:0801.3791
  [astro-ph]}}.

\bibitem{Weissenborn2011}
S.~Weissenborn, I.~Sagert, G.~Pagliara, M.~Hempel, and J.~Schaffner-Bielich
  {\em Astrophys. J.} {\bfseries 740} (2011) L14,
  \href{http://arxiv.org/abs/1102.2869}{{\ttfamily arXiv:1102.2869
  [astro-ph.HE]}}.

\bibitem{Madsen1988}
J.~Madsen {\em Phys. Rev. Lett.} {\bfseries 61} (1988) 2909.

\bibitem{Madsen2005}
J.~Madsen {\em Phys. Rev. D} {\bfseries 71} (2005) 014026,
  \href{http://arxiv.org/abs/astro-ph/0411538}{{\ttfamily
  arXiv:astro-ph/0411538 [astro-ph]}}.

\bibitem{Oaknin2005}
D.~Oaknin and A.~Zhitnitsky {\em Phys. Rev. D} {\bfseries 71} (2005) 023519,
  \href{http://arxiv.org/abs/hep-ph/0309086}{{\ttfamily arXiv:hep-ph/0309086
  [hep-ph]}}.

\bibitem{Forbes2008}
M.~Forbes and A.~Zhitnitsky {\em JCAP} {\bfseries 0801} (2008) 023,
  \href{http://arxiv.org/abs/astro-ph/0611506}{{\ttfamily
  arXiv:astro-ph/0611506 [astro-ph]}}.

\bibitem{Forbes2010}
M.~Forbes, K.~Lawson, and A.~Zhitnitsky {\em Phys. Rev. D} {\bfseries 82}
  (2010) 083510, \href{http://arxiv.org/abs/0910.4541}{{\ttfamily
  arXiv:0910.4541 [astro-ph.GA]}}.

\bibitem{Markevitch2004}
M.~Markevitch, A.~Gonzalez, D.~Clowe, A.~Vikhlinin, L.~David, W.~Forman,
  C.~Jones, S.~Murray, and W.~Tucker {\em Astrophys. J.} {\bfseries 606} (2004)
  819, \href{http://arxiv.org/abs/astro-ph/0309303}{{\ttfamily
  arXiv:astro-ph/0309303 [astro-ph]}}.

\bibitem{Colpi1993}
M.~Colpi, S.~Shapiro, and S.~Teukolsky {\em Astrophys. J.} {\bfseries 414}
  (1993) 717.

\bibitem{Narain2006}
G.~Narain, J.~Schaffner-Bielich, and I.~Mishustin {\em Phys. Rev. D} {\bfseries
  74} (2006) 063003, \href{http://arxiv.org/abs/astro-ph/0605724}{{\ttfamily
  arXiv:astro-ph/0605724 [astro-ph]}}.

\bibitem{Dietl2012}
C.~Dietl, L.~Labun, and J.~Rafelski {\em Phys. Lett. B} {\bfseries 709} (2012)
  123, \href{http://arxiv.org/abs/1110.0551}{{\ttfamily arXiv:1110.0551
  [astro-ph.CO]}}.

\bibitem{Hansson2005}
J.~Hansson and F.~Sandin {\em Phys. Lett. B} {\bfseries 616} (2005) 1,
  \href{http://arxiv.org/abs/astro-ph/0410417}{{\ttfamily
  arXiv:astro-ph/0410417 [astro-ph]}}.

\bibitem{Horvath2007}
J.~Horvath {\em Astrophys. Space Sci.} {\bfseries 307} (2007) 419,
  \href{http://arxiv.org/abs/astro-ph/0702288}{{\ttfamily
  arXiv:astro-ph/0702288 [ASTRO-PH]}}.

\bibitem{Sandin2005}
F.~Sandin {\em Eur. Phys. J. C} {\bfseries 40} (2005) 15,
  \href{http://arxiv.org/abs/astro-ph/0410407}{{\ttfamily
  arXiv:astro-ph/0410407 [astro-ph]}}.

\bibitem{Sandin2007}
F.~Sandin and J.~Hansson {\em Phys. Rev. D} {\bfseries 76} (2007) 125006,
  \href{http://arxiv.org/abs/astro-ph/0701768}{{\ttfamily
  arXiv:astro-ph/0701768 [astro-ph]}}.

\bibitem{Nishimura1987}
H.~Nishimura and Y.~Hayashi {\em Phys. Rev. D} {\bfseries 35} (1987) 3151.

\bibitem{Curie1894}
P.~Curie {\em Seances Sic. Phys. (Paris)} (1894) 76.

\bibitem{Dirac1931}
P.~Dirac {\em Proc. Roy. Soc.} {\bfseries A 133} (1931) 60.

\bibitem{Dirac1948}
P.~Dirac {\em Phys. Rev.} {\bfseries 74} (1948) 817.

\bibitem{tHooft1974}
G.~'t~Hooft {\em Nucl. Phys. B} {\bfseries 79} (1974) 276.

\bibitem{Polyakov1974}
A.~Polyakov {\em JETP Lett.} {\bfseries 20} (1974) 194.

\bibitem{Aharonov1959}
Y.~Aharonov and D.~Bohm {\em Phys. Rev.} {\bfseries 115} (1959) 485.

\bibitem{Preskill1984}
J.~Preskill {\em Ann. Rev. Nucl. Part. Sci.} {\bfseries 34} (1984) 461.

\bibitem{tHooft1976}
G.~'t~Hooft {\em Nucl. Phys. B} {\bfseries 105} (1976) 538.

\bibitem{Corrigan1976}
E.~Corrigan and D.~Olive {\em Nucl. Phys. B} {\bfseries 110} (1976) 237.

\bibitem{Schwinger1966}
J.~S. Schwinger {\em Phys. Rev.} {\bfseries 144} (1966) 1087.

\bibitem{Schwinger1968}
J.~Schwinger {\em Phys. Rev.} {\bfseries 173} (1968) 1536.

\bibitem{Schwinger1969}
J.~Schwinger \href{http://dx.doi.org/10.1126/science.165.3895.757}{{\em
  Science} {\bfseries 165} (1969) 757}.

\bibitem{Yock1969}
P.~Yock {\em Int. J. Theor. Phys.} {\bfseries 2} (1969) 247.

\bibitem{DeRujula1978}
A.~De~Rujula, R.~Giles, and R.~Jaffe {\em Phys. Rev. D} {\bfseries 17} (1978)
  285.

\bibitem{Fryberger1981}
D.~Fryberger {\em Hadronic J.} {\bfseries 4} (1981) 1844.

\bibitem{Kibble:1976sj}
T.~Kibble {\em J.Phys.} {\bfseries A9} (1976) 1387.

\bibitem{Georgi1974}
H.~Georgi and S.~Glashow {\em Phys. Rev. Lett.} {\bfseries 32} (1974) 438.

\bibitem{Pati1974}
J.~Pati and A.~Salam {\em Phys. Rev. D} {\bfseries 10} (1974) 275.

\bibitem{Lazarides:1980cc}
G.~Lazarides, M.~Magg, and Q.~Shafi {\em Phys.Lett.} {\bfseries B97} (1980) 87.

\bibitem{Frampton1989}
P.~Frampton and B.-H. Lee {\em Phys. Rev. Lett.} {\bfseries 64} (1990) 619.

\bibitem{Frampton1990}
P.~Frampton and T.~Kephart {\em Phys. Rev. D} {\bfseries 42} (1990) 3892.

\bibitem{Kephart2007}
T.~Kephart, C.-A. Lee, and Q.~Shafi {\em JHEP} {\bfseries 0701} (2007) 088,
  \href{http://arxiv.org/abs/hep-ph/0602055}{{\ttfamily arXiv:hep-ph/0602055
  [hep-ph]}}.

\bibitem{Vachaspati1992}
T.~Vachaspati and M.~Barriola {\em Phys. Rev. Lett.} {\bfseries 69} (1992)
  1867.

\bibitem{Barriola1994}
M.~Barriola, T.~Vachaspati, and M.~Bucher {\em Phys. Rev. D} {\bfseries 50}
  (1994) 2819, \href{http://arxiv.org/abs/hep-th/9306120}{{\ttfamily
  arXiv:hep-th/9306120 [hep-th]}}.

\bibitem{Cho1997}
Y.~Cho and D.~Maison {\em Phys. Lett. B} {\bfseries 391} (1997) 360,
  \href{http://arxiv.org/abs/9601028}{{\ttfamily arXiv:9601028 [hep-th]}}.

\bibitem{Yang1998}
Y.~Yang {\em Proc. Roy. Soc. Lond. A} {\bfseries 454} (1998) 155.

\bibitem{Bruemmer:2009ky}
F.~Brummer, J.~Jaeckel, and V.~V. Khoze {\em JHEP} {\bfseries 0906} (2009) 037,
\href{http://arxiv.org/abs/0905.0633}{{\ttfamily arXiv:0905.0633 [hep-ph]}}.

\bibitem{Khoze:2014woa}
V.~V. Khoze and G.~Ro
\href{http://arxiv.org/abs/1406.2291}{{\ttfamily arXiv:1406.2291 [hep-ph]}}.

\bibitem{Dokos1980}
C.~Dokos and T.~Tomaras {\em Phys. Rev. D} {\bfseries 21} (1980) 2940.

\bibitem{Rubakov1981}
V.~Rubakov {\em JETP Lett.} {\bfseries 33} (1981) 644.

\bibitem{Rubakov1982}
V.~Rubakov {\em Nucl. Phys. B} {\bfseries 203} (1982) 311.

\bibitem{Callan1982}
C.~Callan {\em Phys. Rev. D} {\bfseries 26} (1982) 2058.

\bibitem{Callan1983}
C.~Callan {\em Nucl. Phys. B} {\bfseries 212} (1983) 391.

\bibitem{Walsh1984}
T.~Walsh, P.~Weisz, and T.~Wu {\em Nucl. Phys. B} {\bfseries 232} (1984) 349.

\bibitem{Rubakov1984}
V.~Rubakov and M.~Serebryakov {\em Nucl. Phys. B} {\bfseries 237} (1984) 329.

\bibitem{Arafune1983}
J.~Arafune and M.~Fukugita {\em Phys. Rev. Lett.} {\bfseries 50} (1983) 1901.

\bibitem{Kibble1980}
T.~Kibble {\em Phys. Rept.} {\bfseries 67} (1980) 183.

\bibitem{Guth1981}
A.~Guth {\em Phys. Rev. D} {\bfseries 23} (1981) 347.

\bibitem{Planck2014}
{Planck Collaboration} {\em Astron. Astrophys.} (2014) ,
  \href{http://arxiv.org/abs/1303.5076}{{\ttfamily arXiv:1303.5076
  [astro-ph.CO]}}.

\bibitem{BICEP22014}
{BICEP2 Collaboration} {\em Phys. Rev. Lett.} {\bfseries 112} (2014) 241101,
  \href{http://arxiv.org/abs/1403.3985}{{\ttfamily arXiv:1403.3985
  [astro-ph.CO]}}.

\bibitem{Sorkin1983}
R.~Sorkin {\em Phys. Rev. Lett.} {\bfseries 51} (1983) 87.

\bibitem{Gross1983}
D.~Gross and M.~Perry {\em Nucl. Phys. B} {\bfseries 226} (1983) 29.

\bibitem{Banks1988}
T.~Banks, M.~Dine, H.~Dykstra, and W.~Fischler {\em Phys. Lett. B} {\bfseries
  212} (1988) 45.

\bibitem{Callan1991}
C.~Callan, J.~Harvey, and A.~Strominger {\em Nucl. Phys. B} {\bfseries 367}
  (1991) 60.

\bibitem{Gauntlett1993}
J.~Gauntlett, J.~Harvey, and J.~Liu {\em Nucl. Phys. B} {\bfseries 409} (1993)
  363, \href{http://arxiv.org/abs/hep-th/9211056}{{\ttfamily
  arXiv:hep-th/9211056 [hep-th]}}.

\bibitem{Ellis2000}
J.~Ellis, N.~Mavromatos, and D.~Nanopoulos {\em Gen. Rel. Grav.} {\bfseries 32}
  (2000) 943, \href{http://arxiv.org/abs/gr-qc/9810086}{{\ttfamily
  arXiv:gr-qc/9810086 [gr-qc]}}.

\bibitem{Ellis2004}
J.~Ellis, N.~Mavromatos, and M.~Westmuckett {\em Phys. Rev. D} {\bfseries 70}
  (2004) 044036, \href{http://arxiv.org/abs/gr-qc/0405066}{{\ttfamily
  arXiv:gr-qc/0405066 [gr-qc]}}.

\bibitem{Ellis2005}
J.~Ellis, N.~Mavromatos, and M.~Westmuckett {\em Phys. Rev. D} {\bfseries 71}
  (2005) 106006, \href{http://arxiv.org/abs/gr-qc/0501060}{{\ttfamily
  arXiv:gr-qc/0501060 [gr-qc]}}.

\bibitem{Ellis2008}
J.~Ellis, N.~Mavromatos, and D.~Nanopoulos {\em Phys. Lett. B} {\bfseries 665}
  (2008) 412, \href{http://arxiv.org/abs/0804.3566}{{\ttfamily arXiv:0804.3566
  [hep-th]}}.

\bibitem{Shiu2004}
G.~Shiu and L.-T. Wang {\em Phys. Rev. D} {\bfseries 69} (2004) 126007,
  \href{http://arxiv.org/abs/hep-ph/0311228}{{\ttfamily arXiv:hep-ph/0311228
  [hep-ph]}}.

\bibitem{Fairbairn2002}
M.~Fairbairn and L.~Griffiths {\em JHEP} {\bfseries 0202} (2002) 024,
  \href{http://arxiv.org/abs/hep-ph/0111435}{{\ttfamily arXiv:hep-ph/0111435
  [hep-ph]}}.

\bibitem{Goto1963}
E.~Goto, H.~Kolm, and K.~Ford {\em Phys. Rev.} {\bfseries 132} (1963) 387.

\bibitem{Osborne1970}
W.~Osborne {\em Phys. Rev. Lett.} {\bfseries 24} (1970) 1441.

\bibitem{Wick2003}
S.~Wick, T.~Kephart, T.~Weiler, and P.~Biermann {\em Astropart. Phys.}
  {\bfseries 18} (2003) 663,
  \href{http://arxiv.org/abs/astro-ph/0001233}{{\ttfamily
  arXiv:astro-ph/0001233 [astro-ph]}}.

\bibitem{Parker1970}
E.~Parker {\em Astrophys. J.} {\bfseries 160} (1970) 383.

\bibitem{Parker1971}
E.~Parker {\em Astrophys. J.} {\bfseries 163} (1971) 255.

\bibitem{Turner1982}
M.~Turner, E.~Parker, and T.~Bogdan {\em Phys. Rev. D} {\bfseries 26} (1982)
  1296.

\bibitem{Adams1993}
F.~Adams, M.~Fatuzzo, K.~Freese, G.~Tarl{\'e}, R.~Watkins, and M.~Turner {\em
  Phys. Rev. Lett.} {\bfseries 70} (1993) 2511.

\bibitem{Jansson2012}
R.~Jansson and G.~Farrar {\em Astrophys. J.} {\bfseries 761} (2012) L11,
  \href{http://arxiv.org/abs/1210.7820}{{\ttfamily arXiv:1210.7820
  [astro-ph.GA]}}.

\bibitem{Jansson2012b}
R.~Jansson and G.~Farrar {\em Astrophys. J.} {\bfseries 757} (2012) 14,
  \href{http://arxiv.org/abs/1204.3662}{{\ttfamily arXiv:1204.3662
  [astro-ph.GA]}}.

\bibitem{GF1}
G.~Farrar {\em private communication} .

\bibitem{Bracci1984}
L.~Bracci, G.~Fiorentini, G.~Mezzorani, and P.~Quarati {\em Phys. Lett. B}
  {\bfseries 143} (1984) 357.

\bibitem{Hill1983}
C.~Hill {\em Nucl. Phys. B} {\bfseries 224} (1983) 469.

\bibitem{Bhattacharjee1995}
P.~Bhattacharjee and G.~Sigl {\em Phys. Rev. D} {\bfseries 51} (1995) 4079,
  \href{http://arxiv.org/abs/astro-ph/9412053}{{\ttfamily
  arXiv:astro-ph/9412053 [astro-ph]}}.

\bibitem{Lee1974}
T.~Lee and G.~Wick {\em Phys. Rev. D} {\bfseries 9} (1974) 2291.

\bibitem{Friedberg1976}
R.~Friedberg, T.~Lee, and A.~Sirlin {\em Phys. Rev. D} {\bfseries 13} (1976)
  2739.

\bibitem{Friedberg1977}
R.~Friedberg and T.~Lee {\em Phys. Rev. D} {\bfseries 15} (1977) 1694.

\bibitem{Coleman1985}
S.~Coleman {\em Nucl. Phys. B} {\bfseries 262} (1985) 263.

\bibitem{Kasuya2000}
S.~Kasuya and M.~Kawasaki {\em Phys. Rev. D} {\bfseries 62} (2000) 023512,
  \href{http://arxiv.org/abs/hep-ph/0002285}{{\ttfamily arXiv:hep-ph/0002285
  [hep-ph]}}.

\bibitem{Kasuya2000b}
S.~Kasuya and M.~Kawasaki {\em Phys. Rev. Lett.} {\bfseries 85} (2000) 2677,
  \href{http://arxiv.org/abs/hep-ph/0006128}{{\ttfamily arXiv:hep-ph/0006128
  [hep-ph]}}.

\bibitem{Kusenko1998}
A.~Kusenko and M.~Shaposhnikov {\em Phys. Lett. B} {\bfseries 418} (1998) 46,
  \href{http://arxiv.org/abs/9709492}{{\ttfamily arXiv:9709492 [hep-ph]}}.

\bibitem{Kusenko1998b}
A.~Kusenko, V.~Kuzmin, M.~Shaposhnikov, and P.~Tinyakov {\em Phys. Rev. Lett.}
  {\bfseries 80} (1998) 3185,
  \href{http://arxiv.org/abs/hep-ph/9712212}{{\ttfamily arXiv:hep-ph/9712212
  [hep-ph]}}.

\bibitem{Zeldovich1967}
Z.~Y.B. and I.~Novikov {\em Sov. Astron.} {\bfseries 10} (1967) 602.

\bibitem{Carr1974}
B.~Carr and S.~Hawking {\em Mon. Not. R. Astron. Soc} {\bfseries 168} (1974)
  399.

\bibitem{Carr1975}
B.~Carr {\em Astrophys. J.} {\bfseries 201} (1975) 1.

\bibitem{Carr2010}
B.~Carr, K.~Kohri, Y.~Sendouda, and J.~Yokoyama {\em Phys. Rev. D} {\bfseries
  81} (2010) 104019, \href{http://arxiv.org/abs/0912.5297}{{\ttfamily
  arXiv:0912.5297 [astro-ph.CO]}}.

\bibitem{Frampton2010}
P.~Frampton, M.~Kawasaki, F.~Takahashi, and T.~Yanagida {\em JCAP} {\bfseries
  1004} (2010) 023, \href{http://arxiv.org/abs/1001.2308}{{\ttfamily
  arXiv:1001.2308 [hep-ph]}}.

\bibitem{Carr1993}
B.~Carr and J.~Lidsey \href{http://dx.doi.org/10.1103/PhysRevD.48.543}{{\em
  Phys. Rev. D} {\bfseries 48} (1993) 543}.

\bibitem{Ivanov1994}
P.~Ivanov, P.~Naselsky, and I.~Novikov
  \href{http://dx.doi.org/10.1103/PhysRevD.50.7173}{{\em Phys. Rev. D}
  {\bfseries 50} (1994) 7173}.

\bibitem{Yokoyama1998}
Y.~Yokoyama \href{http://dx.doi.org/10.1016/S0370-1573(98)00044-1}{{\em Phys.
  Rept.} {\bfseries 307} (1998) 133}.

\bibitem{Kanazawa2000}
T.~Kanazawa, M.~Kawasaki, and T.~Yanagida {\em Phys. Lett. B} {\bfseries 482}
  (2000) 174, \href{http://arxiv.org/abs/hep-ph/0002236}{{\ttfamily
  arXiv:hep-ph/0002236 [hep-ph]}}.

\bibitem{Leach2000}
S.~Leach, I.~Grivell, and A.~Liddle
  \href{http://dx.doi.org/10.1103/PhysRevD.62.043516}{{\em Phys. Rev. D}
  {\bfseries 62} (2000) 043516},
  \href{http://arxiv.org/abs/astro-ph/0004296}{{\ttfamily
  arXiv:astro-ph/0004296 [astro-ph]}}.

\bibitem{Chongchitnan2007}
S.~Chongchitnan and G.~Efstathiou
  \href{http://dx.doi.org/10.1088/1475-7516/2007/01/011}{{\em J. Cosmol.
  Astropart. Phys.} {\bfseries 01} (2007) 011},
  \href{http://arxiv.org/abs/astro-ph/0611818}{{\ttfamily
  arXiv:astro-ph/0611818 [astro-ph]}}.

\bibitem{Nozari2007}
K.~Nozari \href{http://dx.doi.org/10.1016/j.astropartphys.2006.10.001}{{\em
  Astropart. Phys.} {\bfseries 27} (2007) 169}.

\bibitem{Saito2008}
R.~Saito, J.~Yokoyama, and R.~Nagata
  \href{http://dx.doi.org/10.1088/1475-7516/2008/06/024}{{\em J. Cosmol.
  Astropart. Phys.} {\bfseries 06} (2008) 024},
  \href{http://arxiv.org/abs/0804.3470}{{\ttfamily arXiv:0804.3470
  [astro-ph]}}.

\bibitem{Lin2013}
C.-M. Lin and K.-W. Ng {\em Phys. Lett. B} {\bfseries 718} (2013) 1181,
  \href{http://arxiv.org/abs/1206.1685}{{\ttfamily arXiv:1206.1685 [hep-ph]}}.

\bibitem{Hawking1982}
S.~Hawking, I.~Moss, and J.~Stewart
  \href{http://dx.doi.org/10.1103/PhysRevD.26.2681}{{\em Phys. Rev. D}
  {\bfseries 26} (1982) 2681}.

\bibitem{Kodama1982}
H.~Kodama, M.~Sasaki, and K.~Sato {\em Prog. Theor. Phys.} {\bfseries 68}
  (1982) 1979.

\bibitem{Berezin1983}
V.~Berezin, V.~Kuzmin, and I.~Tkachev {\em Phys. Lett. B} {\bfseries 120}
  (1983) 91.

\bibitem{La1989}
D.~La and P.~Steinhardt {\em Phys. Lett. B} {\bfseries 220} (1989) 375.

\bibitem{Moss1994}
I.~Moss {\em Phys. Rev. D} {\bfseries 50} (1994) 676.

\bibitem{Konoplich1998}
R.~Konoplich, S.~Rubin, A.~Sakharov, and M.~Khlopov {\em Astron. Lett.}
  {\bfseries 24} (1998) 413.

\bibitem{Hogan1984}
C.~Hogan {\em Phys. Lett. B} {\bfseries 143} (1984) 87.

\bibitem{Hawking1989}
S.~Hawking {\em Phys. Lett. B} {\bfseries 231} (1989) 237.

\bibitem{Polnarev1991}
A.~Polnarev and R.~Zembowicz {\em Phys. Rev. D} {\bfseries 43} (1991) 1106.

\bibitem{Caldwell1996}
R.~Caldwell and P.~Casper {\em Phys. Rev. D} {\bfseries 53} (1996) 3002,
  \href{http://arxiv.org/abs/gr-qc/9509012}{{\ttfamily arXiv:gr-qc/9509012
  [gr-qc]}}.

\bibitem{MacGibbon1998}
J.~MacGibbon, R.~Brandenberger, and U.~Wichoski {\em Phys. Rev. D} {\bfseries
  57} (1998) 2158, \href{http://arxiv.org/abs/astro-ph/9707146}{{\ttfamily
  arXiv:astro-ph/9707146 [astro-ph]}}.

\bibitem{Hansen2000}
R.~Hansen, M.~Christensen, and A.~Larsen {\em Int. J. Mod. Phys. A} {\bfseries
  15} (2000) 4433, \href{http://arxiv.org/abs/gr-qc/9902048}{{\ttfamily
  arXiv:gr-qc/9902048 [gr-qc]}}.

\bibitem{Nagasawa2005}
M.~Nagasawa {\em Gen. Rel. Grav.} {\bfseries 37} (2005) 1635.

\bibitem{Matsuda2006}
T.~Matsuda {\em JHEP} {\bfseries 0604} (2006) 017,
  \href{http://arxiv.org/abs/hep-ph/0509062}{{\ttfamily arXiv:hep-ph/0509062
  [hep-ph]}}.

\bibitem{Lake2009}
M.~Lake, S.~Thomas, and J.~Ward {\em JHEP} {\bfseries 0912} (2009) 033,
  \href{http://arxiv.org/abs/0906.3695}{{\ttfamily arXiv:0906.3695 [hep-ph]}}.

\bibitem{Hawking1974}
S.~Hawking {\em Nature} {\bfseries 248} (1974) 30.

\bibitem{Hawking1975}
S.~Hawking {\em Commun. Math. Phys.} {\bfseries 43} (1975) 199.

\bibitem{Bekenstein1973}
J.~Bekenstein {\em Phys. Rev. D} {\bfseries 7} (1973) 2333.

\bibitem{Turner1979}
M.~Turner {\em Phys. Lett. B} {\bfseries 89} (1979) 155.

\bibitem{Barrow1991}
J.~Barrow, E.~Copeland, E.~Kolb, and A.~Liddle {\em Phys. Rev. D} {\bfseries
  43} (1991) 984.

\bibitem{Upadhyay1999}
N.~Upadhyay, P.~Das~Gupta, and R.~Saxena {\em Phys. Rev. D} {\bfseries 60}
  (1999) 063513, \href{http://arxiv.org/abs/astro-ph/9903253}{{\ttfamily
  arXiv:astro-ph/9903253 [astro-ph]}}.

\bibitem{Bugaev2003}
E.~Bugaev, M.~Elbakidze, and K.~Konishchev {\em Phys. Atom. Nucl.} {\bfseries
  66} (2003) 476, \href{http://arxiv.org/abs/astro-ph/0110660}{{\ttfamily
  arXiv:astro-ph/0110660 [astro-ph]}}.

\bibitem{Bugaev2002}
E.~Bugaev and K.~Konishchev
  \href{http://dx.doi.org/10.1103/PhysRevD.66.084004}{{\em Phys. Rev. D}
  {\bfseries 66} (2002) 084004}.

\bibitem{Bugaev2009}
E.~Bugaev and P.~Klimai {\em Phys. Rev. D} {\bfseries 79} (2009) 103511,
  \href{http://arxiv.org/abs/0812.4247}{{\ttfamily arXiv:0812.4247
  [astro-ph]}}.

\bibitem{Green1999}
A.~Green {\em Phys. Rev. D} {\bfseries 60} (1999) 063516,
  \href{http://arxiv.org/abs/astro-ph/9903484}{{\ttfamily
  arXiv:astro-ph/9903484 [astro-ph]}}.

\bibitem{Lemoine2000}
M.~Lemoine {\em Phys. Lett. B} {\bfseries 481} (2000) 333,
  \href{http://arxiv.org/abs/hep-ph/0001238}{{\ttfamily arXiv:hep-ph/0001238
  [hep-ph]}}.

\bibitem{Khlopov2006}
M.~Khlopov, A.~Barrau, and J.~Grain {\em Class. Quant. Grav.} {\bfseries 23}
  (2006) 1875, \href{http://arxiv.org/abs/astro-ph/0406621}{{\ttfamily
  arXiv:astro-ph/0406621 [astro-ph]}}.

\bibitem{Izawa1984}
M.~Izawa and K.~Sato {\em Prog. Theor. Phys.} {\bfseries 72} (1984) 768.

\bibitem{Stojkovic2005a}
D.~Stojkovic and K.~Freese {\em Phys. Lett. B} {\bfseries 606} (2005) 251,
  \href{http://arxiv.org/abs/hep-ph/0403248}{{\ttfamily arXiv:hep-ph/0403248
  [hep-ph]}}.

\bibitem{Stojkovic2005b}
D.~Stojkovic, K.~Freese, and G.~Starkman {\em Phys. Rev. D} {\bfseries 72}
  (2005) 045012, \href{http://arxiv.org/abs/hep-ph/0505026}{{\ttfamily
  arXiv:hep-ph/0505026 [hep-ph]}}.

\bibitem{Page1976b}
D.~Page {\em Phys. Rev. D} {\bfseries 13} (1976) 198.

\bibitem{Islam1980}
J.~Islam and B.~Schutz {\em Gen. Rel. Grav.} {\bfseries 12} (1980) 881.

\bibitem{Mack2007}
K.~Mack, J.~Ostriker, and M.~Ricotti {\em Astrophys. J.} {\bfseries 665} (2007)
  1277, \href{http://arxiv.org/abs/astro-ph/0608642}{{\ttfamily
  arXiv:astro-ph/0608642 [astro-ph]}}.

\bibitem{Pani2013}
P.~Pani and A.~Loeb {\em Phys. Rev. D} {\bfseries 88} (2013) 041301,
  \href{http://arxiv.org/abs/1307.5176}{{\ttfamily arXiv:1307.5176
  [astro-ph.CO]}}.

\bibitem{Carr1999}
B.~Carr and M.~Sakellariadou {\em Astrophys. J.} {\bfseries 516} (1999) 195.

\bibitem{Capela2013}
F.~Capela, M.~Pshirkov, and P.~Tinyakov {\em Phys. Rev. D} {\bfseries 87}
  (2013) 123524, \href{http://arxiv.org/abs/1301.4984}{{\ttfamily
  arXiv:1301.4984 [astro-ph.CO]}}.

\bibitem{Adler2001}
R.~Adler, P.~Chen, and D.~Santiago {\em Gen. Rel. Grav.} {\bfseries 33} (2001)
  2101, \href{http://arxiv.org/abs/gr-qc/0106080}{{\ttfamily
  arXiv:gr-qc/0106080 [gr-qc]}}.

\bibitem{Carr2013}
B.~Carr {\em Mod. Phys. Lett. A} {\bfseries 28} (2013) 1340011.

\bibitem{Bowick1988}
M.~Bowick, S.~Giddings, J.~Harvey, G.~Horowitz, and A.~Strominger {\em Phys.
  Rev. Lett.} {\bfseries 61} (1988) 2823.

\bibitem{Lee1992}
K.-M. Lee, V.~Nair, and E.~Weinberg {\em Phys. Rev. Lett.} {\bfseries 68}
  (1992) 1100, \href{http://arxiv.org/abs/hep-th/9111045}{{\ttfamily
  arXiv:hep-th/9111045 [hep-th]}}.

\bibitem{Coleman1991}
S.~Coleman, J.~Preskill, and F.~Wilczek {\em Mod. Phys. Lett. A} {\bfseries 6}
  (1991) 1631.

\bibitem{Torii1993}
T.~Torii and K.~Maeda {\em Phys. Rev. D} {\bfseries 48} (1993) 1643.

\bibitem{MacGibbon1987}
J.~MacGibbon {\em Nature} {\bfseries 329} (1987) 308.

\bibitem{Barrow1992}
J.~Barrow, E.~Copeland, and A.~Liddle {\em Phys. Rev. D} {\bfseries 46} (1992)
  645.

\bibitem{Alexeyev2002}
S.~Alexeyev, A.~Barrau, G.~Boudoul, O.~Khovanskaya, and M.~Sazhin {\em Class.
  Quant. Grav.} {\bfseries 19} (2002) 4431,
  \href{http://arxiv.org/abs/gr-qc/0201069}{{\ttfamily arXiv:gr-qc/0201069
  [gr-qc]}}.

\bibitem{Barrau2004}
A.~Barrau, D.~Blais, G.~Boudoul, and D.~Polarski {\em Annalen Phys.} {\bfseries
  13} (2004) 115, \href{http://arxiv.org/abs/astro-ph/0303330}{{\ttfamily
  arXiv:astro-ph/0303330 [astro-ph]}}.

\bibitem{Chen2005}
P.~Chen {\em New Astron. Rev.} {\bfseries 49} (2005) 233,
  \href{http://arxiv.org/abs/astro-ph/0406514}{{\ttfamily
  arXiv:astro-ph/0406514 [astro-ph]}}.

\bibitem{Nozari2005}
K.~Nozari and S.~Mehdipour {\em Mod. Phys. Lett. A} {\bfseries 20} (2005) 2937,
  \href{http://arxiv.org/abs/0809.3144}{{\ttfamily arXiv:0809.3144 [gr-qc]}}.

\bibitem{Carr1994}
B.~Carr, J.~Gilbert, and J.~Lidsey {\em Phys. Rev. D} {\bfseries 50} (1994)
  4853, \href{http://arxiv.org/abs/astro-ph/9405027}{{\ttfamily
  arXiv:astro-ph/9405027 [astro-ph]}}.

\bibitem{Chuzhoy2009}
L.~Chuzhoy and E.~Kolb {\em JCAP} {\bfseries 0907} (2009) 014,
  \href{http://arxiv.org/abs/0809.0436}{{\ttfamily arXiv:0809.0436
  [astro-ph]}}.

\bibitem{Lewin1977}
J.~Lewin {\em Rutherford Laboratory Technical Report RL-77-126/A} (1977) .

\bibitem{Lindhard1961}
J.~Lindhard and M.~Scharff {\em Phys. Rev.} {\bfseries 124} (1961) 128.

\bibitem{Ormrod1963}
J.~Ormrod and H.~Duckworth {\em Can. J. Phys.} {\bfseries 41} (1963) 1424.

\bibitem{Ormrod1965}
J.~Ormrod, J.~Mcdonald, and H.~Duckworth {\em Can. J. Phys.} {\bfseries 43}
  (1965) 275.

\bibitem{Kraan2004}
A.~Kraan {\em Eur. Phys. J. C} {\bfseries 37} (2004) 91,
  \href{http://arxiv.org/abs/hep-ex/0404001}{{\ttfamily arXiv:hep-ex/0404001
  [hep-ex]}}.

\bibitem{deBoer2008}
Y.~de~Boer, A.~Kaidalov, D.~Milstead, and O.~Piskounova {\em J. Phys. G}
  {\bfseries 35} (2008) 075009,
  \href{http://arxiv.org/abs/0710.3930}{{\ttfamily arXiv:0710.3930 [hep-ph]}}.

\bibitem{Mackeprang2010}
R.~Mackeprang and D.~Milstead {\em Eur. Phys. J. C} {\bfseries 66} (2010) 493,
  \href{http://arxiv.org/abs/0908.1868}{{\ttfamily arXiv:0908.1868 [hep-ph]}}.

\bibitem{Farrar2011}
G.~Farrar, R.~Mackeprang, D.~Milstead, and J.~Roberts {\em JHEP} {\bfseries
  1102} (2011) 018, \href{http://arxiv.org/abs/1011.2964}{{\ttfamily
  arXiv:1011.2964 [hep-ph]}}.

\bibitem{Heiselberg:1993dc}
H.~Heiselberg
{\em Phys.Rev.} {\bfseries D48} (1993) 1418.

\bibitem{Wilk1996}
G.~Wilk and Z.~Wlodarczyk {\em Heavy Ion Phys.} {\bfseries 4} (1996) 395,
  \href{http://arxiv.org/abs/hep-ph/9606401}{{\ttfamily arXiv:hep-ph/9606401
  [hep-ph]}}.

\bibitem{Banerjee2000}
S.~Banerjee, S.~Ghosh, S.~Raha, and D.~Syam {\em Phys. Rev. Lett.} {\bfseries
  85} (2000) 1384, \href{http://arxiv.org/abs/hep-ph/0006286}{{\ttfamily
  arXiv:hep-ph/0006286 [hep-ph]}}.

\bibitem{SuperKamiokande2007}
{Super-Kamiokande Collaboration} {\em Phys. Lett. B} {\bfseries 647} (2007) 18,
  \href{http://arxiv.org/abs/hep-ex/0608057}{{\ttfamily arXiv:hep-ex/0608057
  [hep-ex]}}.

\bibitem{MACRO2000}
{MACRO Collaboration} {\em Eur. Phys. J. C} {\bfseries 13} (2000) 453,
  \href{http://arxiv.org/abs/hep-ex/9904031}{{\ttfamily arXiv:hep-ex/9904031
  [hep-ex]}}.

\bibitem{SLIM2008b}
{SLIM Collaboration}
  \href{http://dx.doi.org/10.1140/epjc/s10052-008-0747-7}{{\em Eur. Phys. J. C}
  {\bfseries 57} (2008) 525}, \href{http://arxiv.org/abs/0805.1797}{{\ttfamily
  arXiv:0805.1797 [hep-ex]}}.

\bibitem{Greenstein1984}
G.~Greenstein and J.~Burns {\em Am. J. Phys.} {\bfseries 52} (1984) 531.

\bibitem{Khriplovich2008}
I.~Khriplovich, A.~Pomeransky, N.~Produit, and G.~Y. Ruban {\em Phys. Rev. D}
  {\bfseries 77} (2008) 064017,
  \href{http://arxiv.org/abs/0710.3438}{{\ttfamily arXiv:0710.3438
  [astro-ph]}}.

\bibitem{Ahlen1975}
S.~Ahlen {\em Phys. Rev. D} {\bfseries 14} (1975) 2935.

\bibitem{Ahlen1978}
S.~Ahlen {\em Phys. Rev. D} {\bfseries 17} (1978) 229.

\bibitem{Sternheimer1971}
R.~Sternheimer and R.~Peierls {\em Phys. Rev. B} {\bfseries 3} (1971) 3681.

\bibitem{Ahlen1982}
S.~Ahlen and K.~Kinoshita
  \href{http://dx.doi.org/10.1103/PhysRevD.26.2347}{{\em Phys. Rev. D}
  {\bfseries 26} (1982) 2347}.

\bibitem{Ahlen1983}
S.~Ahlen and G.~Tarl{\'e} {\em Phys. Rev. D} {\bfseries 27} (1983) 688.

\bibitem{Ficenec1987}
D.~Ficenec, S.~Ahlen, A.~Marin, J.~Musser, and G.~Tarl{\'e} {\em Phys. Rev. D}
  {\bfseries 36} (1987) 311.

\bibitem{Drell1983}
S.~Drell, N.~Kroll, M.~Mueller, S.~Parke, and M.~Ruderman {\em Phys. Rev.
  Lett.} {\bfseries 50} (1983) 644.

\bibitem{Bracci1983}
L.~Bracci and G.~Fiorentini {\em Phys. Lett. B} {\bfseries 124} (1983) 493.

\bibitem{Derkaoui1998}
J.~Derkaoui, G.~Giacomelli, T.~Lari, A.~Margiotta, M.~Ouchrif, L.~Patrizii,
  V.~Popa, and V.~Togo {\em Astropart. Phys.} {\bfseries 9} (1998) 173.

\bibitem{RICE2008}
D.~P. Hogan, D.~Z. Besson, J.~P. Ralston, I.~Kravchenko, and D.~Seckel {\em
  Phys. Rev. D} {\bfseries 78} (2008) 075031,
  \href{http://arxiv.org/abs/0806.2129}{{\ttfamily arXiv:0806.2129
  [astro-ph]}}.

\bibitem{ANITA2011}
{ANITA-II Collaboration} {\em Phys. Rev. D} {\bfseries 83} (2011) 023513,
  \href{http://arxiv.org/abs/1008.1282}{{\ttfamily arXiv:1008.1282
  [astro-ph]}}.

\bibitem{Dutta2001}
S.~Dutta, M.~Reno, I.~Sarcevic, and D.~Seckel {\em Phys. Rev. D} {\bfseries 63}
  (2001) 094020, \href{http://arxiv.org/abs/hep-ph/0012350}{{\ttfamily
  arXiv:hep-ph/0012350 [hep-ph]}}.

\bibitem{Kim1973}
K.~Kim and Y.-S. Tsai {\em Phys. Rev. D} {\bfseries 8} (1973) 3109.

\bibitem{Kittel1977}
C.~Kittel and A.~Manoliu \href{http://dx.doi.org/10.1103/PhysRevB.15.333}{{\em
  Phys. Rev. B} {\bfseries 15} (1977) 333}.

\bibitem{Ruijgrok1983}
T.~Ruijgrok, J.~Tjon, and T.~Wu {\em Phys. Lett. B} {\bfseries 129} (1983) 209.

\bibitem{Ruijgrok1984}
T.~Ruijgrok {\em Acta Phys. Polon. B} {\bfseries 15} (1984) 305.

\bibitem{Sivers1970}
D.~Sivers {\em Phys. Rev. D} {\bfseries 2} (1970) 2048.

\bibitem{Milton2006}
K.~Milton {\em Rep. Prog. Phys.} {\bfseries 69} (2006) 1637,
  \href{http://arxiv.org/abs/0602040}{{\ttfamily arXiv:0602040 [hep-ex]}}.

\bibitem{Olaussen1985}
K.~Olaussen and R.~Sollie {\em Nucl. Phys. B} {\bfseries 255} (1985) 465.

\bibitem{Alvager1967}
T.~{Alv\"ager} and R.~Naumann {\em Phys. Lett. B} {\bfseries 24} (1967) 647.

\bibitem{Muller1977}
R.~Muller, L.~Alvarez, W.~Holley, and E.~Stephenson {\em Science} {\bfseries
  196} (1977) 521.

\bibitem{Boyd1978b}
R.~Boyd, D.~Elmore, D.~Nitz, S.~Olsen, E.~Sugarbaker, and G.~Warren {\em Phys.
  Lett. B} {\bfseries 72} (1978) 484.

\bibitem{Smith1979}
P.~Smith and J.~Bennett {\em Nuclear Physics B} {\bfseries 149} (1979) 525.

\bibitem{Smith1982}
P.~Smith, J.~Bennett, G.~Homer, J.~Lewin, H.~Walford, and W.~Smith {\em Nucl.
  Phys. B} {\bfseries 206} (1982) 333.

\bibitem{Hemmick1990}
T.~Hemmick, D.~Elmore, T.~Gentile, P.~Kubik, S.~Olsen, D.~Ciampa, D.~Nitz,
  H.~Kagan, P.~Haas, P.~Smith, B.~McInteer, and J.~Bigeleisen {\em Phys. Rev.
  D} {\bfseries 41} (1990) 2074.

\bibitem{Verkerk1992}
P.~Verkerk, G.~Grynberg, B.~Pichard, M.~Spiro, S.~Zylberajch, M.~Goldberg, and
  P.~Fayet \href{http://dx.doi.org/10.1103/PhysRevLett.68.1116}{{\em Phys. Rev.
  Lett.} {\bfseries 68} (1992) 1116}.

\bibitem{Yamagata1993}
T.~Yamagata, Y.~Takamori, and H.~Utsunomiya
  \href{http://dx.doi.org/10.1103/PhysRevD.47.1231}{{\em Phys. Rev. D}
  {\bfseries 47} (1993) 1231}.

\bibitem{Lu:2004pk}
Z.~Lu, R.~Holt, P.~Mueller, T.~O'Connor, J.~Schiffer, {\em et~al.} {\em Nucl.
  Phys. A} {\bfseries 754} (2005) 361--368,
\href{http://arxiv.org/abs/nucl-ex/0402015}{{\ttfamily arXiv:nucl-ex/0402015
  [nucl-ex]}}.

\bibitem{Boyd1991}
R.~Boyd and M.~Caffee
  \href{http://dx.doi.org/10.1016/0920-5632(91)90324-8}{{\em Nucl. Phys. B
  (Proc. Suppl.)} {\bfseries 24} (1991) 195}.

\bibitem{Holt1976}
R.~Holt, J.~Schiffer, J.~Specht, L.~Bollinger, G.~Thomas, S.~Fried, J.~Hines,
  and A.~Friedman \href{http://dx.doi.org/10.1103/PhysRevLett.36.183}{{\em
  Phys. Rev. Lett.} {\bfseries 36} (1976) 183}.

\bibitem{Middleton1979}
R.~Middleton, R.~{Zurm\"uhle}, J.~Klein, and R.~Kollarits
  \href{http://dx.doi.org/10.1103/PhysRevLett.43.429}{{\em Phys. Rev. Lett.}
  {\bfseries 43} (1979) 429}.

\bibitem{Dick1984}
W.~Dick, G.~Greenlees, and S.~Kaufman
  \href{http://dx.doi.org/10.1103/PhysRevLett.53.431}{{\em Phys. Rev. Lett.}
  {\bfseries 53} (1984) 431}.

\bibitem{Dick1986}
W.~Dick, G.~Greenlees, and S.~Kaufman
  \href{http://dx.doi.org/10.1103/PhysRevD.33.32}{{\em Phys. Rev. D} {\bfseries
  33} (1986) 32}.

\bibitem{Nitz1986}
D.~Nitz, D.~Ciampa, T.~Hemmick, D.~Elmore, P.~Kubik, S.~Olsen, T.~Gentile,
  H.~Kagan, P.~Haas, and P.~Smith {\em AIP Conf. Proc.} {\bfseries 150} (1986)
  1143.

\bibitem{Norman1987}
E.~Norman, S.~Gazes, and D.~Bennett
  \href{http://dx.doi.org/10.1103/PhysRevLett.58.1403}{{\em Phys. Rev. Lett.}
  {\bfseries 58} (1987) 1403}.

\bibitem{Norman1989}
E.~Norman, R.~Chadwick, K.~Lesko, R.-M. Larimer, and D.~Hoffman
  \href{http://dx.doi.org/10.1103/PhysRevD.39.2499}{{\em Phys. Rev. D}
  {\bfseries 39} (1989) 2499}.

\bibitem{Stevens1976}
C.~Stevens, J.~Schiffer, and W.~Chupka {\em Phys. Rev. D} {\bfseries 14} (1976)
  716.

\bibitem{Han2009}
K.~Han, J.~Ashenfelter, A.~Chikanian, W.~Emmet, L.~Finch, A.~Heinz, J.~Madsen,
  R.~Majka, B.~Monreal, and J.~Sandweiss {\em Phys. Rev. Lett.} {\bfseries 103}
  (2009) 092302, \href{http://arxiv.org/abs/0903.5055}{{\ttfamily
  arXiv:0903.5055 [nucl-ex]}}.

\bibitem{Perillo1998}
M.~Perillo~Isaac, Y.~Chan, R.~Clark, M.~Deleplanque, M.~Dragowsky, P.~Fallon,
  I.~Goldman, R.-M. Larimer, I.~Lee, A.~Macchiavelli, R.~MacLeod,
  K.~Nishiizumi, E.~Norman, L.~Schroeder, and F.~Stephens
  \href{http://dx.doi.org/10.1103/PhysRevLett.81.2416}{{\em Phys. Rev. Lett.}
  {\bfseries 81} (1998) 2416}, \href{http://arxiv.org/abs/9806147}{{\ttfamily
  arXiv:9806147 [astro-ph]}}.

\bibitem{Jones1989}
W.~Jones, P.~Smith, G.~Homer, J.~Lewin, and H.~Walford {\em Z. Phys. C}
  {\bfseries 43} (1989) 349.

\bibitem{Vandegriff1996}
J.~Vandegriff, G.~Raimann, R.~Boyd, M.~Caffee, and B.~Ruiz {\em Phys. Lett. B}
  {\bfseries 365} (1996) 418, \href{http://arxiv.org/abs/9510003}{{\ttfamily
  arXiv:9510003 [nucl-ex]}}.

\bibitem{Javorsek2001a}
D.~Javorsek, D.~Elmore, E.~Fischbach, D.~Granger, T.~Miller, D.~Oliver, and
  V.~Teplitz \href{http://dx.doi.org/10.1103/PhysRevD.64.012005}{{\em Phys.
  Rev. D} {\bfseries 64} (2001) 012005}.

\bibitem{Mueller2004}
P.~Mueller, L.-B. Wang, R.~Holt, Z.-T. Lu, T.~O’Connor, and J.~Schiffer
  \href{http://dx.doi.org/10.1103/PhysRevLett.92.022501}{{\em Phys. Rev. Lett.}
  {\bfseries 92} (2004) 022501}, \href{http://arxiv.org/abs/0302025}{{\ttfamily
  arXiv:0302025 [nucl-ex]}}.

\bibitem{Lu2005}
Z.~Lu, R.~Holt, P.~Mueller, T.~O'Connor, J.~Schiffer, and L.-B. Wang {\em Nucl.
  Phys. A} {\bfseries 754} (2005) 361,
  \href{http://arxiv.org/abs/0402015}{{\ttfamily arXiv:0402015 [nucl-ex]}}.

\bibitem{Turkevich1984}
A.~Turkevich, K.~Wielgoz, and T.~Economou
  \href{http://dx.doi.org/10.1103/PhysRevD.30.1876}{{\em Phys. Rev. D}
  {\bfseries 30} (1984) 1876}.

\bibitem{Polikanov1991}
S.~Polikanov, C.~Sastri, G.~Herrmann, K.~{L\"utzenkirchen}, M.~Overbeck,
  N.~Trautmann, A.~Breskin, R.~Chechik, and Z.~Fraenkel
  \href{http://dx.doi.org/10.1007/BF01288200}{{\em Z. Phys. A} {\bfseries 338}
  (1991) 357}.

\bibitem{Farhi1985}
E.~Farhi and R.~Jaffe {\em Phys. Rev. D} {\bfseries 32} (1985) 2452.

\bibitem{Blackman1989}
E.~Blackman and R.~Jaffe {\em Nucl. Phys. B} {\bfseries 324} (1989) 205.

\bibitem{Klingenberg2001}
R.~Klingenberg {\em J. Phys. G} {\bfseries 27} (2001) 475.

\bibitem{Finch2006}
E.~Finch {\em J. Phys. G} {\bfseries 32} (2006) S251,
  \href{http://arxiv.org/abs/nucl-ex/0605010}{{\ttfamily arXiv:nucl-ex/0605010
  [nucl-ex]}}.

\bibitem{Carrigan1983}
R.~Carrigan and W.~Trowers {\em Nature} {\bfseries 305} (1983) 673.

\bibitem{Carrigan1973}
R.~Carrigan, F.~Nezrick, and B.~Strauss
  \href{http://dx.doi.org/10.1103/PhysRevD.8.3717}{{\em Phys. Rev. D}
  {\bfseries 8} (1973) 3717}.

\bibitem{Kalbfleisch2000}
G.~Kalbfleisch, K.~Milton, M.~Strauss, L.~Gamberg, E.~Smith, and W.~Luo
  \href{http://dx.doi.org/10.1103/PhysRevLett.85.5292}{{\em Phys. Rev. Lett.}
  {\bfseries 85} (2000) 5292}, \href{http://arxiv.org/abs/0005005}{{\ttfamily
  arXiv:0005005 [hep-ex]}}.

\bibitem{Kalbfleisch2004}
G.~Kalbfleisch, W.~Luo, K.~Milton, E.~Smith, and M.~Strauss
  \href{http://dx.doi.org/10.1103/PhysRevD.69.052002}{{\em Phys. Rev. D}
  {\bfseries 69} (2004) 052002}, \href{http://arxiv.org/abs/0306045}{{\ttfamily
  arXiv:0306045 [hep-ex]}}.

\bibitem{H12005}
{H1 Collaboration} \href{http://dx.doi.org/10.1140/epjc/s2005-02201-6}{{\em
  Eur. Phys. J. C} {\bfseries 41} (2005) 133},
  \href{http://arxiv.org/abs/0501039}{{\ttfamily arXiv:0501039 [hep-ex]}}.

\bibitem{Bendtz2013a}
K.~Bendtz, D.~Milstead, H.-P. Hachler, A.~Hirt, P.~Mermod, P.~Michael,
  T.~Sloan, C.~Tegner, and S.~Thorarinsson {\em Phys. Rev. Lett.} {\bfseries
  110} (2013) 121803, \href{http://arxiv.org/abs/1301.6530}{{\ttfamily
  arXiv:1301.6530 [hep-ex]}}.

\bibitem{Petukhov1963}
V.~Petukhov and M.~Yakimenko
  \href{http://dx.doi.org/10.1016/0029-5582(63)90076-2}{{\em Nucl. Phys.}
  {\bfseries 49} (1963) 87}.

\bibitem{Fleischer1969a}
R.~Fleischer, I.~Jacobs, W.~Schwarz, and P.~Price {\em Phys. Rev.} {\bfseries
  177} (1969) 2029.

\bibitem{Fleischer1969b}
R.~Fleischer, H.~Hart, I.~Jacobs, P.~Price, W.~Schwarz, and F.~Aumento
  \href{http://dx.doi.org/10.1103/PhysRev.184.1393}{{\em Phys. Rev.} {\bfseries
  184} (1969) 1393}.

\bibitem{Kolm1971}
H.~Kolm, F.~Villa, and A.~Odian {\em Phys. Rev. D} {\bfseries 4} (1971) 1285.

\bibitem{Carrigan1976}
R.~Carrigan, F.~Nezrick, and B.~Strauss
  \href{http://dx.doi.org/10.1103/PhysRevD.13.1823}{{\em Phys. Rev. D}
  {\bfseries 13} (1976) 1823}.

\bibitem{Malkus1951}
W.~Malkus {\em Phys. Rev.} {\bfseries 83} (1951) 899.

\bibitem{Carithers1966}
W.~Carithers, R.~Stefanski, and R.~Adair
  \href{http://dx.doi.org/10.1103/PhysRev.149.1070}{{\em Phys. Rev.} {\bfseries
  149} (1966) 1070}.

\bibitem{Bartlett1981}
D.~Bartlett, D.~Soo, R.~Fleischer, H.~Hart~Jr., and A.~Mogro-Campero
  \href{http://dx.doi.org/10.1103/PhysRevD.24.612}{{\em Phys. Rev. D}
  {\bfseries 24} (1981) 612}.

\bibitem{Goto1958}
E.~Goto {\em J. Phys. Soc. Jpn.} {\bfseries 13} (1958) 1413.

\bibitem{Jeon1995}
H.~Jeon and M.~Longo \href{http://dx.doi.org/10.1103/PhysRevLett.75.1443}{{\em
  Phys. Rev. Lett.} {\bfseries 75} (1995) 1443},
  \href{http://arxiv.org/abs/9508003}{{\ttfamily arXiv:9508003 [hep-ex]}}.

\bibitem{Kovalik1986}
J.~Kovalik and J.~Kirschvink {\em Phys. Rev. A} {\bfseries 33} (1986) 1183.

\bibitem{Eberhard1971}
P.~H. Eberhard, R.~Ross, L.~W. Alvarez, and R.~Watt {\em Phys. Rev. D}
  {\bfseries 4} (1971) 3260.

\bibitem{Ross1973}
R.~Ross, P.~Eberhard, L.~Alvarez, and R.~Watt {\em Phys. Rev. D} {\bfseries 8}
  (1973) 698.

\bibitem{Ebisu1987}
T.~Ebisu and T.~Watanabe \href{http://dx.doi.org/10.1103/PhysRevD.36.3359}{{\em
  Phys. Rev. D} {\bfseries 36} (1987) 3359}.

\bibitem{Carrigan1980}
R.~Carrigan \href{http://dx.doi.org/10.1038/288348a0}{{\em Nature} {\bfseries
  288} (1980) 348}.

\bibitem{Schatten1983}
K.~Schatten \href{http://dx.doi.org/10.1103/PhysRevD.27.1525}{{\em Phys. Rev.
  D} {\bfseries 27} (1983) 1525}.

\bibitem{Groom1986}
D.~E. Groom {\em Phys. Rept.} {\bfseries 140} (1986) 323.

\bibitem{Incandela1984}
J.~Incandela, M.~Campbell, H.~Frisch, S.~Somalwar, M.~Kuchnir, and H.~Gustafson
  \href{http://dx.doi.org/10.1103/PhysRevLett.53.2067}{{\em Phys. Rev. Lett.}
  {\bfseries 53} (1984) 2067}.

\bibitem{Ebisu1985}
T.~Ebisu and T.~Watanabe {\em J. Phys. G} {\bfseries 11} (1985) 883.

\bibitem{Incandela1986}
J.~Incandela, H.~Frisch, S.~Somalwar, M.~Kuchnir, and H.~Gustafson {\em Phys.
  Rev. D} {\bfseries 34} (1986) 2637.

\bibitem{Cabrera1983a}
B.~Cabrera, M.~Taber, R.~Gardner, and J.~Bourg
  \href{http://dx.doi.org/10.1103/PhysRevLett.51.1933}{{\em Phys. Rev. Lett.}
  {\bfseries 51} (1983) 1933}.

\bibitem{Caplin1985}
A.~Caplin, C.~Guy, M.~Hardiman, J.~Park, and J.~Schouten {\em Nature}
  {\bfseries 317} (1985) 234.

\bibitem{Caplin1986}
A.~Caplin, M.~Hardiman, M.~Koratzinos, and J.~Schouten {\em Nature} {\bfseries
  321} (1986) 402.

\bibitem{Cromar1986}
M.~Cromar, A.~Clark, and F.~Fickett {\em Phys. Rev. Lett.} {\bfseries 56}
  (1986) 2561.

\bibitem{Gardner1991}
R.~Gardner, B.~Cabrera, M.~Huber, and M.~Taber {\em Phys. Rev. D} {\bfseries
  44} (1991) 622.

\bibitem{Huber1990}
M.~Huber, B.~Cabrera, M.~Taber, and R.~Gardner
  \href{http://dx.doi.org/10.1103/PhysRevLett.64.835}{{\em Phys. Rev. Lett.}
  {\bfseries 64} (1990) 835}.

\bibitem{Huber1991}
M.~Huber, B.~Cabrera, M.~Taber, and R.~Gardner {\em Phys. Rev. D} {\bfseries
  44} (1991) 636.

\bibitem{Bermon1985}
S.~Bermon, P.~Chaudhari, C.~Chi, C.~Tesche, and C.~Tsuei {\em Phys. Rev. Lett.}
  {\bfseries 55} (1985) 1850.

\bibitem{Bermon1990}
S.~Bermon, C.~Chi, C.~Tsuei, J.~Rozen, P.~Chaudhari, M.~McElfresh, and
  A.~Prodell {\em Phys. Rev. Lett.} {\bfseries 64} (1990) 839.

\bibitem{Jesse1964}
P.~Jesse \href{http://dx.doi.org/10.1063/1.1726206}{{\em J. Chem. Phys.}
  {\bfseries 41} (1964) 2060}.

\bibitem{MACRO1994}
{MACRO Collaboration} {\em Phys. Rev. Lett.} {\bfseries 72} (1994) 608.

\bibitem{MACRO1997}
{MACRO Collaboration} {\em Phys. Lett. B} {\bfseries 406} no.~3, (1997) 249.

\bibitem{MACRO2002a}
{MACRO Collaboration} {\em Eur. Phys. J. C} {\bfseries 25} (2002) 511,
  \href{http://arxiv.org/abs/hep-ex/0207020}{{\ttfamily arXiv:hep-ex/0207020
  [hep-ex]}}.

\bibitem{Mashimo1982}
T.~Mashimo, K.~Kawagoe, and M.~Koshiba {\em JPSJ} {\bfseries 49} (1982) 1114.

\bibitem{Bonarelli1982}
R.~Bonarelli, P.~Capiluppi, I.~D'Antone, G.~Giacomelli, G.~Mandrioli, C.~Merli,
  and A.~Rossi {\em Phys. Lett. B} {\bfseries 112} (1982) 100.

\bibitem{Bonarelli1983}
R.~Bonarelli, P.~Capiluppi, I.~D'antone, G.~Giacomelli, G.~Mandrioli, C.~Merli,
  and A.~Rossi {\em Phys. Lett. B} {\bfseries 126} (1983) 137.

\bibitem{Kajino1984a}
F.~Kajino, S.~Matsuno, Y.~Yuan, and T.~Kitamura
  \href{http://dx.doi.org/10.1103/PhysRevLett.52.1373}{{\em Phys. Rev. Lett.}
  {\bfseries 52} (1984) 1373}.

\bibitem{Tarle1984}
G.~Tarl{\'e}, S.~Ahlen, and T.~Liss
  \href{http://dx.doi.org/10.1103/PhysRevLett.52.90}{{\em Phys. Rev. Lett.}
  {\bfseries 52} (1984) 90}.

\bibitem{Liss1984}
T.~Liss, S.~Ahlen, and G.~Tarl{\'e} {\em Phys. Rev. D} {\bfseries 30} (1984)
  884.

\bibitem{Barish1987}
B.~Barish, G.~Liu, and C.~Lane
  \href{http://dx.doi.org/10.1103/PhysRevD.36.2641}{{\em Phys. Rev. D}
  {\bfseries 36} (1987) 2641}.

\bibitem{Alekseev1982}
E.~Alekseev, M.~Boliev, A.~Chudakov, B.~Makoev, S.~Mikheev, and Y.~Stenkin {\em
  Lett. Nuovo Cim.} {\bfseries 35} (1982) 413.

\bibitem{Mashimo1983}
T.~Mashimo, S.~Orito, K.~Kawagoe, S.~Nakamura, and M.~Nozaki {\em Phys. Lett.
  B} {\bfseries 128} (1983) 327.

\bibitem{Groom1983}
D.~Groom, E.~Loh, H.~Nelson, and D.~Ritson
  \href{http://dx.doi.org/10.1103/PhysRevLett.50.573}{{\em Phys. Rev. Lett.}
  {\bfseries 50} (1983) 573}.

\bibitem{Kawagoe1984}
K.~Kawagoe, T.~Mashimo, S.~Nakamura, M.~Nozaki, and S.~Orito {\em Lett. Nuovo
  Cim.} {\bfseries 41} (1984) 315.

\bibitem{Tsukamoto1987}
T.~Tsukamoto, K.~Nagano, K.~Anraku, M.~Imori, K.~Kawagoe, S.~Nakamura,
  M.~Nozaki, S.~Orito, K.~Yamamoto, and T.~Yoshida {\em Europhys. Lett.}
  {\bfseries 3} (1987) 39.

\bibitem{Shepko1987}
M.~Shepko, C.~Gagliardi, P.~Green, P.~Mcintyre, T.~Meyer, R.~Tribble, and
  R.~Webb {\em Phys. Rev. D} {\bfseries 35} (1987) 2917.

\bibitem{Bartelt1983}
J.~Bartelt, H.~Courant, K.~Heller, T.~Joyce, M.~Marshak, E.~Peterson,
  K.~Ruddick, M.~Shupe, D.~Ayres, J.~Dawson, T.~Fields, E.~May, and L.~Price
  \href{http://dx.doi.org/10.1103/PhysRevLett.50.655}{{\em Phys. Rev. Lett.}
  {\bfseries 50} (1983) 655}.

\bibitem{Krishnaswamy1984}
M.~Krishnaswamy, M.~Menon, N.~Mondal, V.~Narasimham, B.~Sreekantan, Y.~Hayashi,
  N.~Ito, S.~Kawakami, and S.~Miyake, ``{Limits on the flux of monopoles from
  the Kolar Gold Mine experiments},''
  \href{http://dx.doi.org/10.1016/0370-2693(84)91143-2}{{\em Phys. Lett. B}
  {\bfseries 142} (1984) 99}.

\bibitem{Soudan21992}
{Soudan 2 Collaboration} \href{http://dx.doi.org/10.1103/PhysRevD.46.4846}{{\em
  Phys. Rev. D} {\bfseries 46} (1992) 4846}.

\bibitem{L3C2009}
{L3+C Collaboration} {\em Chin. Phys. C} {\bfseries 33} (2009) 603.

\bibitem{Ullman1981}
J.~Ullman \href{http://dx.doi.org/10.1103/PhysRevLett.47.289}{{\em Phys. Rev.
  Lett.} {\bfseries 47} (1981) 289}.

\bibitem{Hara1986}
T.~Hara, M.~Honda, Y.~Ohno, N.~Hayashida, K.~Kamata, T.~Kifune, G.~Tanahashi,
  M.~Mori, Y.~Matsubara, M.~Teshima, M.~Kobayashi, T.~Kondo, K.~Nishijima, and
  Y.~Totsuka {\em Phys. Rev. Lett.} {\bfseries 56} (1986) 553.

\bibitem{Masek1987}
G.~Masek, L.~Knapp, E.~Miller, J.~Stronski, W.~Vernon, and J.~White {\em Phys.
  Rev. D} {\bfseries 35} (1987) 2758.

\bibitem{Buckland1990}
K.~Buckland, G.~Masek, W.~Vernon, L.~Knapp, and J.~Stronski {\em Phys. Rev. D}
  {\bfseries 41} (1990) 2726.

\bibitem{Battistoni1983}
G.~Battistoni, E.~Bellotti, G.~Bologna, P.~Campana, C.~Castagnoli,
  V.~Chiarella, A.~Ciocio, D.~Cundy, B.~D'Ettorre-Piazzoli, E.~Fiorini,
  P.~Galeotti, E.~Iarocci, C.~Liguori, G.~Mannocchi, G.~Murtas, P.~Negri,
  G.~Nicoletti, P.~Picchi, M.~Price, A.~Pullia, S.~Ragazzi, M.~Rollier,
  O.~Saavedra, L.~Satta, L.~Trasatti, and L.~Zanotti {\em Phys. Lett. B}
  {\bfseries 133} (1983) 454.

\bibitem{MACRO2000b}
{MACRO Collaboration} {\em Phys. Rev. D} {\bfseries 62} (2000) 052003,
  \href{http://arxiv.org/abs/hep-ex/0002029}{{\ttfamily arXiv:hep-ex/0002029
  [hep-ex]}}.

\bibitem{MACRO2004}
{MACRO Collaboration} \href{http://arxiv.org/abs/hep-ex/0402006}{{\ttfamily
  arXiv:hep-ex/0402006 [hep-ex]}}.

\bibitem{Aglietta1994}
M.~Aglietta, P.~Antonioli, G.~Badino, C.~Castagnoli, A.~Castellina, V.~Dadykin,
  W.~Fulgione, P.~Galeotti, F.~Khalchukov, E.~Korolkova, P.~Kortchaguin,
  V.~Kortchaguin, V.~Kudryavtsev, A.~Malguin, G.~Marchetti, L.~Periale,
  V.~Ryassnii, O.~Ryazhskaya, O.~Saavedra, G.~Trinchero, S.~Vernetto,
  V.~Yakushev, and G.~Zatsepin {\em Astropart. Phys.} {\bfseries 2} (1994) 29.

\bibitem{DeLise1965}
D.~De~Lise and T.~Bowen {\em Phys. Rev.} {\bfseries 140} (1965) B458.

\bibitem{MACRO1992}
{MACRO Collaboration} {\em Phys. Rev. Lett.} {\bfseries 69} (1992) 1860.

\bibitem{Starkman1990}
G.~Starkman, A.~Gould, R.~Esmailzadeh, and S.~Dimopoulos {\em Phys. Rev. D}
  {\bfseries 41} (1990) 3594.

\bibitem{Bernabei1999}
R.~Bernabei, P.~Belli, R.~Cerulli, F.~Montecchia, M.~Amato, G.~Ignesti,
  A.~Incicchitti, D.~Prosperi, C.~Dai, H.~He, H.~Kuang, J.~Ma, G.~Sun, and
  Z.~Ye {\em Phys. Rev. Lett.} {\bfseries 83} (1999) 4918.

\bibitem{Orito1991}
S.~Orito, H.~Ichinose, S.~Nakamura, K.~Kuwahara, T.~Doke, K.~Ogura, H.~Tawara,
  M.~Imori, K.~Yamamoto, H.~Yamakawa, T.~Suzuki, K.~Anraku, M.~Nozaki,
  M.~Sasaki, and T.~Yoshida
  \href{http://dx.doi.org/10.1103/PhysRevLett.66.1951}{{\em Phys. Rev. Lett.}
  {\bfseries 66} (1991) 1951}.

\bibitem{SLIM2008a}
{SLIM Collaboration}
  \href{http://dx.doi.org/10.1140/epjc/s10052-008-0597-3}{{\em Eur. Phys. J. C}
  {\bfseries 55} (2008) 57}, \href{http://arxiv.org/abs/0801.4913}{{\ttfamily
  arXiv:0801.4913 [hep-ex]}}.

\bibitem{Derkaoui1999}
J.~Derkaoui, G.~Giacomelli, T.~Lari, G.~Mandrioli, M.~Ouchrif, L.~Patrizii, and
  V.~Popa {\em Astropart. Phys.} {\bfseries 10} (1999) 339.

\bibitem{Fleischer1971}
R.~Fleischer, H.~Hart, G.~Nichols, and P.~Price {\em Phys. Rev. D} {\bfseries
  4} (1971) 24.

\bibitem{Doke1983}
T.~Doke, T.~Hayashi, R.~Hamasaki, T.~Akioka, T.~Naito, K.~Ito, T.~Yanagimachi,
  S.~Kobayashi, T.~Takenaka, M.~Ohe, K.~Nagata, and T.~Takahashi {\em Phys.
  Lett. B} {\bfseries 129} (1983) 370.

\bibitem{Nakamura1987}
S.~Nakamura, K.~Kawagoe, K.~Yamamoto, S.~Orito, H.~Ichinose, T.~Doke,
  T.~Hayashi, H.~Tawara, and K.~Ogura {\em Phys. Lett. B} {\bfseries 183}
  (1987) 395.

\bibitem{Price1975}
P.~Price, E.~Shirk, W.~Osborne, and L.~Pinsky
  \href{http://dx.doi.org/10.1103/PhysRevLett.35.487}{{\em Phys. Rev. Lett.}
  {\bfseries 35} (1975) 487}.

\bibitem{Price1978}
P.~Price, E.~Shirk, R.~Hagstrom, and W.~Osborne {\em Phys. Rev. D} {\bfseries
  18} (1978) 1382.

\bibitem{Barwick1983}
S.~Barwick, K.~Kinoshita, and P.~Price
  \href{http://dx.doi.org/10.1103/PhysRevD.28.2338}{{\em Phys. Rev. D}
  {\bfseries 28} (1983) 2338}.

\bibitem{Price1984b}
P.~Price {\em Phys. Lett. B} {\bfseries 140} (1984) 112.

\bibitem{Kinoshita1981}
K.~Kinoshita and P.~Price {\em Phys. Rev. D} {\bfseries 24} (1981) 1707.

\bibitem{Shirk1978}
E.~Shirk and P.~Price {\em Astrophys. J.} {\bfseries 220} (1978) 719.

\bibitem{Weaver1998}
B.~Weaver, A.~Westphal, P.~Price, V.~Afanasyev, and V.~Akimov
  \href{http://dx.doi.org/10.1016/S0168-583X(98)00403-0}{{\em Nucl. Instrum.
  Meth. B} {\bfseries 145} (1998) 409}.

\bibitem{Nakamura1990}
S.~Nakamura, S.~Orito, T.~Suzuki, T.~Doke, H.~Ichinose, K.~Kuwahara, and
  K.~Ogura {\em Phys. Lett. B} {\bfseries 263} (1991) 529.

\bibitem{Arafune2000}
J.~Arafune, T.~Yoshida, S.~Nakamura, and K.~Ogure {\em Phys. Rev. D} {\bfseries
  62} (2000) 105013, \href{http://arxiv.org/abs/hep-ph/0005103}{{\ttfamily
  arXiv:hep-ph/0005103 [hep-ph]}}.

\bibitem{Fleischer1965}
R.~Fleischer, P.~Price, and R.~Walker
  \href{http://dx.doi.org/10.1126/science.149.3682.383}{{\em Science}
  {\bfseries 149} (1965) 383}.

\bibitem{Uzgiris1971}
E.~Uzgiris and R.~Fleischer {\em Nature} {\bfseries 234} (1971) 28.

\bibitem{Fleischer1964a}
R.~Fleischer, P.~Price, E.~Symes, and D.~Miller
  \href{http://dx.doi.org/10.1126/science.143.3604.349}{{\em Science}
  {\bfseries 143} (1964) 349}.

\bibitem{Fleischer1967}
R.~Fleischer, P.~Price, R.~Walker, and E.~Hubbard {\em Phys. Rev.} {\bfseries
  156} (1967) 353.

\bibitem{Fleischer1969c}
R.~Fleischer, P.~Price, and R.~Woods
  \href{http://dx.doi.org/10.1103/PhysRev.184.1398}{{\em Phys. Rev.} {\bfseries
  184} (1969) 1398}.

\bibitem{Price1984a}
P.~Price, S.~Guo, S.~Ahlen, and R.~Fleischer
  \href{http://dx.doi.org/10.1103/PhysRevLett.52.1265}{{\em Phys. Rev. Lett.}
  {\bfseries 52} (1984) 1265}.

\bibitem{Price1986}
P.~Price and M.~Salamon {\em Phys. Rev. Lett.} {\bfseries 56} (1986) 1226.

\bibitem{Ghosh1990}
D.~Ghosh and S.~Chatterjea {\em Europhys. Lett.} {\bfseries 12} (1990) 25.

\bibitem{Price1988}
P.~Price {\em Phys. Rev. D} {\bfseries 38} (1988) 3813.

\bibitem{KM3NeT2011}
{KM3NeT Collaboration} {\em Nucl. Instrum. Meth. A} {\bfseries 630} (2011) 125.

\bibitem{ANTARES2012}
{ANTARES Collaboration} {\em Astropart. Phys.} {\bfseries 35} (2012) 634,
  \href{http://arxiv.org/abs/1110.2656}{{\ttfamily arXiv:1110.2656
  [astro-ph]}}.

\bibitem{AMANDA2010}
{IceCube Collaboration} {\em Eur. Phys. J. C} {\bfseries 69} (2010) 361.

\bibitem{BAIKAL2008}
{BAIKAL Collaboration} {\em Astropart. Phys.} {\bfseries 29} (2008) 366.

\bibitem{BAIKAL1995}
{BAIKAL Collaboration} {\em Proceedings of the 2nd Workshop on ``The dark side
  of the Universe: experimental efforts and theoretical framework''} (1995) ,
  \href{http://arxiv.org/abs/astro-ph/9601160}{{\ttfamily
  arXiv:astro-ph/9601160 [astro-ph]}}.

\bibitem{IceCube2013}
{IceCube Collaboration} {\em Phys. Rev. D} {\bfseries 87} (2013) 022001,
  \href{http://arxiv.org/abs/1208.4861}{{\ttfamily arXiv:1208.4861
  [astro-ph.HE]}}.

\bibitem{Kamiokande1991}
{Kamiokande-II Collaboration} {\em Phys. Rev. D} {\bfseries 43} (1991) 2843.

\bibitem{Pavalas2010}
G.~Pavalas {\em AIP Conf. Proc.} {\bfseries 1304} (2010) 454,
  \href{http://arxiv.org/abs/1010.2071}{{\ttfamily arXiv:1010.2071
  [astro-ph.HE]}}.

\bibitem{ask1}
G.~Askaryan {\em JETP} {\bfseries 14} 441.

\bibitem{ask2}
G.~Askaryan {\em JETP} {\bfseries 21} 658.

\bibitem{Lehtinen2004}
N.~Lehtinen, P.~Gorham, A.~Jacobson, and R.~Roussel-Dupre {\em Phys. Rev. D}
  {\bfseries 69} (2004) 013008,
  \href{http://arxiv.org/abs/astro-ph/0309656}{{\ttfamily
  arXiv:astro-ph/0309656 [astro-ph]}}.

\bibitem{Jaeger2010}
T.~Jaeger, R.~Mutel, and K.~Gayley {\em Astropart. Phys.} {\bfseries 34} (2010)
  293, \href{http://arxiv.org/abs/0910.5949}{{\ttfamily arXiv:0910.5949
  [astro-ph.IM]}}.

\bibitem{Buitink2010}
S.~Buitink, O.~Scholten, J.~Bacelar, R.~Braun, A.~de~Bruyn, H.~Falcke,
  K.~Singh, B.~Stappers, R.~Strom, and R.~Yahyaoui {\em Astron. Astrophys.}
  {\bfseries 521} (2010) A47,
\href{http://arxiv.org/abs/1004.0274}{{\ttfamily arXiv:1004.0274
  [astro-ph.HE]}}.

\bibitem{ANITA2010}
{ANITA Collaboration} \href{http://dx.doi.org/10.1103/PhysRevD.82.022004}{{\em
  Phys. Rev. D} {\bfseries 82} (2010) 022004},
  \href{http://arxiv.org/abs/1106.1164}{{\ttfamily arXiv:1106.1164
  [astro-ph.HE]}}.

\bibitem{Kravchenko2012}
I.~Kravchenko, S.~Hussain, D.~Seckel, D.~Besson, E.~Fensholt, J.~Ralston,
  J.~Taylor, K.~Ratzlaff, and R.~Young {\em Phys. Rev. D} {\bfseries 85} (2012)
  062004, \href{http://arxiv.org/abs/1106.1164}{{\ttfamily arXiv:1106.1164
  [astro-ph.HE]}}.

\bibitem{Yamamoto2008}
K.~Yamamoto, H.~Hayakawa, A.~Okada, T.~Uchiyama, S.~Miyoki, M.~Ohashi, and
  K.~Kuroda {\em Phys. Rev. D} {\bfseries 78} (2008) 022004,
  \href{http://arxiv.org/abs/0805.2387}{{\ttfamily arXiv:0805.2387 [gr-qc]}}.

\bibitem{Allega1983}
A.~Allega and N.~Cabibbo {\em Lett. Nuovo Cim.} {\bfseries 38} (1983) 263.

\bibitem{Bernard1984}
C.~Bernard, A.~De~Rujula, and B.~Lautrup
  \href{http://dx.doi.org/http://dx.doi.org/10.1016/0550-3213(84)90136-6}{{\em
  Nucl. Phys. B} {\bfseries 242} (1984) 93}.

\bibitem{Liu1988}
G.~Liu and B.~Barish {\em Phys. Rev. Lett.} {\bfseries 61} (1988) 271.

\bibitem{Astone1993}
P.~Astone, M.~Bassan, P.~Bonifazi, E.~Coccia, C.~Cosmelli, V.~Fafone,
  S.~Frasca, E.~Majorana, I.~Modena, G.~Pallottino, G.~Pizzella, P.~Rapagnani,
  F.~Ricci, F.~Ronga, and M.~Visco {\em Phys. Rev. D} {\bfseries 47} (1993)
  4770.

\bibitem{Astone2013}
P.~Astone, M.~Bassan, E.~Coccia, S.~D'Antonio, V.~Fafone, G.~Giordano,
  A.~Marini, Y.~Minenkov, I.~Modena, A.~Moleti, G.~Pallottino, G.~Pizzella,
  A.~Rocchi, F.~Ronga, and M.~Visco {\em proceedings of 33rd International
  Cosmic Ray Conference} (2013) ,
  \href{http://arxiv.org/abs/1306.5164}{{\ttfamily arXiv:1306.5164
  [astro-ph.HE]}}.

\bibitem{Nahnhauer2012}
R.~Nahnhauer {\em Nucl. Instrum. Meth. A} {\bfseries 662} (2012) S20,
  \href{http://arxiv.org/abs/1010.3082}{{\ttfamily arXiv:1010.3082
  [astro-ph.IM]}}.

\bibitem{Kurahashi2010}
N.~Kurahashi, J.~Vandenbroucke, and G.~Gratta {\em Phys. Rev. D} {\bfseries 82}
  (2010) 073006, \href{http://arxiv.org/abs/1007.5517}{{\ttfamily
  arXiv:1007.5517 [hep-ex]}}.

\bibitem{Danaher2012}
S.~Danaher, W.~Ooppakaew, and T.~Sloan {\em Nucl. Instrum. Meth. A} {\bfseries
  662} (2012) S198.

\bibitem{HiRes2004}
{HiRes Collaboration} {\em Phys. Rev. Lett.} {\bfseries 92} (2004) 151101,
  \href{http://arxiv.org/abs/astro-ph/0208243}{{\ttfamily
  arXiv:astro-ph/0208243 [astro-ph]}}.

\bibitem{Albuquerque1998}
I.~Albuquerque, G.~Farrar, and E.~Kolb {\em Phys. Rev. D} {\bfseries 59} (1999)
  015021, \href{http://arxiv.org/abs/hep-ph/9805288}{{\ttfamily
  arXiv:hep-ph/9805288 [hep-ph]}}.

\bibitem{Berezinsky2001}
V.~Berezinsky, M.~Kachelriess, and S.~Ostapchenko {\em Phys. Rev. D} {\bfseries
  65} (2002) 083004, \href{http://arxiv.org/abs/astro-ph/0109026}{{\ttfamily
  arXiv:astro-ph/0109026 [astro-ph]}}.

\bibitem{Gonzalez2005}
J.~Gonzalez, S.~Reucroft, and J.~Swain {\em Phys. Rev. D} {\bfseries 74} (2006)
  027701, \href{http://arxiv.org/abs/hep-ph/0504260}{{\ttfamily
  arXiv:hep-ph/0504260 [hep-ph]}}.

\bibitem{Anchordoqui2007}
L.~Anchordoqui, A.~Delgado, C.~Garcia~Canal, and S.~Sciutto {\em Phys. Rev. D}
  {\bfseries 77} (2008) 023009,
  \href{http://arxiv.org/abs/0710.0525}{{\ttfamily arXiv:0710.0525 [hep-ph]}}.

\bibitem{Albuquerque2009}
I.~Albuquerque and W.~Carvalho {\em Phys. Rev. D} {\bfseries 80} (2009) 023006,
  \href{http://arxiv.org/abs/0901.3572}{{\ttfamily arXiv:0901.3572 [hep-ph]}}.

\bibitem{Auger2012}
 {\em Bulletin of the American Physical Society} (2012) .
  {\tt{http://meetings.aps.org/link/BAPS.2012.APR.T7.7}}.

\bibitem{Fowler1987}
P.~Fowler, R.~Walker, M.~Masheder, R.~Moses, A.~Worley, and A.~Gay
  \href{http://dx.doi.org/10.1086/165101}{{\em Astrophys. J.} {\bfseries 314}
  (1987) 739}.

\bibitem{Binns1989}
W.~R. Binns, T.~Garrard, P.~Gibner, M.~Israel, M.~P. Kertzman, J.~Klarmann,
  B.~Newport, E.~Stone, and C.~Waddington
  \href{http://dx.doi.org/10.1086/168082}{{\em Astrophys. J.} {\bfseries 346}
  (1989) 997}.

\bibitem{Westphal1998}
A.~Westphal, P.~Price, B.~Weaver, and V.~Afanasiev
  \href{http://dx.doi.org/10.1038/23887}{{\em Nature} {\bfseries 396} (1998)
  50}.

\bibitem{Sandweiss2004}
J.~Sandweiss {\em J. Phys. G} {\bfseries 30} (2004) S51.

\bibitem{BESS2008}
{BESS Collaboration} {\em Adv. Space Res.} {\bfseries 41} (2008) 2050.

\bibitem{Sbarra2003}
C.~Sbarra, D.~Casadei, L.~Brocco, A.~Contin, G.~Levi, and F.~Palmonari {\em
  proceedings of 18th European Cosmic Rays Symposium} (2003) ,
  \href{http://arxiv.org/abs/astro-ph/0304192}{{\ttfamily
  arXiv:astro-ph/0304192 [astro-ph]}}.

\bibitem{Rubakov1983}
V.~Rubakov and M.~Serebryakov {\em Nucl. Phys. B} {\bfseries 218} (1983) 240.

\bibitem{Bosetti1983}
P.~Bosetti, P.~Gorham, F.~Harris, J.~Learned, M.~McMurdo, D.~O'Connor,
  V.~Stenger, S.~Thompson, and K.~Mitsui {\em Phys. Lett. B} {\bfseries 133}
  (1983) 265.

\bibitem{Errede1983}
S.~Errede, J.~Stone, J.~van~der Velde, R.~Bionta, G.~Blewitt, C.~Bratton,
  B.~Cortez, G.~Foster, W.~Gajewski, M.~Goldhaber, J.~Greenberg, T.~Haines,
  T.~Jones, D.~Kielczewska, W.~Kropp, J.~Learned, E.~Lehmann, J.~LoSecco, P.~R.
  Murthy, H.~Park, F.~Reines, J.~Schultz, E.~Shumard, D.~Sinclair, D.~Smith,
  H.~Sobel, L.~Sulak, R.~Svoboda, and C.~Wuest {\em Phys. Rev. Lett.}
  {\bfseries 51} (1983) 245.

\bibitem{Kajita1985}
T.~Kajita, K.~Arisaka, M.~Koshiba, M.~Nakahata, Y.~Oyama, A.~Suzuki, M.~Takita,
  Y.~Totsuka, T.~Kifune, T.~Suda, K.~Takahashi, and K.~Miyano {\em J. Phys.
  Soc. Jap.} {\bfseries 54} (1985) 4065.

\bibitem{Stone1985}
J.~Stone, H.~Park, G.~Blewitt, B.~Cortez, G.~Forter, W.~Gajewski, T.~Haines,
  D.~Kielczewska, J.~Losecco, R.~Bionta, C.~Bratton, D.~Casper,
  P.~Chrysicopoulou, R.~Claus, S.~Errede, K.~Ganezer, M.~Goldhaber, T.~Jones,
  W.~Kropp, J.~Learned, E.~Lehmann, F.~Reines, J.~Schultz, S.~Seidel,
  E.~Shumard, D.~Sinclair, H.~Sobel, J.~Stone, L.~Sulak, R.~Svoboda, J.~Van
  Der~Velde, and C.~Wuest
  \href{http://dx.doi.org/10.1016/0550-3213(85)90441-9}{{\em Nuclear Physics B}
  {\bfseries 252} (1985) 261}.

\bibitem{Bartelt1987}
J.~Bartelt, H.~Courant, K.~Heller, T.~Joyce, M.~Marshak, E.~Peterson,
  K.~Ruddick, M.~Shupe, D.~Ayres, J.~Dawson, T.~Fields, E.~May, and L.~Price
  \href{http://dx.doi.org/10.1103/PhysRevD.36.1990}{{\em Phys. Rev. D}
  {\bfseries 36} (1987) 1990}.

\bibitem{IMB1994}
R.~Becker-Szendy, C.~Bratton, J.~Breault, D.~Casper, S.~Dye, K.~Ganezer,
  W.~Gajewski, M.~Goldhaber, T.~Haines, P.~Halverson, D.~Kielczewska, W.~Kropp,
  J.~Learned, J.~LoSecco, S.~Matsuno, G.~McGrath, C.~McGrew, R.~Miller,
  L.~Price, F.~Reines, J.~Schultz, H.~Sobel, J.~Stone, L.~Sulak, and R.~Svoboda
  \href{http://dx.doi.org/10.1103/PhysRevD.49.2169}{{\em Phys. Rev. D}
  {\bfseries 49} (1994) 2169}.

\bibitem{BAIKAL1998}
{Baikal Collaboration} {\em Prog. Part. Nucl. Phys.} {\bfseries 40} (1998) 391,
  \href{http://arxiv.org/abs/astro-ph/9801044}{{\ttfamily
  arXiv:astro-ph/9801044 [astro-ph]}}.

\bibitem{MACRO2002b}
{MACRO Collaboration} {\em Eur. Phys. J. C} {\bfseries 26} (2002) 163,
  \href{http://arxiv.org/abs/hep-ex/0207024}{{\ttfamily arXiv:hep-ex/0207024
  [hep-ex]}}.

\bibitem{IceCube2014}
{IceCube Collaboration} {\em Eur. Phys. J. C} {\bfseries 74} (2014) 2938,
  \href{http://arxiv.org/abs/1402.3460}{{\ttfamily arXiv:1402.3460
  [astro-ph.CO]}}.

\bibitem{Meyer1985}
N.~Meyer-Vernet {\em Astrophys. J.} {\bfseries 290} (1985) 21.

\bibitem{SuperKamiokande2012}
{Super-Kamiokande Collaboration} {\em Astropart. Phys.} {\bfseries 36} (2012)
  131, \href{http://arxiv.org/abs/1203.0940}{{\ttfamily arXiv:1203.0940
  [hep-ex]}}.

\bibitem{Kolb1982}
E.~Kolb, S.~Colgate, and J.~Harvey
  \href{http://dx.doi.org/10.1103/PhysRevLett.49.1373}{{\em Phys. Rev. Lett.}
  {\bfseries 49} (1982) 1373}.

\bibitem{Dimopoulos1982b}
S.~Dimopoulos, J.~Preskill, and F.~Wilczek {\em Phys. Lett. B} {\bfseries 119}
  (1982) 320.

\bibitem{Kolb1984}
E.~Kolb and M.~Turner {\em Astrophys. J.} {\bfseries 286} (1984) 702.

\bibitem{Freese1983b}
K.~Freese, M.~Turner, and D.~Schramm {\em Phys. Rev. Lett.} {\bfseries 51}
  (1983) 1625.

\bibitem{Cabrera1982}
B.~Cabrera \href{http://dx.doi.org/10.1103/PhysRevLett.48.1378}{{\em Phys. Rev.
  Lett.} {\bfseries 48} (1982) 1378}.

\bibitem{Guy1987}
J.~Guy {\em Nature} {\bfseries 325} (1987) 463.

\bibitem{Shaulov1996}
S.~Shaulov {\em Acta Phys. Hung. New Series Heavy Ion Phys.} {\bfseries 4}
  (1996) 403.

\bibitem{Chen1997}
H.~Chen, C.~Dai, L.~Ding, Y.~Guo, A.~Huo, C.~Jing, H.~Kuang, H.~Liu, J.~Ma,
  C.~Shen, H.~Sheng, Z.~Yao, Z.~Yu, Q.~Zhu, C.~Ching, T.~Ho, and C.~Gao {\em
  Phys. Rept.} {\bfseries 282} (1997) 1.

\bibitem{Lattes1980}
C.~Lattes, Y.~Fujimoto, and S.~Hasegawa {\em Phys. Rept.} {\bfseries 65} (1980)
  151.

\bibitem{Kopenkin2003}
V.~Kopenkin, Y.~Fujimoto, and T.~Sinzi {\em Phys. Rev. D} {\bfseries 68} (2003)
  052007.

\bibitem{Ohsawa2004}
A.~Ohsawa, E.~Shibuya, and M.~Tamada {\em Phys. Rev. D} {\bfseries 70} (2004)
  074028.

\bibitem{Kopenkin2006}
V.~Kopenkin and Y.~Fujimoto {\em Phys. Rev. D} {\bfseries 73} (2006) 082001.

\bibitem{Kopenkin2008}
V.~Kopenkin, Y.~Fujimoto, and T.~Sinzi {\em Phys. Rev. D} {\bfseries 77} (2008)
  128301.

\bibitem{Anderson1983}
S.~Anderson, J.~Lord, S.~Strausz, and R.~Wilkes
  \href{http://dx.doi.org/10.1103/PhysRevD.28.2308}{{\em Phys. Rev. D}
  {\bfseries 28} (1983) 2308}.

\bibitem{Saito1990}
T.~Saito, Y.~Hatano, Y.~Fukada, and H.~Oda {\em Phys. Rev. Lett.} {\bfseries
  65} (1990) 2094.

\bibitem{Ichimura1993}
M.~Ichimura, E.~Kamioka, M.~Kitazawa, T.~Kobayashi, T.~Shibata, S.~Somemiya,
  M.~Kogawa, S.~Kuramata, H.~Matsutani, T.~Murabayashi, H.~Nanjyo, Z.~Watanabe,
  H.~Sugimoto, and K.~Nakazawa {\em Nuovo Cim. A} {\bfseries 106} (1993) 843.

\bibitem{Bertani1990}
M.~Bertani, G.~Giacomelli, M.~Mondardini, B.~Pal, L.~Patrizii, F.~Predieri,
  P.~Serra-Lugaresi, G.~Sini, M.~Spurio, V.~Togo, and S.~Zucchelli
  \href{http://dx.doi.org/10.1209/0295-5075/12/7/007}{{\em Europhys. Lett.}
  {\bfseries 12} (1990) 613}.

\bibitem{CDF2006}
{CDF Collaboration} \href{http://dx.doi.org/10.1103/PhysRevLett.96.201801}{{\em
  Phys. Rev. Lett.} {\bfseries 96} (2006) 201801},
  \href{http://arxiv.org/abs/0509015}{{\ttfamily arXiv:0509015 [hep-ex]}}.

\bibitem{Capela2013b}
F.~Capela, M.~Pshirkov, and P.~Tinyakov {\em Phys .Rev. D} {\bfseries 87}
  (2013) 023507, \href{http://arxiv.org/abs/1209.6021}{{\ttfamily
  arXiv:1209.6021 [astro-ph.CO]}}.

\bibitem{Dalcanton1994}
J.~Dalcanton, C.~Canizares, A.~Granados, C.~Steidel, and J.~Stocke {\em
  Astrophysical Journal} {\bfseries 424} (1994) 550.

\bibitem{MACHO1998}
{MACHO and EROS Collaborations} {\em Astrophys. J. Lett.} {\bfseries 499}
  (1998) L9, \href{http://arxiv.org/abs/astro-ph/9803082}{{\ttfamily
  arXiv:astro-ph/9803082 [astro-ph]}}.

\bibitem{Griest2013}
K.~Griest, A.~Cieplak, and M.~Lehner {\em Phys. Rev. Lett.} {\bfseries 111}
  (2013) 181302, \href{http://arxiv.org/abs/1307.5798}{{\ttfamily
  arXiv:1307.5798 [astro-ph.CO]}}.

\bibitem{Gould1992}
A.~Gould {\em Astrophys. J.} {\bfseries 386} (1992) L5.

\bibitem{Marani1999}
G.~Marani, R.~Nemiroff, J.~Norris, K.~Hurley, and J.~Bonnell {\em Astrophys. J.
  Lett.} {\bfseries 512} (1999) L13,
  \href{http://arxiv.org/abs/astro-ph/9810391}{{\ttfamily
  arXiv:astro-ph/9810391 [astro-ph]}}.

\bibitem{Barnacka2012}
A.~Barnacka, J.~Glicenstein, and R.~Moderski {\em Phys. Rev. D} {\bfseries 86}
  (2012) 043001, \href{http://arxiv.org/abs/1204.2056}{{\ttfamily
  arXiv:1204.2056 [astro-ph.CO]}}.

\bibitem{Luo2012}
Y.~Luo, S.~Hanasoge, J.~Tromp, and F.~Pretorius {\em Astrophys. J.} {\bfseries
  751} (2012) 16, \href{http://arxiv.org/abs/1203.3806}{{\ttfamily
  arXiv:1203.3806 [astro-ph.CO]}}.

\bibitem{Rafelski2013}
J.~Rafelski, L.~Labun, and J.~Birrell {\em Phys. Rev. Lett.} {\bfseries 110}
  no.~11, (2013) 111102, \href{http://arxiv.org/abs/1104.4572}{{\ttfamily
  arXiv:1104.4572 [astro-ph.EP]}}.

\bibitem{Herrin1996}
E.~Herrin and V.~Teplitz {\em Phys. Rev. D} {\bfseries 53} (1996) 6762.

\bibitem{Herrin2006}
E.~Herrin, D.~Rosenbaum, and V.~Teplitz {\em Phys. Rev. D} {\bfseries 73}
  (2006) 043511, \href{http://arxiv.org/abs/astro-ph/0505584}{{\ttfamily
  arXiv:astro-ph/0505584 [astro-ph]}}.

\bibitem{Banerdt2006}
W.~Banerdt, T.~Chui, E.~Herrin, D.~Rosenbaum, and V.~Teplitz
  \href{http://dx.doi.org/10.1016/j.asr.2005.06.034}{{\em Advances in Space
  Research} {\bfseries 37} (2006) 1889}.

\bibitem{Herrin2007}
E.~Herrin, D.~Rosenbaum, and V.~Teplitz {\em Nucl. Phys. Proc. Suppl.}
  {\bfseries 173} (2007) 72.

\bibitem{Jackson1973}
A.~Jackson and M.~Ryan \href{http://dx.doi.org/10.1038/245088a0}{{\em Nature}
  {\bfseries 245} (1973) 88}.

\bibitem{Beasley1974}
W.~Beasley and B.~Tinsley {\em Nature} {\bfseries 250} (1974) 555.

\bibitem{Collar1999}
J.~Collar and K.~Zioutas {\em Phys. Rev. Lett.} {\bfseries 83} (1999) 3097,
  \href{http://arxiv.org/abs/astro-ph/9902310}{{\ttfamily
  arXiv:astro-ph/9902310 [astro-ph]}}.

\bibitem{Porter1985}
N.~Porter, D.~Fegan, G.~MacNeill, and T.~Weekes {\em Nature} {\bfseries 316}
  (1985) 49.

\bibitem{Gorham2012}
P.~Gorham {\em Phys. Rev. D} {\bfseries 86} (2012) 123005,
  \href{http://arxiv.org/abs/1208.3697}{{\ttfamily arXiv:1208.3697
  [astro-ph.CO]}}.

\bibitem{Bertaina2014}
M.~Bertaina, A.~Cellino, and F.~Ronga {\em Exp. Astron.} (2014) .

\bibitem{Levison2002}
H.~Levison, A.~Morbidelli, L.~Dones, R.~Jedicke, P.~Wiegert, and W.~Bottke~Jr.
  \href{http://dx.doi.org/10.1126/science.1070226}{{\em Science} {\bfseries
  296} (2002) 2212}.

\bibitem{Knight2014}
M.~Knight and K.~Battams {\em Astrophys. J. Lett.} {\bfseries 782} (2014) L37,
  \href{http://arxiv.org/abs/1401.7028}{{\ttfamily arXiv:1401.7028
  [astro-ph.EP]}}.

\bibitem{AHearn2007}
M.~A'Hearn and M.~Combi, ``Deep impact at comet tempel 1,''
  \href{http://dx.doi.org/http://dx.doi.org/10.1016/j.icarus.2006.12.011}{{\em
  Icarus} {\bfseries 187} (2007) 1}.

\bibitem{Thomas2005}
P.~Thomas and M.~Robinson \href{http://dx.doi.org/10.1038/nature03855}{{\em
  Nature} {\bfseries 436} (2005) 366}.

\bibitem{OBrien2009}
D.~O’Brien, ``The yarkovsky effect is not responsible for small crater
  depletion on eros and itokawa,''
  \href{http://dx.doi.org/10.1016/j.icarus.2009.03.038}{{\em Icarus} {\bfseries
  203} (2009) 112}.

\bibitem{Lowry2007}
S.~Lowry, A.~Fitzsimmons, P.~Pravec, D.~Vokrouhlicky, H.~Boehnhardt, P.~Taylor,
  J.-L. Margot, A.~Galad, M.~Irwin, J.~Irwin, and P.~Kusnirak {\em Science}
  {\bfseries 316} (2007) 272.

\bibitem{Taylor2007}
P.~Taylor, J.-L. Margot, D.~Vokrouhlicky, D.~Scheeres, P.~Pravec, S.~Lowry,
  A.~Fitzsimmons, M.~Nolan, S.~Ostro, L.~Benner, J.~Giorgini, and C.~Magri {\em
  Science} {\bfseries 316} (2007) 274.

\bibitem{Kaasalainen2007}
M.~Kaasalainen, J.~Dcaronurech, B.~Warner, Y.~Krugly, and N.~Gaftonyuk {\em
  Nature} {\bfseries 446} (2007) 420.

\bibitem{Harris2006}
A.~Harris and P.~Pravec \href{http://dx.doi.org/10.1017/S1743921305006903}{{\em
  Proceedings of IAU Symposium, Asteroids, Comets, Meteors} {\bfseries 229}
  (2006) 439}.

\bibitem{Eubanks2014}
T.~Eubanks {\em proceedings of IAA Space Exploration Conference} (2014) .

\bibitem{GonzalezGarcia2008}
M.~Gonzalez-Garcia, F.~Halzen, M.~Maltoni, and H.~K. Tanaka {\em Phys. Rev.
  Lett.} {\bfseries 100} (2008) 061802,
  \href{http://arxiv.org/abs/0711.0745}{{\ttfamily arXiv:0711.0745 [hep-ph]}}.

\bibitem{Hoshina2012}
K.~Hoshina and H.~Tanaka {\em EGU General Assembly Conference Abstracts}
  {\bfseries 14} (2012) 3246.

\bibitem{HyperKamiokande2011}
{Hyper-Kamiokande Collaboration} {\em letter of intent} (2011) ,
  \href{http://arxiv.org/abs/1109.3262}{{\ttfamily arXiv:1109.3262 [hep-ex]}}.

\bibitem{Olinto1987}
A.~Olinto {\em Phys. Lett. B} {\bfseries 192} (1987) 71.

\bibitem{Alpar1987}
M.~Alpar \href{http://dx.doi.org/10.1103/PhysRevLett.58.2152}{{\em Phys. Rev.
  Lett.} {\bfseries 58} (1987) 2152}.

\bibitem{Friedman1991}
R.~Caldwell and J.~Friedman {\em Phys. Lett. B} {\bfseries 264} (1991) 143.

\bibitem{Bauswein2009}
A.~Bauswein, H.-T. Janka, R.~Oechslin, G.~Pagliara, I.~Sagert,
  J.~Schaffner-Bielich, M.~Hohle, and R.~Neuhauser {\em Phys. Rev. Lett.}
  {\bfseries 103} (2009) 011101,
  \href{http://arxiv.org/abs/0812.4248}{{\ttfamily arXiv:0812.4248
  [astro-ph]}}.

\bibitem{Pani2014}
P.~Pani and A.~Loeb {\em JCAP} {\bfseries 1406} (2014) 026,
  \href{http://arxiv.org/abs/1401.3025}{{\ttfamily arXiv:1401.3025
  [astro-ph.CO]}}.

\bibitem{Kesden2011}
M.~Kesden and S.~Hanasoge {\em Phys. Rev. Lett.} {\bfseries 107} (2011) 111101,
  \href{http://arxiv.org/abs/1106.0011}{{\ttfamily arXiv:1106.0011
  [astro-ph.CO]}}.

\bibitem{Prantzos2011}
N.~Prantzos, C.~Boehm, A.~Bykov, R.~Diehl, K.~Ferriere, N.~Guessoum, P.~Jean,
  J.~Knoedlseder, A.~Marcowith, I.~Moskalenko, A.~Strong, and
  G.~Weidenspointner {\em Rev. Mod. Phys.} {\bfseries 83} (2011) 1001,
  \href{http://arxiv.org/abs/1009.4620}{{\ttfamily arXiv:1009.4620
  [astro-ph.HE]}}.

\bibitem{Lawson2013}
K.~Lawson and A.~Zhitnitsky {\em Phys. Lett. B} {\bfseries 724} (2013) 17,
  \href{http://arxiv.org/abs/1210.2400}{{\ttfamily arXiv:1210.2400
  [astro-ph.CO]}}.

\bibitem{Fornengo2011}
N.~Fornengo, R.~Lineros, M.~Regis, and M.~Taoso {\em Phys. Rev. Lett.}
  {\bfseries 107} (2011) 271302,
  \href{http://arxiv.org/abs/1108.0569}{{\ttfamily arXiv:1108.0569 [hep-ph]}}.

\bibitem{Page1976}
D.~Page and S.~Hawking {\em Astrophys. J.} {\bfseries 206} (1976) 1.

\bibitem{Lehoucq2009}
R.~Lehoucq, M.~Casse, J.-M. Casandjian, and I.~Grenier {\em Astron. Astrophys.}
  {\bfseries 502} (2009) 37, \href{http://arxiv.org/abs/0906.1648}{{\ttfamily
  arXiv:0906.1648 [astro-ph.HE]}}.

\bibitem{Carr1976}
B.~Carr {\em Astrophys. J.} {\bfseries 206} (1976) 8.

\bibitem{Turner1982b}
M.~Turner {\em Nature} {\bfseries 297} (1982) 379.

\bibitem{Halzen1991}
F.~Halzen, E.~Zas, J.~MacGibbon, and T.~Weekes {\em Nature} {\bfseries 353}
  (1991) 807.

\bibitem{AMS2013}
{AMS Collaboration} {\em Phys. Rev. Lett.} {\bfseries 110} (2013) 141102.

\bibitem{Cline2011}
D.~Cline \href{http://dx.doi.org/10.4236/ijaa.2011.13021}{{\em Int. J. Astron.
  Astroph.} {\bfseries 1} (2011) 164}.

\bibitem{Rees1977}
M.~Rees \href{http://dx.doi.org/10.1038/266333a0}{{\em Nature} {\bfseries 266}
  (1977) 333}.

\bibitem{Okele1980}
P.~Okele and M.~Rees {\em Astron. Astrophys.} {\bfseries 81} (1980) 263.

\bibitem{Barrau2000}
A.~Barrau {\em Astropart. Phys.} {\bfseries 12} (2000) 269,
  \href{http://arxiv.org/abs/astro-ph/9907347}{{\ttfamily
  arXiv:astro-ph/9907347 [astro-ph]}}.

\bibitem{Heckler1997}
A.~Heckler {\em Phys. Rev. D} {\bfseries 55} (1997) 480,
  \href{http://arxiv.org/abs/astro-ph/9601029}{{\ttfamily
  arXiv:astro-ph/9601029 [astro-ph]}}.

\bibitem{MacGibbon2008}
J.~MacGibbon, B.~Carr, and D.~Page {\em Phys. Rev. D} {\bfseries 78} (2008)
  064043, \href{http://arxiv.org/abs/0709.2380}{{\ttfamily arXiv:0709.2380
  [astro-ph]}}.

\bibitem{Greisen1966}
K.~Greisen {\em Phys. Rev. Lett.} {\bfseries 16} (1966) 748.

\bibitem{Zatsepin1966}
G.~Zatsepin and V.~Kuzmin {\em JETP Lett.} {\bfseries 4} (1966) 78.

\bibitem{Auger2013}
{Pierre Auger Collaboration} {\em Adv. Space Res.} {\bfseries 53} (2013) 1476,
  \href{http://arxiv.org/abs/1309.1249}{{\ttfamily arXiv:1309.1249
  [astro-ph.HE]}}.

\bibitem{Alexandreas1993}
D.~E. Alexandreas, G.~E. Allen, D.~Berley, S.~Biller, R.~L. Burman,
  M.~Cavalli-Sforza, C.~Y. Chang, M.~L. Chen, P.~Chumney, D.~Coyne, C.~Dion,
  G.~M. Dion, D.~Dorfan, R.~W. Ellsworth, J.~A. Goodman, T.~J. Haines,
  M.~Harmon, C.~M. Hoffman, L.~Kelley, S.~Klein, D.~E. Nagle, D.~M. Schmidt,
  R.~Schnee, C.~Sinnis, A.~Shoup, M.~J. Stark, D.~D. Weeks, D.~A. Williams,
  J.~P. Wu, T.~Yang, G.~B. Yodh, and W.~P. Zhang {\em Phys. Rev. Lett.}
  {\bfseries 71} (1993) 2524.

\bibitem{Krennrich2000}
F.~Krennrich, S.~Le~Bohec, and T.~Weekes {\em Astrophys. J.} {\bfseries 529}
  (2000) 506, \href{http://arxiv.org/abs/astro-ph/9909078}{{\ttfamily
  arXiv:astro-ph/9909078 [astro-ph]}}.

\bibitem{Alam1999}
J.-E. Alam, S.~Raha, and B.~Sinha {\em Astrophys. J.} {\bfseries 513} (1999)
  572, \href{http://arxiv.org/abs/astro-ph/9704226}{{\ttfamily
  arXiv:astro-ph/9704226 [astro-ph]}}.

\end{thebibliography}\endgroup
